\documentclass[10pt,letterpaper]{article}
\usepackage[top=0.85in,left=1in,footskip=0.75in,marginparwidth=1in]{geometry}
\usepackage[utf8]{inputenc}
\usepackage{natbib}
\usepackage{cite}
\usepackage{nameref,hyperref}
\usepackage[right]{lineno}
\usepackage{microtype}
\DisableLigatures[f]{encoding = *, family = * }
\usepackage{changepage}
\usepackage[aboveskip=1pt,labelfont=bf,labelsep=period,singlelinecheck=off]{caption}
\makeatletter
\renewcommand{\@biblabel}[1]{\quad#1.}
\makeatother
\usepackage{lastpage,fancyhdr,graphicx}
\usepackage{epstopdf, epsfig}
\fancyheadoffset[L]{2.25in}
\fancyfootoffset[L]{2.25in}
\usepackage{color}
\definecolor{Gray}{gray}{.25}
\usepackage{sidecap}
\usepackage{wrapfig}
\usepackage[pscoord]{eso-pic}
\usepackage[fulladjust]{marginnote}
\reversemarginpar

\usepackage{mathabx}
\usepackage{morefloats}
\usepackage{upgreek}
\usepackage{ragged2e}

\input{definitions.sty}
\def\minf {\mathrm{M}_{\infty}}
\def\rec {\mathrm{Re}_c}

\begin{document}
\vspace*{0.35in}

\begin{flushleft}

{\Large
\textbf\newline{Bursting and reformation cycle of the laminar separation bubble over a NACA-0012 aerofoil: Characterisation of the f\/low-f\/ield}
}
\newline\\
Eltayeb M. ElJack\textsuperscript{1,*},
Ibraheem M. AlQadi\textsuperscript{2}, and
Julio Soria\textsuperscript{2,3}\\
\bigskip
\bf{1} Mechanical Engineering Department, University of Khartoum, Khartoum, Sudan\\
\bf{2} Aeronautical Engineering Department, King Abdulaziz University, Jeddah, Saudi Arabia\\
\bf{3} Laboratory for Turbulence Research in Aerospace and Combustion, Department of Mechanical and Aerospace Engineering, Monash University, Melbourne, Australia\\
\bigskip
* emeljack@uofk.edu

\end{flushleft}

\section*{Abstract}
This study examines the effects of angle of attack on the characteristics of the laminar separation bubble (LSB), its associated low-frequency f\/low oscillation (LFO), and the f\/low-f\/ield around a NACA-0012 aerofoil at $\rec = 5\times10^4$ and $\minf = 0.4$. In the range of the investigated angles of attack, statistics of the f\/low-f\/ield suggest the existence of three distinct angle-of-attack regimes: 1) the f\/low and aerodynamic characteristics are not much affected by the LFO for angles of attack $\alpha < 9.25^{\circ}$; 2) the f\/low-f\/ield undergoes a transition regime in which the bubble bursting and reformation cycle, LFO, develops until it reaches a quasi-periodic switching between separated and attached f\/low in the range of angles of attack $9.25^{\circ} \leq \alpha \leq 9.6^{\circ}$; and 3) for the angles of attack $\alpha > 9.6^{\circ}$, the f\/low-f\/ield and the aerodynamic characteristics are overwhelmed by a quasi-periodic and self-sustained LFO until the aerofoil approaches the angle of full stall. The length of the bubble, in the mean sense, decreases to a minimum size of $33.5\%$ of the aerofoil chord at $\alpha = 9.0^{\circ}$ then it grows in size when the angle of attack is further increased. A trailing-edge bubble (TEB) forms at $\alpha > 9.25^{\circ}$ and grows with the angle of attack. The LSB and TEB merge and continue to deform until they form an open bubble at $\alpha = 10.5^{\circ}$. On the suction surface of the aerofoil, the pressure distribution shows that the presence of the LSB induces a gradual and continues adverse pressure gradient (APG) when the f\/low is attached. The bursting of the bubble causes a gradual and continues favourable pressure gradient (FPG) when the f\/low is separated. This is indicative that a natural forcing mechanism keeps the f\/low attached against the APG and separated despite the FPG. It is shown that perturbations of the wall-normal and the spanwise velocity components are extracted exclusively by the local velocity gradient inside the shear layer via the Kelvin--Helmholtz instability. Whereas, the f\/luctuations in the streamwise velocity component and the pressure are due to a global oscillation in the f\/low-f\/ield in addition to the velocity gradient across the shear layer. The variance of the pressure f\/luctuations has a signif\/icant magnitude in the laminar portion of the separated shear layer. This is indicative that the instability that generates and sustains the LFO originates at this location. The present investigation suggests that most of the observations reported in the literature about the LSB and its associated LFO are neither thresholds nor indicators for the inception of the instability, but rather are consequences of it.

\section{Introduction}\label{sec:intorudction}
\begin{figure}
\begin{center}
\includegraphics[width=250pt, trim={0mm 0mm 0mm 0mm}, clip]{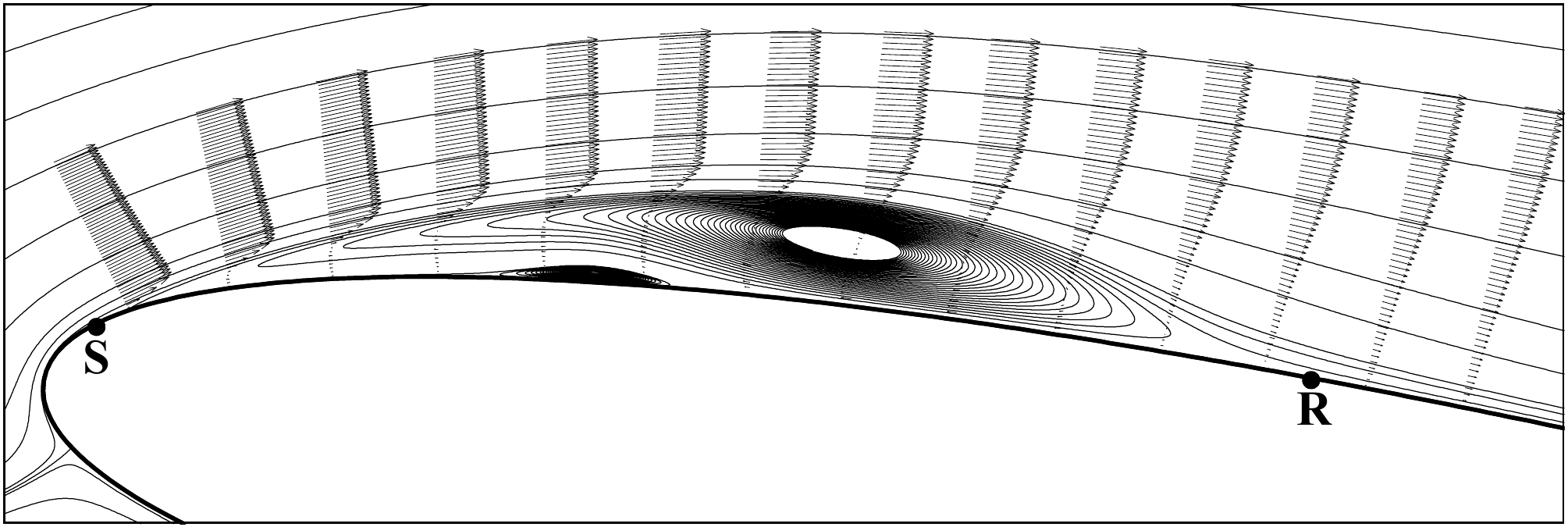}
\caption{A laminar separation bubble over a NACA-0012 aerofoil at the angle of attack of $9.25^{\circ}$.}
\label{bubble}
\end{center}
\end{figure}

Interest in low Reynolds number aerodynamics, $\rec = 10^4$ to $10^6$ based on the free-stream velocity and aerofoil chord, has signif\/icantly increased over the past three decades. This is because of the steady increase in applications that operate at low Reynolds numbers due to high altitude f\/light, where the kinematic viscosity is relatively higher, and/or having small geometrical dimensions. The low Reynolds number aerodynamics of an aerofoil at incidence is characterised by its proclivity to induce a laminar separation bubble (LSB) on the upper surface of the aerofoil. Some of the known conditions for the formation of an LSB are functions of the Reynolds number of the f\/low, pressure distribution, aerofoil geometry, surface roughness, and turbulence intensity of the free-stream. F\/igure~\ref{bubble} shows an LSB over a NACA-0012 aerofoil at Reynolds number of $5\times10^4$ and the angle of attack of $9.25^{\circ}$.
\newline
At the leading-edge of the aerofoil, a laminar boundary layer is established by the incoming f\/low. The boundary layer remains laminar, if not forced into early transition, until the pressure gradient changes from negative to positive. If the adverse pressure gradient is strong, the laminar boundary layer detaches and travels away from the aerofoil surface to create a region of separated f\/low near the surface (point \textbf{S} in f\/igure~\ref{bubble}). Along the separated shear layer, kinetic energy is extracted from the mean f\/low and fed to the perturbations. The process of feeding energy into the $2$D--rolls continue until they break down into three-dimensional structures and transition to turbulence initiates. Consequently, the f\/low reattaches because the boundary layer becomes more energetic (point \textbf{R} in f\/igure~\ref{bubble}). The f\/low is reversed in the region between points \textbf{S} and \textbf{R} in f\/igure~\ref{bubble} which constitutes an LSB. Beneath the LSB lies a small counter-rotating bubble, which is called a \textit{secondary bubble}. The turbulent boundary layer downstream of the bubble has more momentum near the surface to resist the adverse pressure gradient and avoid a new separation.\newline
\citet{owen1953laminar} classif\/ied the LSB into two distinct formats of short and long bubble. The bubble is short if the ratio of the bubble length to the displacement thickness at the point of separation is in the order of $100$. Whereas, the bubble is termed long if the ratio is in the order of $400$. A short bubble has little effect on the external potential f\/low, while the long bubble has a more notable inf\/luence. At certain conditions, short bubble suddenly alters to a long bubble or a fully separated f\/low without any subsequent reattachment which is termed as bubble bursting. Recent experimental and numerical observations have shown an unusual switching between bubble bursting and reattachment at a low frequency on the inception of stall~\citep{rinoie2004oscillating, tanaka2004flow, almutairi2013large, almutairi2015large, eljack2016large, almutairi2017dynamics, eljack2017high}.\newline
\citet{mccullough1951examples} classif\/ied the aerofoil stall into three main categories: 1) leading-edge, 2) thin-aerofoil, and 3) trailing-edge stall. The leading-edge stall results from the f\/low separation near the leading-edge without any subsequent reattachment downstream of the separation. In the thin-aerofoil stall, the f\/low reattaches downstream the separation, and then the reattachment point moves toward the trailing-edge as the angle of attack increases. The trailing-edge stall initiates at the trailing-edge where the f\/low separates, and the separation point moves toward the leading-edge as the angle of attack increases.~\citet{gaster1967structure} who was the f\/irst to systematically investigate the stability of the LSB, found that bubble bursting occurs either by a gradual increment in the bubble length or by a suddenly discontinuous event. The former applies in the present work as will be seen in $\S$~\ref{sec:results}.\newline
The instantaneous shape of the LSB is random, and an instantaneous continues separation or reattachment line in the spanwise direction is not def\/ined. A question was raised as to whether spanwise averaging of the data is justif\/ied, and before that, if the assumption of periodic f\/low in this direction is valid. The most important to us is the shape of the large-scale bubbles; are they two-dimensional? Previous works on two-dimensional aerofoil stalling characteristics have shown a low-frequency and highly unsteady f\/low or a steady large-scale three-dimensional structure. The latter is termed stall cells, and there was considerable amount of research focused on their structure.~\citet{winkelman1980flow} carried out an oil--f\/low visualisations of the f\/low-f\/ield on the suction surface of stalled wings of Clark Y aerofoil section at chord Reynolds numbers of $24.5\times10^4$, $26\times10^4$, and $38.5\times10^4$. The ends of the wing are f\/lush with the tunnel side walls or splitter plates and on plane rectangular wings of inf\/inite aspect ratio. They observed that the mushroom-shaped cells started to form at the onset of stall on the two-dimensional model. The three-dimensional cells coexisted with trailing-edge stall cell on the surface of the wing several degrees above the stalling angle of attack. They tentatively sketched a f\/low-f\/ield model showing the general features of a leading-edge separation bubble and trailing-edge separation.~\citet{winkelman1990flow} took up~\citet{winkelman1980flow} work and measured the spectra of the velocity in the wake of a rectangular wing having the same aerofoil section. The spectra of the wake measurements did not show any indications of a dominant low-frequency mode.~\citet{yon1998study} used f\/ine-thread tuft-f\/low visualisation and high-frequency response pressure transducers measurements to investigate the unsteady features of the stall cells. They studied the f\/low-f\/ield around a wing of aspect ratio ranging from $2$ to $6$ with the NACA-0015 aerofoil section. They imposed the two-dimensionality by using end plates that effectively eliminated the tip f\/low. The authors reported the existence of a low-frequency mode in their pressure spectra, but with considerably smaller amplitudes of oscillations.~\citet{broeren2001spanwise} studied f\/ive different aerofoil conf\/igurations (NACA-2414, NACA-64A010, LRN-1007, E374 and Ultra-Sport) by measuring the wake velocity across the spanwise direction, and using mini-tufts for f\/low visualisation. They found that all the stall types were dependent on the type of the aerofoil. They concluded that the low-frequency f\/low oscillation (LFO) phenomenon always occurs in the aerofoils that exhibits a thin-aerofoil stall or the combination of both thin-aerofoil and trailing-edge stall. Most importantly to us, they found that the low-frequency unsteadiness is essentially two dimensional. Their conclusions were in good agreement with that of~\citet{zaman1989natural} who observed that the LFO takes place with aerofoils exhibiting either the trailing-edge or thin-aerofoil type stalls but does not take place with the leading-edge type stall. To this end, it is evident that the LFO phenomenon is inherently two-dimensional by its nature and neither the imposed boundary condition nor the spanwise averaging affected the results presented in $\S$~\ref{sec:results}.\newline
\citet{eljack2017high} showed that at relatively low angles of attack, the LSB is present on the upper surface of an aerofoil and remains intact. As the angle of attack of the aerofoil is increased to a critical value, the LSB abruptly and intermittently bursts, which causes an oscillation in the f\/low-f\/ield and consequently the aerodynamic forces. As the angle of attack is increased above the critical angle, the bursting becomes more frequent, and the LSB exhibits a quasi-periodic switching between long and short bubbles, which results in a global LFO. As the aerofoil approaches the full stall angle, the f\/low remains separated with intermittent and abrupt random reattachments. The aerofoil eventually undergoes a full stall as the angle of attack is further increased.\newline
The presence of the bubble signif\/icantly deteriorates the aerodynamic performance, such as loss of lift, undesirable change in the moment, and increase in the drag. F\/low oscillations due to bubble shedding and sudden aerofoil stalling due to bubble bursting are direct consequences of the complex and random behaviour of the LSB. The ultimate goal is to control the f\/low-f\/ield around an aerofoil at low Reynolds numbers in the presence of an LSB to improve its performance. However, very little of the general character of the LSB is understood. Revealing the underlying mechanism behind the self-sustained periodic bursting and reformation of the bubble is an essential step towards control of its undesirable effects.\newline
The objective of the present study is to examine the effects of the angle of attack on the characteristics of the LSB, its associated LFO, and the f\/low-f\/ield around a NACA-0012 aerofoil at $\rec = 5\times10^4$ and $\minf = 0.4$. A conditional time-averaging is used to characterise the f\/low-f\/ield. The characteristics of the f\/low-f\/ield, the LSB, and the LFO are provided along with careful comparisons with existing experimental and numerical work. Similarities are discussed in detail and discrepancies are justif\/ied wherever necessary.

\section{Mathematical modelling and computational setup}
The non-dimensional Favre-f\/iltered compressible Navier--Stokes equations in curvilinear coordinates are given by:
\begin{equation}\label{mass}
\frac{\partial\rhob}{\partial t}
+ \frac{\partial}{\partial \xi_{j}} \left[\xixt \left(\rhob\util_j \right)\right]
= 0
\end{equation}
\begin{equation}\label{momentum}
\frac{\partial}{\partial t} (\rhob \util_i)
+ \frac{\partial}{\partial \xi_{j}} \left[\xixt \left(\rhob\util_i\util_j
+ \pb \delta_{ij} - \widetilde{\sigma}_{ij} \right) \right]
= - \underbrace{\frac{\partial}{\partial \xi_{j}} \left(\xixt \tau_{ij}\right)}_{(1)} +
\underbrace{\frac{\partial}{\partial \xi_{j}} \left[\xixt \left(\bar{\sigma}_{ij}- \widetilde{\sigma}_{ij}\right)\right]}_{(2)}
\end{equation}
\begin{equation}\label{energy}
\frac{\partial}{\partial t} (\rhob \Etil)
+ \frac{\partial}{\partial \xi_{j}}
\left[\xixt \left(\left(\rhob \Etil + \pb\right)\util_j + \widetilde{q}_j
- \widetilde{\sigma}_{ij}\util_i\right) \right] =
- \underbrace{\frac{\partial}{\partial \xi_{j}} \left[\xixt \left(Q_j
+ {\textstyle\half}{\cal G}_j
- {\cal D}_j \right)\right]}_{(3)}
\end{equation}
Where, $\util_i$ is the velocity component in the $x_i$ direction, $\widetilde{p}$ is the pressure, $\widetilde{\rho}$ is the density, $\widetilde{T}$ is the temperature, and $\widetilde{E}$ is the total energy per unit mass. The stress tensor is given by:
\begin{equation}
{\sigma}_{ij} = \frac{2{\widetilde{\mu}}}{\mathop{\rm Re}} S_{ij}
- {\frac{2{\widetilde{\mu}}}{3 \mathop{\rm Re}}} \delta_{ij} S_{kk} \ , \
S_{ij}=\frac{1}{2}\left(\frac{\partial u_i}{\partial \xi_j}\frac{\partial {\xi_j}}{\partial {x_i}} + \frac{\partial u_j}{\partial \xi_i}\frac{\partial {\xi_i}}{\partial {x_j}} \right)
\end{equation}
The heat f\/lux vector is given by:
\begin{equation}
{\widetilde{q}_j} = \frac{\widetilde{\mu}}{{(\gamma - 1){\mathop{\rm Re}\nolimits} \Pr {\mathop M\nolimits_\infty ^2}}}\frac{\partial {\widetilde T}}{\partial {\xi_j}} \xix
\end{equation}
where
\begin{equation}
\widetilde{\mu} = {\widetilde T^{\frac{3}{2}}}\frac{{1 + C}}{{\widetilde T + C}} { \ , \ } (C = 0.3686)    \ , \
\widetilde T = \gamma {\mathop M\nolimits_\infty ^2} \frac{{\bar p}}{{\bar \rho }}
\end{equation}
$\grave{\xi _{{ij}}}$ is the transformation metrics tensor given by:
\begin{equation}
\grave{\xi _{{ij}}} = \frac{1}{J}\frac{{\partial {\xi _j}}}{{\partial {x_i}}} \ , \ J = \left| {\frac{{\partial {x_i}}}{{\partial {\xi _j}}}} \right|
\end{equation}
The Favre-f\/iltered equations consist of the terms from the Navier--Stokes equations on the left hand side, in addition to the terms on the right hand side resulting from the f\/iltering operation. The under-bracketed terms represent the effects of the unresolved subgrid-scale (\textsf{sgs}) turbulent structures. $\tij$ is the \textsf{sgs} stress tensor ($\tij = \rhob(\widetilde{u_iu_j}-\util_i\util_j)$), $Q_j$ is the \textsf{sgs} heat f\/lux vector ($Q_j = \rhob \left(\widetilde{u_j T}-\util_j\widetilde{T}\right)$), and $\partial{\cal G}_j/\partial x_j$ is the \textsf{sgs} turbulent diffusion term, where ${\cal G}_j = \rhob\left(\widetilde{u_ju_ku_k} -\util_j\widetilde{u_ku_k} \right)$ is the \textsf{sgs} viscous diffusion term and ${\cal D}_j = \overline{\sigma_{ij}u_i}-\widetilde{\sigma}_{ij}\util_i$. The nonlinear sub-f\/ilter viscous contribution, term (2) in equation~\ref{momentum}, is from the use of $\widetilde{T}$ to evaluate $\mu(\widetilde{T})$. This term is very small and negligible compared to (1) in the same equation. Similarly, the \textsf{sgs} viscous diffusion term $\partial{\cal D}_j/\partial x_j$ is also neglected. The \textsf{sgs} heat f\/lux vector $Q_j$ and \textsf{sgs} turbulent diffusion term $\partial{\cal G}_j/\partial x_j$ are negligible for low Mach number f\/lows~\citet{vreman1995direct}.\newline
The \textsf{sgs} stress tensor $\tau _{ij}$ represents the effect of the small-scale turbulent structures and is def\/ined as:
\begin{equation}
{\tau _{ij}} = \bar \rho \left({\widetilde {{u_i}{u_j}}{\rm{-}}{{\widetilde u}_i}{{\widetilde u}_j}} \right)
\end{equation}
The \textsf{sgs} stress tensor $\tau _{ij}$ is modelled by using the eddy viscosity concept under the low compressibility:
\begin{equation}
{\tau _{ij}} - \frac{1}{3}{\delta _{ij}}{\tau _{kk}} = 2{\nu_{_{\sf{t}}}}{{{\widetilde S}}_{ij}}
\end{equation}
where $\tau _{kk}$ refers to the trace of the \textsf{sgs} Reynolds stress tensor and $\nu_{_{\sf{t}}}$ refers to the turbulent eddy viscosity obtained by the mixed-time-scale (\textsf{mts}) model developed by~\citet{inagaki2005mixed}. The model is turned off automatically in the laminar f\/low region. Thus, it overcomes the drawbacks of using a wall-damping function. In this model, the eddy viscosity is calculated by using the following def\/inition:
\begin{equation}
{\nu_{_{\sf{t}}}} = {{{C}}_{_{{\sf{mts}}}}}{\uptau_{\rm{s}}}{k_{_{{\sf {sgs}}}}}
\end{equation}
where ${C}_{_{{\sf{mts}}}}$ refers to the model f\/ixed parameter. ${\uptau_{\rm{s}}}$ refers to the time scale given by:
\begin{equation}
{\uptau_{\rm{s}}}^{- 1} = {\left({\frac{{\bar \Delta}}{{\sqrt {{k_{_{\sf{sgs}}}}}}}} \right)^{- 1}} + {\left({\frac{{{C_{\uptau}}}}{{\left| {\bar S} \right|}}} \right)^{- 1}}
\end{equation}
where
\begin{equation}
|\bar S|=\sqrt {2 {\bar{S}_{ij}} {\bar{S}_{ij}}} \ \ ,\ \ \bar \Delta = {\left({\Delta x_1\Delta x_2\Delta x_3} \right)^{\frac{1}{3}}}
\end{equation}
${\bar \Delta}$ refers to the f\/ilter size, and $\mathrm{C}_{\uptau}$ is the time-scale parameter. The velocity scale $k_{_{{\sf sgs}}}$ is def\/ined by:
\begin{equation}
{k_{_{\sf sgs}}} = {(\bar u - \widehat{\bar {u}})^2}
\end{equation}
As long as the f\/low is fully resolved, the above def\/inition gives a zero velocity scale in the laminar f\/low region. Consequently, the eddy viscosity ($\nu_{_{\sf{t}}}$) also approaches zero in this region.\newline
The LES code utilised in the present simulations is an LES version of the direct numerical simulation (DNS) code written and validated by~\citet{jones2010stability}. The Navier--Stokes equations were discretised using a fourth-order explicit central difference scheme for spatial discretisation in the interior points. The fourth-order boundary scheme of~\citet{carpenter1999stable} was used to treat points near and at the boundary. To preserve the spatial characteristics, the transformation metrics tensor $\grave{\xi _{{ij}}}$ was evaluated by using the same fourth-order scheme. Temporal discretisation is performed using a low-storage fourth-order Runge--Kutta scheme. The solution stability was improved by implementing an entropy splitting scheme~\citet{sandham2001entropy}. The entropy splitting constant $\beta$ was set equal to $2.0$ \citet{jones2008numerical}. The combination of a relatively coarse grid and high-order scheme results in the generation of spurious high-frequency waves. The non-physical waves will contaminate the solution if they are not eliminated at each time step. Thus, the solution must be f\/iltered to eliminate spurious waves at the high unresolved wavenumber range by damping their amplitude with minimal effect on the resolved waves at lower wavenumbers. The f\/iltering scheme used here is a fourth-order compact scheme developed by~\citet{gaitonde1998high}. The integral characteristic boundary condition is applied at the free-stream and the far-f\/ield boundaries~\citet{sandhu1994boundary}. The zonal characteristic boundary condition is applied at the downstream exit boundary~\citet{sandberg2006nonreflecting} which are considered as non-ref\/lected boundary conditions to overcome the circulation effects at the boundaries. The adiabatic and no-slip conditions are applied at the aerofoil surface. The LSB, its associated LFO, and the f\/low-f\/ield are inherently two dimensional. Therefore, a periodic boundary condition is applied in the spanwise direction for each step of the Runge-Kutta time steps. The internal branch-cut boundary was updated at each step of the Runge--Kutta scheme.\newline
In terms of the aerofoil chord $(c)$, the dimensions of the computational domain were set as follows: $W=5.0c$ in the wake region, from the aerofoil trailing-edge to the outf\/low boundary in the streamwise direction; $L_\eta=7.3c$ in the wall-normal direction (the C-grid radius); and $L_\zeta=0.5c$ in the spanwise direction. Where $\xi, \eta$ and $\zeta$ are the curvilinear coordinates along the aerofoil surface, in the wall-normal direction, and the spanwise direction, respectively.
\begin{table*}
	\begin{center}
		\Large
		\caption{Computational grid parameters}
		\label{table:1}
		\resizebox{\textwidth}{!}{%
			\begin{tabular}{ | p{3cm} | p{1.5cm} | p{1.5cm} | p{1.5cm} | p{1.5cm} | p{1.5cm} | p{1.5cm} | p{3cm} | }
				\hline
				Grid & $y^+$ & $\Delta x^+$ & $\Delta z^+$ & $N_\xi$ & $N_\eta$ & $N_\zeta$ & Total points \\ \hline
				Grid--$1$ & $>1$ & $<50$ & $<50$	& $637$ & $320$ & $86$ & $17,530,240$\\ \hline
			 	Grid--$2$ & $<1$ & $<15$ & $<15$ & $780$ & $320$ & $101$ & $25,209,600$\\ \hline
				Grid--$3$ & $<1$ & $<10$ & $<10$ & $980$ & $320$ & $151$ & $47,353,600$\\ \hline
			\end{tabular}
		}
	\end{center}
\end{table*}
\begin{figure}
\begin{center}
\begin{minipage}{220pt}
\includegraphics[height=150pt , trim={0mm 0mm 0mm 0mm}, clip]{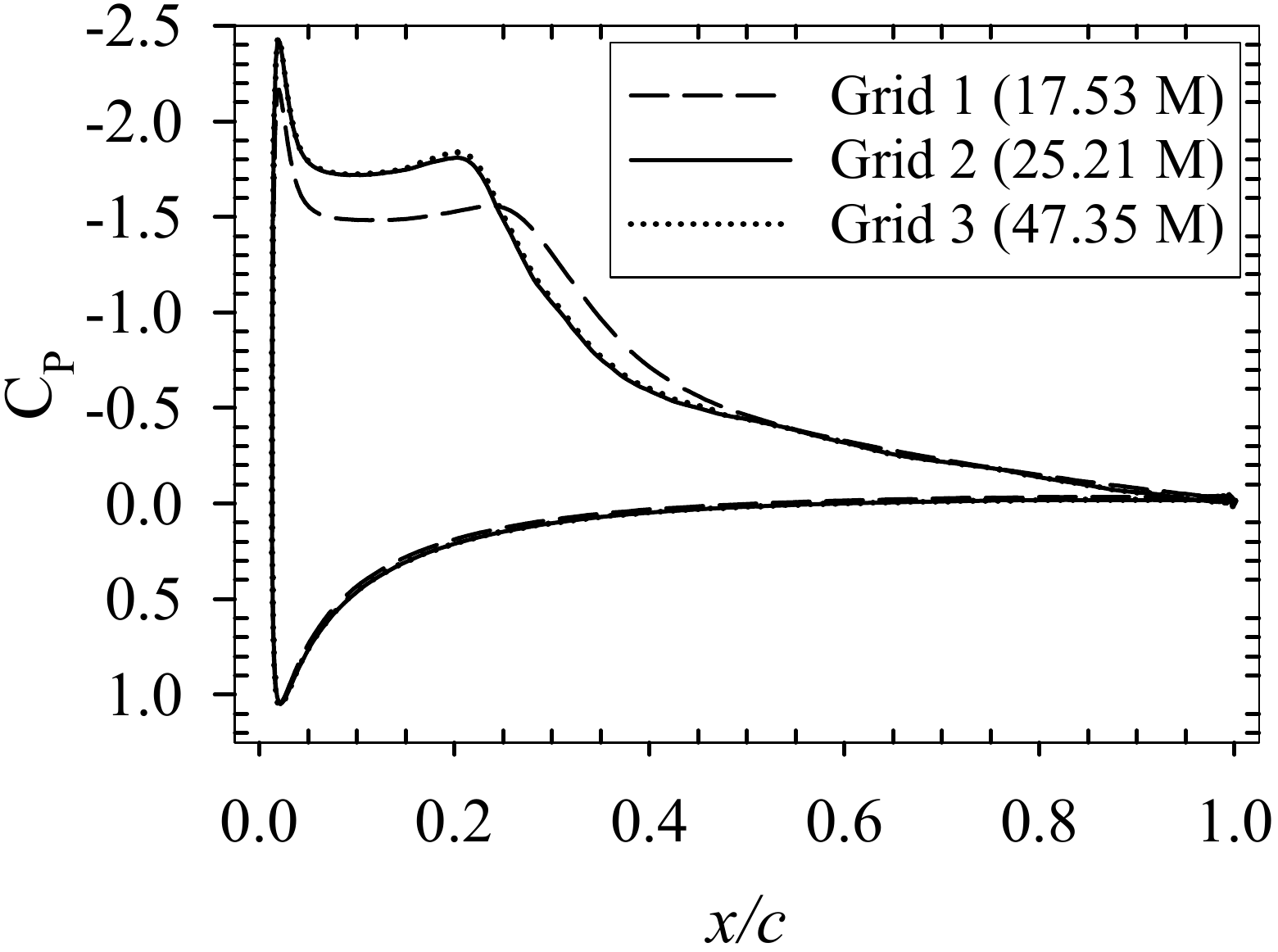}
\end{minipage}
\begin{minipage}{220pt}
\includegraphics[height=150pt , trim={0mm 0mm 0mm 0mm}, clip]{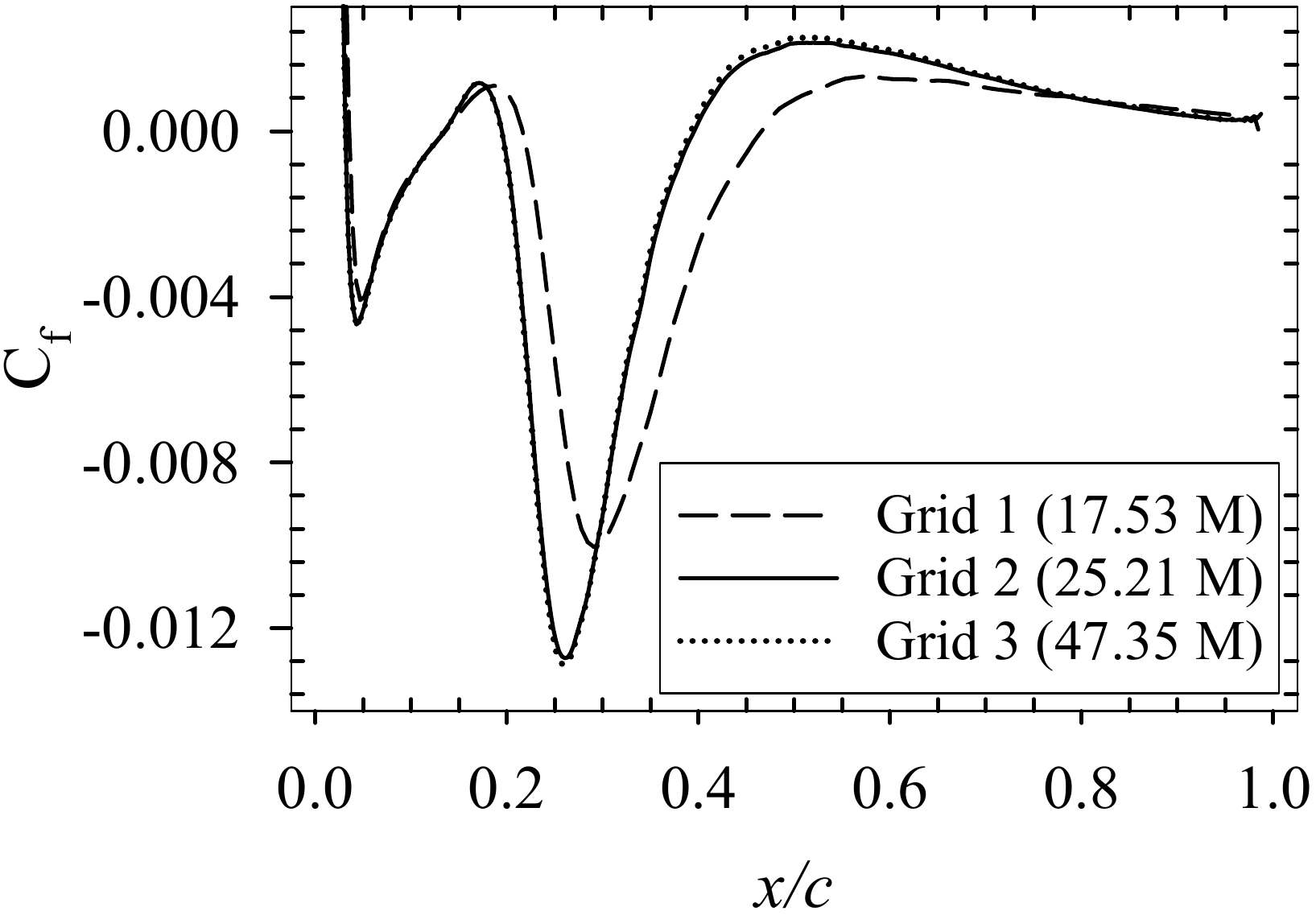}
\end{minipage}
\caption{Sensitivity of the mean pressure coeff\/icient and the mean skin-friction coeff\/icient to grid ref\/inement at the angle of attack $\alpha = 9.25^{\circ}$.}
\label{cp_cf}
\end{center}
\end{figure}
\newline
A grid sensitivity study was performed to assess the effect of the grid distribution on the characteristics of the LSB and its associated LFO. Three different grid distributions were used; a coarse grid (Grid--$1$) with the dimensions of $N_\xi\times N_\eta\times N_\zeta=637\times320\times86$, a f\/ine grid with the dimensions of $780\times320\times 101$ (Grid--$2$), and a f\/iner grid with the dimensions of $980\times320\times 151$ (Grid--$3$). Table~\ref{table:1} shows summary of the three different grids. Grid--$2$ is a ref\/ined and optimised version of Grid--$1$ by redistributing grid points in the $\eta$ direction such that around $60\%$ of the total grid points were within one chord from the aerofoil surface. Grid--$2$ and Grid--$3$ were generated using a hyperbolic grid generator to improve the grid orthogonality and minimise the grid skewness. The wall-normal grid spacing was reduced to ensure a wall-normal spacing of $y^{+}\leq1$. Having inadequate spacing in the $\xi$ direction could affect the development of the evolving Kelvin--Helmholtz instability and consequently the transition process. Equidistant grid spacing with a relatively small $\Delta x$ was adopted in the range $0 \leq x/c \leq 0.5$ for Grid--$2$ and Grid--$3$. The number of grid points in the spanwise direction was increased from $86$ grid points in Grid--$1$ to $101$ grid points in Grid--$2$, and $151$ grid points in Grid--$3$. This has reduced the grid spacing in viscous wall units to less than $15$ and signif\/icantly improved the solution. An adequate grid distribution in the spanwise direction is crucial for the break-down of the evolving Kelvin--Helmholtz instability. Thus, predicting the correct location of transition depends on the grid distribution.
\begin{figure}
\begin{center}
\begin{minipage}{220pt}
\includegraphics[height=140pt , trim={0mm 0mm 0mm 0mm}, clip]{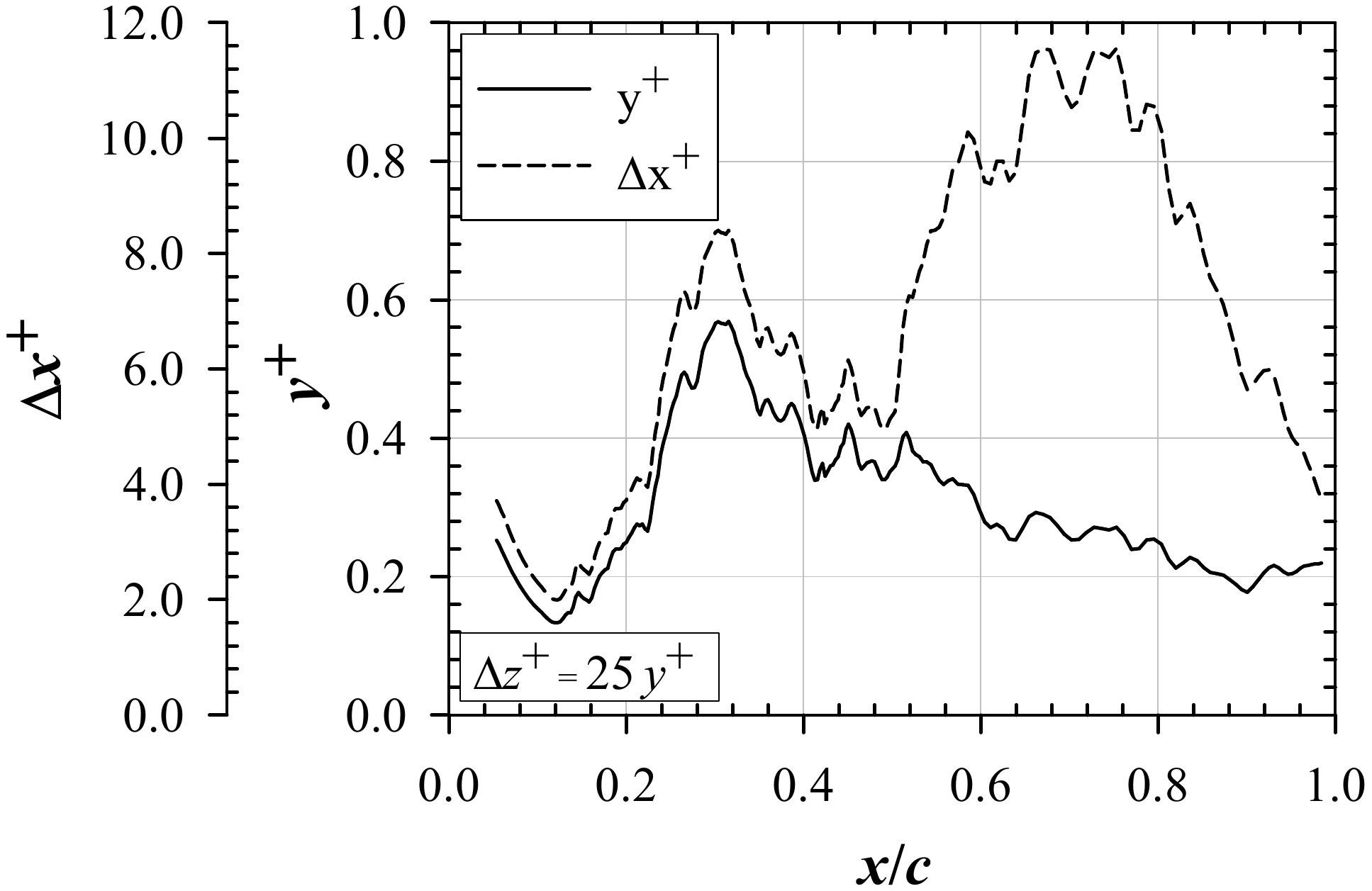}
\caption{Grid resolution, attached-phase.}
\label{yplus_attached}
\end{minipage}
\begin{minipage}{220pt}
\includegraphics[height=140pt , trim={0mm 0mm 0mm 0mm}, clip]{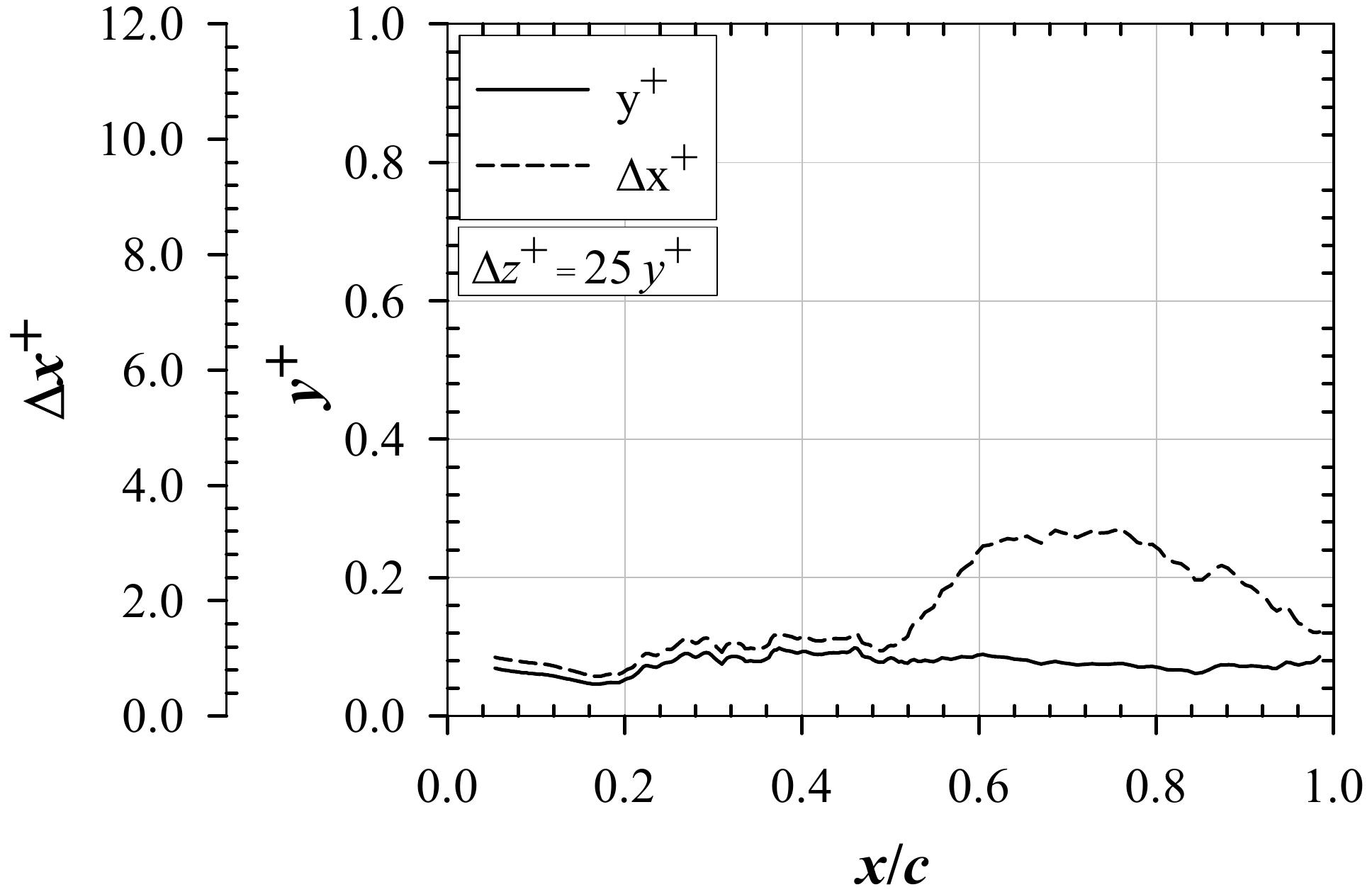}
\caption{Grid resolution, separated-phase.}
\label{yplus_separated}
\end{minipage}
\end{center}
\end{figure}
\newline
F\/igure~\ref{cp_cf} shows the mean pressure and mean skin-friction coeff\/icients for Grid--$1$, Grid--$2$, and Grid--$3$. The averaged pressure coeff\/icient distribution on the aerofoil suction side for Grid--$2$ has a much stronger adverse pressure near the leading edge. For the mean skin-friction coeff\/icient, the bubble size was smaller in Grid--$2$ case. In contrast, it was a bit longer in the coarse grid case (Grid--$1$). The bubble characteristics are independent of the grid at a grid resolution higher than that of Grid--$2$ as seen in the f\/igure. As seen in the f\/igure, the differences in the mean pressure and mean skin-friction distributions for Grid--$1$ and Grid--$2$ are huge compared to the small change in the overall grid points. However, the difference in Grid--$1$ and Grid--$2$ is not only in the number of grid points, but also in the quality of the grid. Grid--$1$ contains high skewed cells, the grid points are not adequately distributed in the wall-normal direction, and the grid is stretched in the streamwise direction. On the contrary, Grid--$2$ is optimized to have orthogonal cells and minimum skewness. Moreover, the grid points are optimally distributed in the wall-normal direction such that the grid points are intensif\/ied near the wall and along the shear layer. Also, an equidistance grid distribution is employed in the transition region. F\/inally, the number of grid points is signif\/icantly increased in the spanwise direction which enhanced the prediction of the break-down of the evolving Kelvin--Helmholtz instability and the transition location. Thus, the mean pressure coeff\/icient and mean skin-friction coeff\/icient have huge difference despite the relatively small change in the number of grid points.\newline
For the aerodynamic forces, the grid sensitivity study showed that the LFO was always captured whether the coarse or f\/ine grid is used. However, when coarse grids is used, the LFO is captured at lower angles of attack. As the grid is ref\/ined, the LFO is captured at slightly higher angles of attack. When an adequate grid distribution is used, the angle of attack at which the LFO takes place becomes independent of the grid resolution. This explains why~\citet{mukai2006large} observed several aspects of the LFO phenomenon despite using a much coarser grid.\newline
An investigation was performed to assess the effect of the domain spanwise width on the characteristics of the LSB and its associated LFO phenomena. F\/ive spanwise widths of $L_\zeta=0.2c$, $0.25c$, $0.5c$, $0.75c$, and $1.0c$ were considered. The mean skin-friction coeff\/icient and the mean pressure coeff\/icient showed that the shape and length of the LSB are independent of the computational domain width for $L_\zeta\geq0.25c$. Therefore, a domain width of $L_{\zeta}=0.5c$ was considered suff\/icient for the current Reynolds number. A computational grid with the dimensions $N_\xi\times N_\eta\times N_\zeta=780\times320\times 101$ was generated around a NACA-0012 aerofoil for each of the sixteen angles of attack. The aerofoil was oriented such that the incoming free-stream was always at a zero angle. F\/igures~\ref{yplus_attached} and~\ref{yplus_separated} show the grid resolution at the aerofoil surface for the attached and separated phases of the LFO cycle at $\alpha=9.8^\circ$. The f\/igures illustrate the variation of $y^+$ in the $\eta$ direction, $\Delta x^+$ in the $\xi$ direction, and $\Delta z^+$ in the $\zeta$ direction on the aerofoil suction side. The maximum grid spacing values for the wall-normal units during the attached phase were $y^+=0.55$, $\Delta x^+=12$, and $\Delta z^+=14$ with 20 grid points within $y^+ \leq 10$. $y^+$ had relatively high values in the range $0.2 \leq x/c \leq 0.4$; this is the region where the transition to turbulence takes place. The maximum grid spacing is smaller for the separated phase because the velocity gradient is much smaller in this case.

\section{Results and discussion} \label{sec:results}
LESs were carried out for the f\/low around the NACA-0012 aerofoil at sixteen angles of attack ($\alpha = 9.0^{\circ}$--$10.1^{\circ}$ at increments of $0.1^{\circ}$ as well as $\alpha = 8.5^{\circ}$, $8.8^{\circ}$, $9.25^{\circ}$, and $10.5^{\circ}$). The simulation Reynolds number and Mach number were $\rec=5\times10^4$ and $\minf = 0.4$, respectively. The free-stream f\/low direction was set parallel to the horizontal axis for all simulations ($u=1$, $v=0$, and $w=0$). The entire domain was initialised using the free-stream conditions. The simulations were performed with a time step of $10^{-4}$ non-dimensional time units. The samples for statistics are collected once transition of the simulations has decayed and the f\/low became stationary in time after $50$ f\/low-through times which is equivalent to $50$ non-dimensional time units. Aerodynamic coeff\/icients (lift coeff\/icient $(\mathrm{C}_{\!_\mathrm{L}})$, drag coeff\/icient $(\mathrm{C}_{\!_\mathrm{D}})$, skin-friction coeff\/icient $(\mathrm{C}_{\!_\mathrm{f}})$, and moment coeff\/icient $(\mathrm{C}_{\!_\mathrm{m}})$) were sampled for each angle of attack at a frequency of $10,000$ to generate two and a half million samples over a time period of $250$ non-dimensional time units. The locally-time-averaged and spanwise ensemble-averaged pressure, velocity components, and Reynolds stresses were sampled every $50$ time steps on the $x$--$y$ plane. A dataset of $20,000$ $x$--$y$ planes was recorded at a frequency of $204$ at each angle of attack.

\subsection{Conditional time-averaging}
\begin{figure}
\begin{center}
\includegraphics[width=300pt, trim={0mm 0mm 0mm 0mm}, clip]{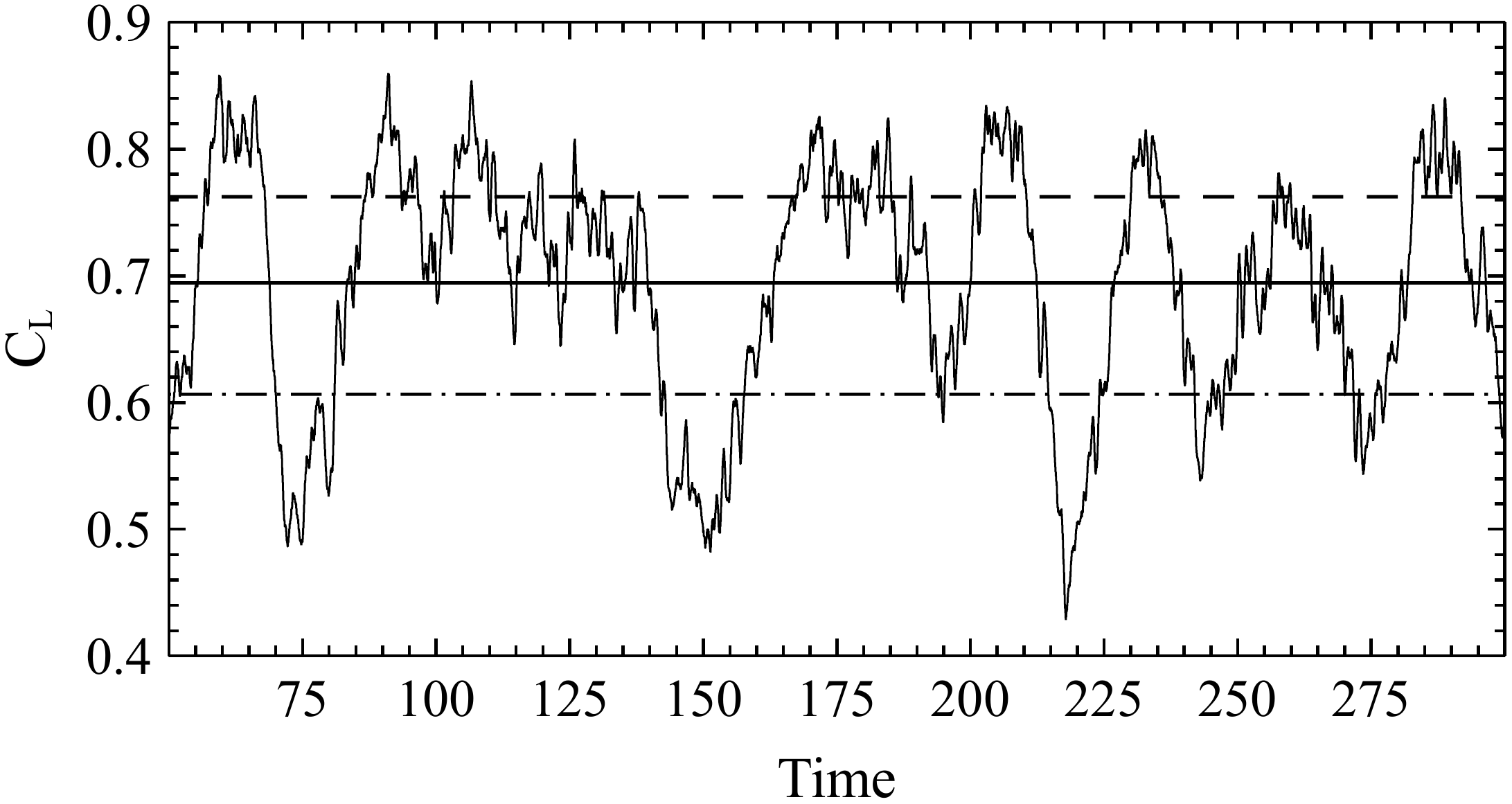}
\caption{Time history of the lift coeff\/icient at the angle of attack $9.80^{\circ}$ plotted versus non-dimensional time. Soild line: mean ($\overline{\mathrm{C}_{\!_\mathrm{L}}}$), dashed line: high-lift mean ($\widehat{\mathrm{C}_{\!_\mathrm{L}}}$), and dash-dot line: low-lift mean ($\widecheck{\mathrm{C}_{\!_\mathrm{L}}}$).}
\label{conditional_average}
\end{center}
\end{figure}

The signal of the lift coeff\/icient at each angle of attack was used as a reference for conditionally time-averaging the f\/low-f\/ield. The time-average of an instantaneous variable of the f\/low-f\/ield, $\psi$, was def\/ined on three different levels: 1) a mean-lift average; 2) a high-lift average in which the f\/low-f\/ield is attached in the statistical sense; and 3) a low-lift average in which the f\/low is statistically separated. The mean-lift average of the variable, $\overline{\psi}$, is simply the time-average of all data samples of the variable. The high-lift average of the variable, $\widehat{\psi}$, is the time-average of all data samples that have values higher than the mean values. The low-lift average of the variable, $\widecheck{\psi}$, is the average of all data samples that have values less than the mean values. The lift coeff\/icient signal is used to identify the data samples that are above or below the mean. It is implemented by taking the mean of the lift coeff\/icient at each angle of attack. The indices of data points of the time-history of the lift coeff\/icient that are above/below the mean of the lift coeff\/icient are then stored in a high-lift/low-lift data f\/iles, respectively. The indices are then used to locate the data of other f\/low variables that are above/below their corresponding mean and consequently used to estimate the low-lift and high-lift time-average for all of the f\/low variables. F\/igure~\ref{conditional_average} illustrates the concept of the conditional time-averaging that is based on the lift coeff\/icient.

\subsection{Mean f\/low}
\begin{figure}
\begin{center}
\begin{minipage}{220pt}
\centering
\includegraphics[width=220pt, trim={0mm 0mm 0mm 0mm}, clip]{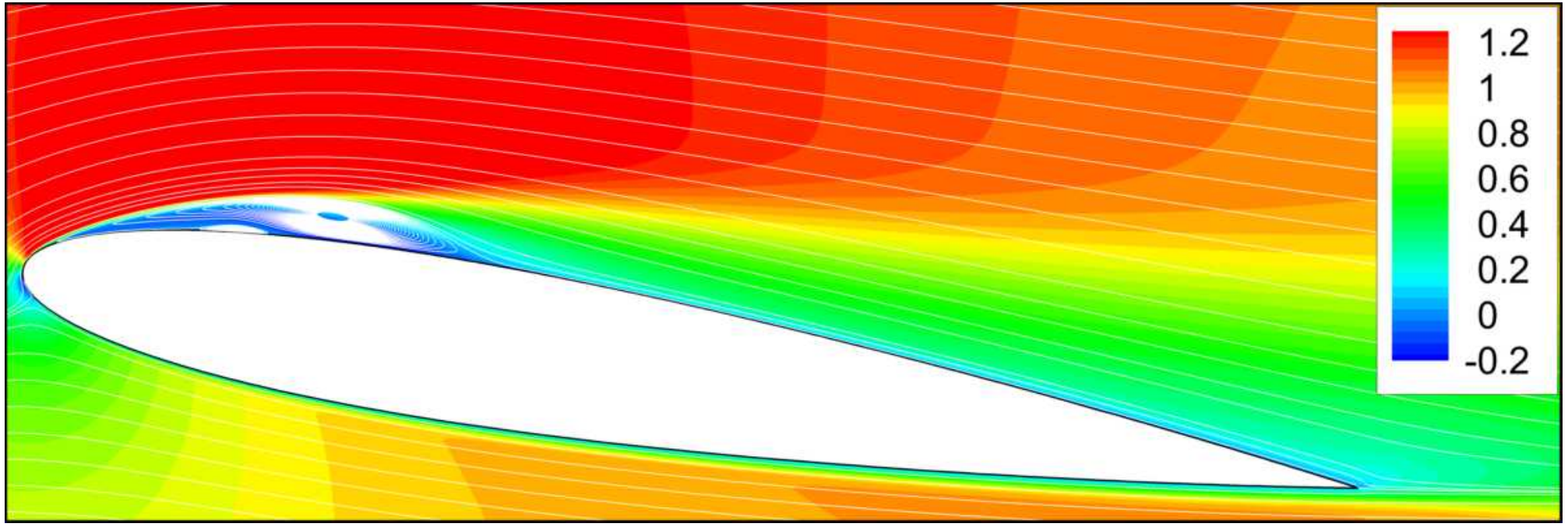}
\textit{$\alpha = 9.25^{\circ}$}
\end{minipage}
\medskip
\begin{minipage}{220pt}
\centering
\includegraphics[width=220pt, trim={0mm 0mm 0mm 0mm}, clip]{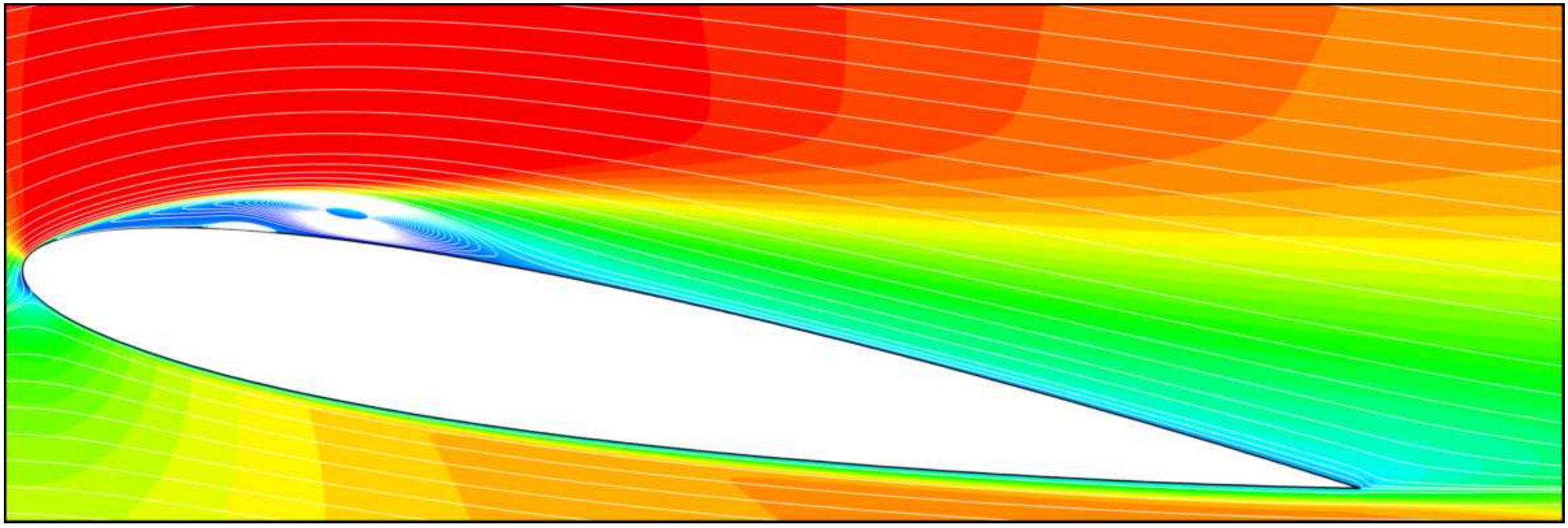}
\textit{$\alpha = 9.40^{\circ}$}
\end{minipage}
\medskip
\begin{minipage}{220pt}
\centering
\includegraphics[width=220pt, trim={0mm 0mm 0mm 0mm}, clip]{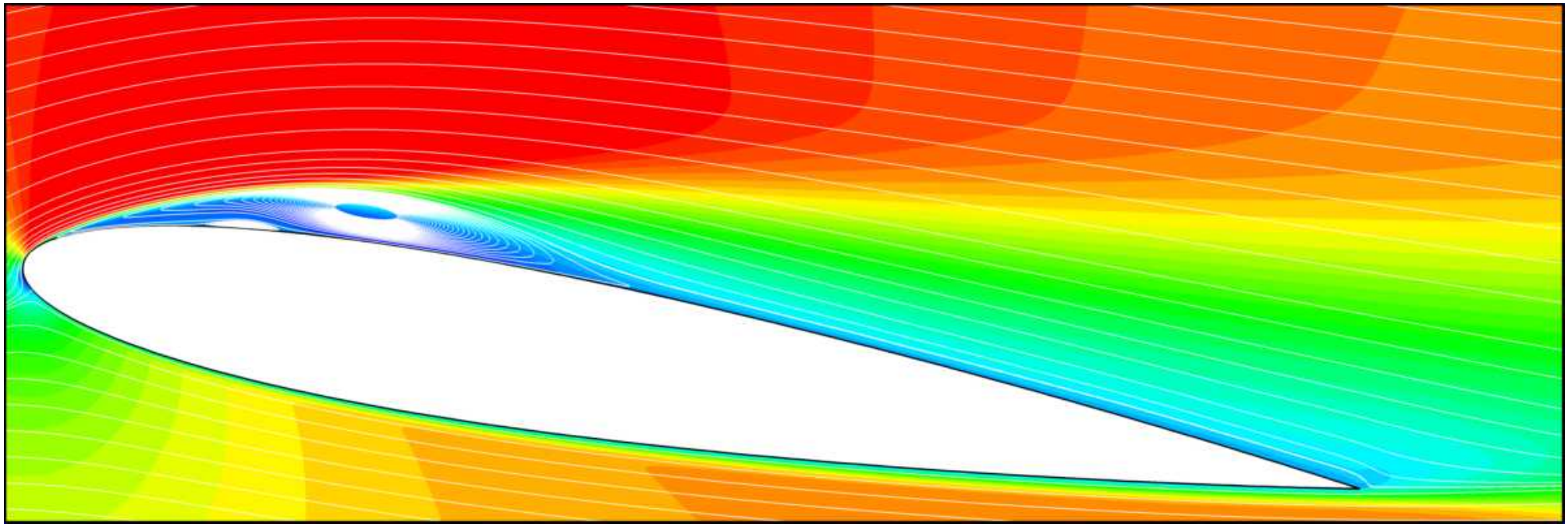}
\textit{$\alpha = 9.50^{\circ}$}
\end{minipage}
\medskip
\begin{minipage}{220pt}
\centering
\includegraphics[width=220pt, trim={0mm 0mm 0mm 0mm}, clip]{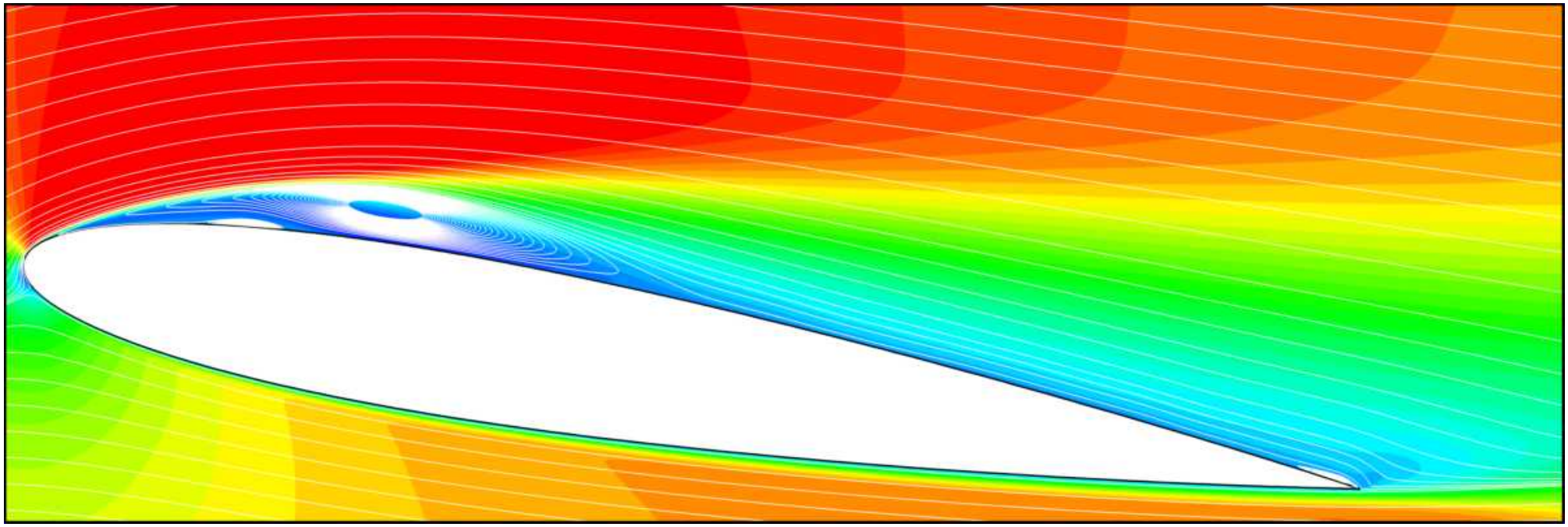}
\textit{$\alpha = 9.60^{\circ}$}
\end{minipage}
\medskip
\begin{minipage}{220pt}
\centering
\includegraphics[width=220pt, trim={0mm 0mm 0mm 0mm}, clip]{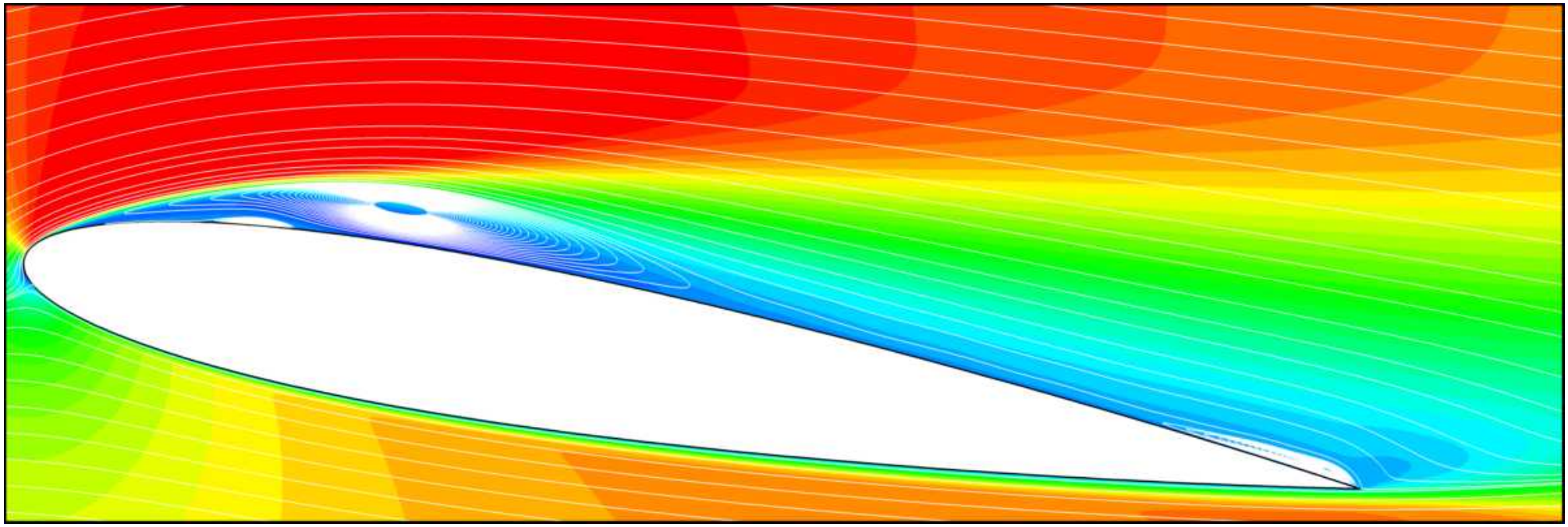}
\textit{$\alpha = 9.70^{\circ}$}
\end{minipage}
\medskip
\begin{minipage}{220pt}
\centering
\includegraphics[width=220pt, trim={0mm 0mm 0mm 0mm}, clip]{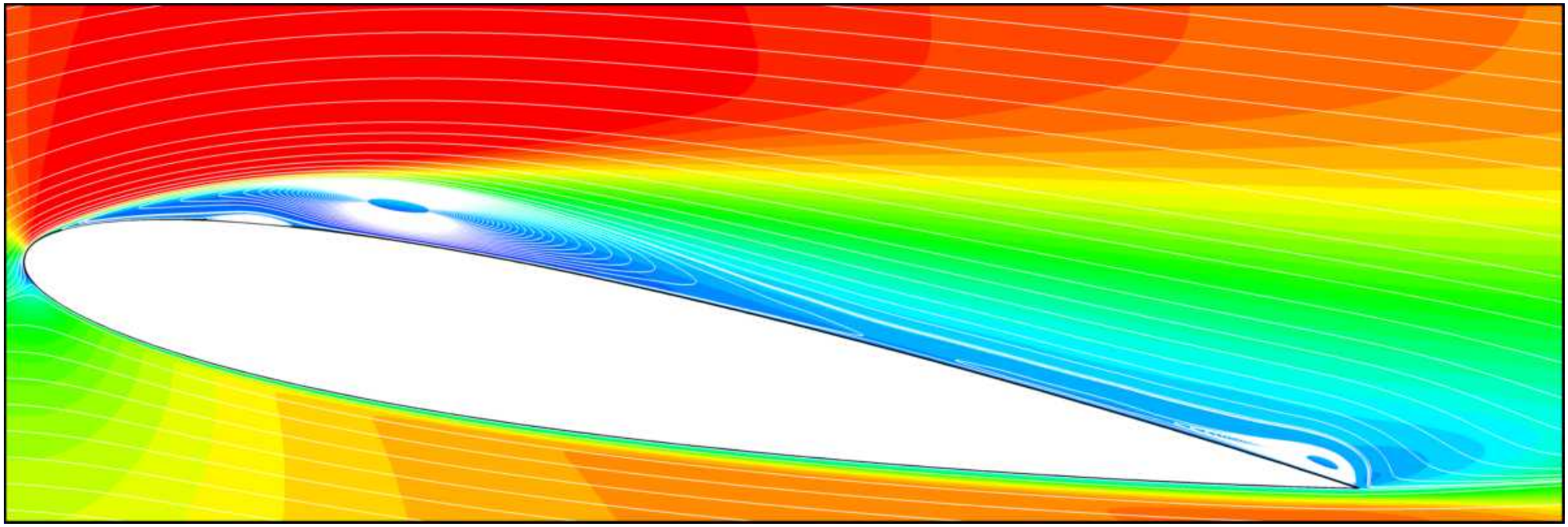}
\textit{$\alpha = 9.80^{\circ}$}
\end{minipage}
\medskip
\begin{minipage}{220pt}
\centering
\includegraphics[width=220pt, trim={0mm 0mm 0mm 0mm}, clip]{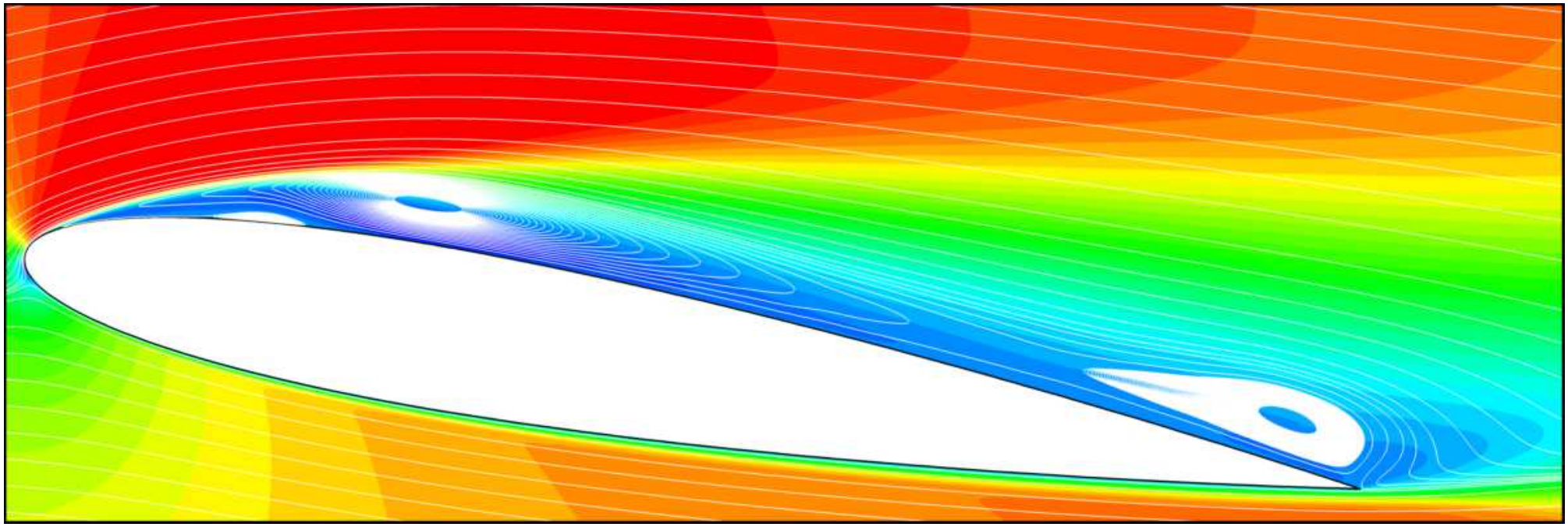}
\textit{$\alpha = 9.90^{\circ}$}
\end{minipage}
\medskip
\begin{minipage}{220pt}
\centering
\includegraphics[width=220pt, trim={0mm 0mm 0mm 0mm}, clip]{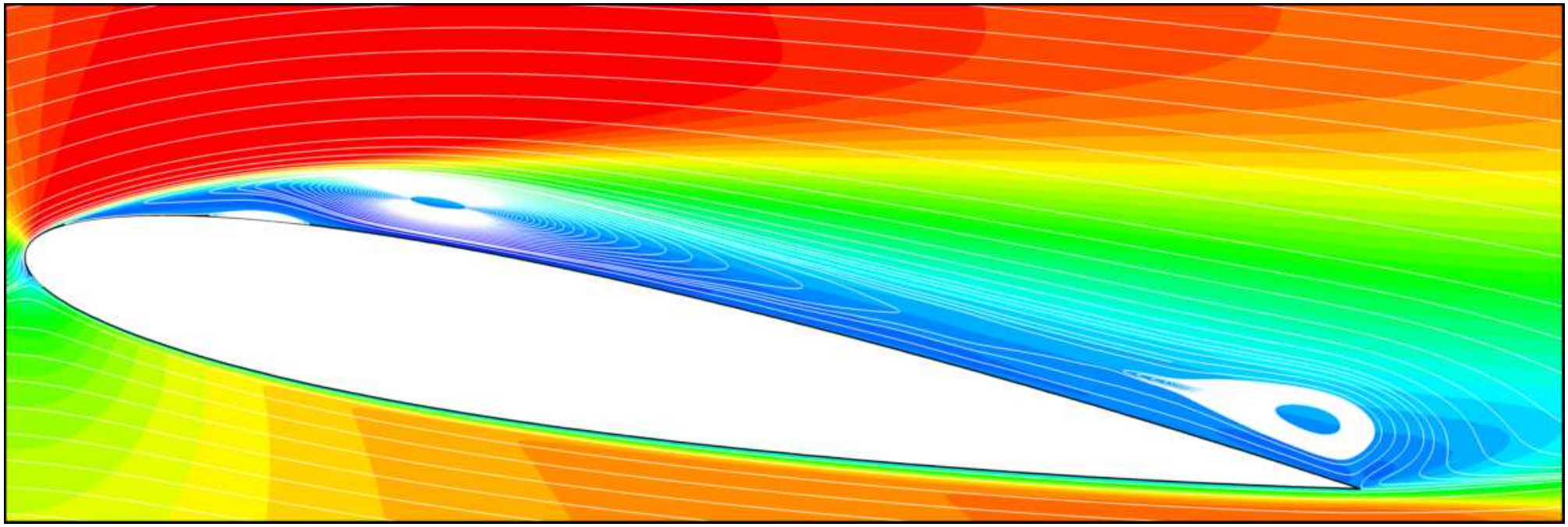}
\textit{$\alpha = 10.0^{\circ}$}
\end{minipage}
\medskip
\begin{minipage}{220pt}
\centering
\includegraphics[width=220pt, trim={0mm 0mm 0mm 0mm}, clip]{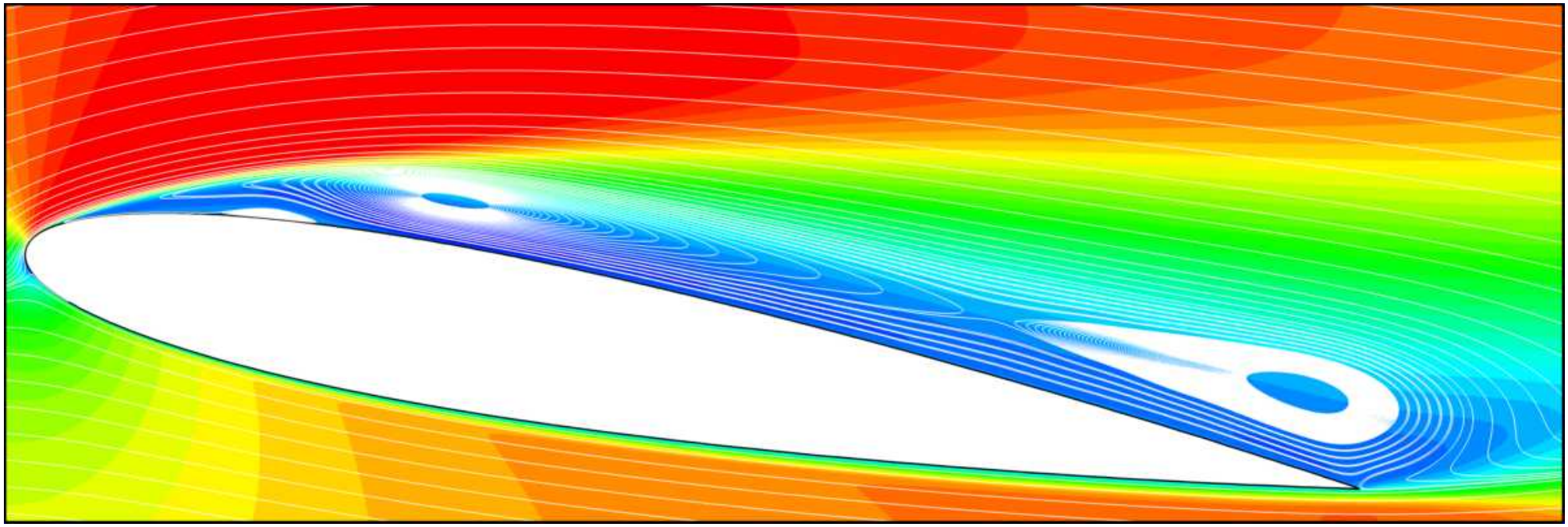}
\textit{$\alpha = 10.1^{\circ}$}
\end{minipage}
\begin{minipage}{220pt}
\centering
\includegraphics[width=220pt, trim={0mm 0mm 0mm 0mm}, clip]{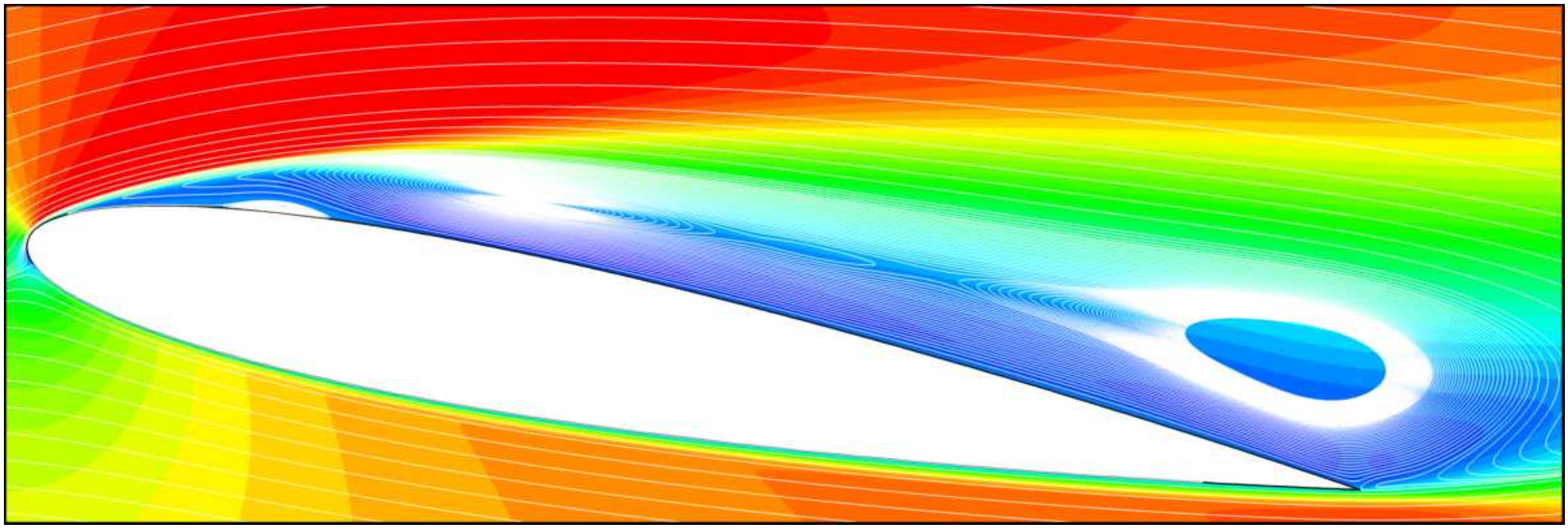}
\textit{$\alpha = 10.5^{\circ}$}
\end{minipage}
\caption{Streamlines patterns of the mean f\/low-f\/ield superimposed on colour maps of the mean streamwise velocity component, $\overline{U}$, for the angles of attack $\alpha = 9.25^{\circ}$--$10.5^{\circ}$.}
\label{bubble_mean}
\end{center}
\end{figure}

\begin{figure}
\begin{center}
\begin{minipage}{220pt}
\centering
\includegraphics[width=220pt, trim={0mm 0mm 0mm 0mm}, clip]{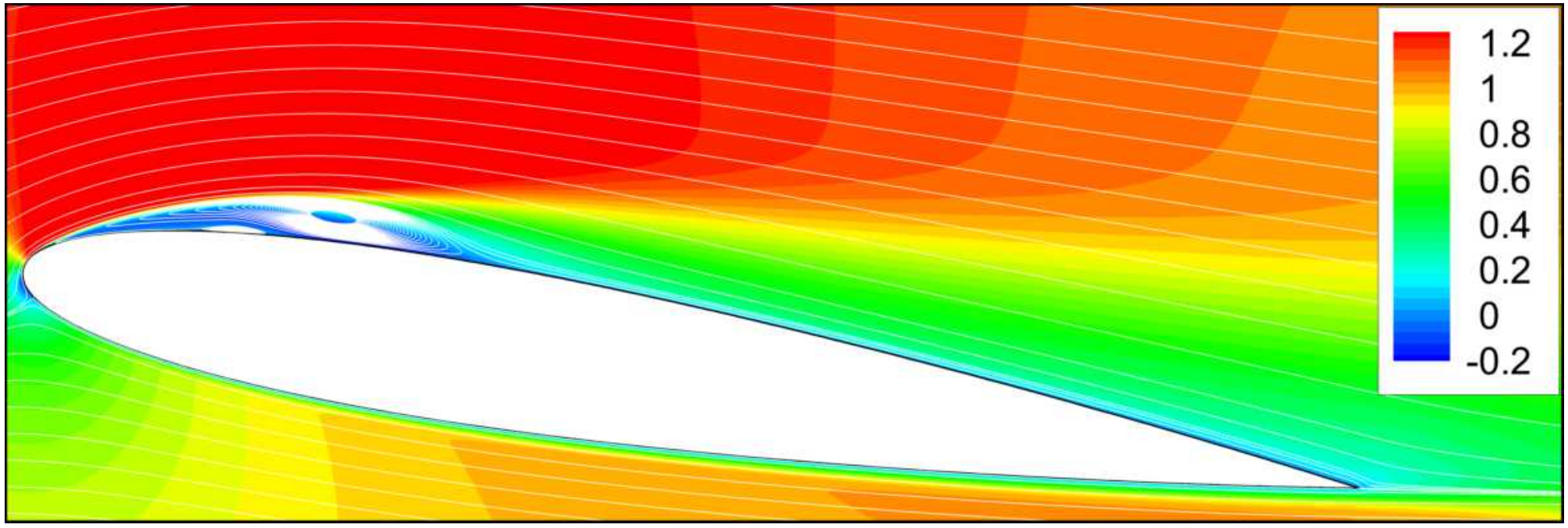}
\textit{$\alpha = 9.25^{\circ}$}
\end{minipage}
\begin{minipage}{220pt}
\centering
\includegraphics[width=220pt, trim={0mm 0mm 0mm 0mm}, clip]{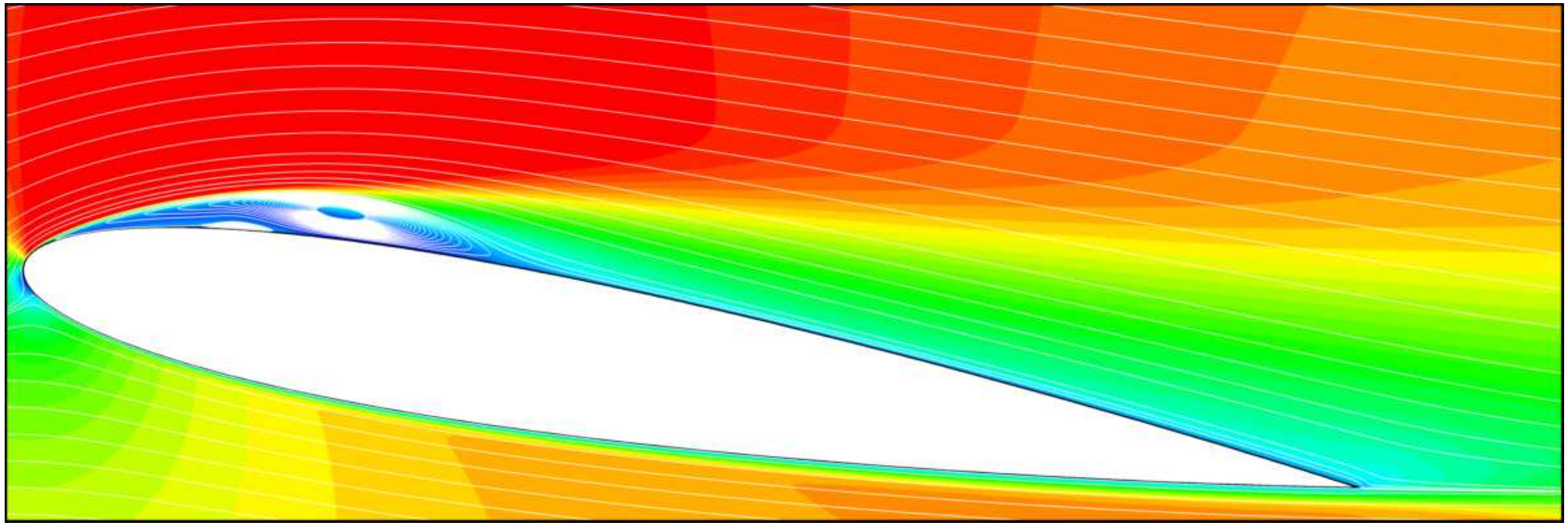}
\textit{$\alpha = 9.40^{\circ}$}
\end{minipage}
\begin{minipage}{220pt}
\centering
\includegraphics[width=220pt, trim={0mm 0mm 0mm 0mm}, clip]{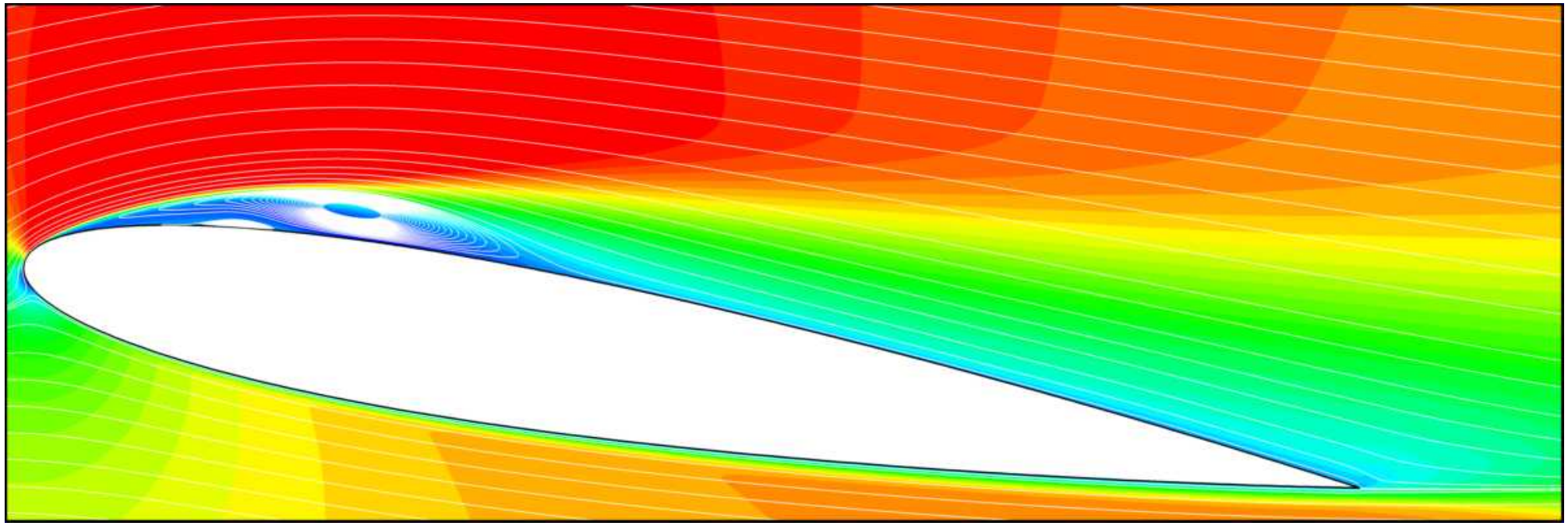}
\textit{$\alpha = 9.50^{\circ}$}
\end{minipage}
\begin{minipage}{220pt}
\centering
\includegraphics[width=220pt, trim={0mm 0mm 0mm 0mm}, clip]{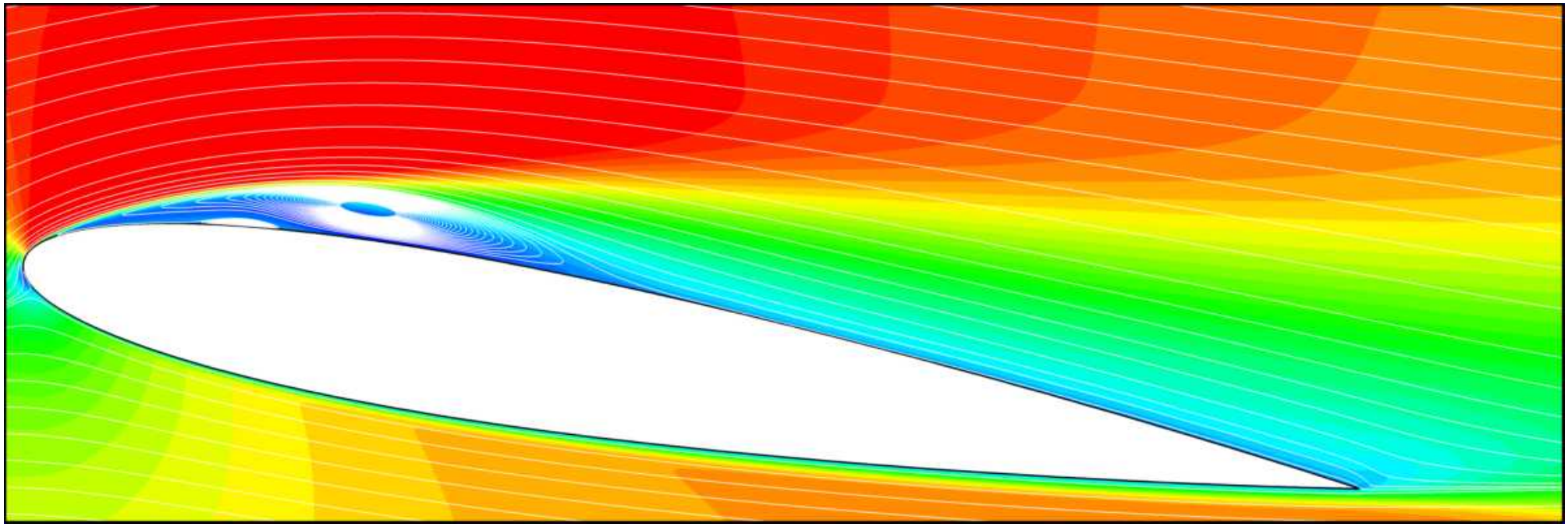}
\textit{$\alpha = 9.60^{\circ}$}
\end{minipage}
\begin{minipage}{220pt}
\centering
\includegraphics[width=220pt, trim={0mm 0mm 0mm 0mm}, clip]{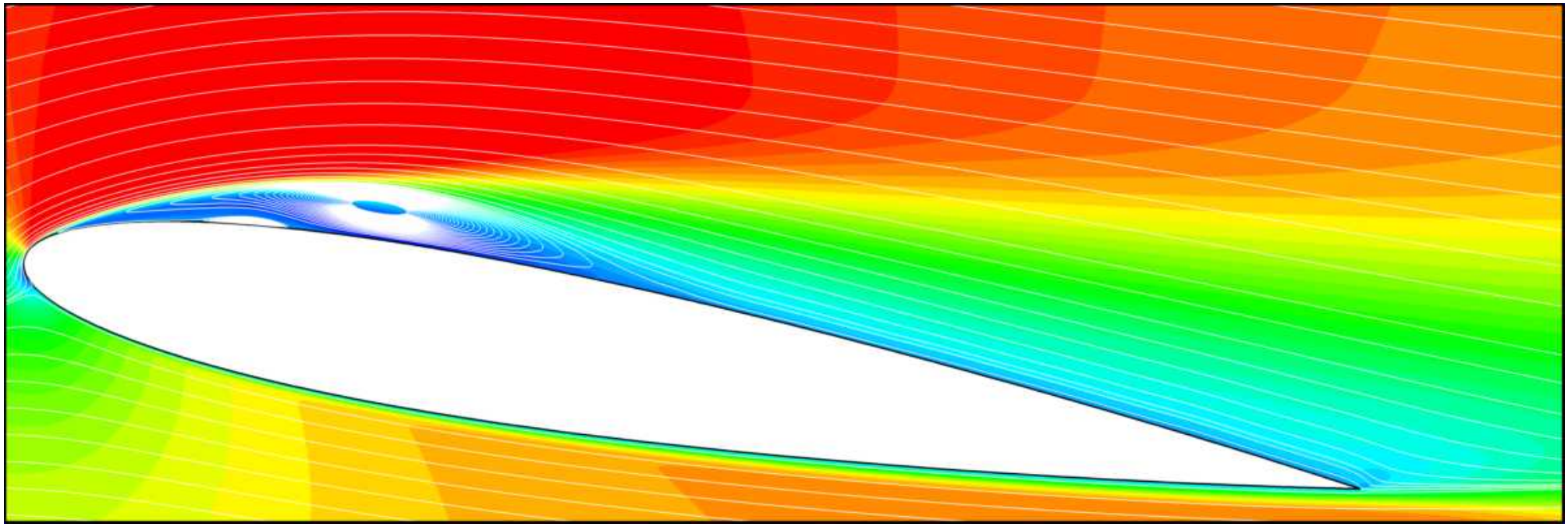}
\textit{$\alpha = 9.70^{\circ}$}
\end{minipage}
\begin{minipage}{220pt}
\centering
\includegraphics[width=220pt, trim={0mm 0mm 0mm 0mm}, clip]{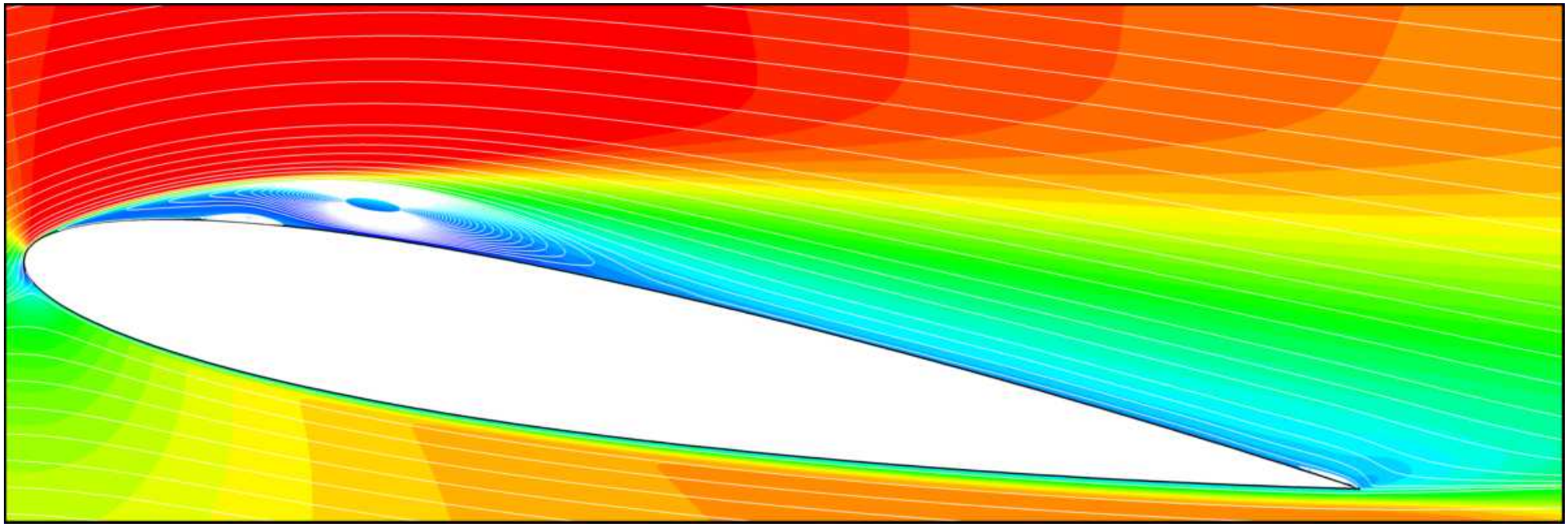}
\textit{$\alpha = 9.80^{\circ}$}
\end{minipage}
\begin{minipage}{220pt}
\centering
\includegraphics[width=220pt, trim={0mm 0mm 0mm 0mm}, clip]{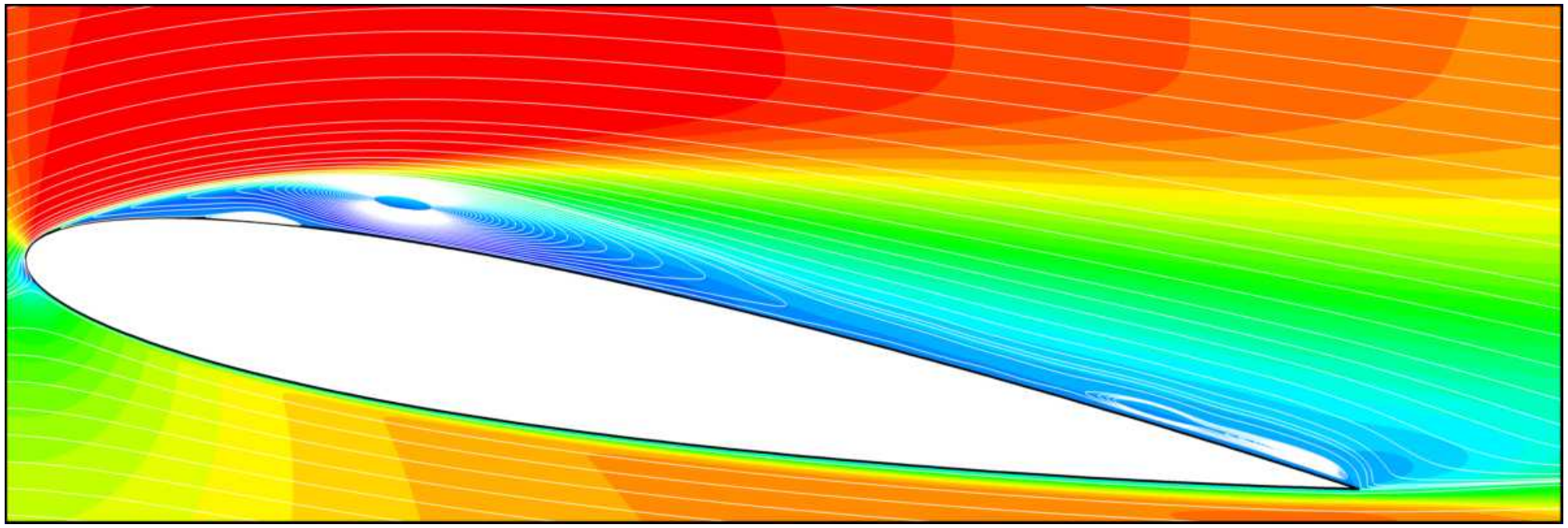}
\textit{$\alpha = 9.90^{\circ}$}
\end{minipage}
\begin{minipage}{220pt}
\centering
\includegraphics[width=220pt, trim={0mm 0mm 0mm 0mm}, clip]{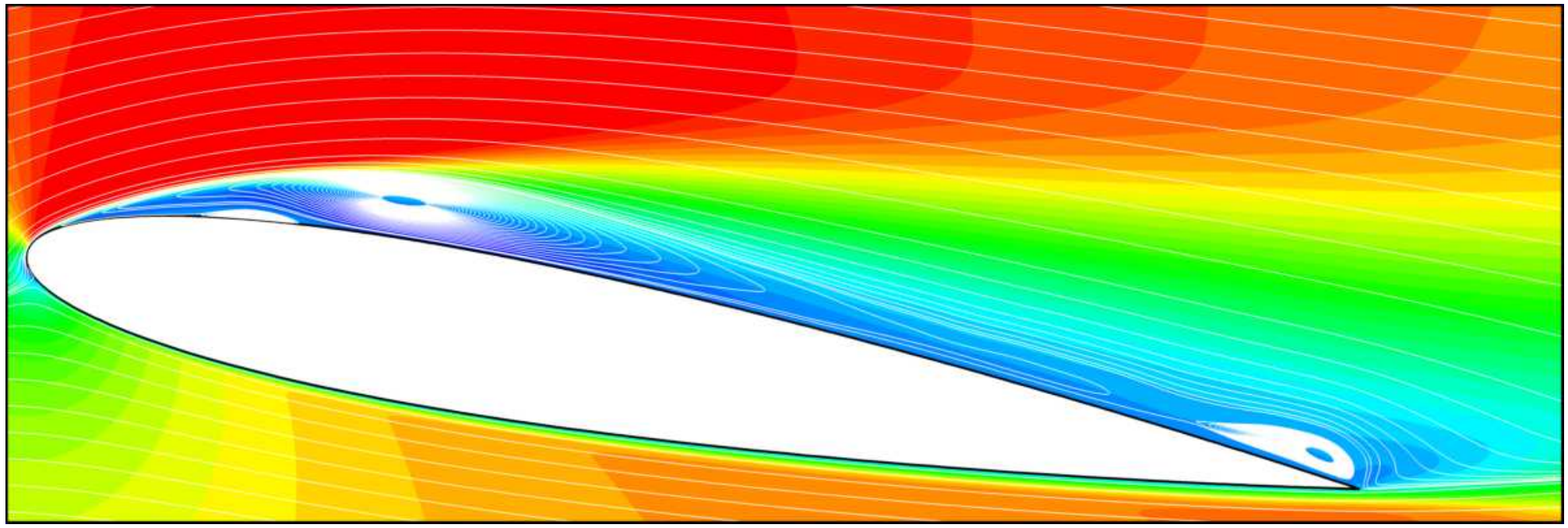}
\textit{$\alpha = 10.0^{\circ}$}
\end{minipage}
\begin{minipage}{220pt}
\centering
\includegraphics[width=220pt, trim={0mm 0mm 0mm 0mm}, clip]{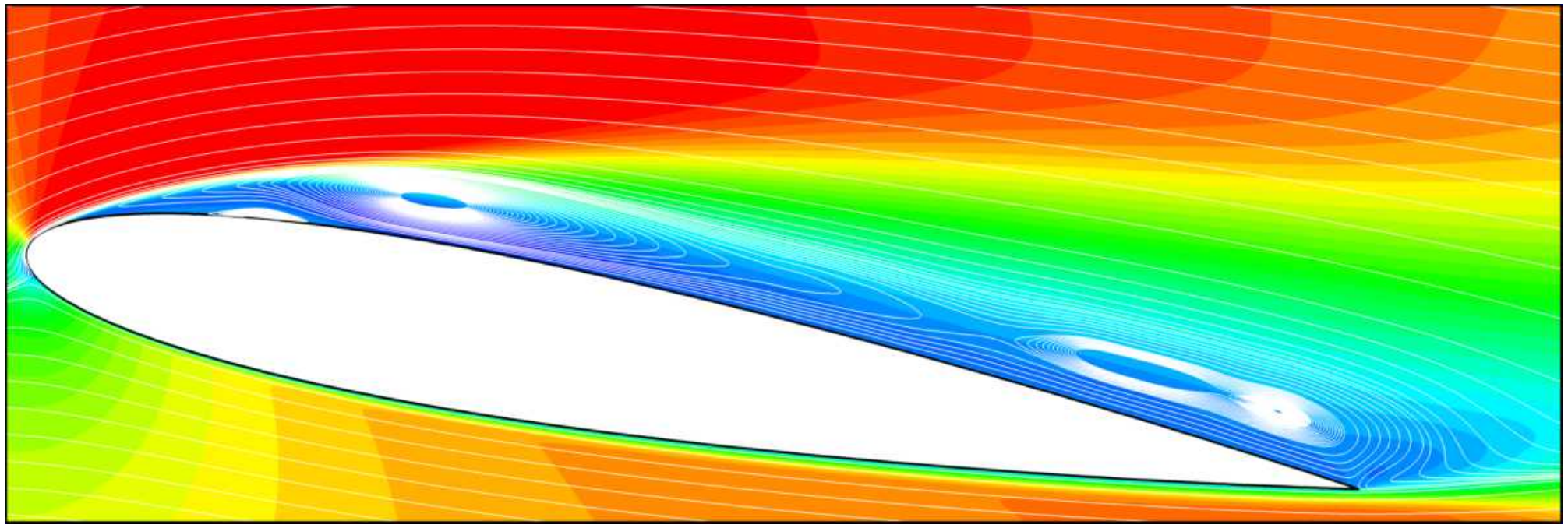}
\textit{$\alpha = 10.1^{\circ}$}
\end{minipage}
\begin{minipage}{220pt}
\centering
\includegraphics[width=220pt, trim={0mm 0mm 0mm 0mm}, clip]{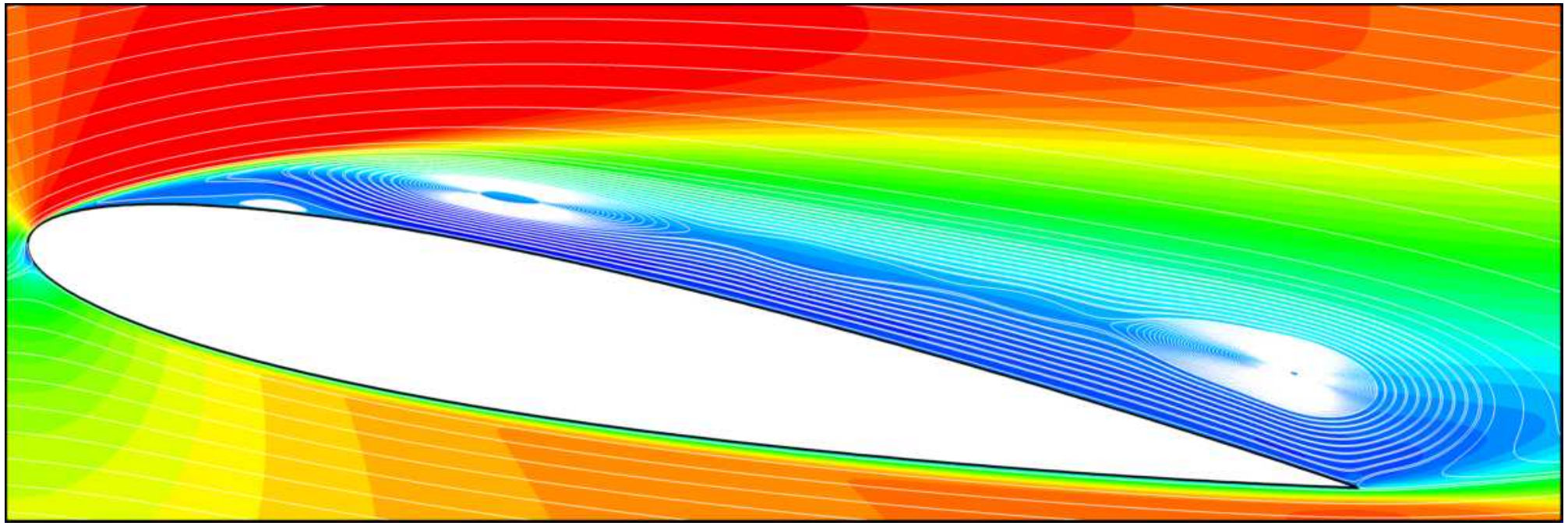}
\textit{$\alpha = 10.5^{\circ}$}
\end{minipage}
\caption{Streamlines patterns of the high-lift f\/low-f\/ield superimposed on colour maps of the high-lift streamwise velocity component, $\widehat{U}$, for the angles of attack $\alpha = 9.25^{\circ}$--$10.5^{\circ}$.}
\label{bubble_above}
\end{center}
\end{figure}

\begin{figure}
\begin{center}
\begin{minipage}{220pt}
\includegraphics[height=125pt , trim={0mm 0mm 0mm 0mm}, clip]{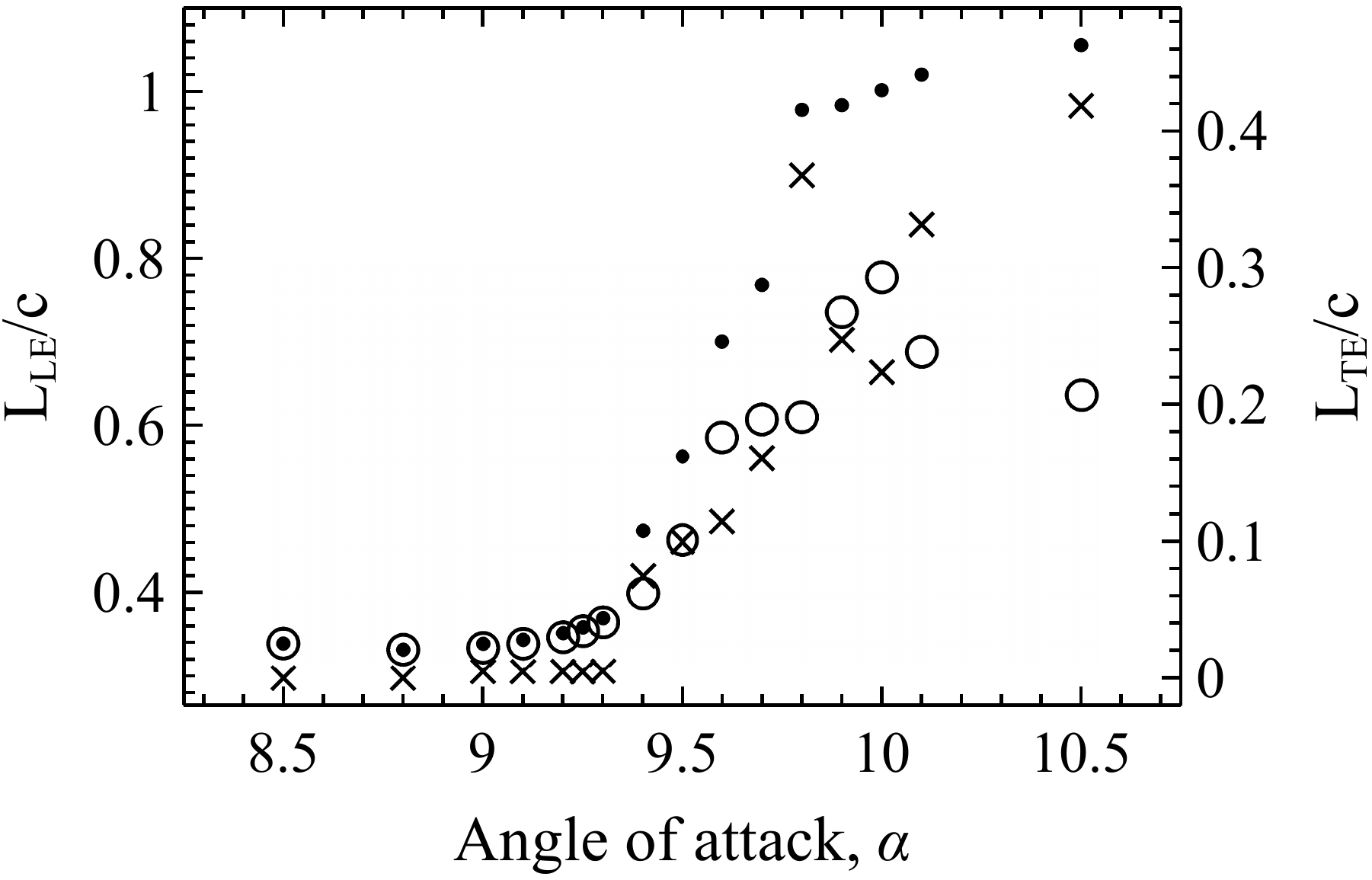}
\end{minipage}
\begin{minipage}{220pt}
\includegraphics[height=125pt , trim={0mm 0mm 0mm 0mm}, clip]{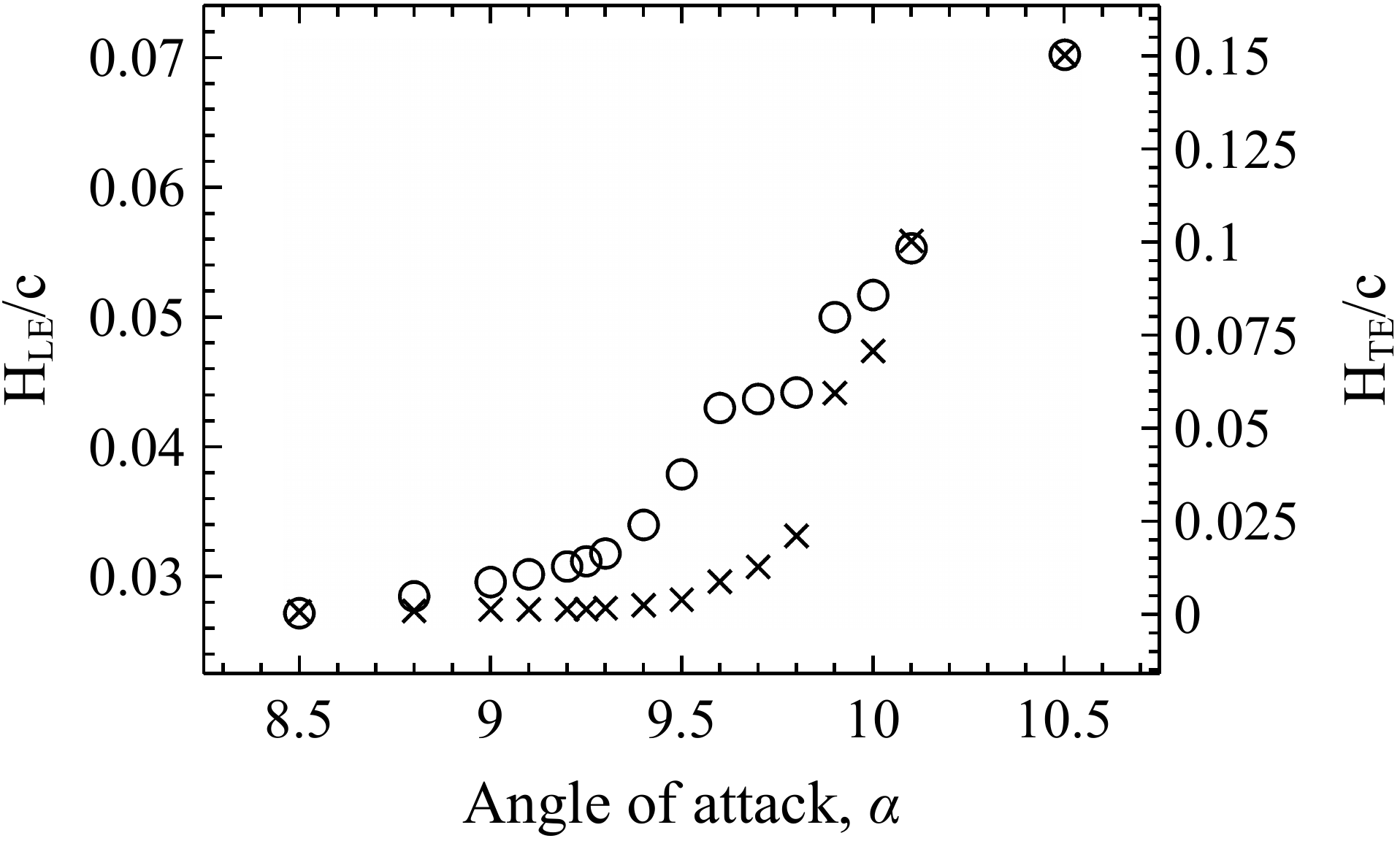}
\end{minipage}
\caption{Left: the bubble length plotted versus the angle of attack $\alpha$. Circles: the length of the leading-edge bubble ($L_{LE}$); $\times$'s: the length of the trailing-edge bubble ($L_{TE}$); black dots: $L_{LE} + L_{TE}$. Right: the bubble height plotted versus the angle of attack $\alpha$. Circles: the height of the leading-edge bubble ($H_{LE}$); $\times$'s: the height of the trailing-edge bubble ($H_{TE}$).}
\label{bubble_size}
\end{center}
\end{figure}

\begin{figure}
\begin{center}
\begin{minipage}{220pt}
\centering
\includegraphics[width=220pt, trim={0mm 0mm 0mm 0mm}, clip]{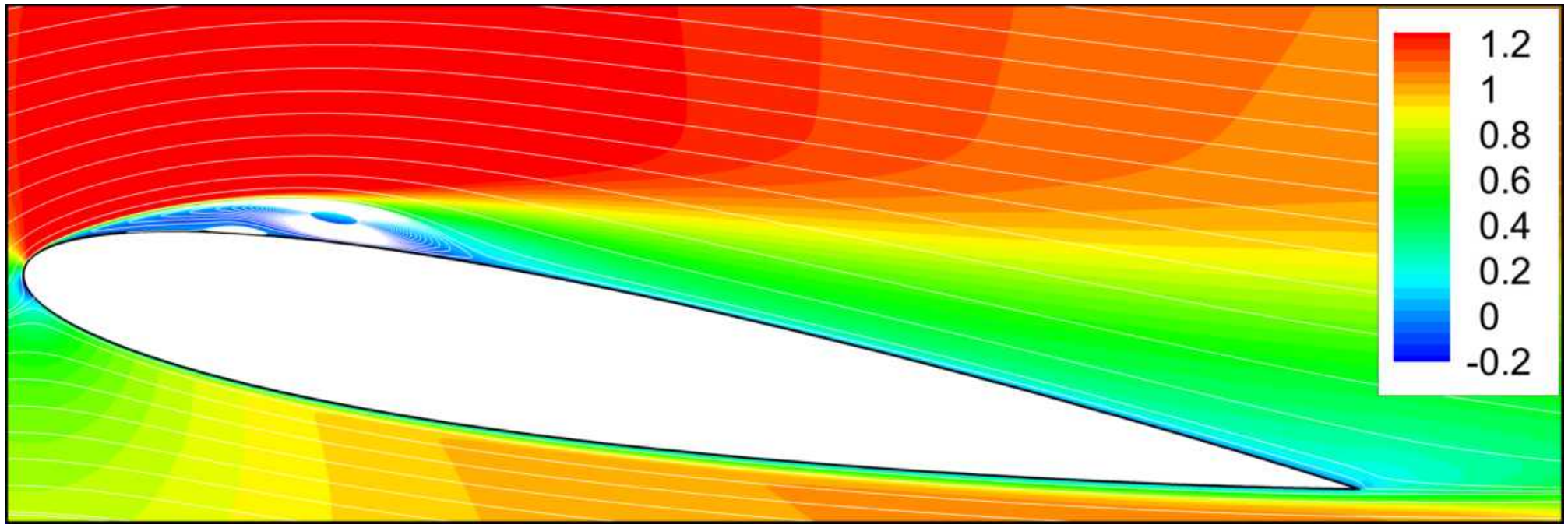}
\textit{$\alpha = 9.25^{\circ}$}
\end{minipage}
\begin{minipage}{220pt}
\centering
\includegraphics[width=220pt, trim={0mm 0mm 0mm 0mm}, clip]{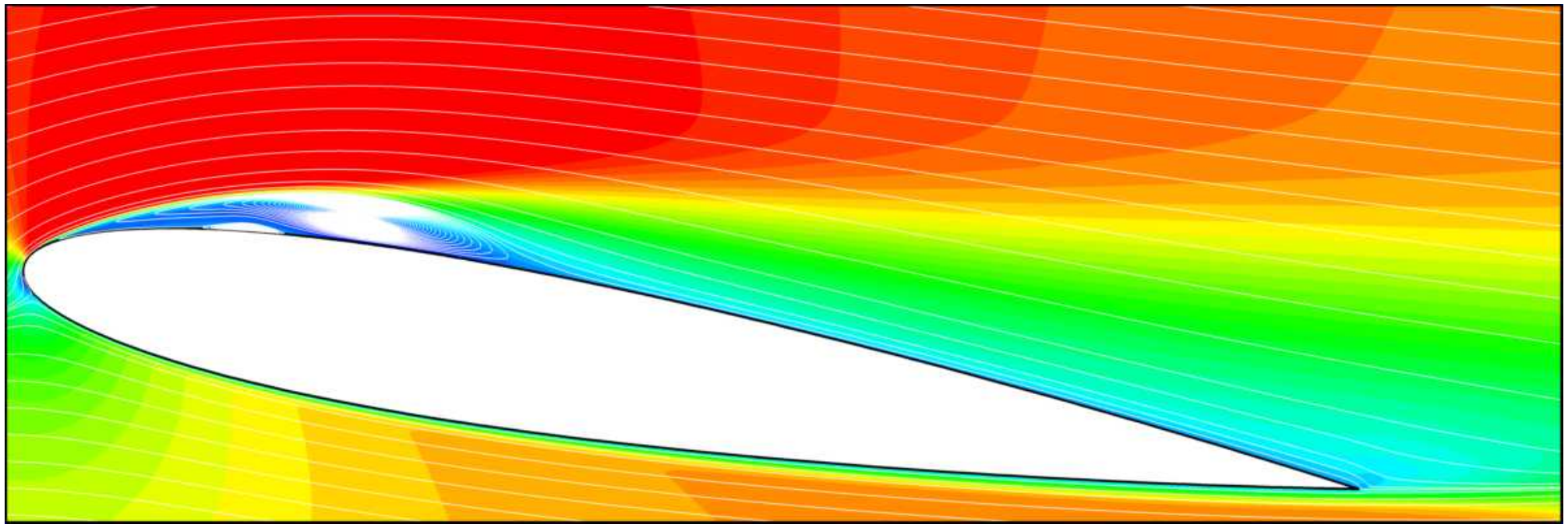}
\textit{$\alpha = 9.40^{\circ}$}
\end{minipage}
\begin{minipage}{220pt}
\centering
\includegraphics[width=220pt, trim={0mm 0mm 0mm 0mm}, clip]{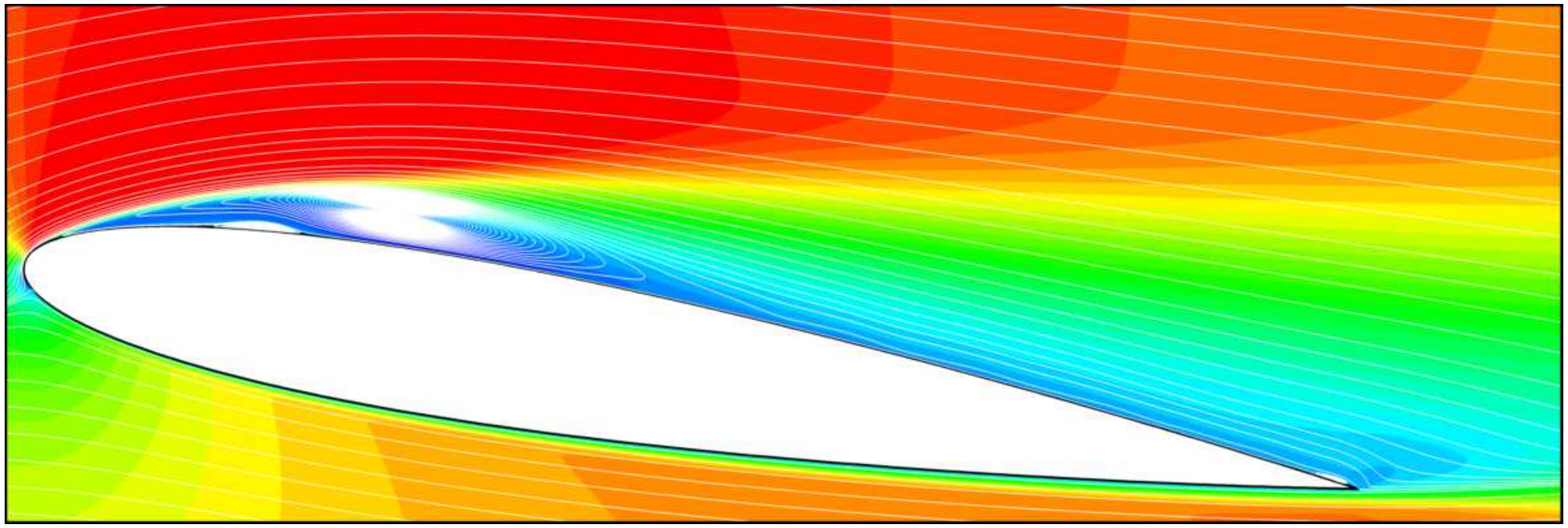}
\textit{$\alpha = 9.50^{\circ}$}
\end{minipage}
\begin{minipage}{220pt}
\centering
\includegraphics[width=220pt, trim={0mm 0mm 0mm 0mm}, clip]{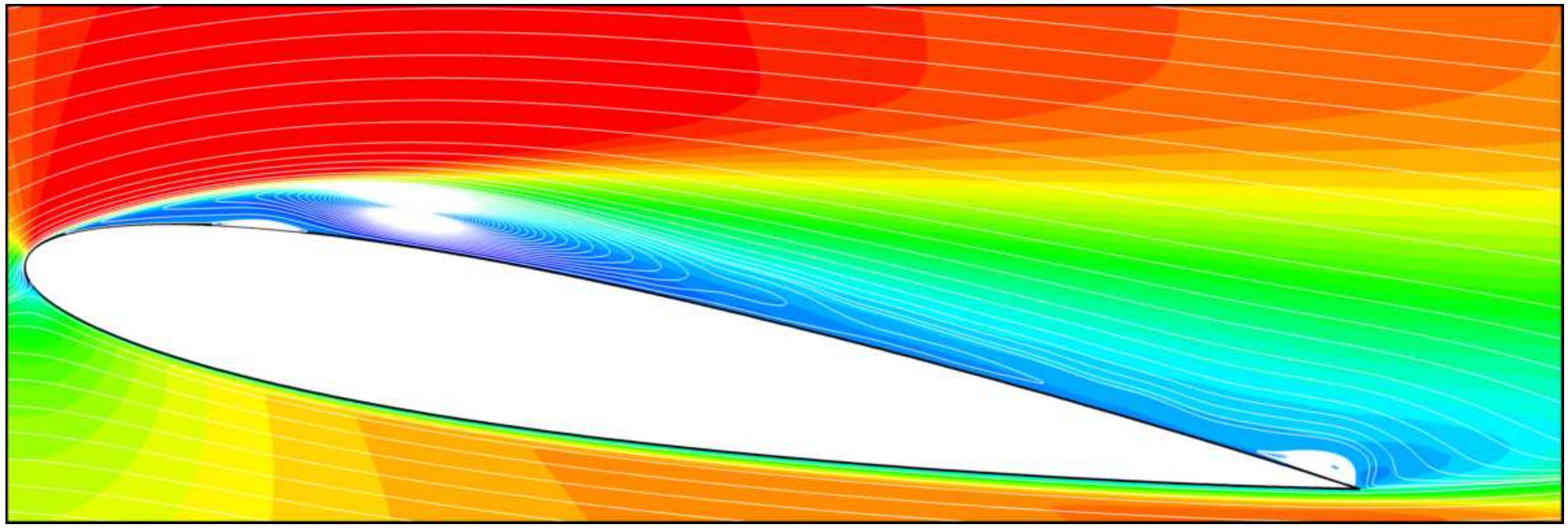}
\textit{$\alpha = 9.60^{\circ}$}
\end{minipage}
\begin{minipage}{220pt}
\centering
\includegraphics[width=220pt, trim={0mm 0mm 0mm 0mm}, clip]{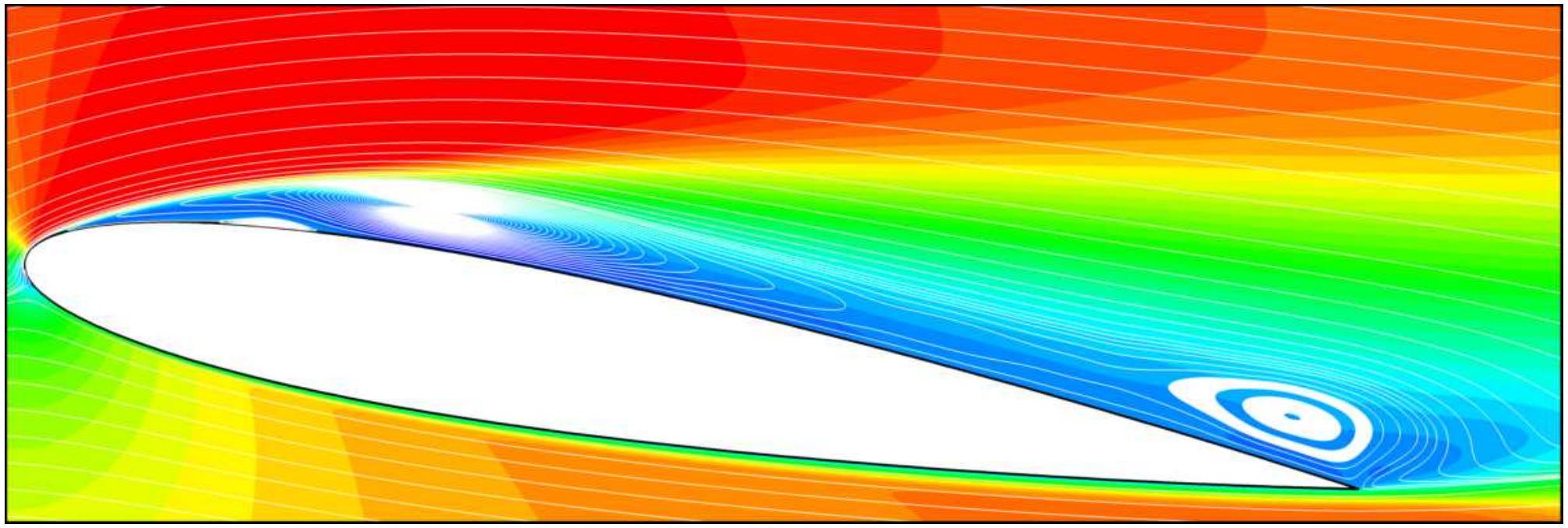}
\textit{$\alpha = 9.70^{\circ}$}
\end{minipage}
\begin{minipage}{220pt}
\centering
\includegraphics[width=220pt, trim={0mm 0mm 0mm 0mm}, clip]{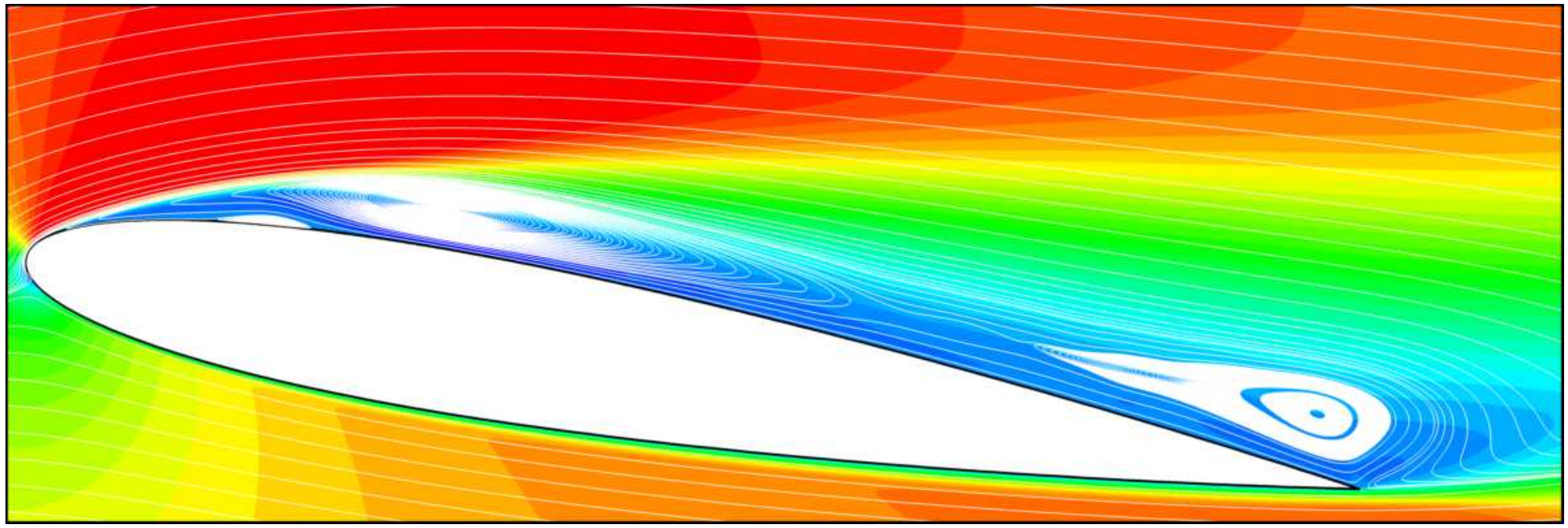}
\textit{$\alpha = 9.80^{\circ}$}
\end{minipage}
\begin{minipage}{220pt}
\centering
\includegraphics[width=220pt, trim={0mm 0mm 0mm 0mm}, clip]{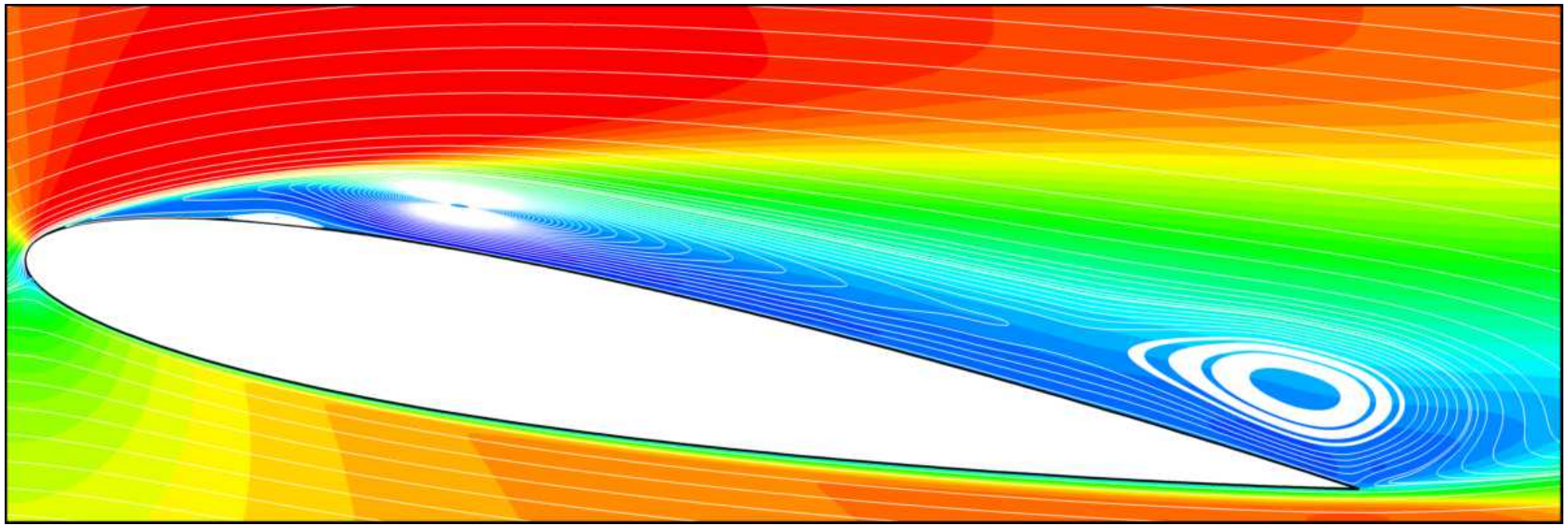}
\textit{$\alpha = 9.90^{\circ}$}
\end{minipage}
\begin{minipage}{220pt}
\centering
\includegraphics[width=220pt, trim={0mm 0mm 0mm 0mm}, clip]{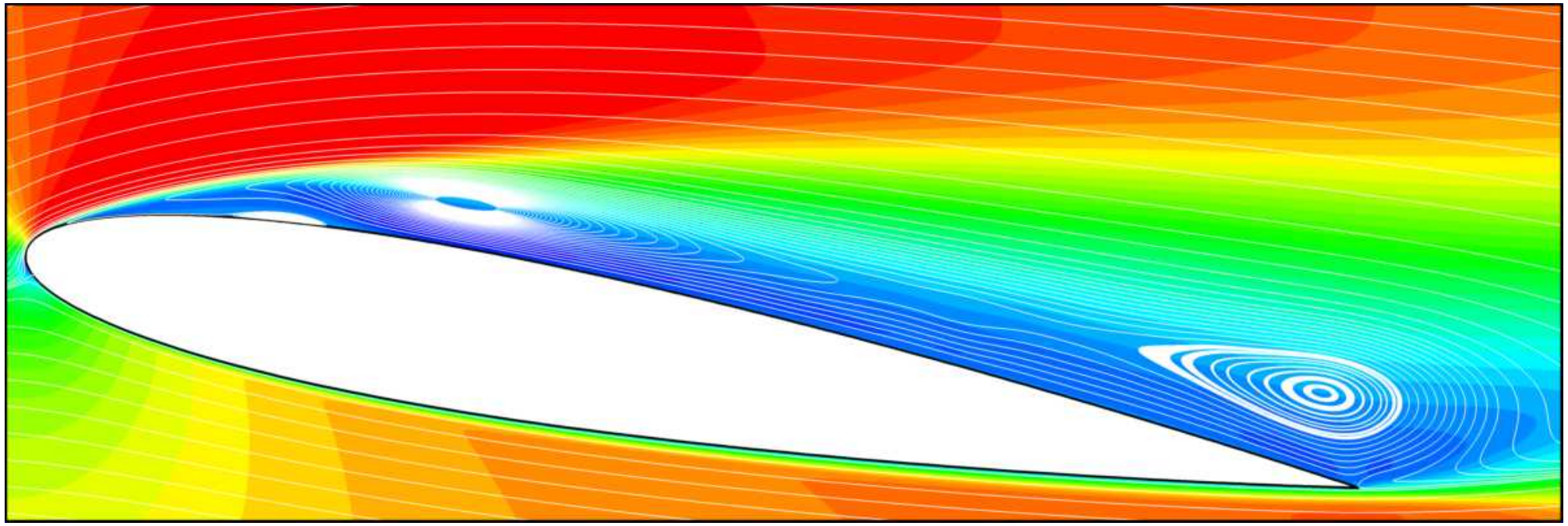}
\textit{$\alpha = 10.0^{\circ}$}
\end{minipage}
\begin{minipage}{220pt}
\centering
\includegraphics[width=220pt, trim={0mm 0mm 0mm 0mm}, clip]{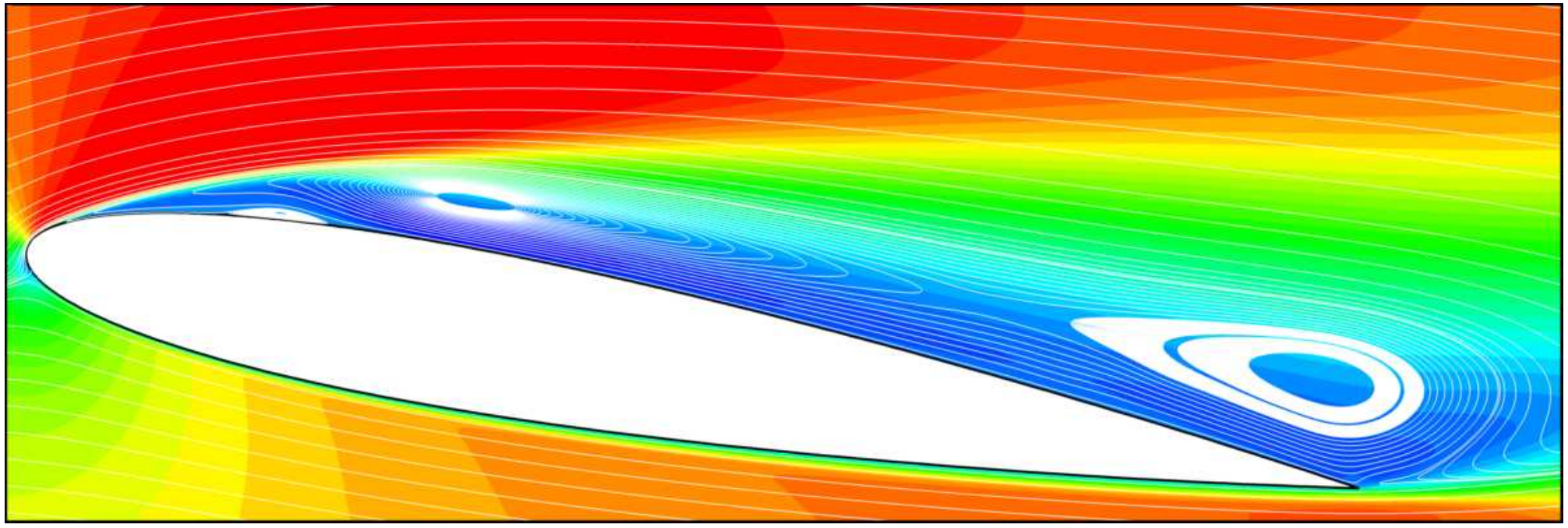}
\textit{$\alpha = 10.1^{\circ}$}
\end{minipage}
\begin{minipage}{220pt}
\centering
\includegraphics[width=220pt, trim={0mm 0mm 0mm 0mm}, clip]{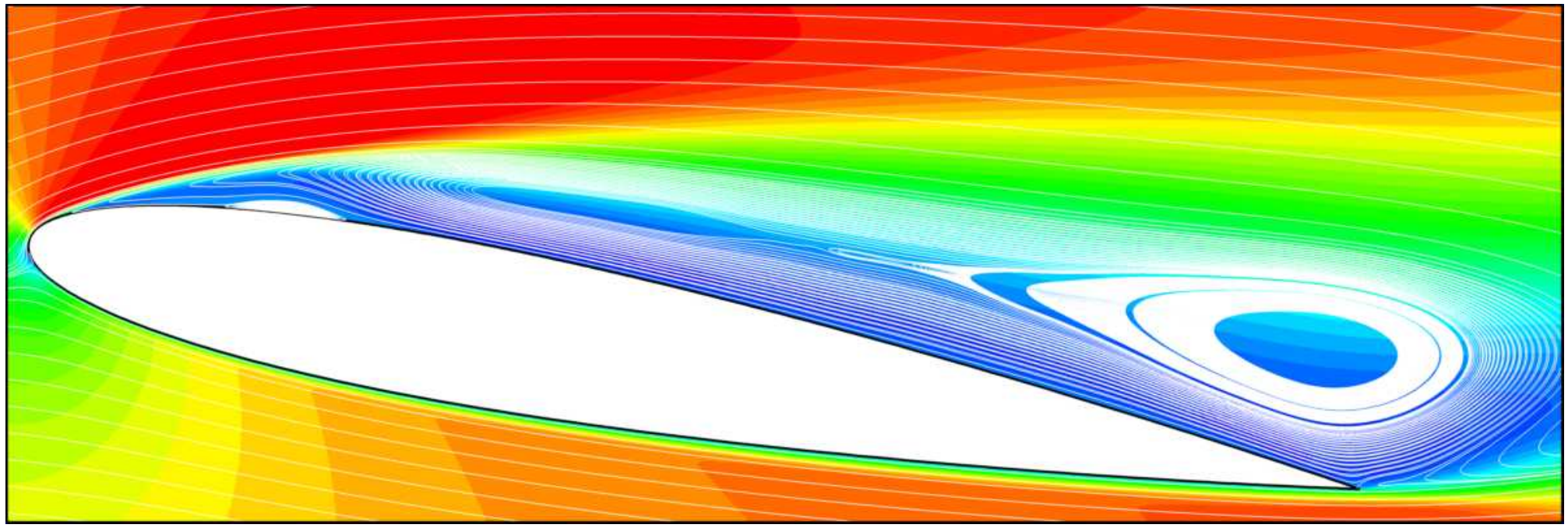}
\textit{$\alpha = 10.5^{\circ}$}
\end{minipage}
\caption{Streamlines patterns of the low-lift f\/low-f\/ield superimposed on colour maps of the low-lift streamwise velocity component, $\widecheck{U}$, for the angles of attack $\alpha = 9.25^{\circ}$--$10.5^{\circ}$.}
\label{bubble_below}
\end{center}
\end{figure}

\begin{figure}
\begin{center}
\begin{minipage}{220pt}
\includegraphics[height=125pt , trim={0mm 0mm 0mm 0mm}, clip]{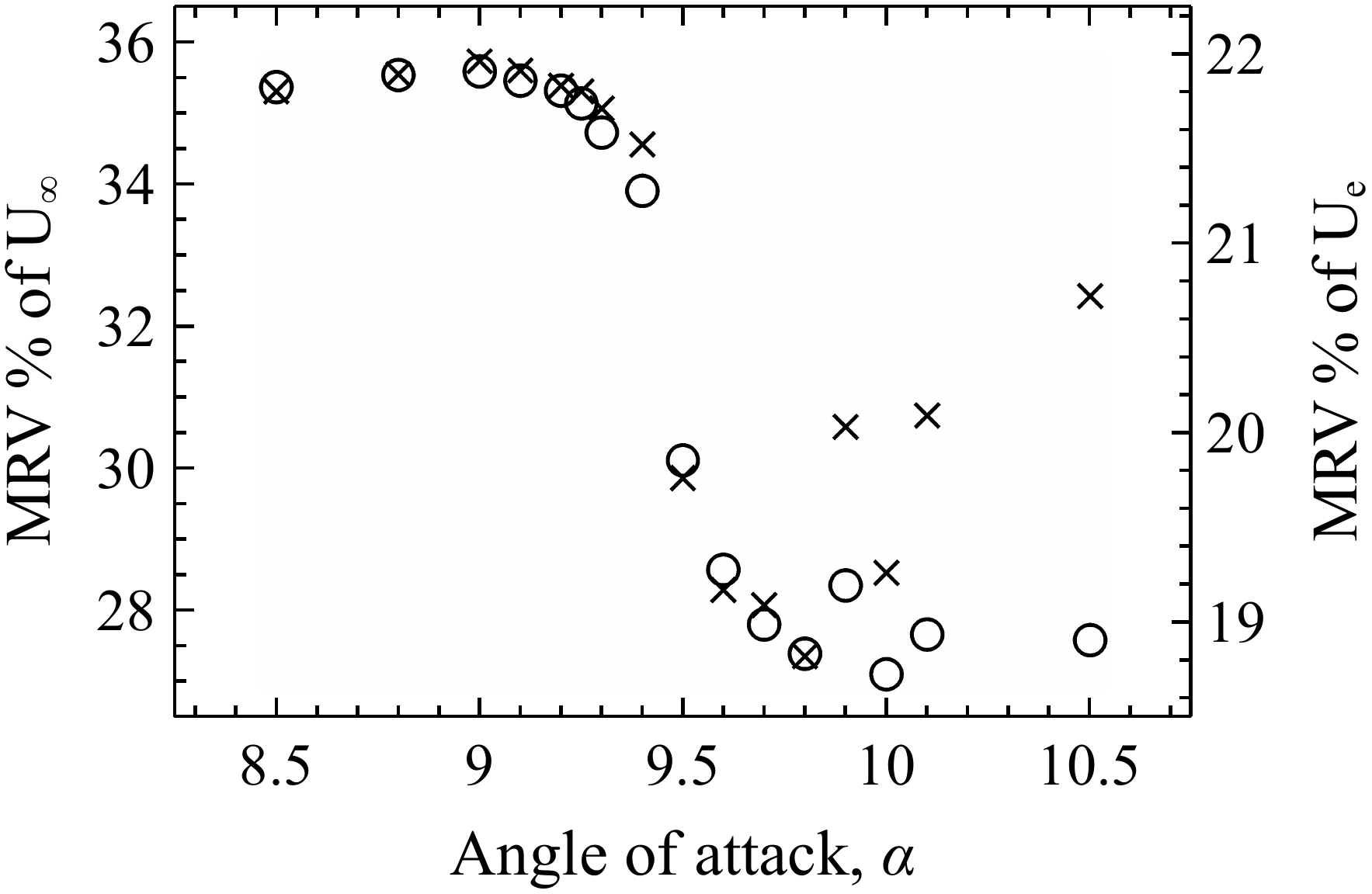}
\end{minipage}
\begin{minipage}{220pt}
\includegraphics[height=125pt , trim={0mm 0mm 0mm 0mm}, clip]{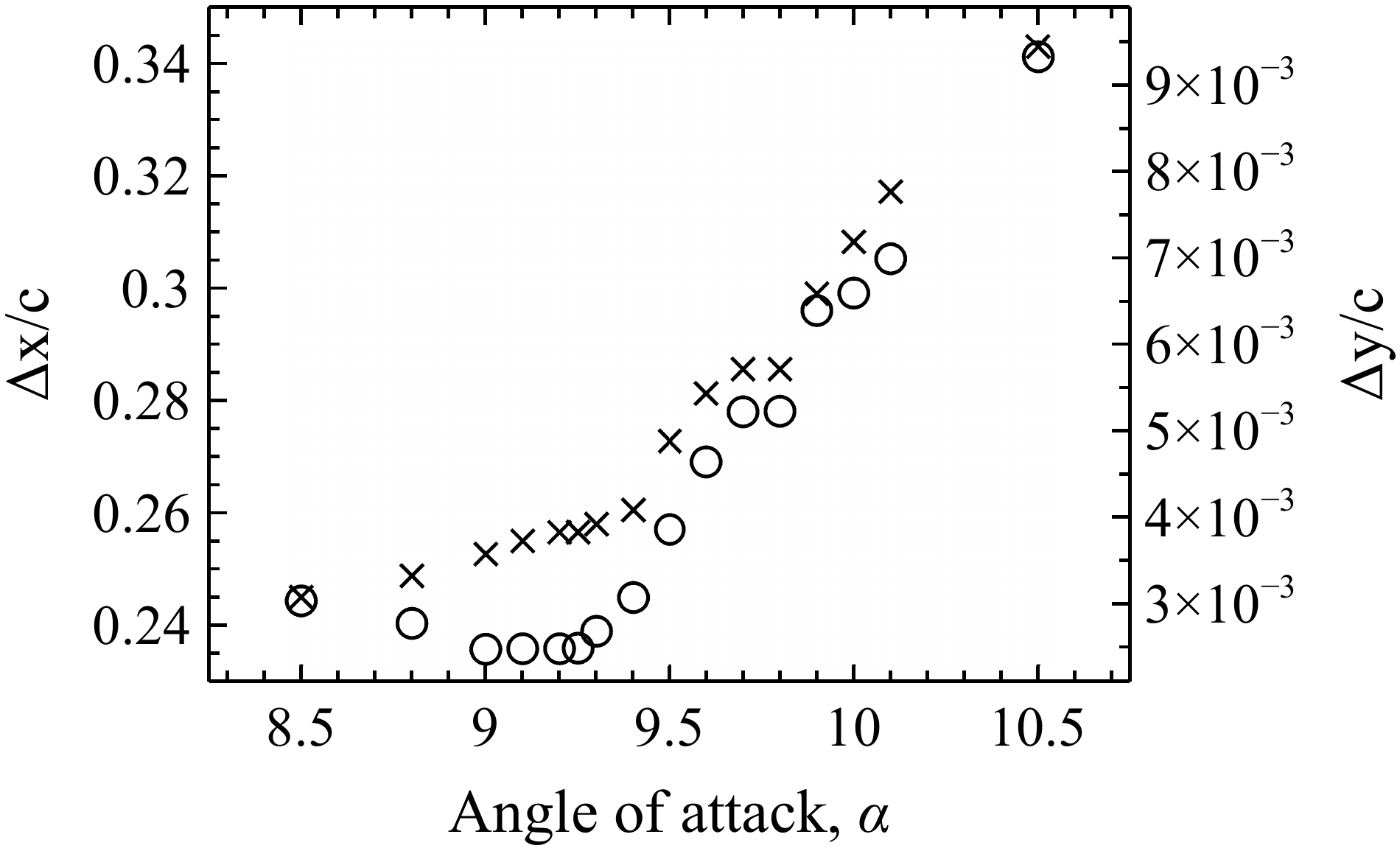}
\end{minipage}
\caption{Left: the maximum reverse velocity ($\mathrm{MRV}$) plotted versus the angle of attack $\alpha$. Circles: the $\mathrm{MRV}$ as a percentage of the free-stream velocity $U_\infty$; $\times$'s: the $\mathrm{MRV}$ as a percentage of the local free-stream velocity $U_e$. Right: the location of the $\mathrm{MRV}$ plotted versus the angle of attack $\alpha$. Circles: $\Delta x/c$ measured from the aerofoil leading-edge; $\times$'s: $\Delta y/c$ measured from the aerofoil surface.}
\label{MRV}
\end{center}
\end{figure}

\begin{figure}
\begin{center}
\begin{minipage}{220pt}
\centering
\includegraphics[width=220pt, trim={0mm 0mm 0mm 0mm}, clip]{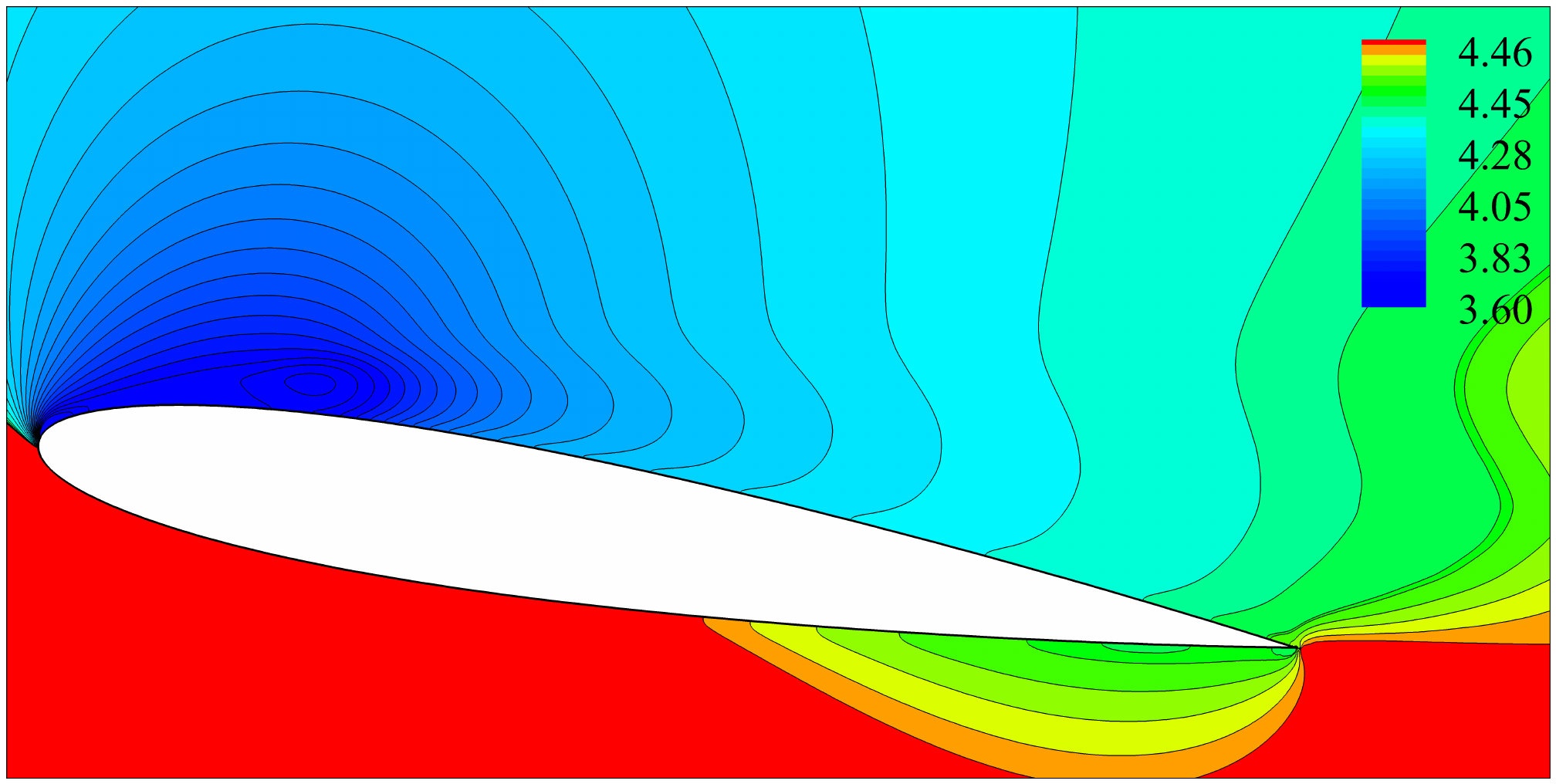}
\textit{$\alpha = 9.25^{\circ}$}
\end{minipage}
\begin{minipage}{220pt}
\centering
\includegraphics[width=220pt, trim={0mm 0mm 0mm 0mm}, clip]{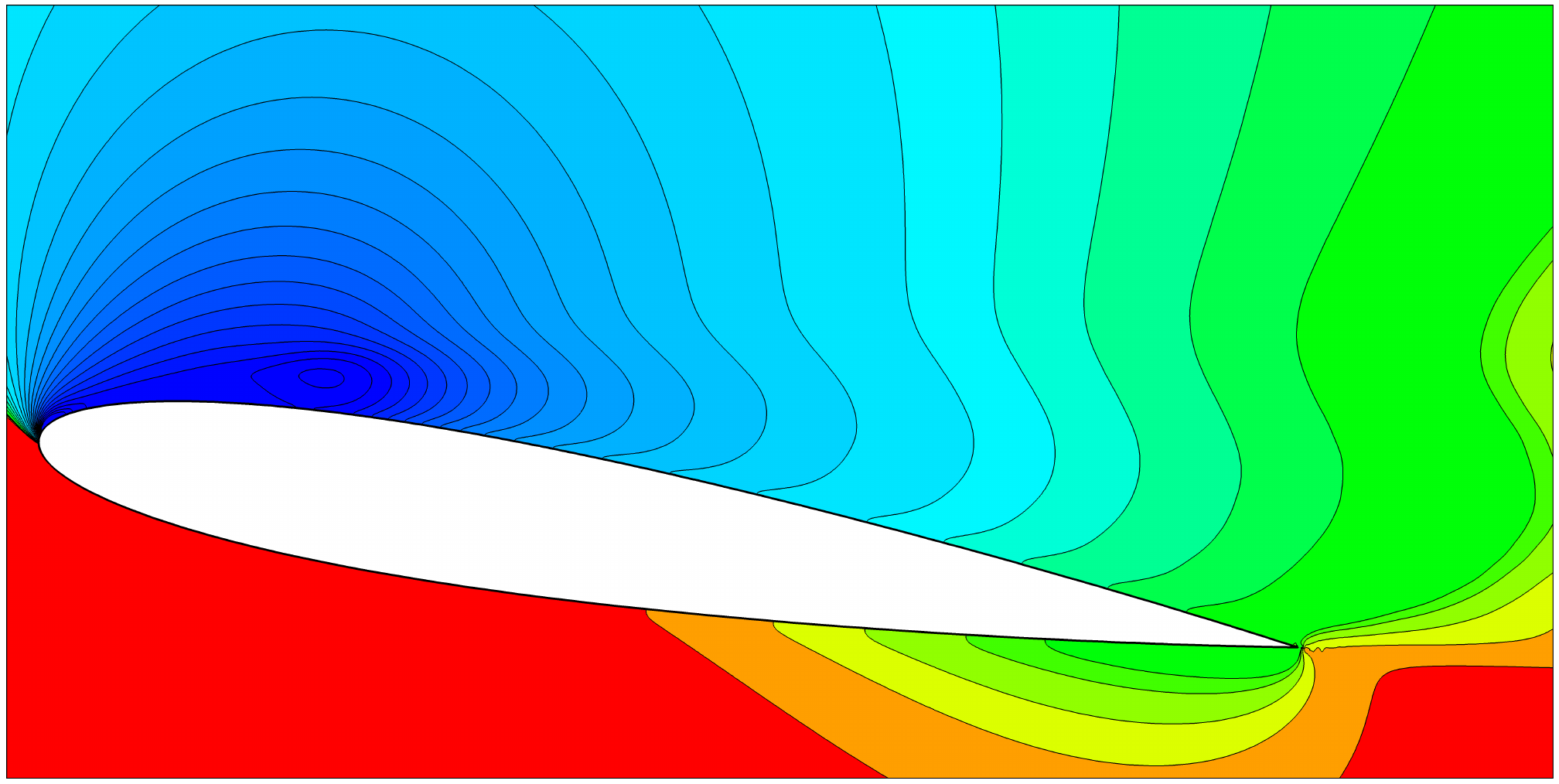}
\textit{$\alpha = 9.40^{\circ}$}
\end{minipage}
\begin{minipage}{220pt}
\centering
\includegraphics[width=220pt, trim={0mm 0mm 0mm 0mm}, clip]{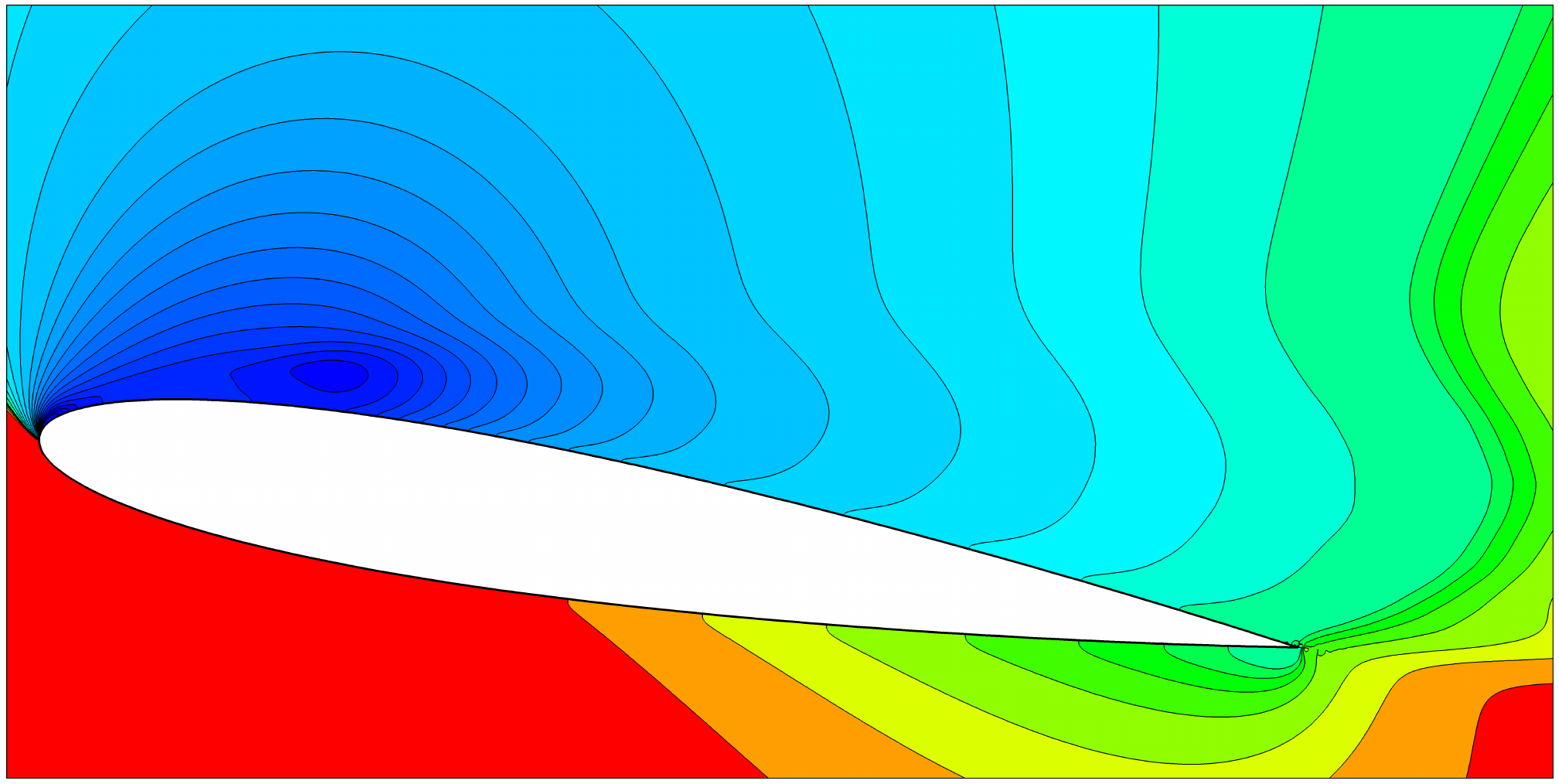}
\textit{$\alpha = 9.50^{\circ}$}
\end{minipage}
\begin{minipage}{220pt}
\centering
\includegraphics[width=220pt, trim={0mm 0mm 0mm 0mm}, clip]{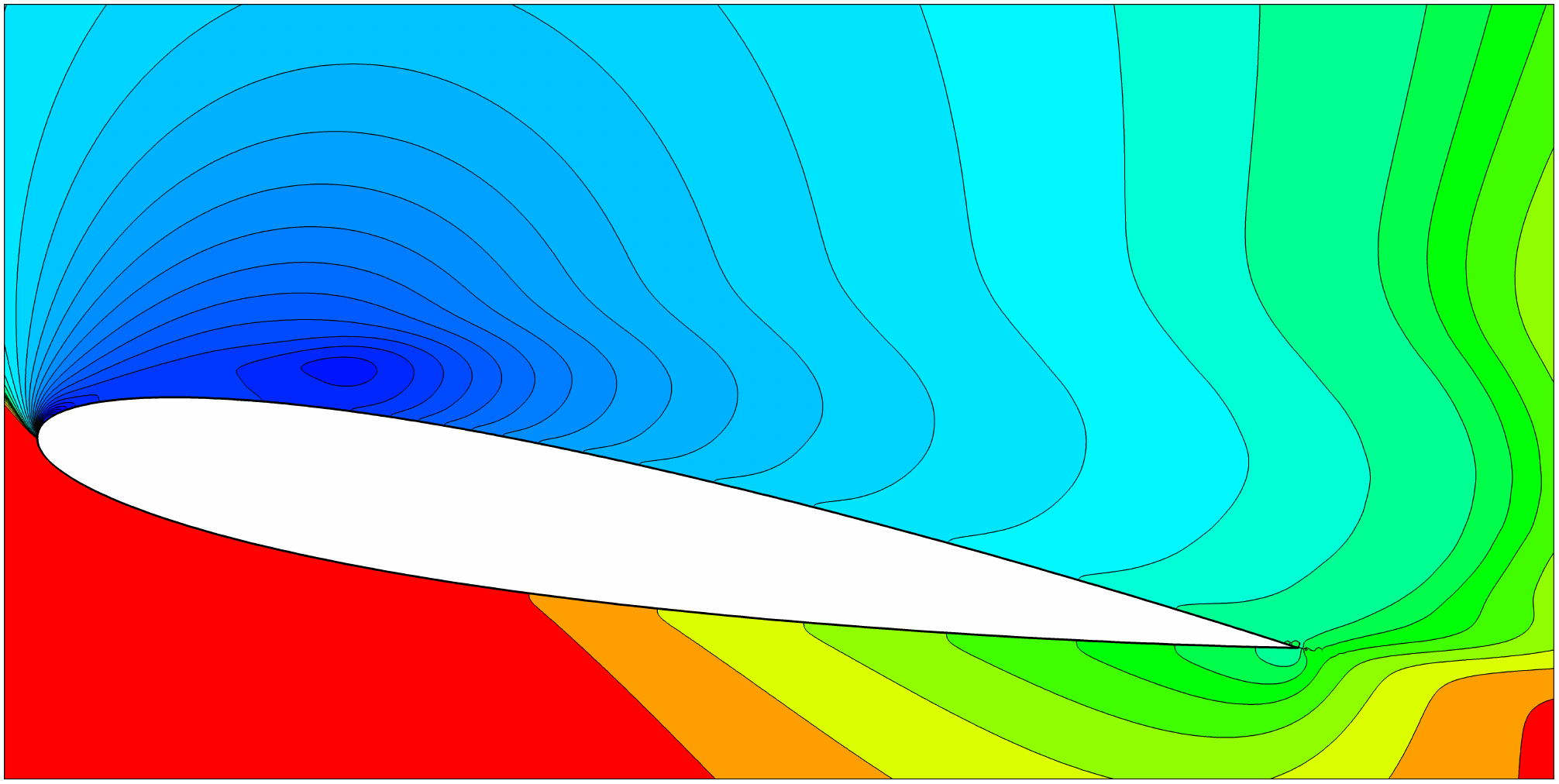}
\textit{$\alpha = 9.60^{\circ}$}
\end{minipage}
\begin{minipage}{220pt}
\centering
\includegraphics[width=220pt, trim={0mm 0mm 0mm 0mm}, clip]{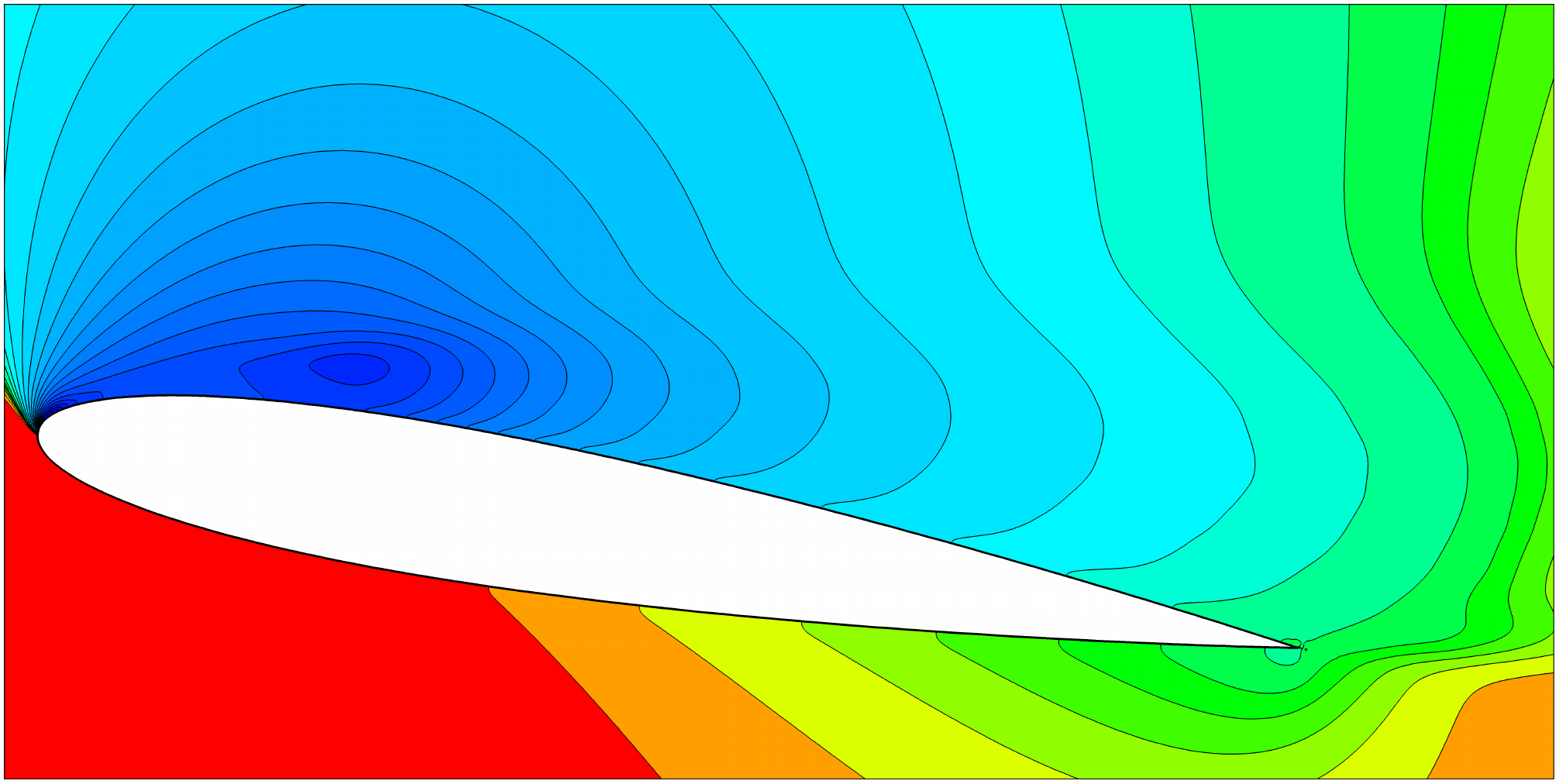}
\textit{$\alpha = 9.70^{\circ}$}
\end{minipage}
\begin{minipage}{220pt}
\centering
\includegraphics[width=220pt, trim={0mm 0mm 0mm 0mm}, clip]{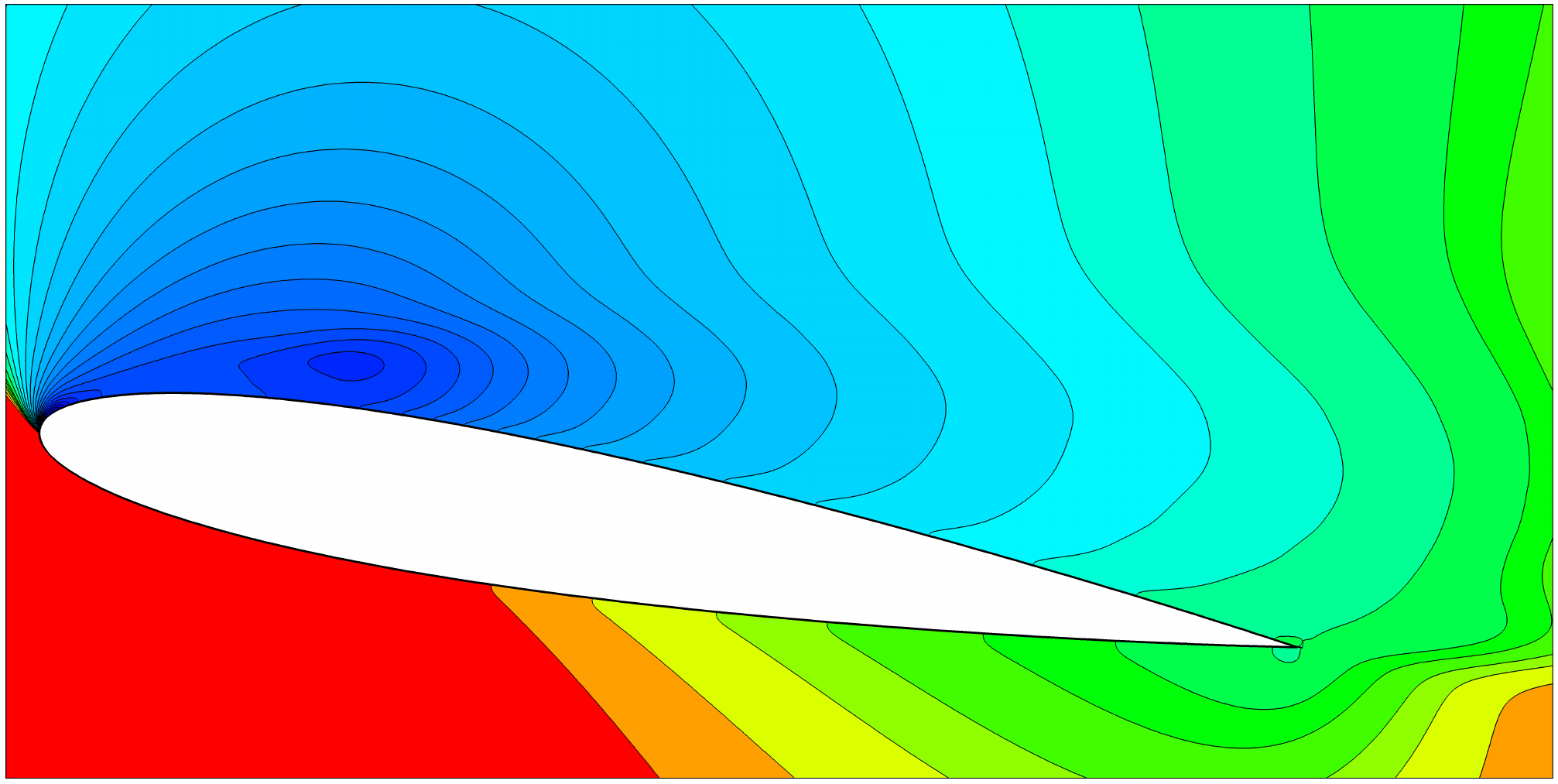}
\textit{$\alpha = 9.80^{\circ}$}
\end{minipage}
\begin{minipage}{220pt}
\centering
\includegraphics[width=220pt, trim={0mm 0mm 0mm 0mm}, clip]{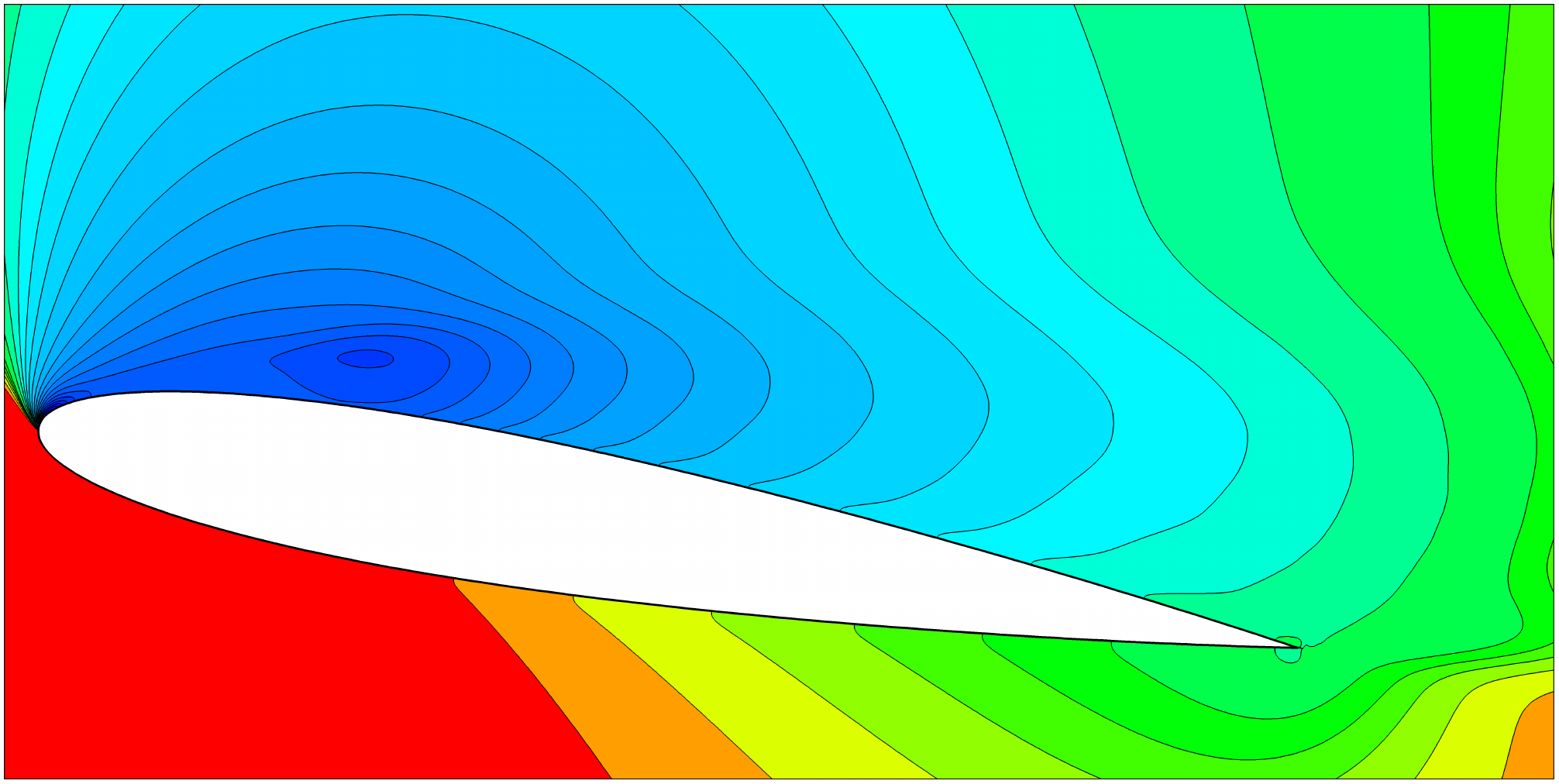}
\textit{$\alpha = 9.90^{\circ}$}
\end{minipage}
\begin{minipage}{220pt}
\centering
\includegraphics[width=220pt, trim={0mm 0mm 0mm 0mm}, clip]{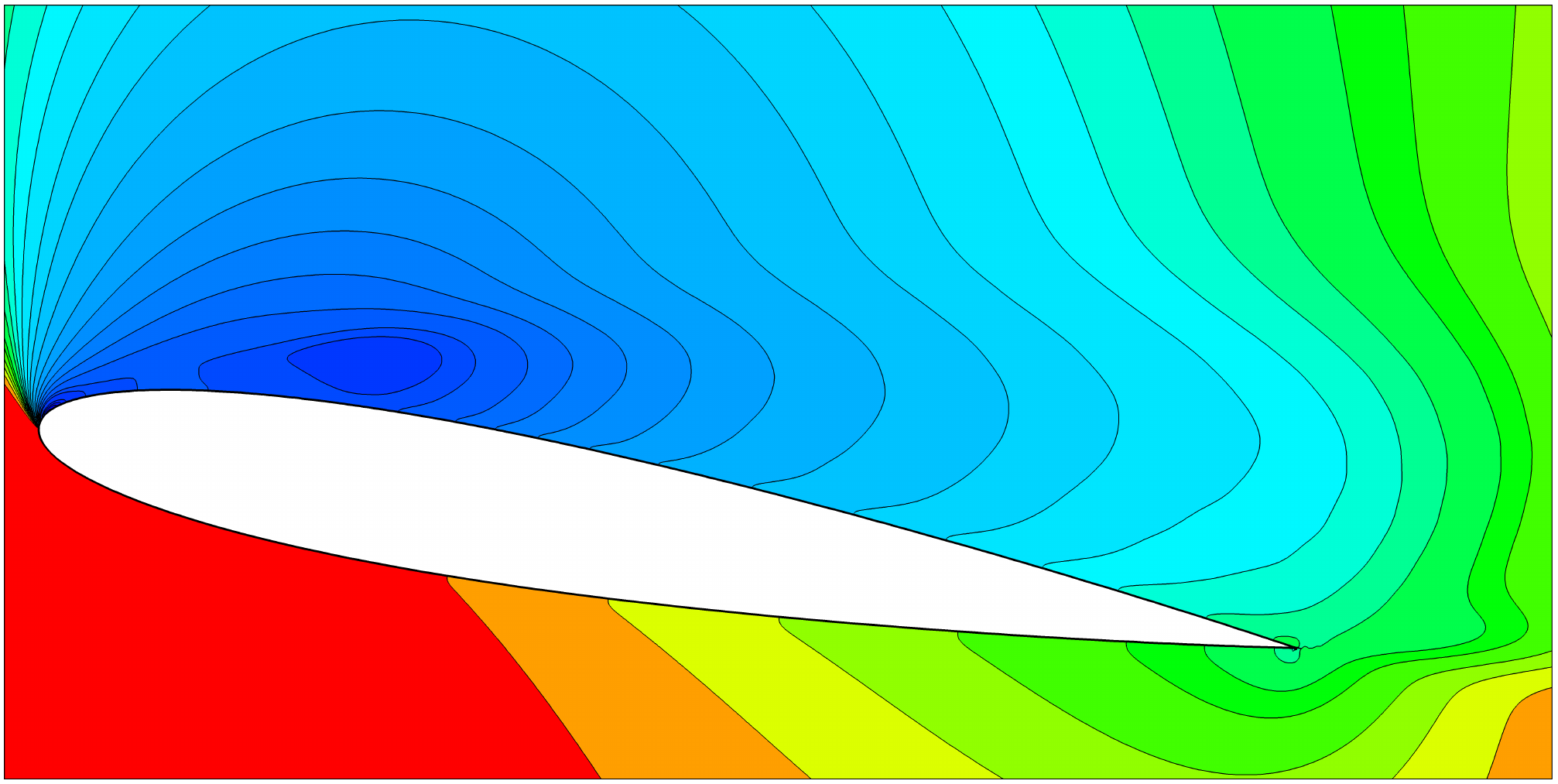}
\textit{$\alpha = 10.0^{\circ}$}
\end{minipage}
\begin{minipage}{220pt}
\centering
\includegraphics[width=220pt, trim={0mm 0mm 0mm 0mm}, clip]{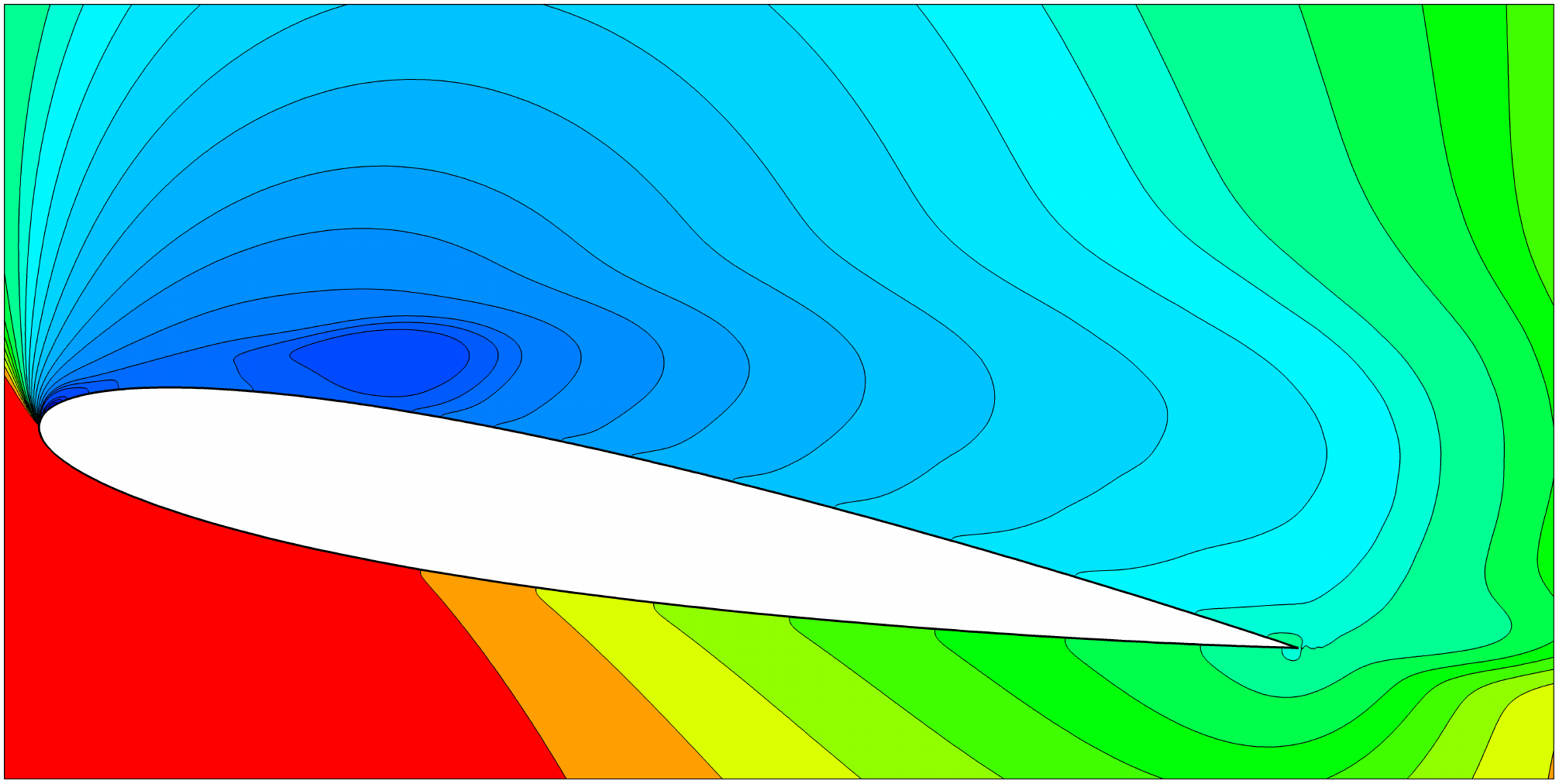}
\textit{$\alpha = 10.1^{\circ}$}
\end{minipage}
\begin{minipage}{220pt}
\centering
\includegraphics[width=220pt, trim={0mm 0mm 0mm 0mm}, clip]{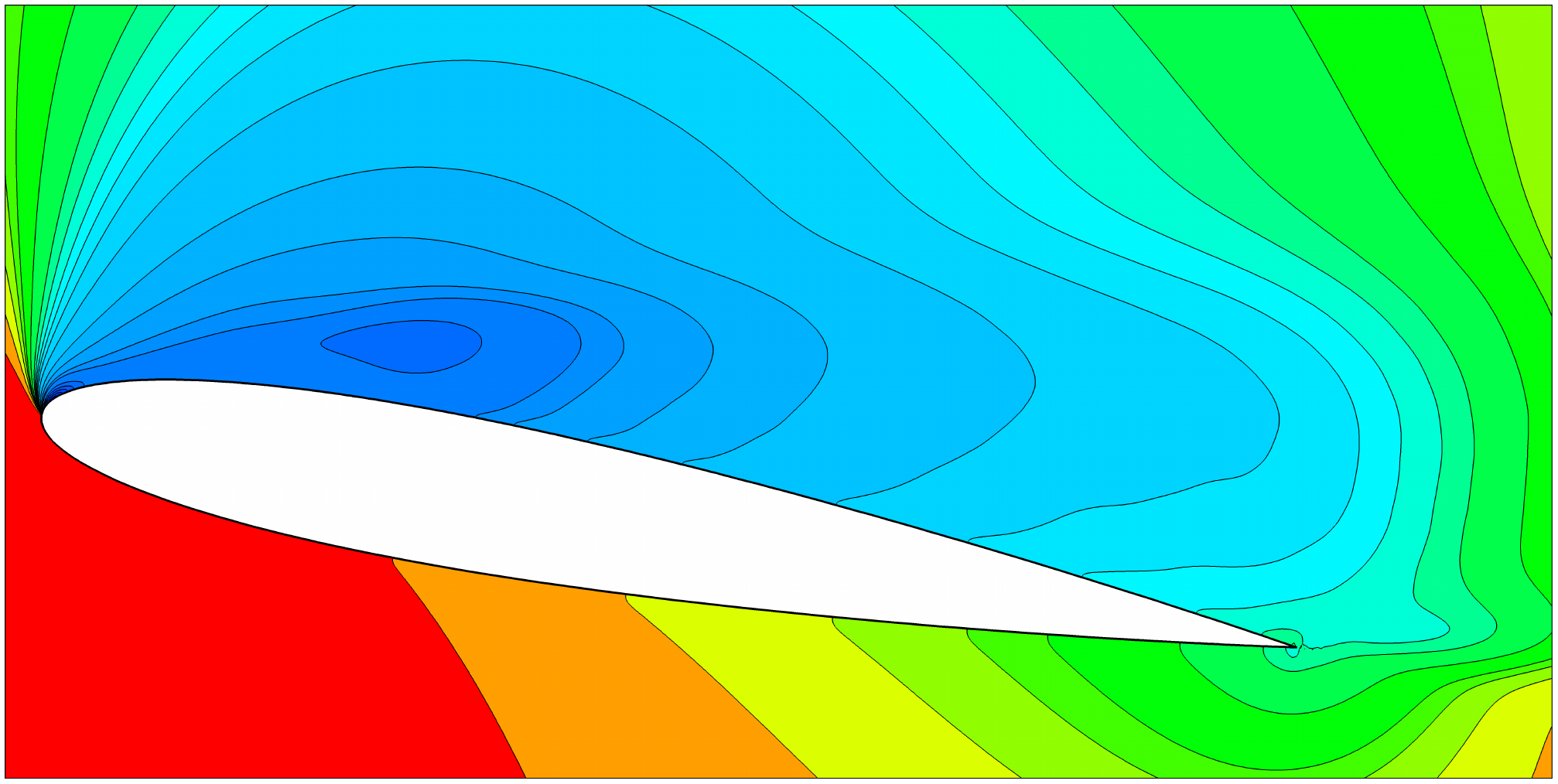}
\textit{$\alpha = 10.5^{\circ}$}
\end{minipage}
\caption{Contours plot of the mean pressure, $\overline{P}$, for the angles of attack $\alpha = 9.25^{\circ}$--$10.5^{\circ}$.}
\label{p_mean}
\end{center}
\end{figure}

The f\/irst observations and descriptions of the LSB were reported by~\citet{jones1934experimental} and~\citet{jones1934stalling}. After that, the structure of the LSB was investigated in the work of~\citet{young1966some} and its averaged shape was described by~\citet{horton1968laminar}. Since then, many experimental measurements were conducted on the suction surface of different aerofoils. In most of the previous works, it was observed that the length of the LSB decreases as the angle of attack is increased. At a critical angle of attack, the shortest bubble is achieved, and the LSB length increases when the angle of attack is increased above the critical angle.\newline
F\/igures~\ref{bubble_mean},~\ref{bubble_above}, and~\ref{bubble_below} show streamlines patterns superimposed on colour maps of the streamwise velocity component of the mean, high-lift, and low-lift f\/low-f\/ields, respectively, at the angles of attack $\alpha = 9.25^{\circ}$--$10.5^{\circ}$. As seen in the f\/igures, the LSB is formed on the upper surface of the aerofoil. The bubble length, height, and shape varies as the angle of attack is increased. It is noted that the f\/low is fully attached, in the mean sense, in the three f\/igures at the angle of attack of $9.25^{\circ}$. The f\/low is massively separated, open bubble, in the three f\/igures at the angle of attack of $10.5^{\circ}$. This is indicative that these two angles set the limits of the angles of attack of interest.\newline
F\/igure~\ref{bubble_below} shows that a trailing-edge separation region that constitutes a trailing-edge bubble (TEB) is formed at angles of attack $\alpha > 9.25^{\circ}$ and continues to grow as the angle of attack is increased. At the angle of attack of $9.7^{\circ}$ the leading-edge and the trailing-edge bubbles merge. The merged bubbles continue to deform until an open bubble is formed at the angle of attack of $10.5^{\circ}$. This type of stall is a combination of thin-aerofoil and trailing-edge stall as classif\/ied by~\citet{mccullough1951examples}. This process occurs at the same sequence of events but at higher angles of attack for the high-lift mean, f\/igure~\ref{bubble_above}, and at lower angles of attack for the mean, f\/igure~\ref{bubble_mean}.\newline
\citet{broeren1999flowfield} studied the f\/low-f\/ield around an LRN$(1)$--$1007$ aerofoil at $\rec = 3\times10^5$ and an angle of attack $\alpha=15^{\circ}$. Two velocity components were measured at $687$ locations in the $x$--$y$ plane at the aerofoil mid-span using laser Doppler velocimeter (LDV). A Conditional-averaging method was used and $24$ time slots within one cycle resolved the oscillation in $15^{\circ}$ intervals. They showed four snapshots of contours of the streamwise velocity at four different phases. The snapshots show small leading-edge and trailing-edge separated zones, they grow in the second snapshot, merge in the third one, and become an open bubble in the fourth snapshot. Their experimental observations are in good agreement with the present discussion inequities of how limited and coarse their data is.\newline
It is interesting to note that the secondary bubble exists at the same location when the f\/low is attached and when it is massively separated. This is indicative that the secondary bubble plays a profound role in the underlying mechanism behind the self-sustained quasi-periodic low-frequency switching between attached and separated f\/low. Furthermore, the current observation being statistical implies that the secondary bubble and the conditions for its formation are permanent in space and stationary in time. To the best of the authors' knowledge, this has never been observed in any of the previous work.\newline
F\/igure~\ref{bubble_size} illustrates the mean length and the mean height of the LSB and the TEB plotted versus the angle of attack. The mean length and height of the bubbles were obtained by approximating the locations of separation and reattachment for each bubble from the streamlines patterns. At relatively low angles of attack, before the merging of the two bubbles, the LSB and TEB are bounded by four half-saddle points. The f\/irst half-saddle point is the separation point at the vicinity of the leading-edge (\textbf{SP1}). The second half-saddle point is the reattachment point located just downstream the LSB (\textbf{SP2}). The third half-saddle is the separation point near the trailing-edge (\textbf{SP3}). The fourth half-saddle point is the reattachment point located at the trailing-edge (\textbf{SP4}). The f\/irst two half-saddle points constitute the LSB, and the latter two half-saddle points constitute the TEB. At relatively high angles of attack, when the two bubbles start to merge, the two half-saddle points \textbf{SP2} and \textbf{SP3} move away from the wall and form a full-saddle point (\textbf{SP}). The LSB and TEB are bounded by two half-saddle points and the newly formed full-saddle point: \textbf{SP1}, \textbf{SP}, and \textbf{SP4}. The length of the LSB is measured from \textbf{SP1} to \textbf{SP} and the length of the TEB is measured from \textbf{SP} to \textbf{SP4}. The size of the mean bubbles grows with the angle of attack. When the LFO become fully developed, the length of the LSB starts to decrease as the angle of attack is increased and the summation of the LSB and TEB length spans the whole aerofoil chord. The height of the bubbles is measured across each bubble from the aerofoil surface, passing through the bubble focus, and to the edge of the bubble. As shown in f\/igure~\ref{bubble_size} (right), the height of the bubbles increases with the angle of attack.\newline
The separated shear layer is highly unstable, but a distinction needs to be drawn between convective instability, where disturbances grow in space, and absolute instability where disturbances grow in time. Most of the work pegs the stability of the bubble to changes in the separated shear layer and its associated Kelvin--Helmholtz instability. Analysis performed by~\citet{niew1993stability} for backward facing step f\/lows suggests absolute instability for f\/low with more than $20$\% reverse f\/low. Whereas,~\citet{hammond1998local} predicted $30$\% for the start of absolute instability in LSBs. F\/igure~\ref{MRV} shows the variation of the maximum reverse velocity ($\mathrm{MRV}$) inside the bubble as a percentage of the free-stream velocity, denoted by circles, and as a percentage of the maximum external velocity outside the bubble, denoted by $\times$'s. The $\mathrm{MRV}$ reach a maximum of $22\%$ of the maximum external velocity outside the bubble and a maximum of $35.6\%$ of the free-stream velocity at the angle of attack of $9.0^{\circ}$. The location of the $\mathrm{MRV}$ is shown on the right-hand side of f\/igure~\ref{MRV}. The circles denote the location of $\mathrm{MRV}$ in $x$ direction measured from the aerofoil leading-edge and the $\times$'s denote the location of the $\mathrm{MRV}$ in the $y$ direction measured from the aerofoil surface. At the angles of attack of $9.0^{\circ} \leq \alpha \leq 9.25^{\circ}$ the location of the $\mathrm{MRV}$ was about $0.236$ chords downstream the leading-edge. At higher angles of attack, the location of the $\mathrm{MRV}$ moved downstream. The location of the $\mathrm{MRV}$ in $y$ direction moves away from the aerofoil surface when the angle of attack is increased. The study by~\citet{alam2000direct} showed that the separation bubble is considered absolutely unstable if the separation bubble sustains a maximum reverse velocity percentage in the range $\mathrm{MRV} = 15\%$ to $20\%$. The $\mathrm{MRV}$ for the current simulation, $22\%$, is larger than the range pointed out by~\citet{alam2000direct} and can be considered as an indication of the presence of an absolute instability in the bubble. However, it is not clear yet whether the threshold of the $\mathrm{MRV}$ is an indication of the inception of the absolute instability or it is a consequence of the absolute instability for the current conf\/iguration. That is, is this threshold a necessary or a suff\/icient condition for the absolute instability. All previous observations, including the present study, reported the threshold as a necessary condition but it was never established if it is a suff\/icient condition. Therefore, it is concluded that the threshold of the $\mathrm{MRV}$ is an observation accompanying the LSB being absolutely unstable, but it is not the root cause that triggers the instability.\newline
Contours plots of the mean pressure distributions around the aerofoil ($\overline{p}$) is illustrated in f\/igure~\ref{p_mean}. The maximum contour level is set equal to the far-f\/ield pressure which is equal to $4.46$ for the current conf\/iguration. It is noted that---on the suction surface of the aerofoil---the high-lift pressure in the vicinity of the LSB is lower than the mean pressure at the same location. Thus, the presence of the bubble induces an APG along the aerofoil chord when the f\/low is attached. It is also noted that, the low-lift pressure is higher than the mean pressure at the region of the LSB. Hence, the pressure gradient is favourable along the aerofoil chord when the f\/low is separated. This changes what is continuously reported in the literature that the bubble bursts because of the APG and the f\/low reattach because the early transition adds more energy to the boundary layer helping it overcome the APG and avoids new separation. In the light of the new observations, it seems that there is a natural forcing mechanism, instability, that forces the f\/low to remain attached against the APG and forces it to separate despite the FPG. Such a natural forcing mechanism seems to add momentum to the boundary layer during the attached f\/low phase and subtract momentum from the boundary layer during the separated f\/low phase.\newline
At zero angle of attack, a reversed f\/low region exists on both sides of the aerofoil in the vicinity of the trailing-edge. These are consequences of the von Karman alternating vortices that shed at the wake of the aerofoil. For symmetric aerofoils at zero angle of attack, the von Karman vortices are mirror symmetric. However, as the angle of attack increases the upper vortex becomes larger and stronger than the lower one. At a critical angle of attack, the lower reversed f\/low region and the APG disappear and is replaced by an FPG. As seen in f\/igure~\ref{p_mean}, at the angle of attack of $9.25^{\circ}$ there is a small APG region on the pressure surface of the aerofoil. At higher angles of attack, this region is replaced by an FPG. The disappearance of the APG region on the pressure surface of the aerofoil might affects the dynamics of the f\/low and alters the oscillating f\/low mode at the trailing-edge. However, further investigation on the dynamics of the f\/low is required before such a conclusion can be drawn.

\subsection{Turbulent kinetic energy}\label{sec:tke}
\begin{figure}
\begin{center}
\begin{minipage}{220pt}
\centering
\includegraphics[width=220pt, trim={0mm 0mm 0mm 0mm}, clip]{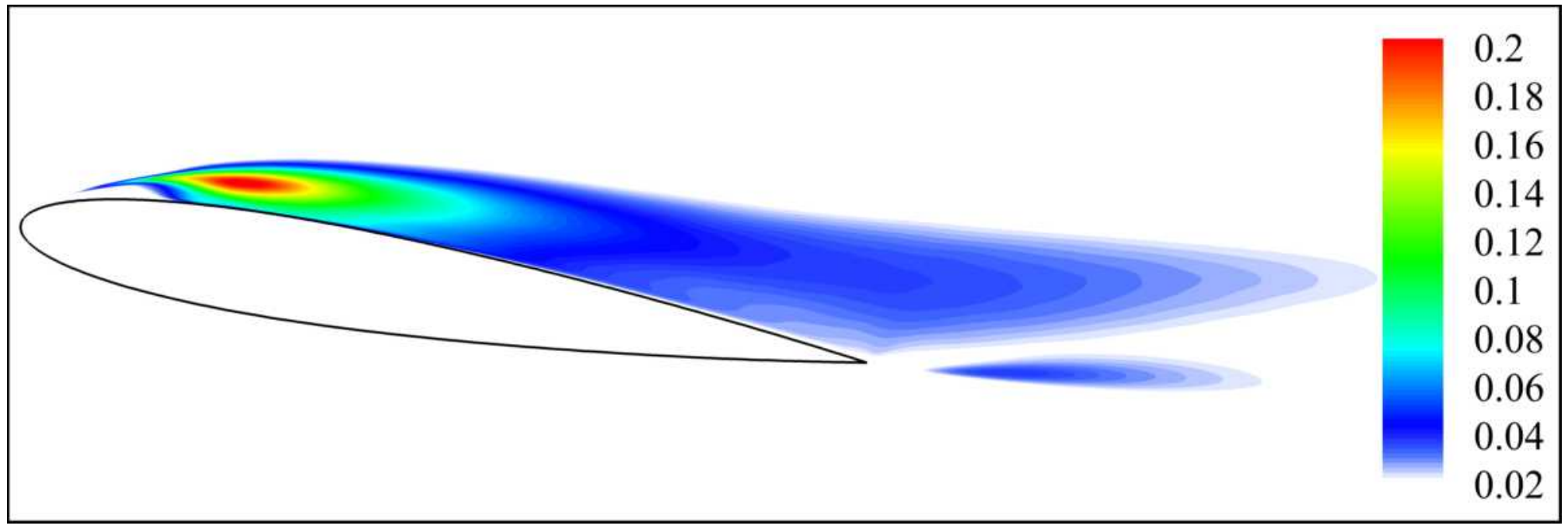}
\textit{$\alpha = 9.25^{\circ}$}
\end{minipage}
\medskip
\begin{minipage}{220pt}
\centering
\includegraphics[width=220pt, trim={0mm 0mm 0mm 0mm}, clip]{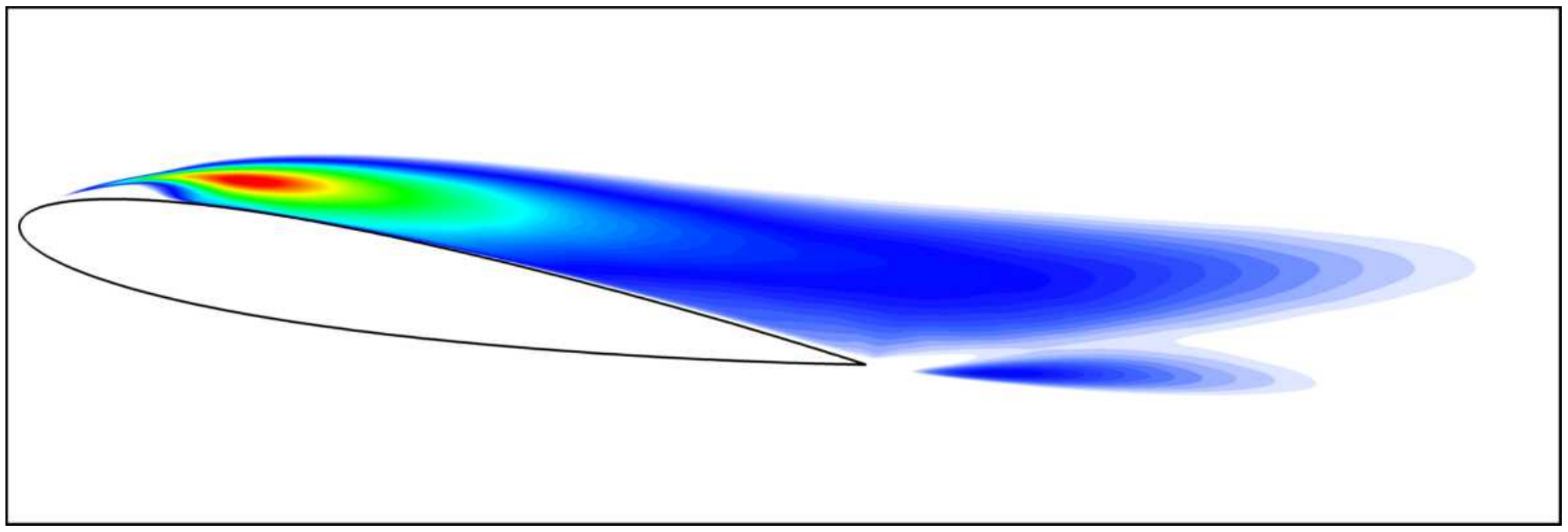}
\textit{$\alpha = 9.40^{\circ}$}
\end{minipage}
\medskip
\begin{minipage}{220pt}
\centering
\includegraphics[width=220pt, trim={0mm 0mm 0mm 0mm}, clip]{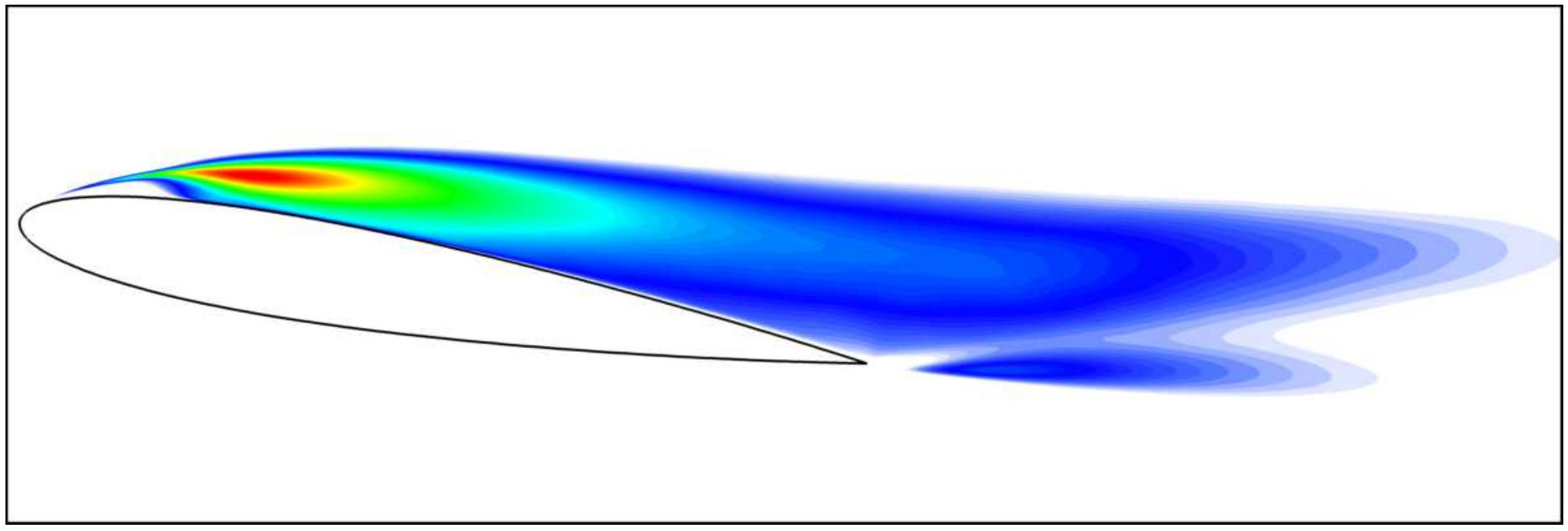}
\textit{$\alpha = 9.50^{\circ}$}
\end{minipage}
\medskip
\begin{minipage}{220pt}
\centering
\includegraphics[width=220pt, trim={0mm 0mm 0mm 0mm}, clip]{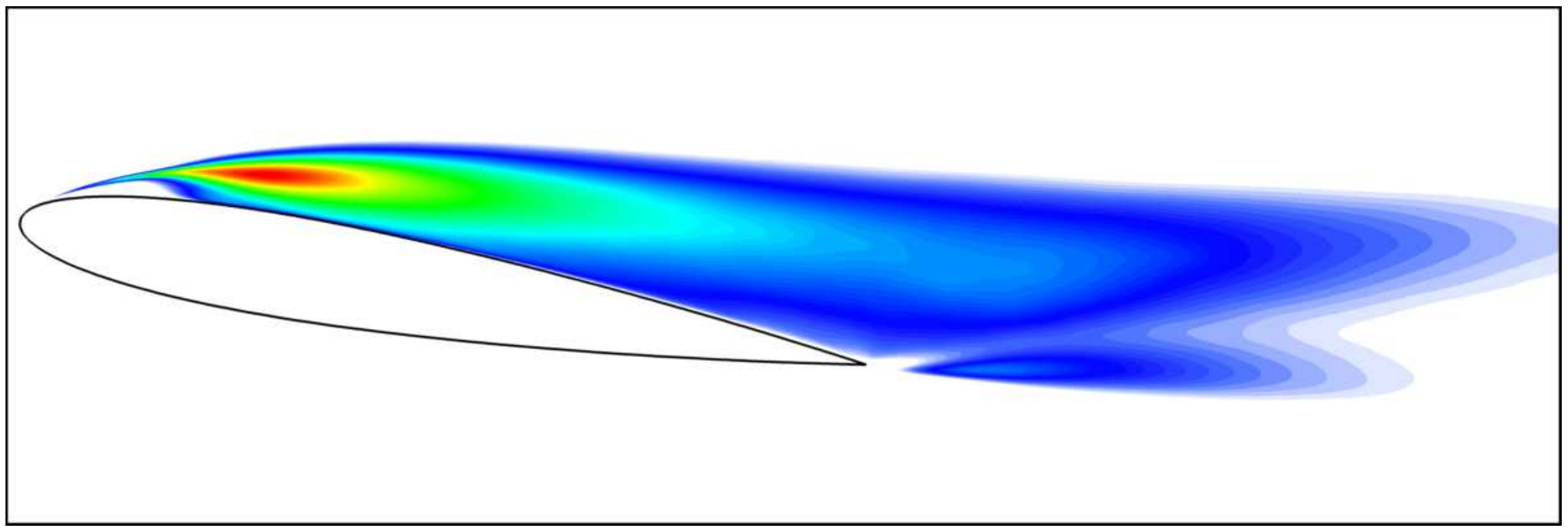}
\textit{$\alpha = 9.60^{\circ}$}
\end{minipage}
\medskip
\begin{minipage}{220pt}
\centering
\includegraphics[width=220pt, trim={0mm 0mm 0mm 0mm}, clip]{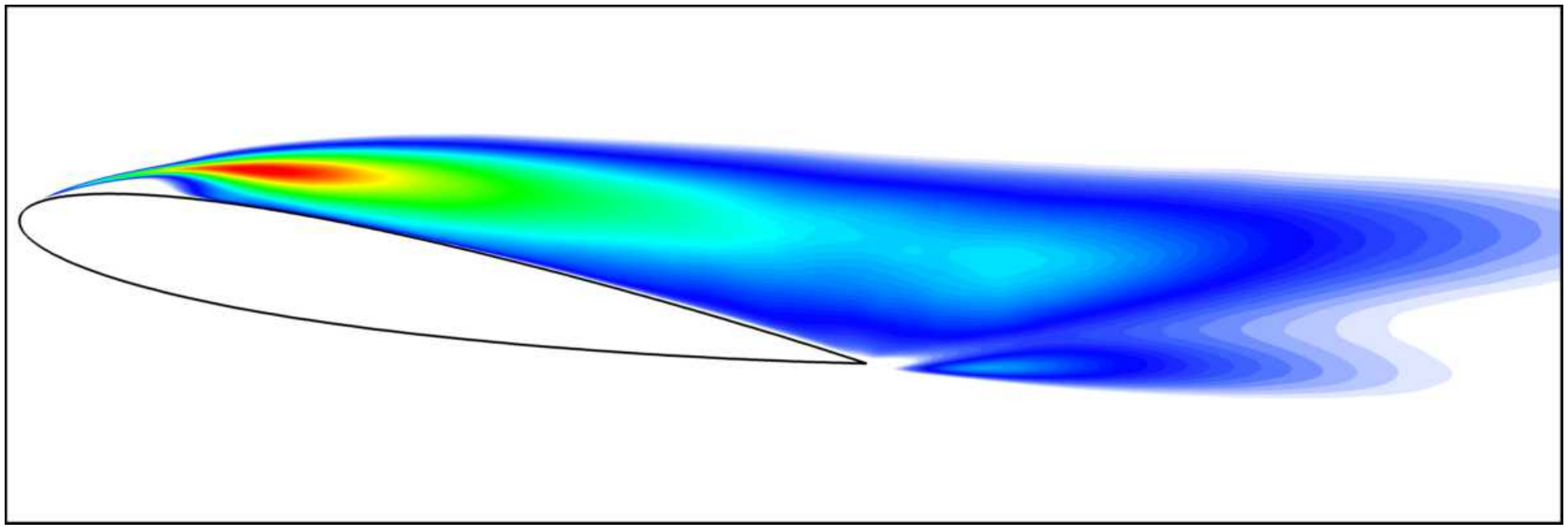}
\textit{$\alpha = 9.70^{\circ}$}
\end{minipage}
\medskip
\begin{minipage}{220pt}
\centering
\includegraphics[width=220pt, trim={0mm 0mm 0mm 0mm}, clip]{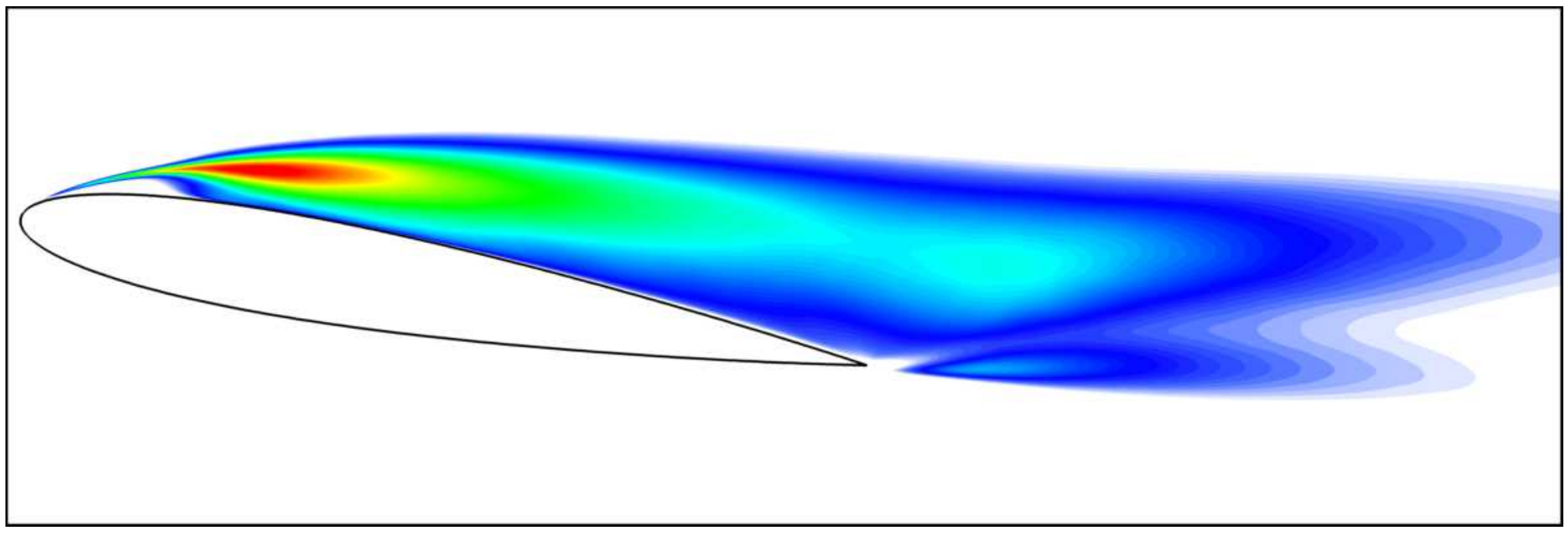}
\textit{$\alpha = 9.80^{\circ}$}
\end{minipage}
\medskip
\begin{minipage}{220pt}
\centering
\includegraphics[width=220pt, trim={0mm 0mm 0mm 0mm}, clip]{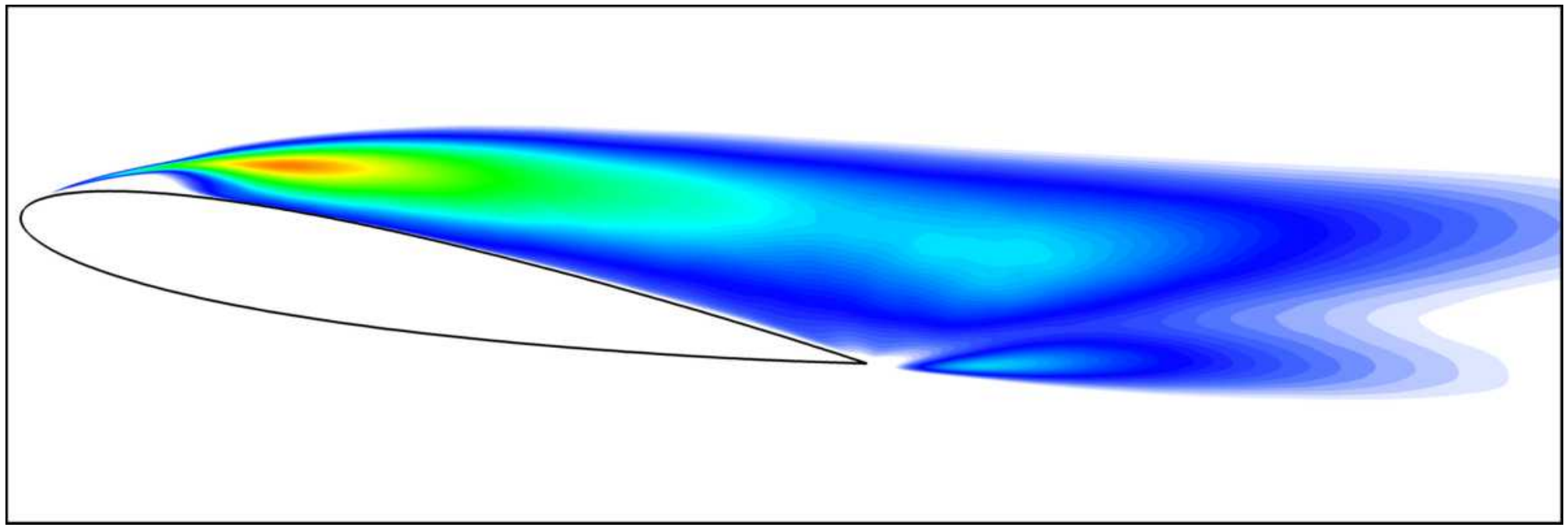}
\textit{$\alpha = 9.90^{\circ}$}
\end{minipage}
\medskip
\begin{minipage}{220pt}
\centering
\includegraphics[width=220pt, trim={0mm 0mm 0mm 0mm}, clip]{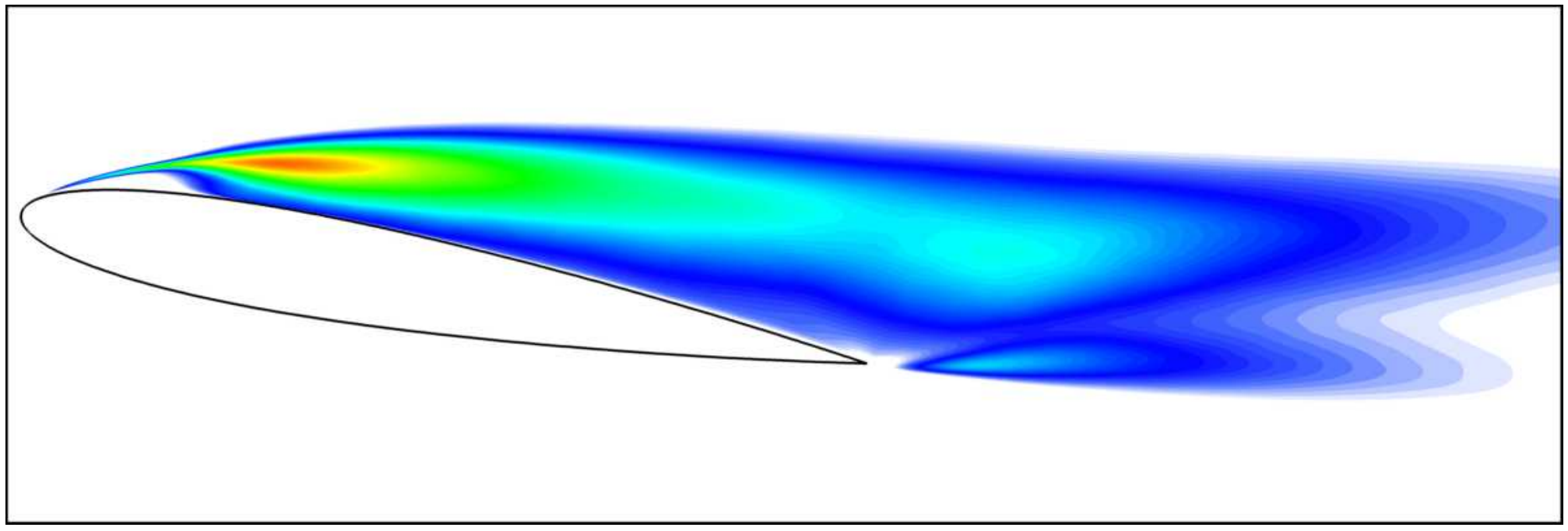}
\textit{$\alpha = 10.0^{\circ}$}
\end{minipage}
\medskip
\begin{minipage}{220pt}
\centering
\includegraphics[width=220pt, trim={0mm 0mm 0mm 0mm}, clip]{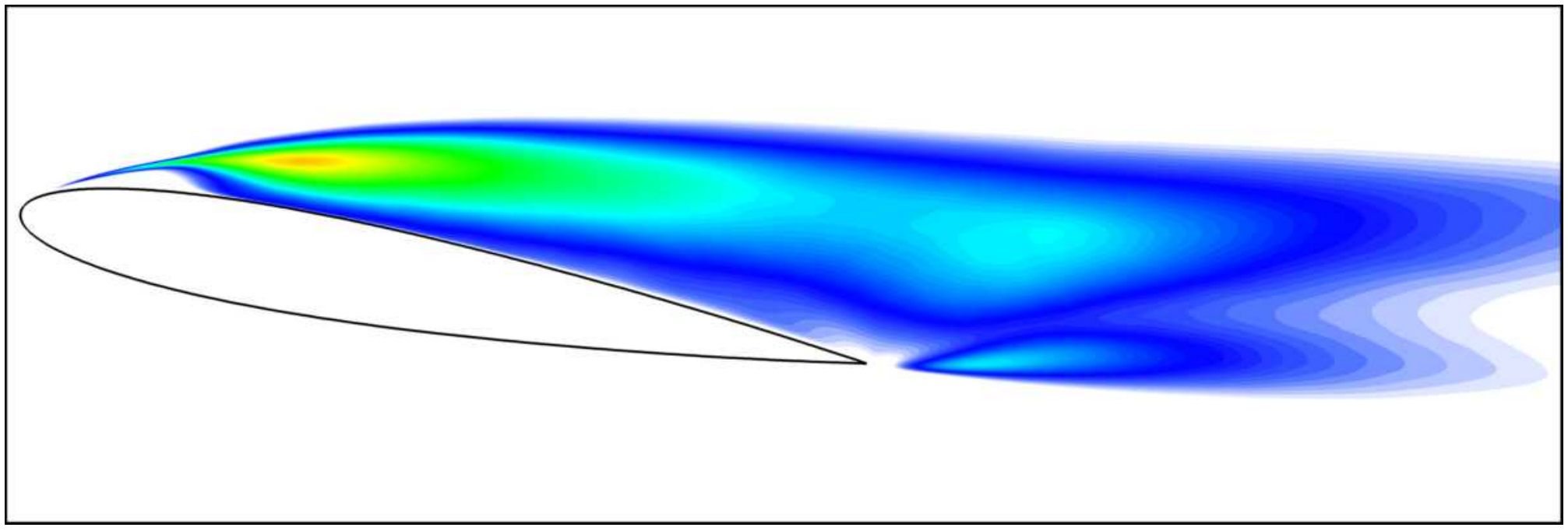}
\textit{$\alpha = 10.1^{\circ}$}
\end{minipage}
\begin{minipage}{220pt}
\centering
\includegraphics[width=220pt, trim={0mm 0mm 0mm 0mm}, clip]{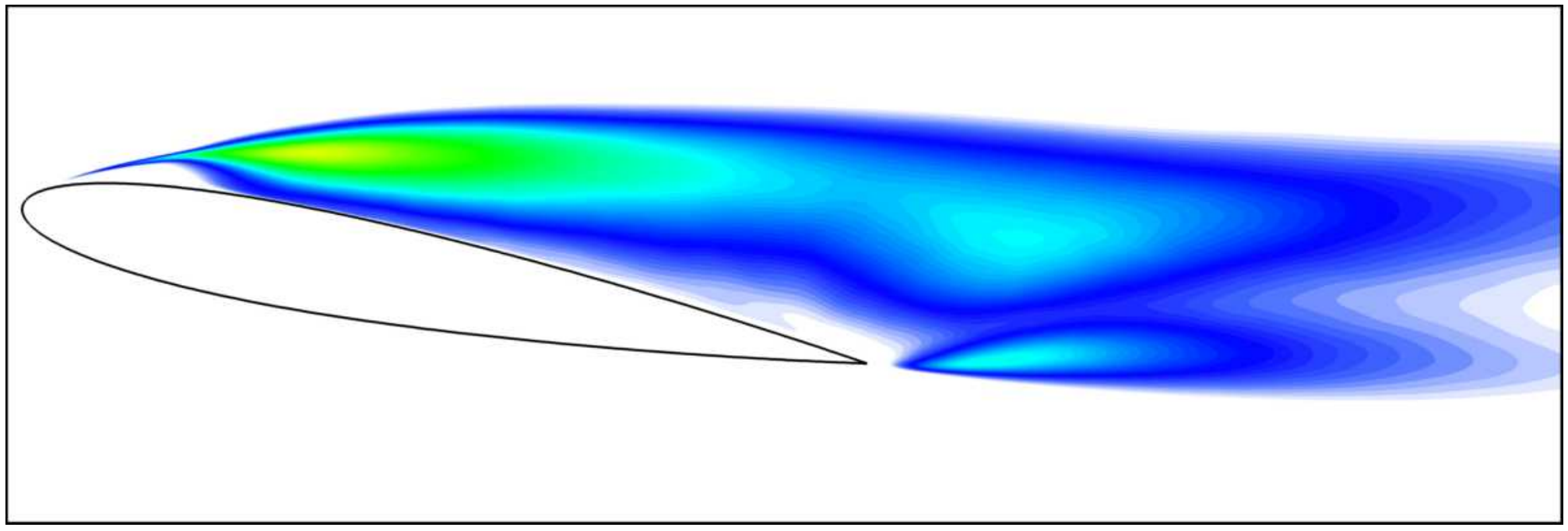}
\textit{$\alpha = 10.5^{\circ}$}
\end{minipage}
\caption{Colours map of the variance of the streamwise velocity component, $\overline{{u\mydprime}^2}$, for the angles of attack $\alpha = 9.25^{\circ}$--$10.5^{\circ}$.}
\label{u2_mean}
\end{center}
\end{figure}

\begin{figure}
\begin{center}
\begin{minipage}{220pt}
\centering
\includegraphics[width=220pt, trim={0mm 0mm 0mm 0mm}, clip]{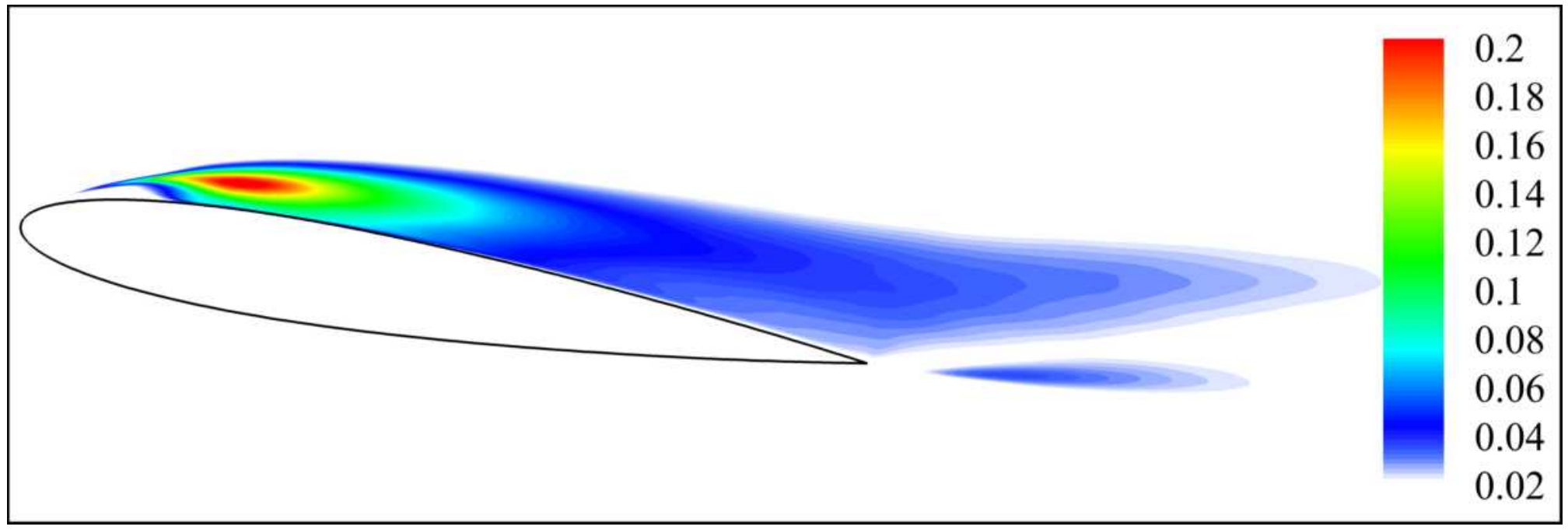}
\textit{$\alpha = 9.25^{\circ}$}
\end{minipage}
\medskip
\begin{minipage}{220pt}
\centering
\includegraphics[width=220pt, trim={0mm 0mm 0mm 0mm}, clip]{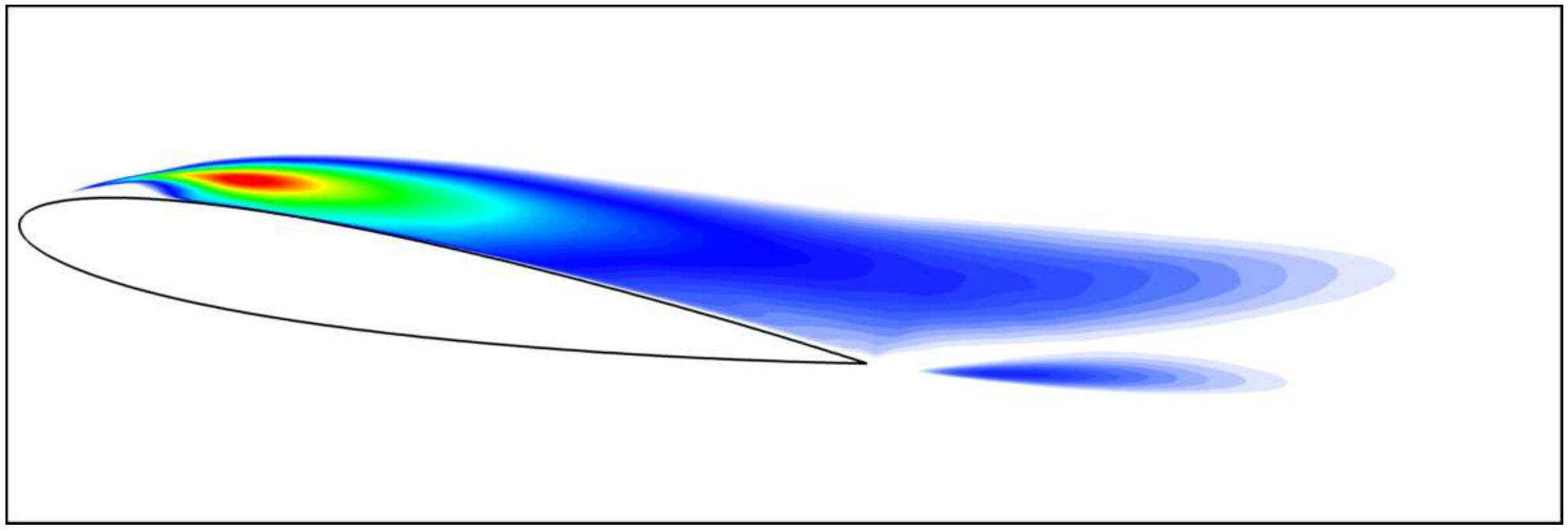}
\textit{$\alpha = 9.40^{\circ}$}
\end{minipage}
\medskip
\begin{minipage}{220pt}
\centering
\includegraphics[width=220pt, trim={0mm 0mm 0mm 0mm}, clip]{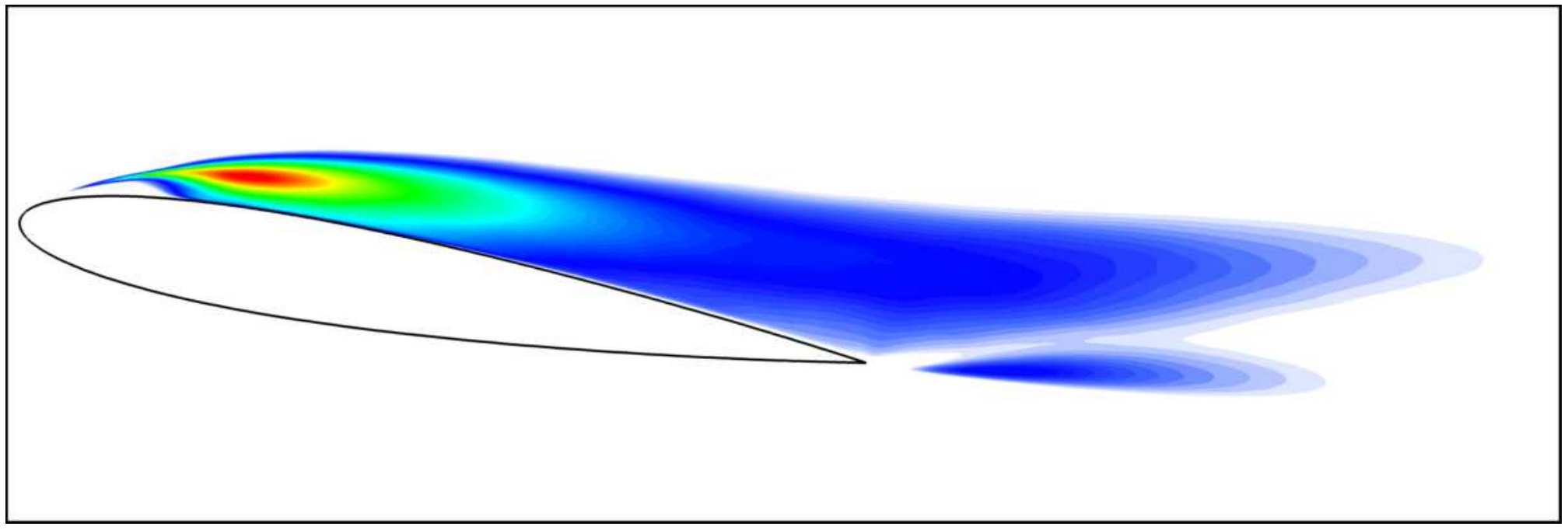}
\textit{$\alpha = 9.50^{\circ}$}
\end{minipage}
\medskip
\begin{minipage}{220pt}
\centering
\includegraphics[width=220pt, trim={0mm 0mm 0mm 0mm}, clip]{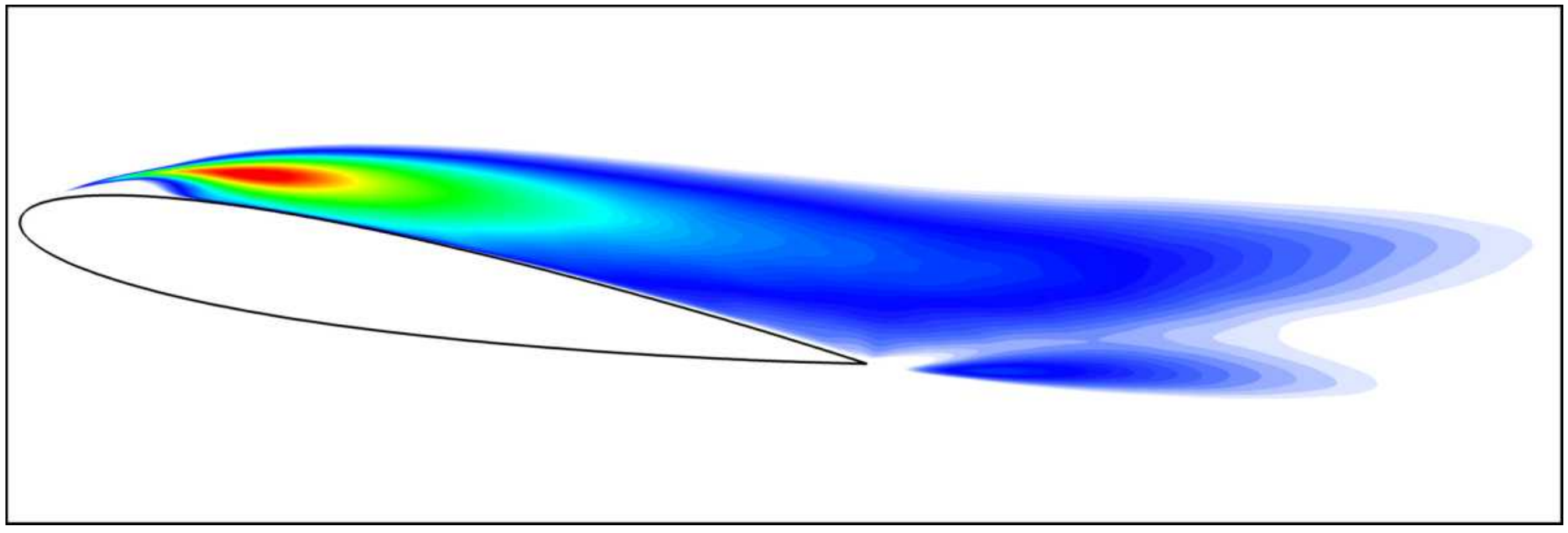}
\textit{$\alpha = 9.60^{\circ}$}
\end{minipage}
\medskip
\begin{minipage}{220pt}
\centering
\includegraphics[width=220pt, trim={0mm 0mm 0mm 0mm}, clip]{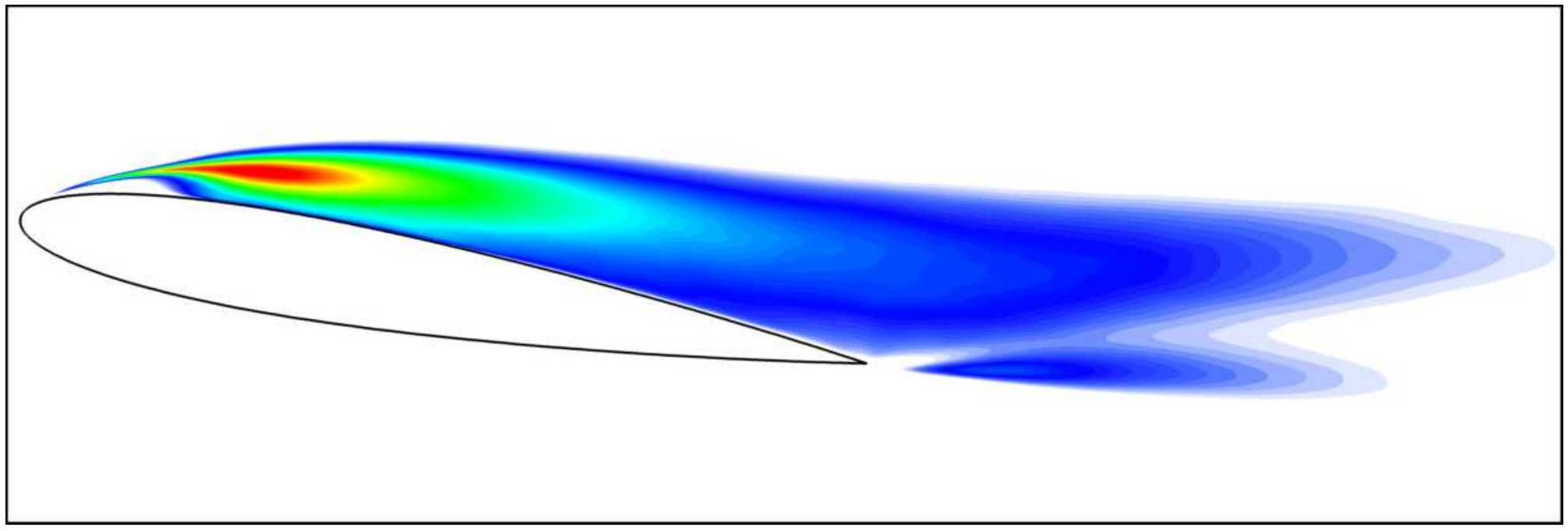}
\textit{$\alpha = 9.70^{\circ}$}
\end{minipage}
\medskip
\begin{minipage}{220pt}
\centering
\includegraphics[width=220pt, trim={0mm 0mm 0mm 0mm}, clip]{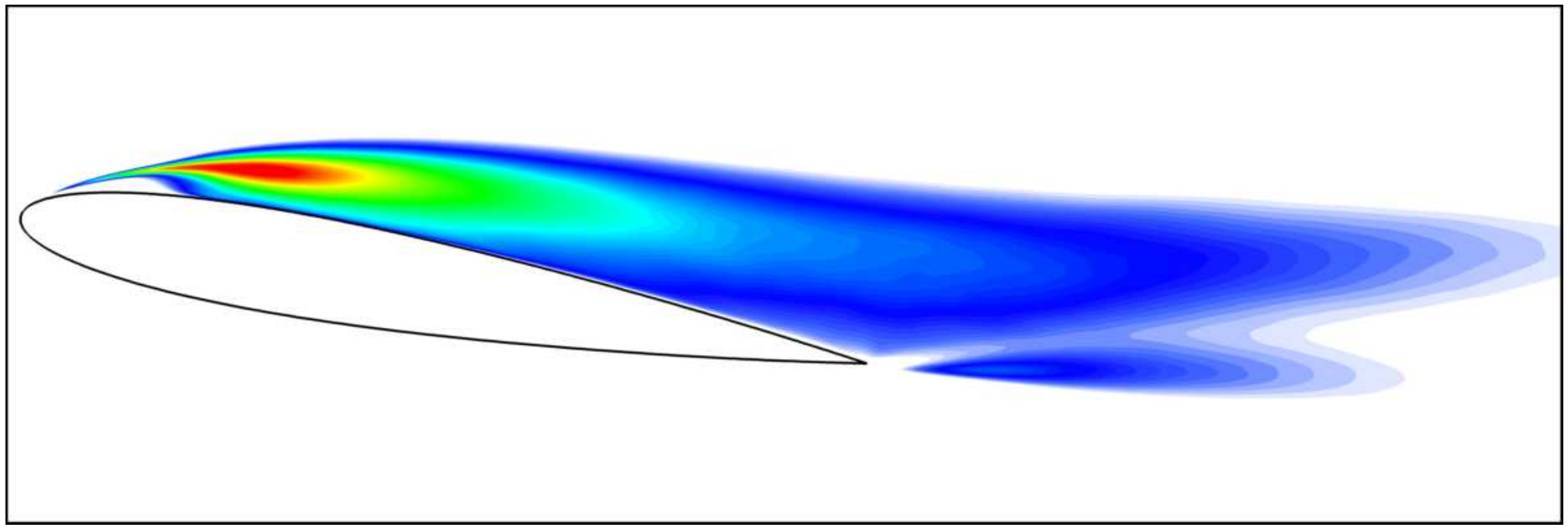}
\textit{$\alpha = 9.80^{\circ}$}
\end{minipage}
\medskip
\begin{minipage}{220pt}
\centering
\includegraphics[width=220pt, trim={0mm 0mm 0mm 0mm}, clip]{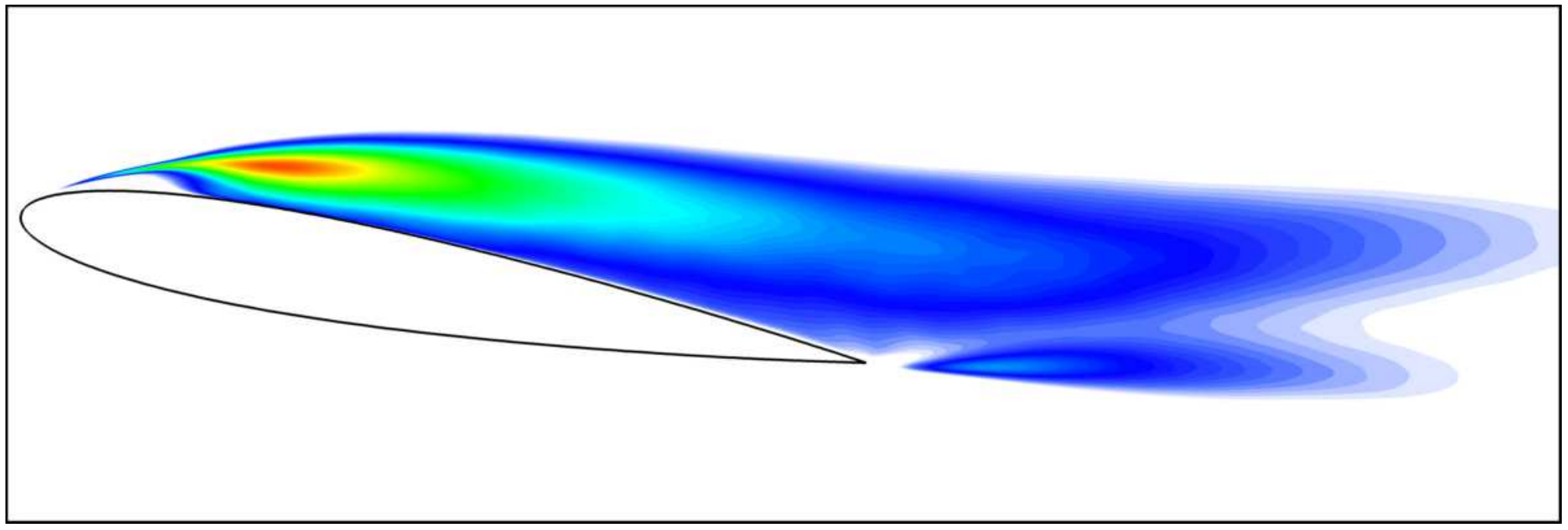}
\textit{$\alpha = 9.90^{\circ}$}
\end{minipage}
\medskip
\begin{minipage}{220pt}
\centering
\includegraphics[width=220pt, trim={0mm 0mm 0mm 0mm}, clip]{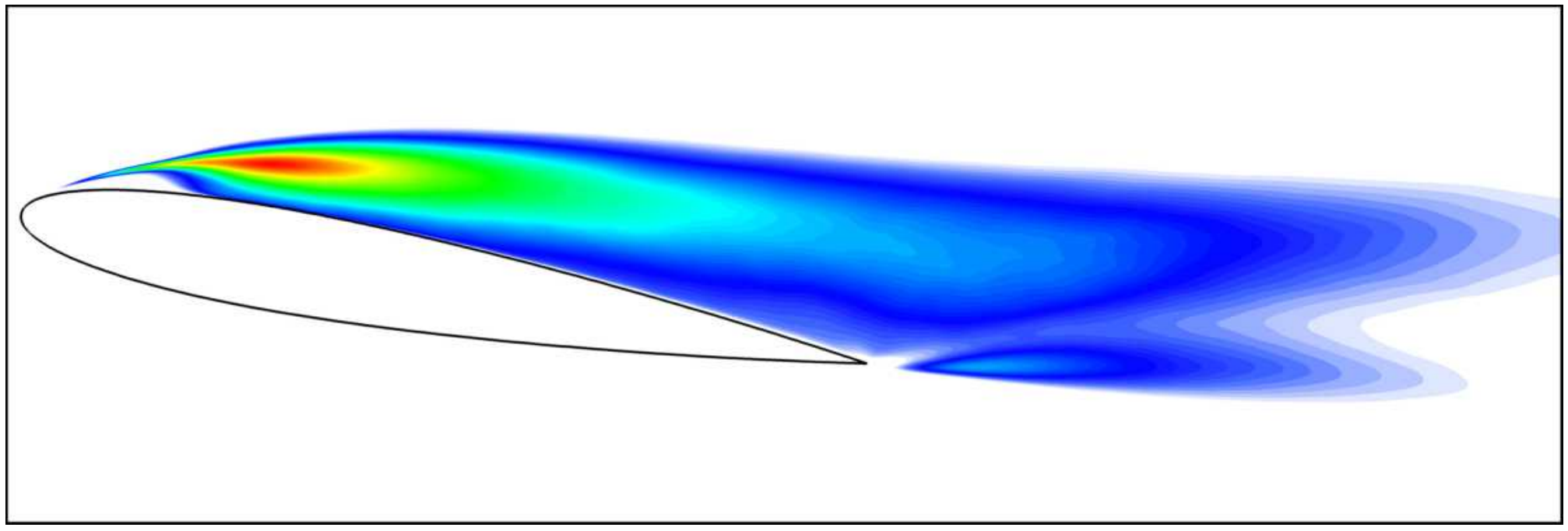}
\textit{$\alpha = 10.0^{\circ}$}
\end{minipage}
\medskip
\begin{minipage}{220pt}
\centering
\includegraphics[width=220pt, trim={0mm 0mm 0mm 0mm}, clip]{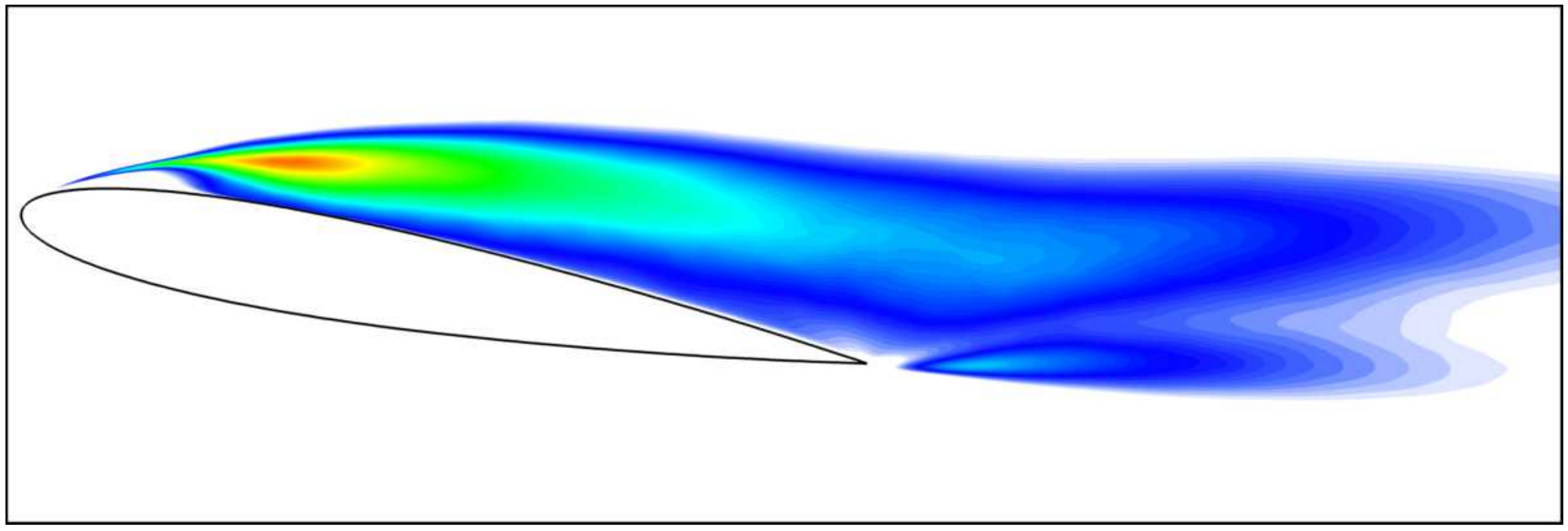}
\textit{$\alpha = 10.1^{\circ}$}
\end{minipage}
\begin{minipage}{220pt}
\centering
\includegraphics[width=220pt, trim={0mm 0mm 0mm 0mm}, clip]{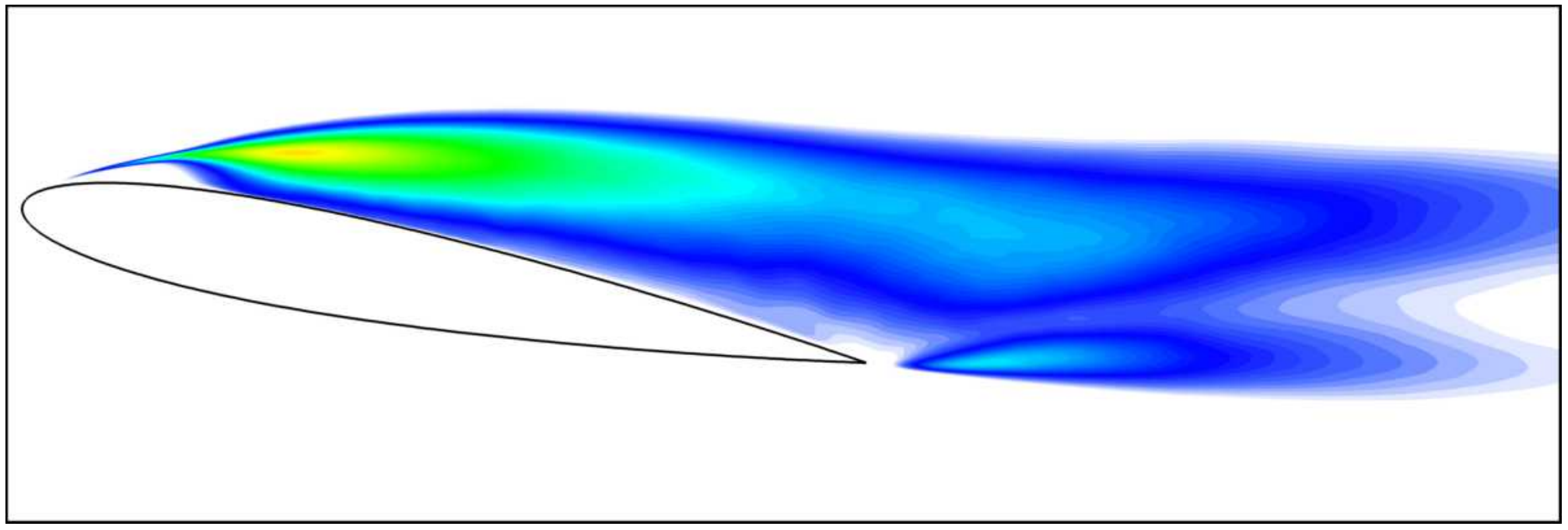}
\textit{$\alpha = 10.5^{\circ}$}
\end{minipage}
\caption{Colours map of the high-lift variance of the streamwise velocity component, $\widehat{{u\mydprime}^2}$, for the angles of attack $\alpha = 9.25^{\circ}$--$10.5^{\circ}$.}
\label{u2_above}
\end{center}
\end{figure}

\begin{figure}
\begin{center}
\begin{minipage}{220pt}
\centering
\includegraphics[width=220pt, trim={0mm 0mm 0mm 0mm}, clip]{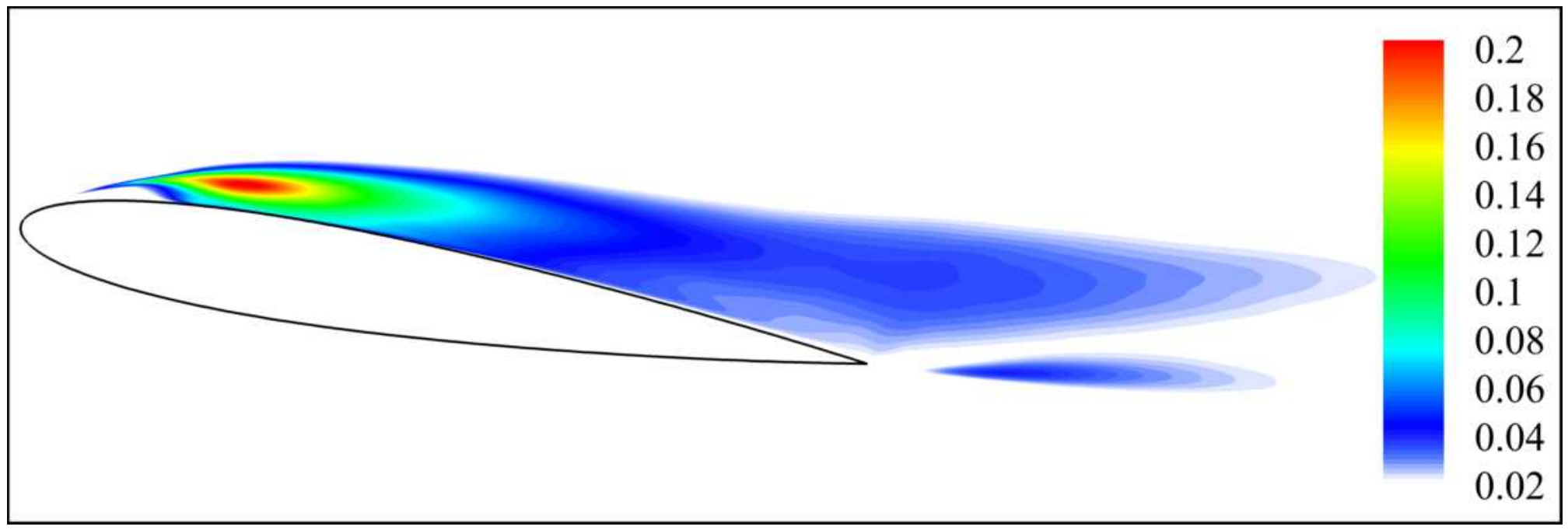}
\textit{$\alpha = 9.25^{\circ}$}
\end{minipage}
\begin{minipage}{220pt}
\centering
\includegraphics[width=220pt, trim={0mm 0mm 0mm 0mm}, clip]{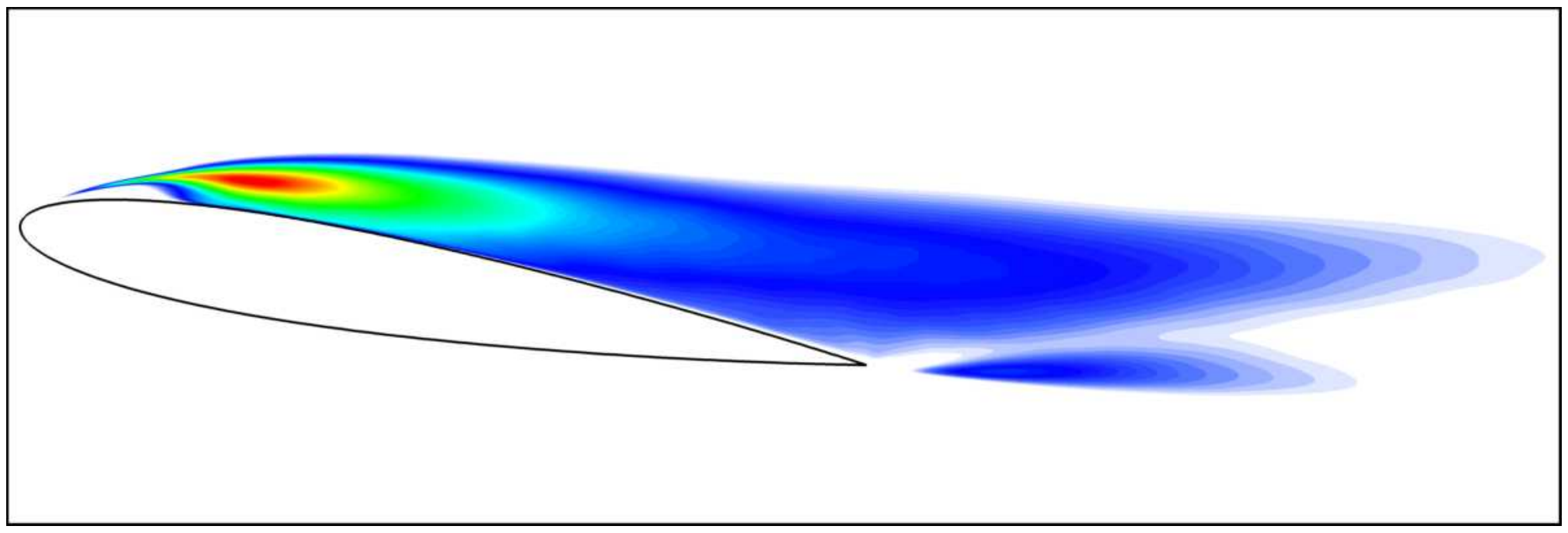}
\textit{$\alpha = 9.40^{\circ}$}
\end{minipage}
\begin{minipage}{220pt}
\centering
\includegraphics[width=220pt, trim={0mm 0mm 0mm 0mm}, clip]{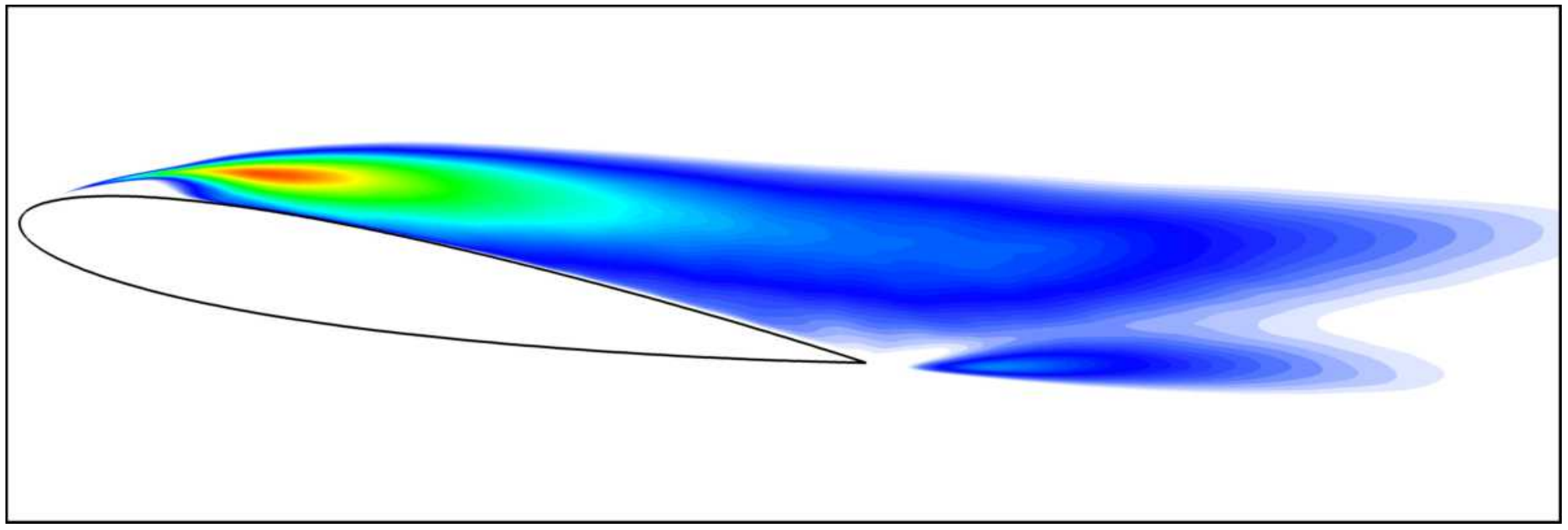}
\textit{$\alpha = 9.50^{\circ}$}
\end{minipage}
\begin{minipage}{220pt}
\centering
\includegraphics[width=220pt, trim={0mm 0mm 0mm 0mm}, clip]{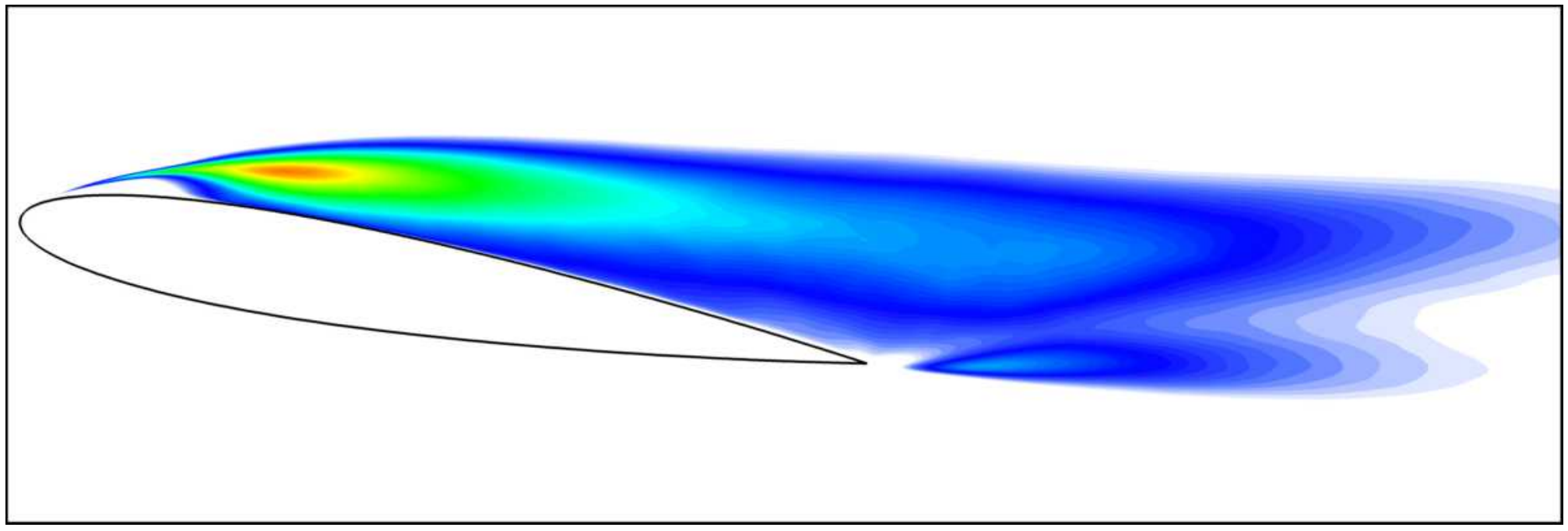}
\textit{$\alpha = 9.60^{\circ}$}
\end{minipage}
\begin{minipage}{220pt}
\centering
\includegraphics[width=220pt, trim={0mm 0mm 0mm 0mm}, clip]{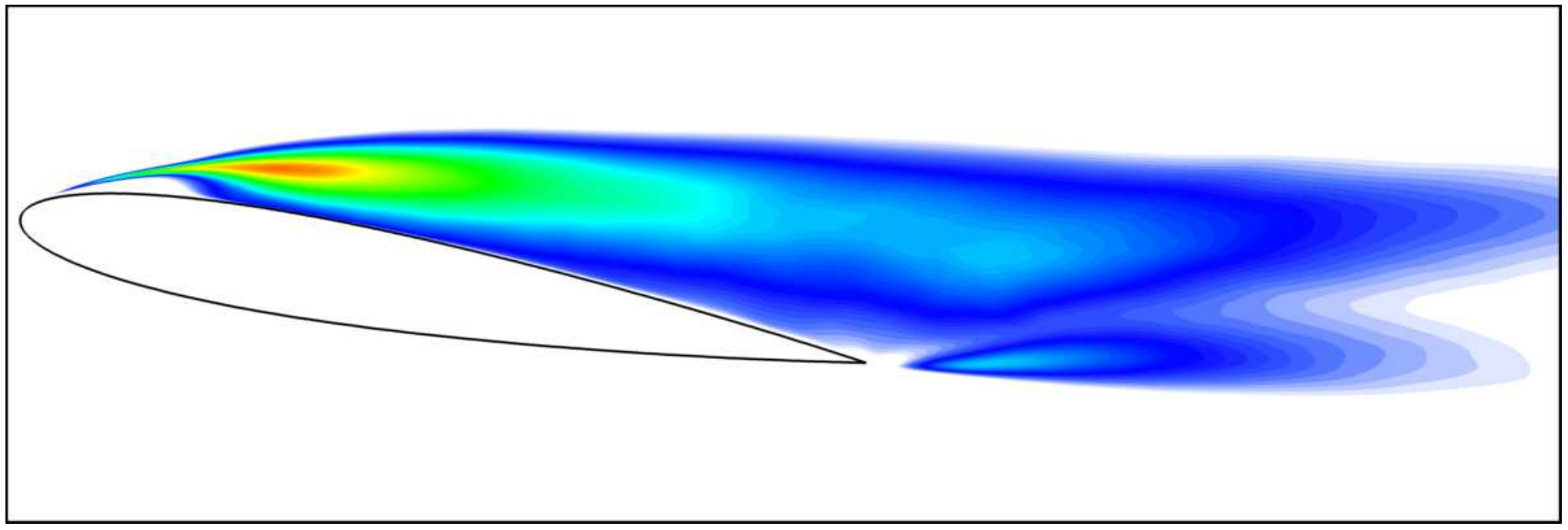}
\textit{$\alpha = 9.70^{\circ}$}
\end{minipage}
\begin{minipage}{220pt}
\centering
\includegraphics[width=220pt, trim={0mm 0mm 0mm 0mm}, clip]{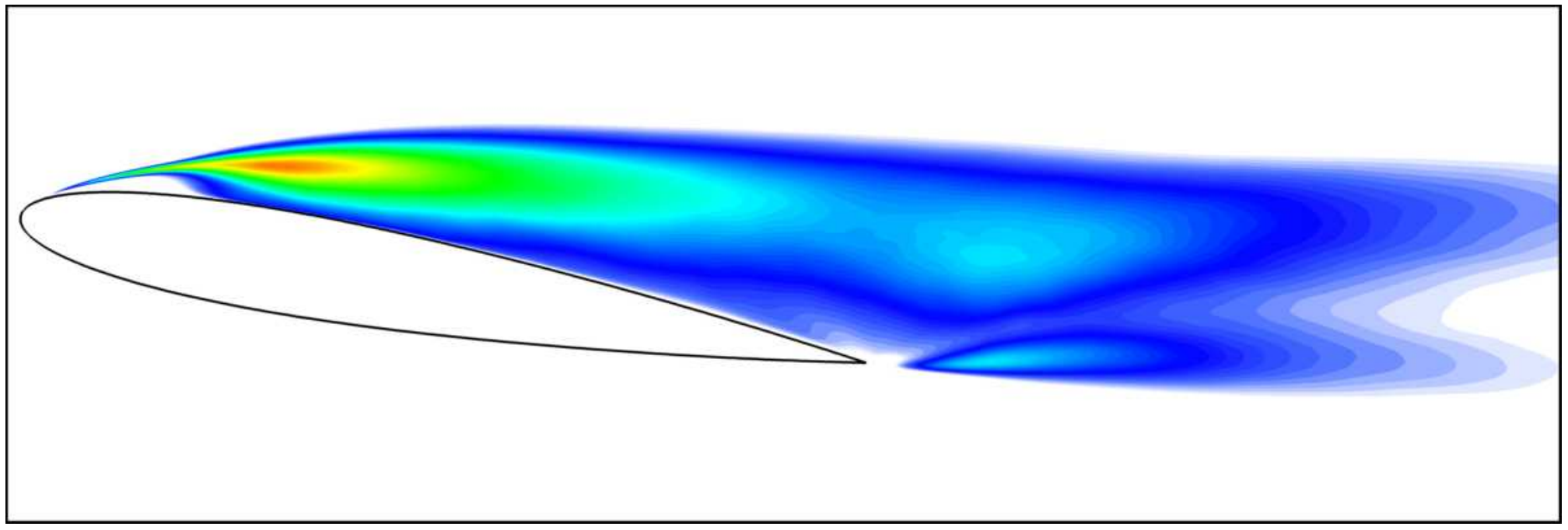}
\textit{$\alpha = 9.80^{\circ}$}
\end{minipage}
\begin{minipage}{220pt}
\centering
\includegraphics[width=220pt, trim={0mm 0mm 0mm 0mm}, clip]{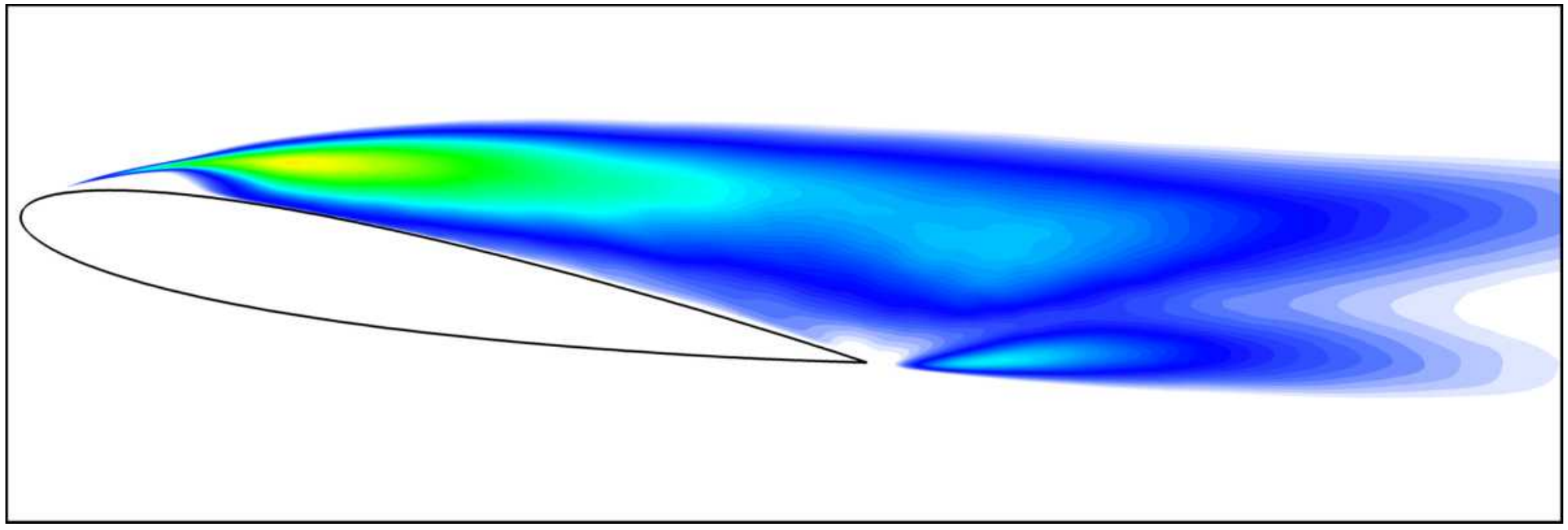}
\textit{$\alpha = 9.90^{\circ}$}
\end{minipage}
\begin{minipage}{220pt}
\centering
\includegraphics[width=220pt, trim={0mm 0mm 0mm 0mm}, clip]{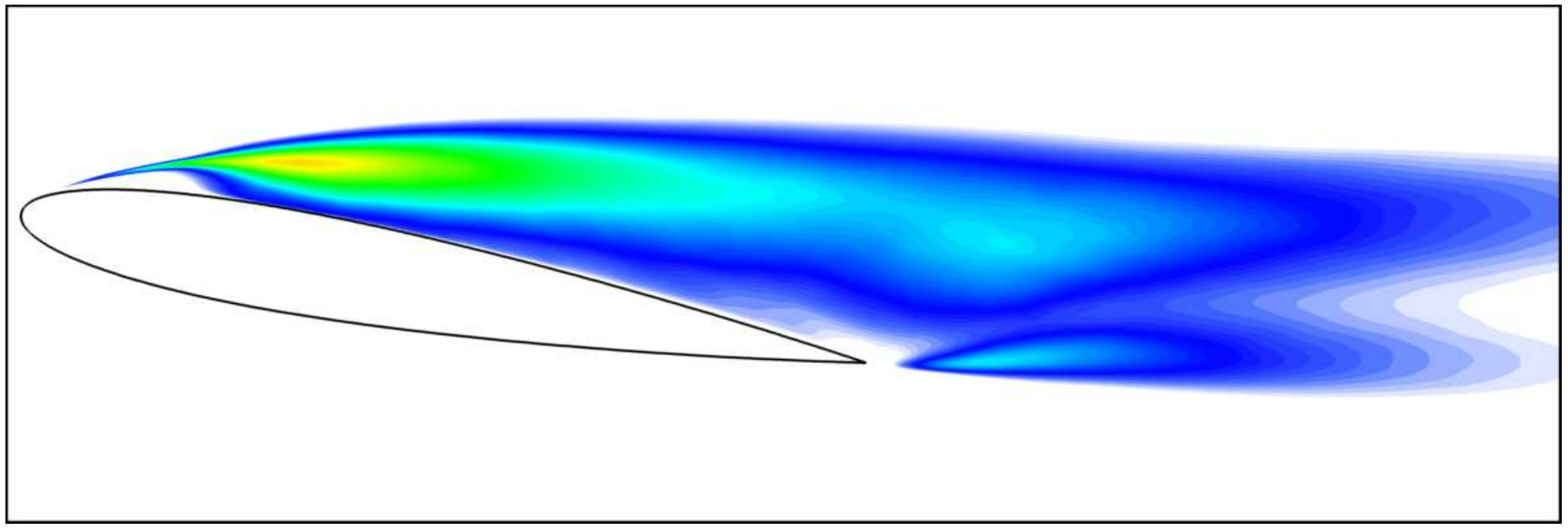}
\textit{$\alpha = 10.0^{\circ}$}
\end{minipage}
\begin{minipage}{220pt}
\centering
\includegraphics[width=220pt, trim={0mm 0mm 0mm 0mm}, clip]{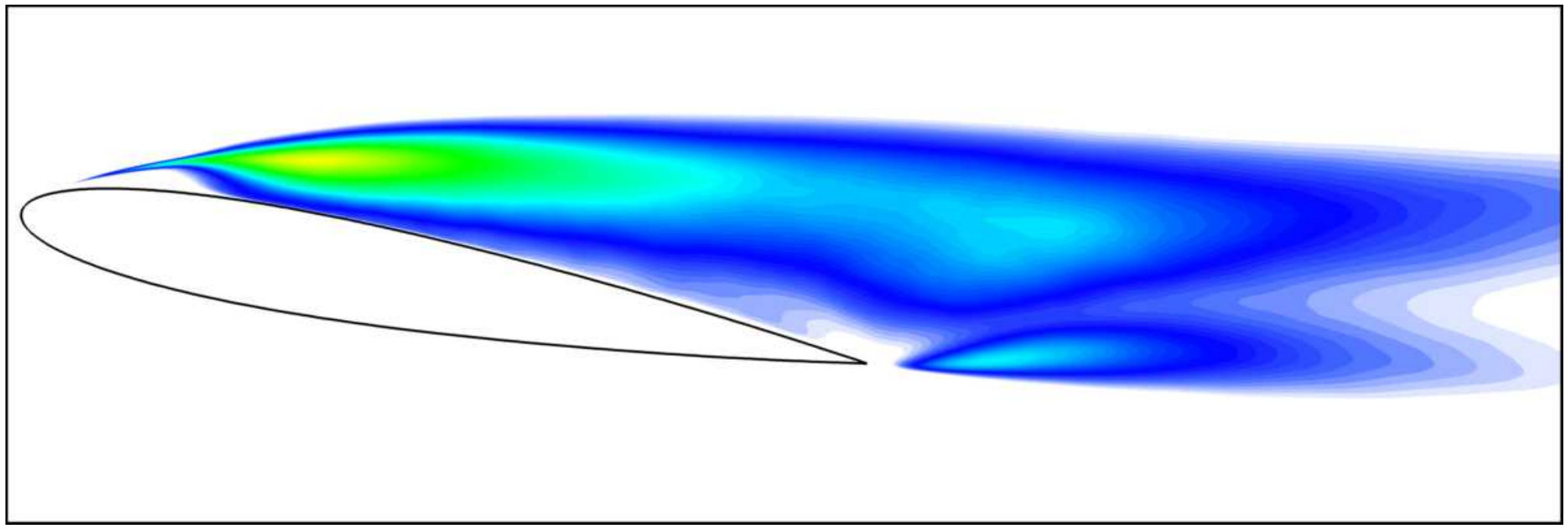}
\textit{$\alpha = 10.1^{\circ}$}
\end{minipage}
\begin{minipage}{220pt}
\centering
\includegraphics[width=220pt, trim={0mm 0mm 0mm 0mm}, clip]{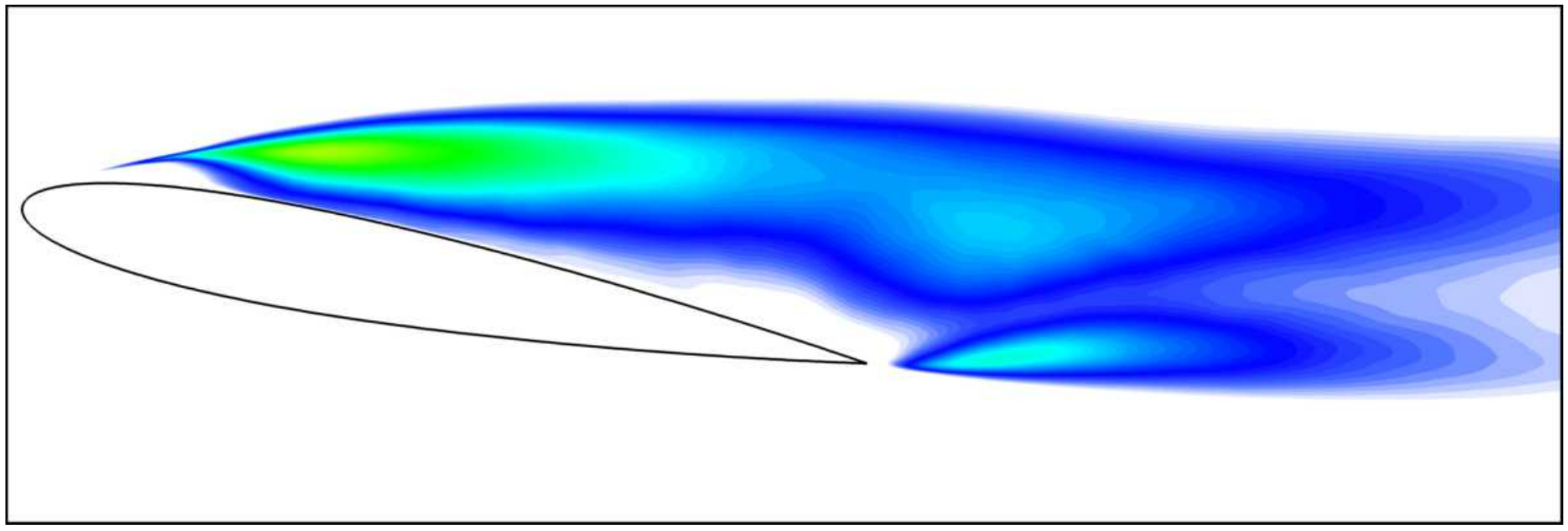}
\textit{$\alpha = 10.5^{\circ}$}
\end{minipage}
\caption{Colours map of the low-lift variance of the streamwise velocity component, $\widecheck{{u\mydprime}^2}$, for the angles of attack $\alpha = 9.25^{\circ}$--$10.5^{\circ}$.}
\label{u2_below}
\end{center}
\end{figure}

\begin{figure}
\begin{center}
\begin{minipage}{220pt}
\includegraphics[height=125pt , trim={0mm 0mm 0mm 0mm}, clip]{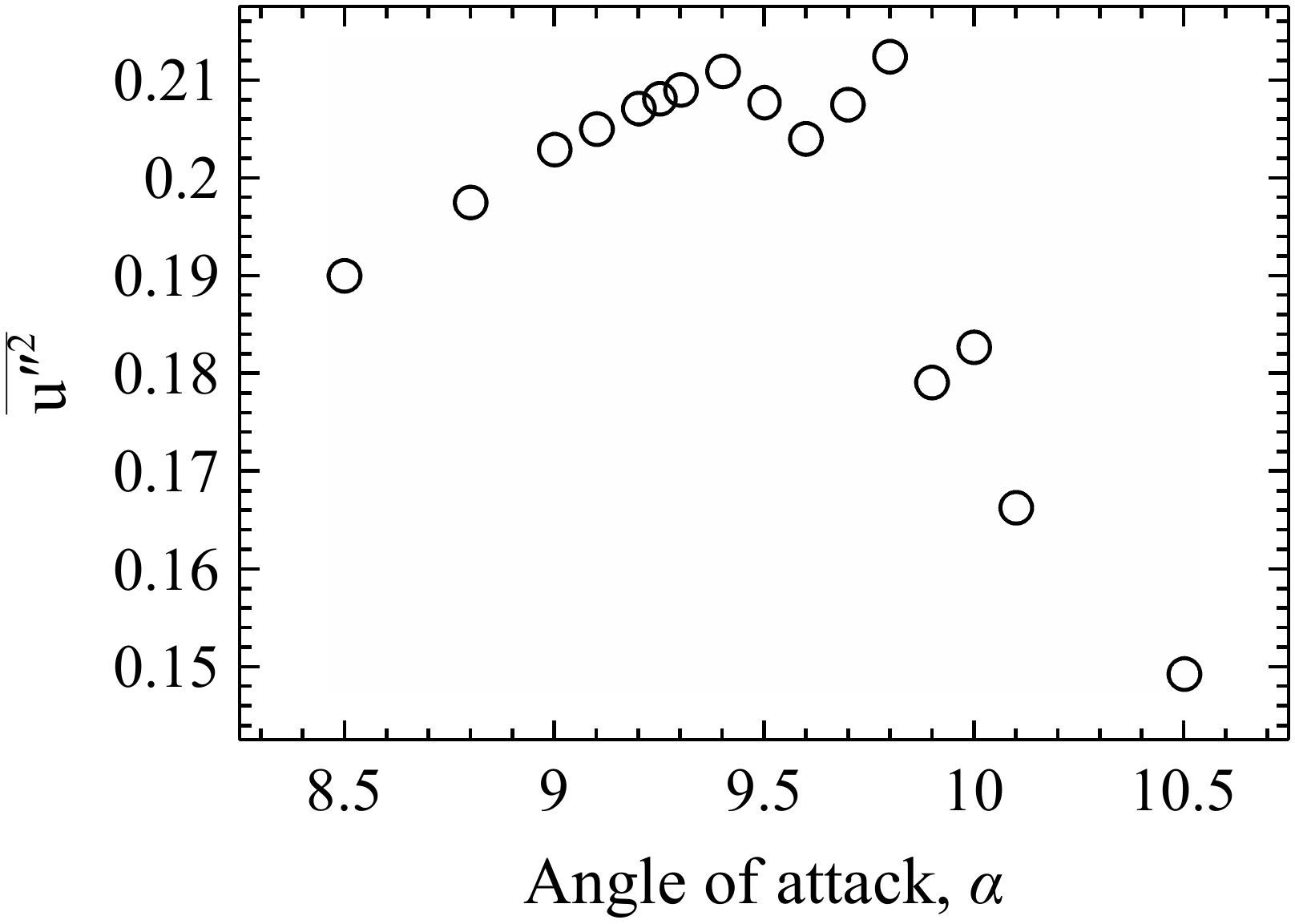}
\end{minipage}
\begin{minipage}{220pt}
\includegraphics[height=125pt , trim={0mm 0mm 0mm 0mm}, clip]{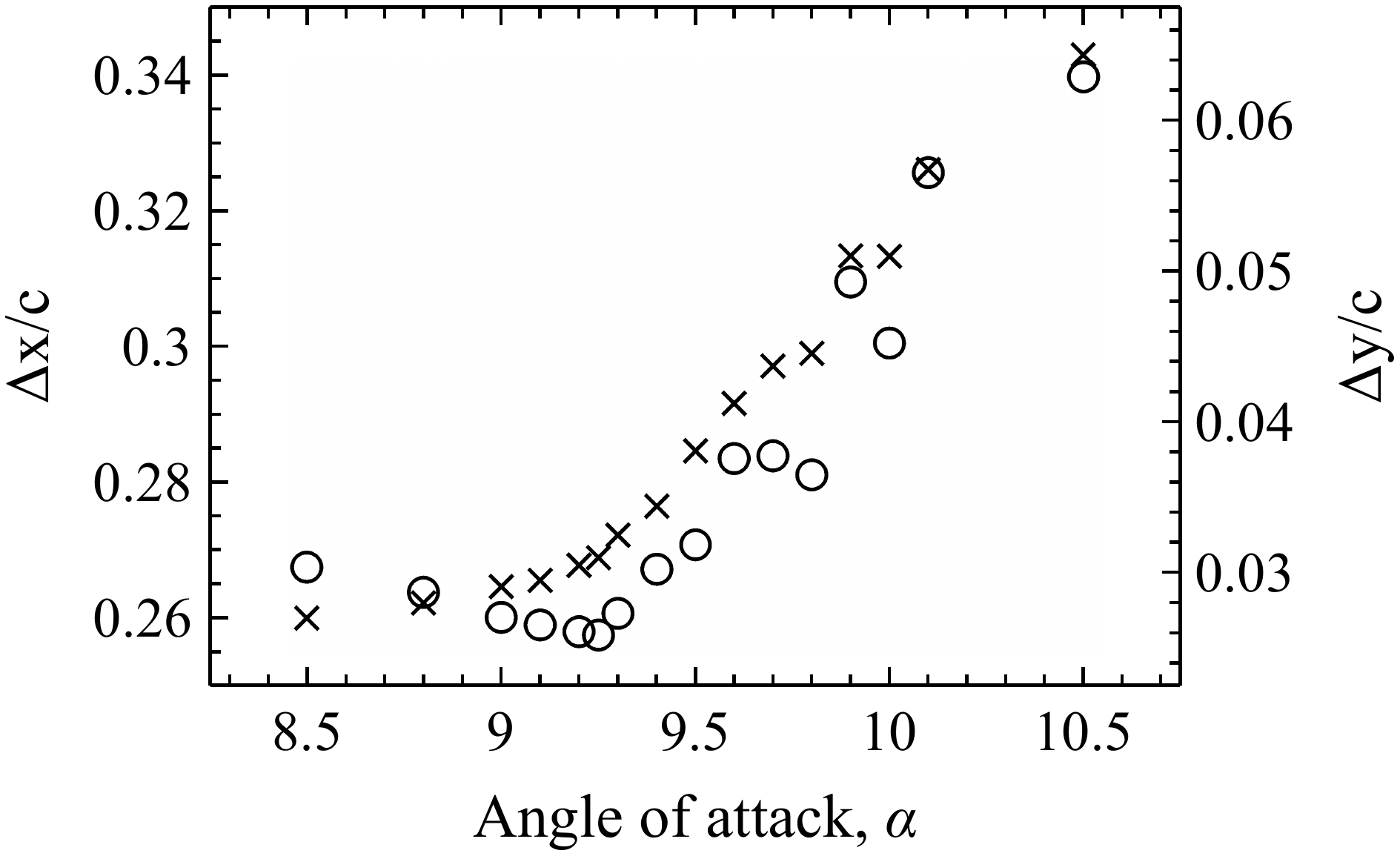}
\end{minipage}
\caption{Left: the maximum $\overline{{u\mydprime}^2}$ plotted versus the angle of attack $\alpha$. Right: the locations of the maximum $\overline{{u\mydprime}^2}$ plotted versus the angle of attack $\alpha$. Circles: $\Delta x/c$ measured from the aerofoil leading-edge, $\times$'s: $\Delta y/c$ measured from the aerofoil surface.}
\label{u2_max}
\end{center}
\end{figure}

\begin{figure}
\begin{center}
\begin{minipage}{220pt}
\centering
\includegraphics[width=220pt, trim={0mm 0mm 0mm 0mm}, clip]{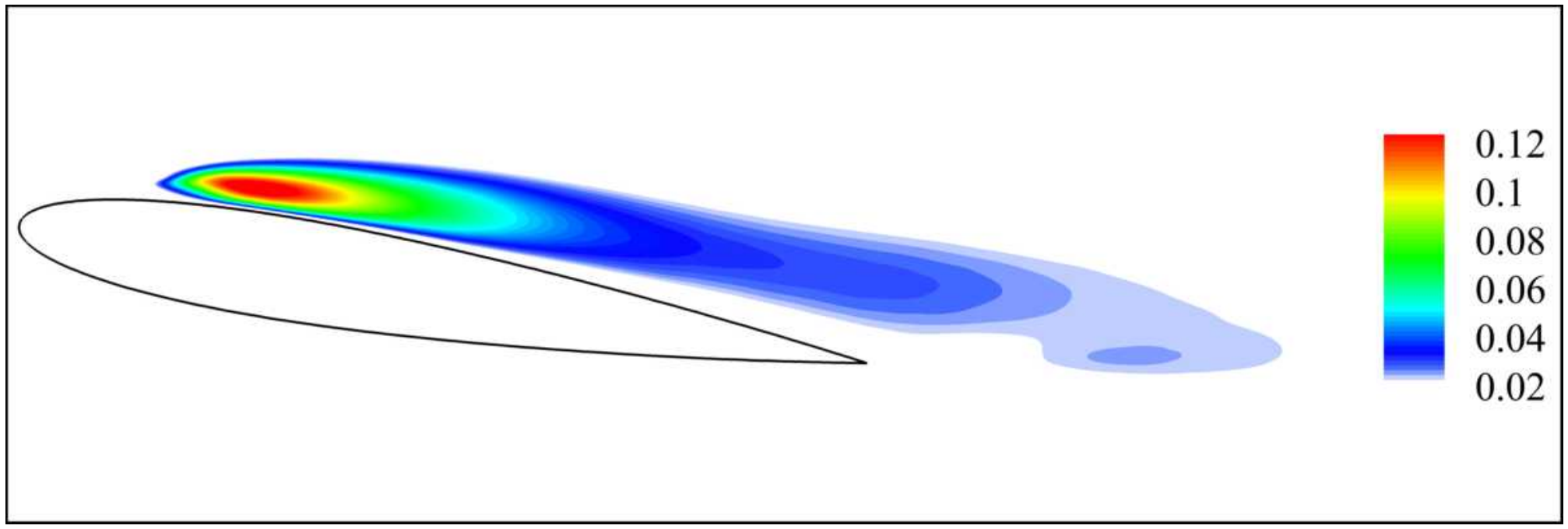}
\textit{$\alpha = 9.25^{\circ}$}
\end{minipage}
\medskip
\begin{minipage}{220pt}
\centering
\includegraphics[width=220pt, trim={0mm 0mm 0mm 0mm}, clip]{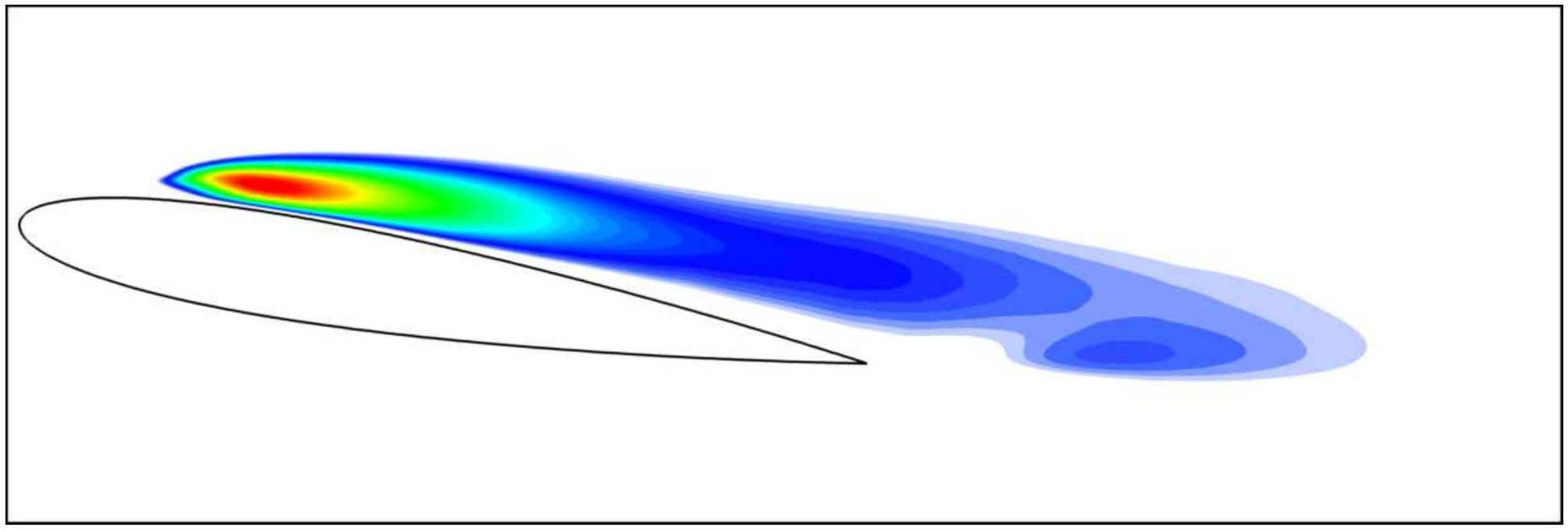}
\textit{$\alpha = 9.40^{\circ}$}
\end{minipage}
\medskip
\begin{minipage}{220pt}
\centering
\includegraphics[width=220pt, trim={0mm 0mm 0mm 0mm}, clip]{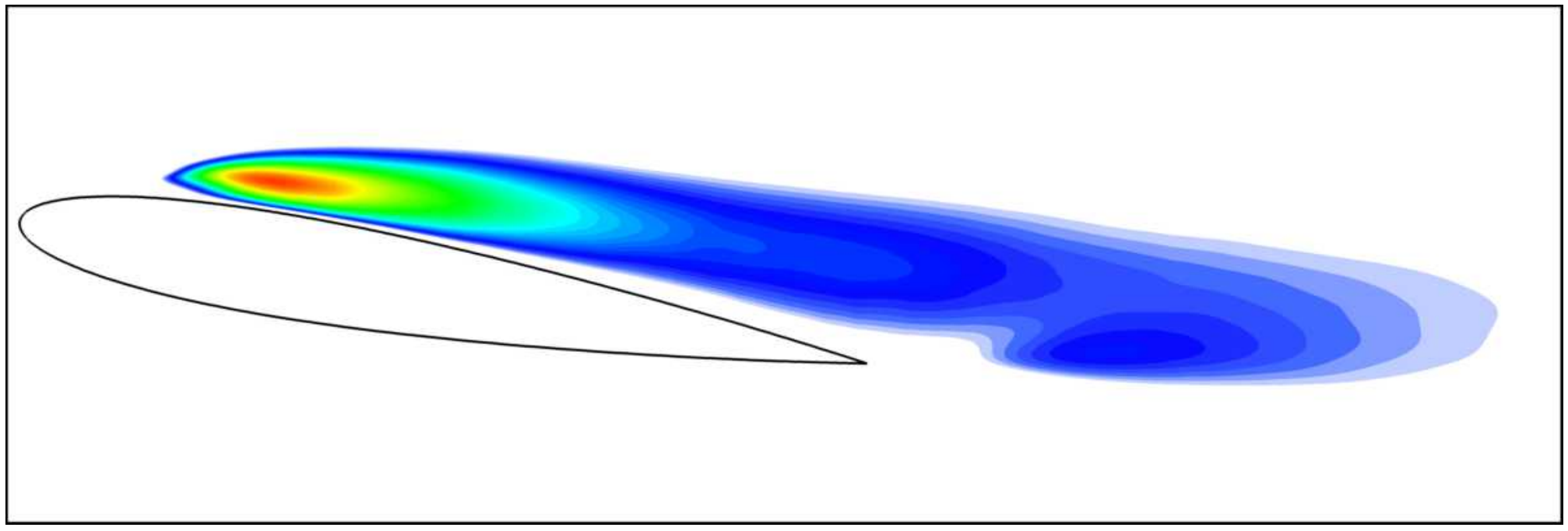}
\textit{$\alpha = 9.50^{\circ}$}
\end{minipage}
\medskip
\begin{minipage}{220pt}
\centering
\includegraphics[width=220pt, trim={0mm 0mm 0mm 0mm}, clip]{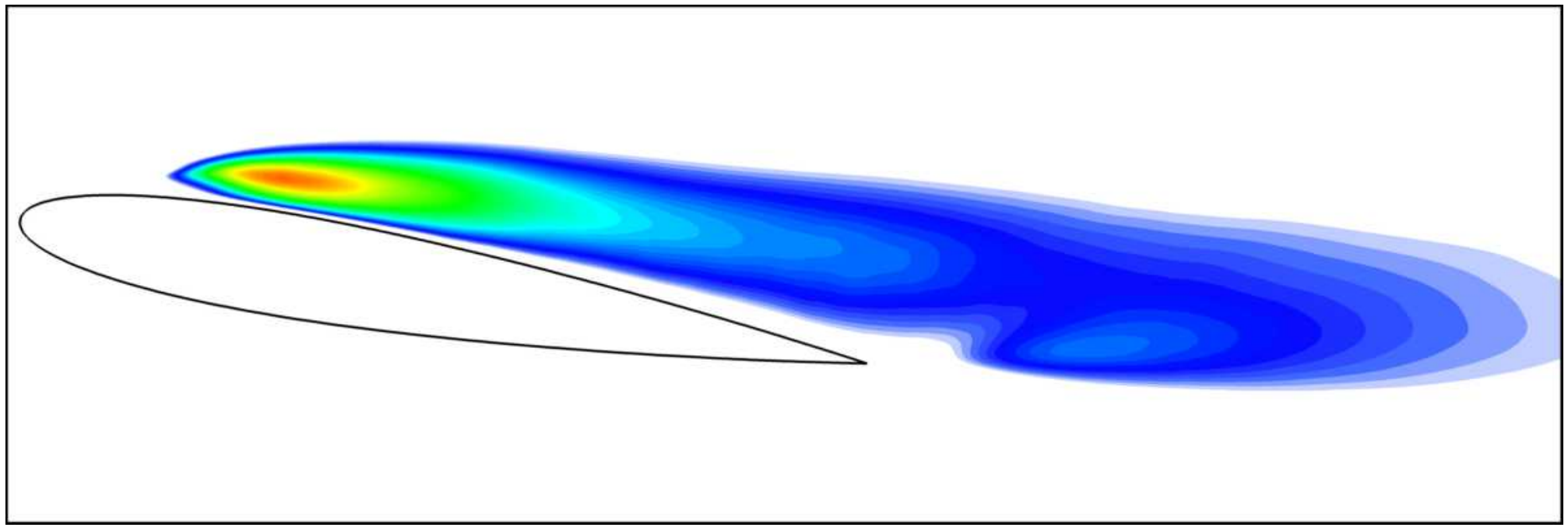}
\textit{$\alpha = 9.60^{\circ}$}
\end{minipage}
\medskip
\begin{minipage}{220pt}
\centering
\includegraphics[width=220pt, trim={0mm 0mm 0mm 0mm}, clip]{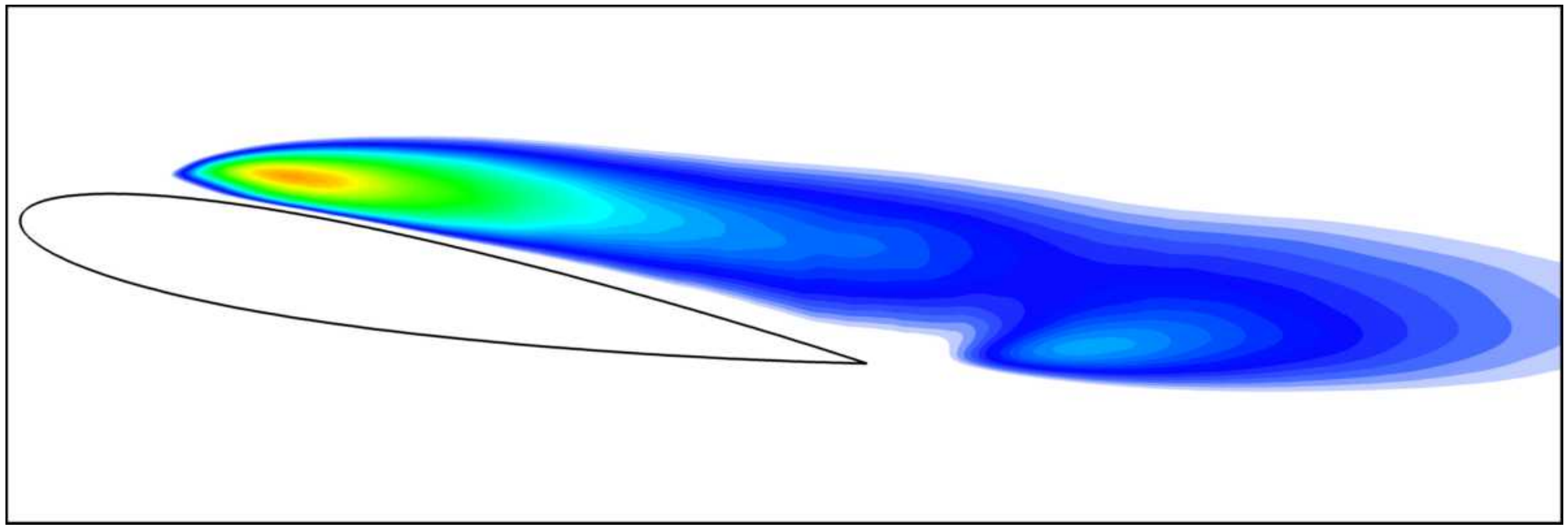}
\textit{$\alpha = 9.70^{\circ}$}
\end{minipage}
\medskip
\begin{minipage}{220pt}
\centering
\includegraphics[width=220pt, trim={0mm 0mm 0mm 0mm}, clip]{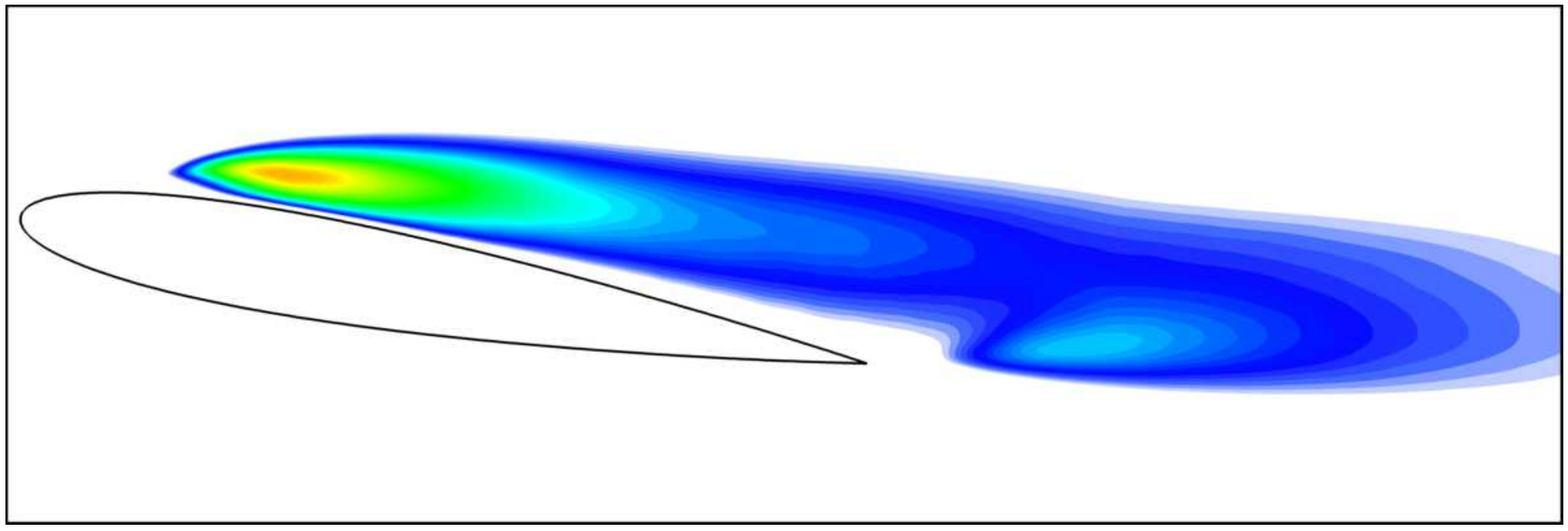}
\textit{$\alpha = 9.80^{\circ}$}
\end{minipage}
\medskip
\begin{minipage}{220pt}
\centering
\includegraphics[width=220pt, trim={0mm 0mm 0mm 0mm}, clip]{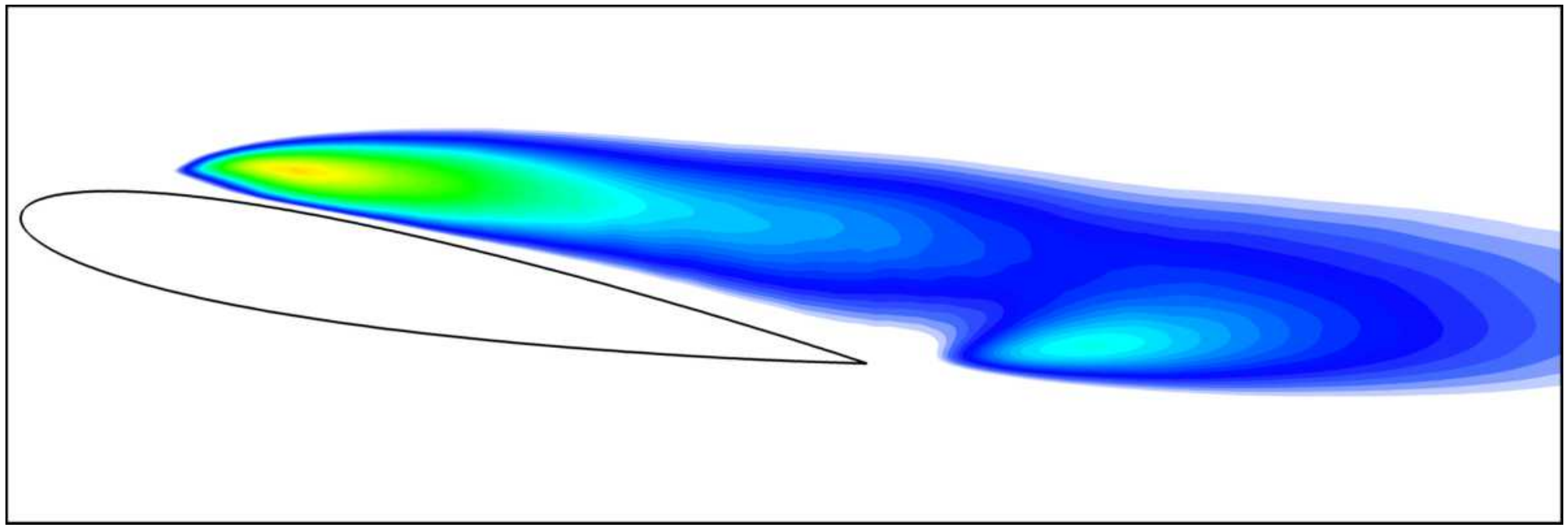}
\textit{$\alpha = 9.90^{\circ}$}
\end{minipage}
\medskip
\begin{minipage}{220pt}
\centering
\includegraphics[width=220pt, trim={0mm 0mm 0mm 0mm}, clip]{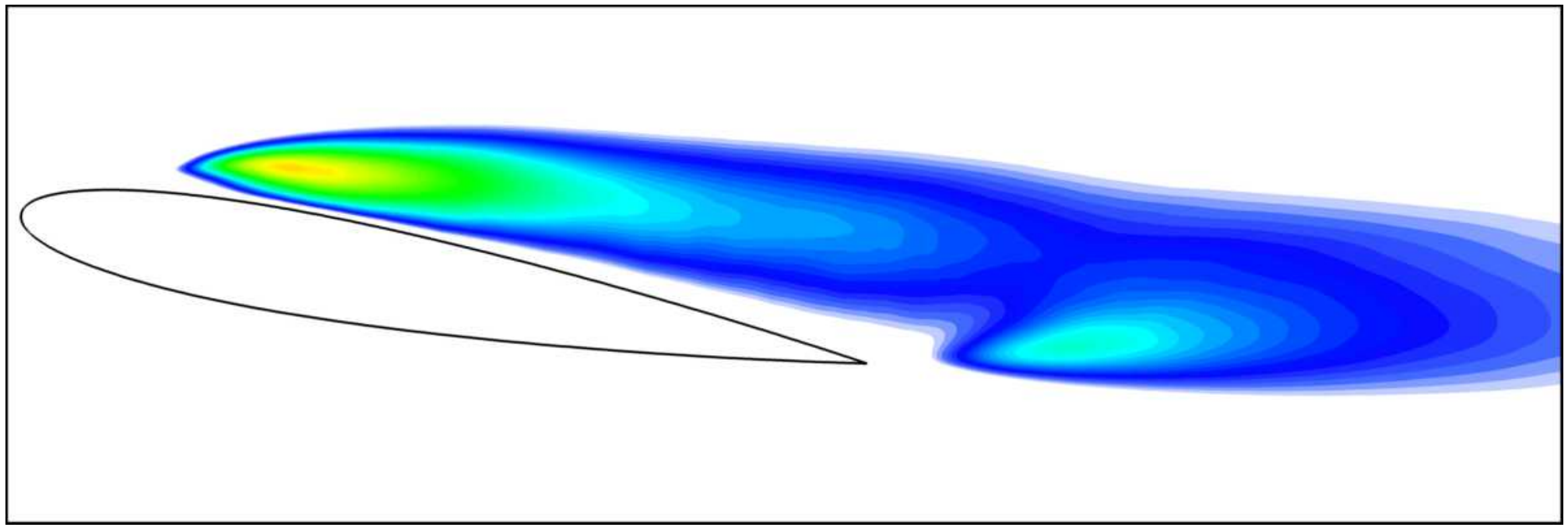}
\textit{$\alpha = 10.0^{\circ}$}
\end{minipage}
\medskip
\begin{minipage}{220pt}
\centering
\includegraphics[width=220pt, trim={0mm 0mm 0mm 0mm}, clip]{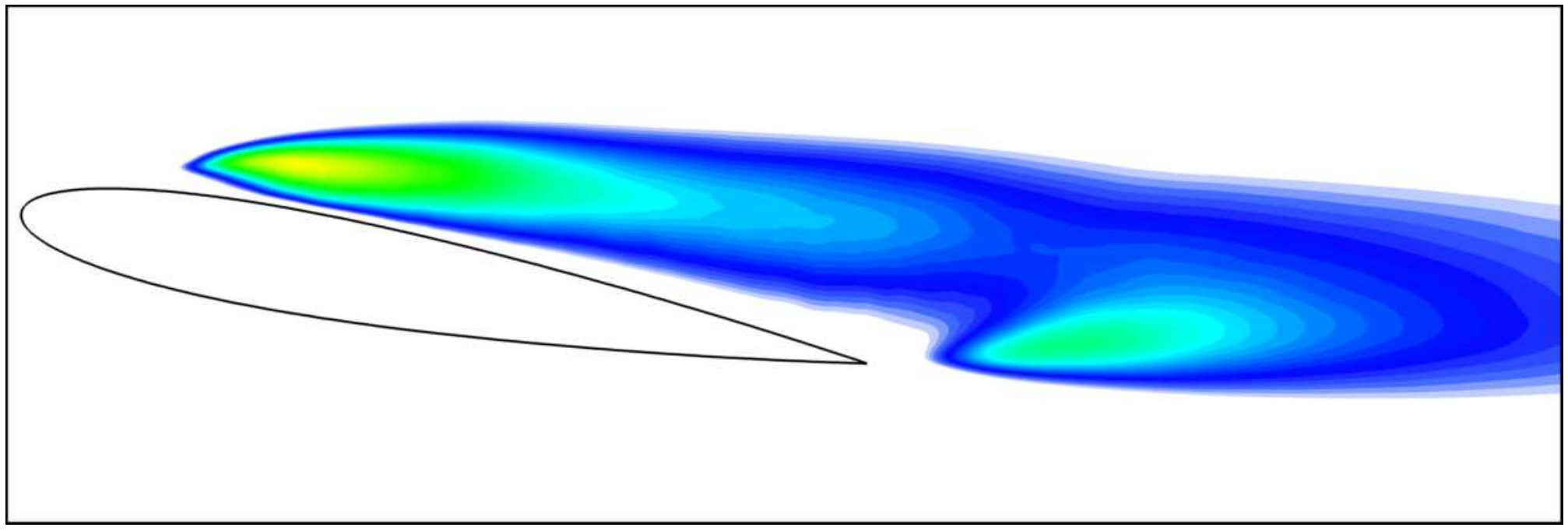}
\textit{$\alpha = 10.1^{\circ}$}
\end{minipage}
\begin{minipage}{220pt}
\centering
\includegraphics[width=220pt, trim={0mm 0mm 0mm 0mm}, clip]{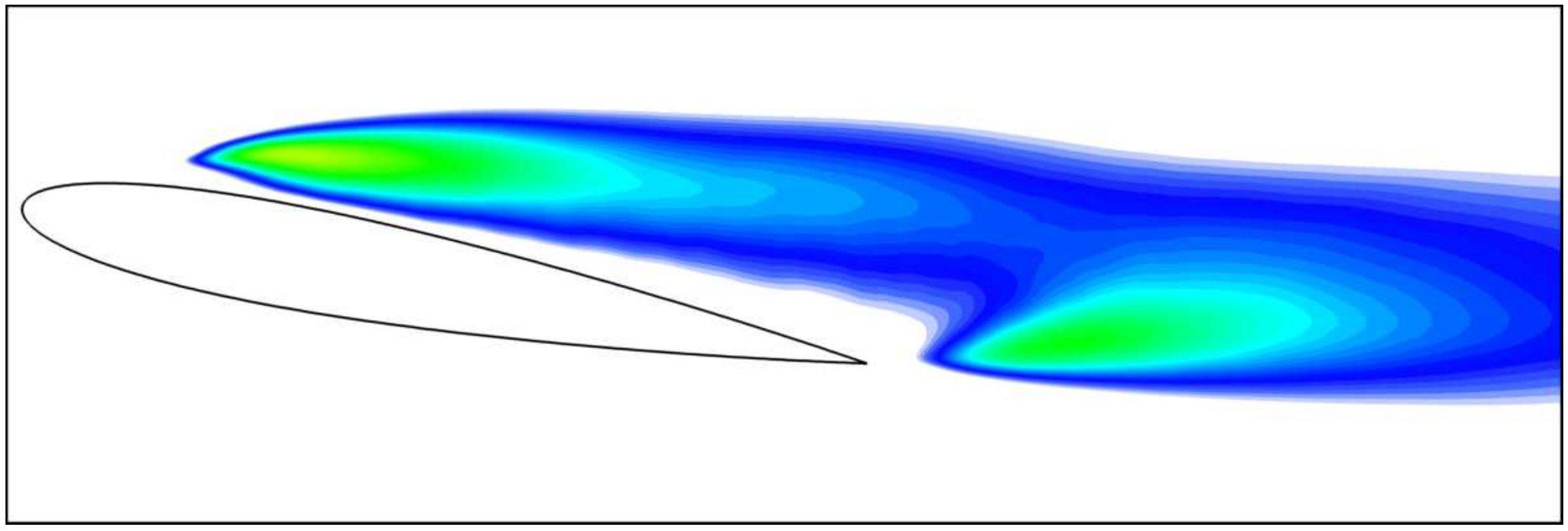}
\textit{$\alpha = 10.5^{\circ}$}
\end{minipage}
\caption{Colours map of the variance of the wall-normal velocity component, $\overline{{v\mydprime}^2}$, for the angles of attack $\alpha = 9.25^{\circ}$--$10.5^{\circ}$.}
\label{v2_mean}
\end{center}
\end{figure}

\begin{figure}
\begin{center}
\begin{minipage}{220pt}
\centering
\includegraphics[width=220pt, trim={0mm 0mm 0mm 0mm}, clip]{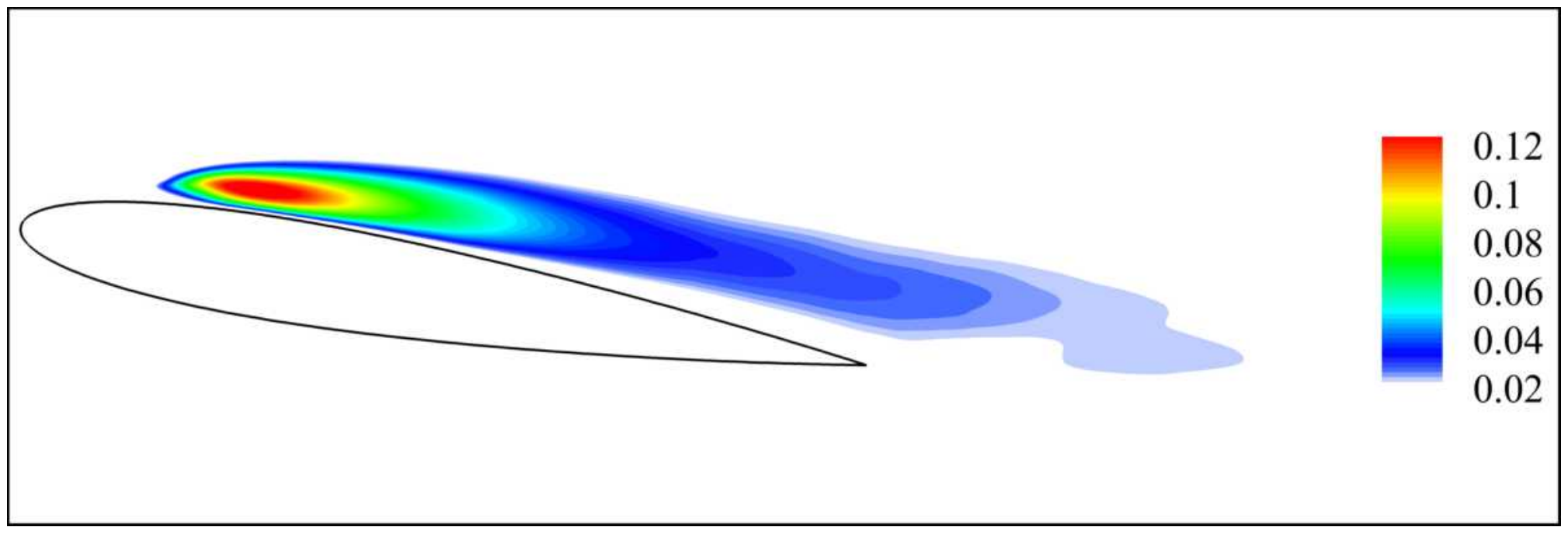}
\textit{$\alpha = 9.25^{\circ}$}
\end{minipage}
\medskip
\begin{minipage}{220pt}
\centering
\includegraphics[width=220pt, trim={0mm 0mm 0mm 0mm}, clip]{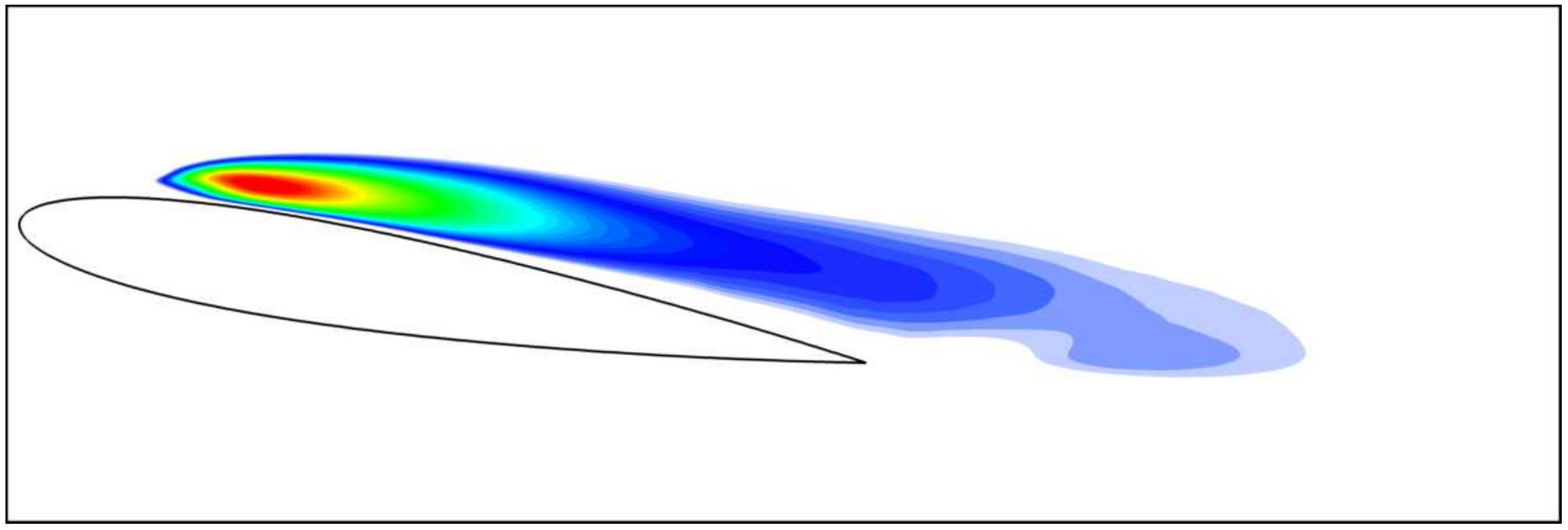}
\textit{$\alpha = 9.40^{\circ}$}
\end{minipage}
\medskip
\begin{minipage}{220pt}
\centering
\includegraphics[width=220pt, trim={0mm 0mm 0mm 0mm}, clip]{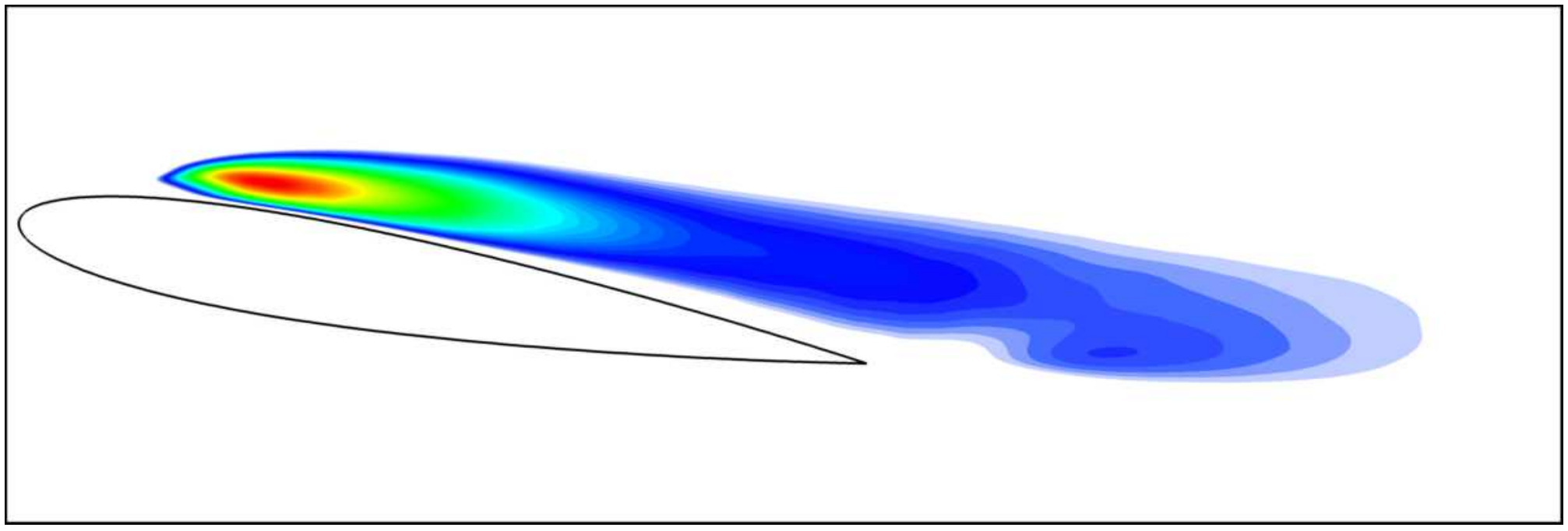}
\textit{$\alpha = 9.50^{\circ}$}
\end{minipage}
\medskip
\begin{minipage}{220pt}
\centering
\includegraphics[width=220pt, trim={0mm 0mm 0mm 0mm}, clip]{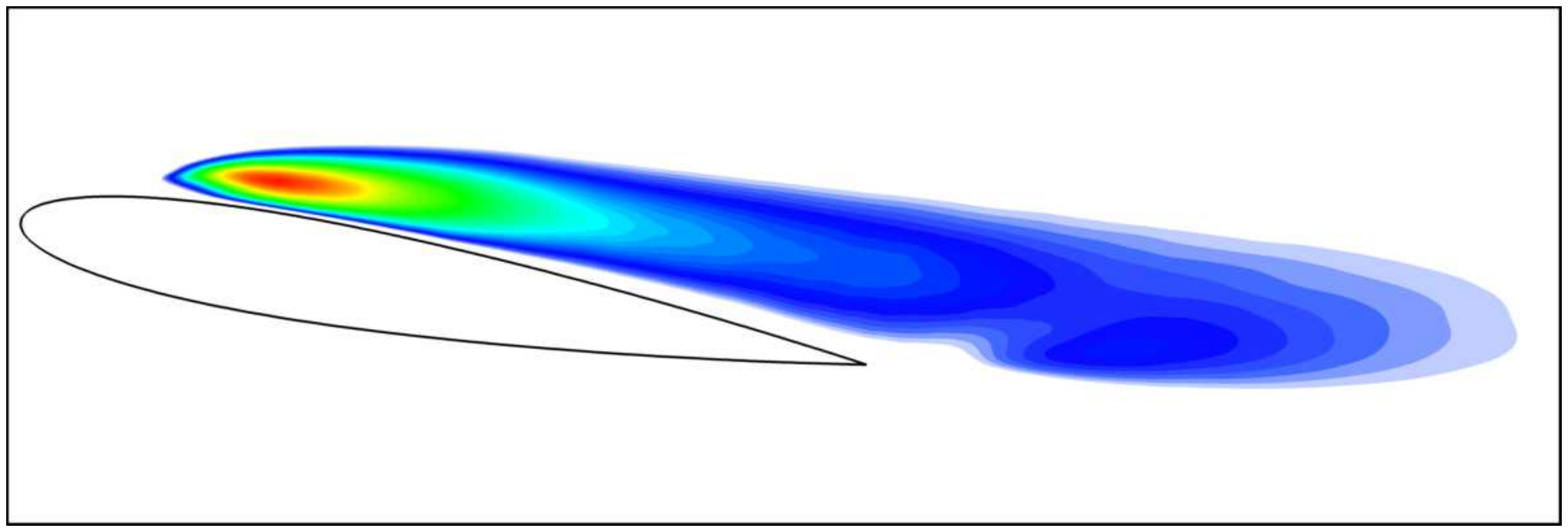}
\textit{$\alpha = 9.60^{\circ}$}
\end{minipage}
\medskip
\begin{minipage}{220pt}
\centering
\includegraphics[width=220pt, trim={0mm 0mm 0mm 0mm}, clip]{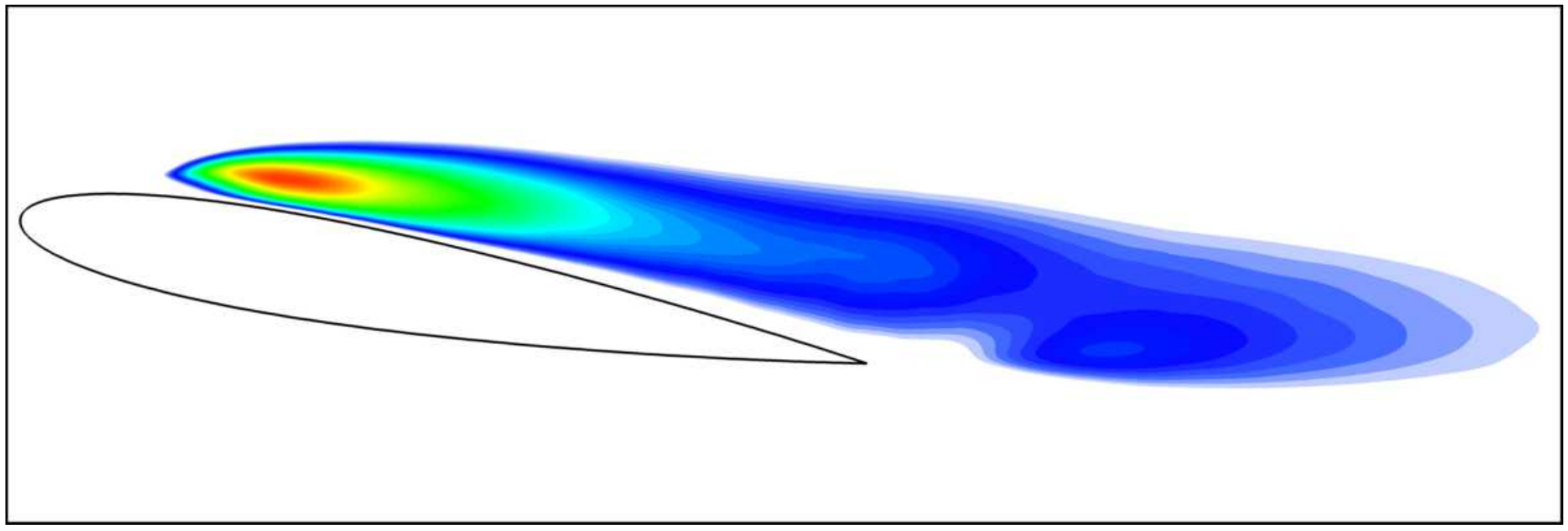}
\textit{$\alpha = 9.70^{\circ}$}
\end{minipage}
\medskip
\begin{minipage}{220pt}
\centering
\includegraphics[width=220pt, trim={0mm 0mm 0mm 0mm}, clip]{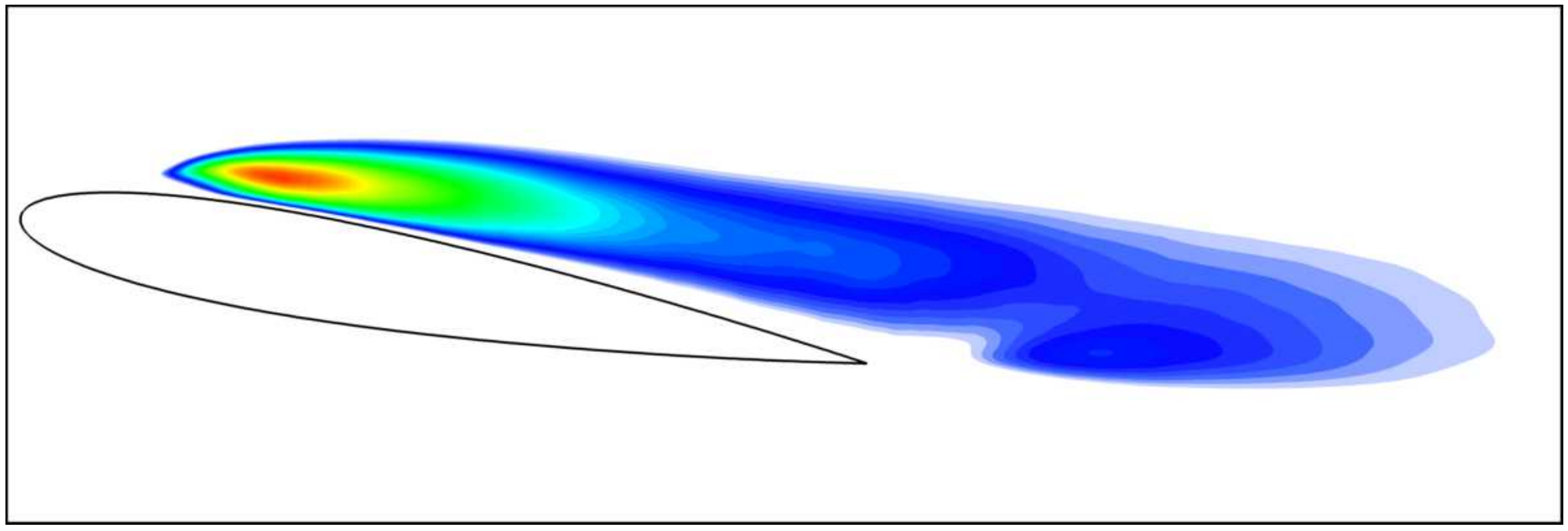}
\textit{$\alpha = 9.80^{\circ}$}
\end{minipage}
\medskip
\begin{minipage}{220pt}
\centering
\includegraphics[width=220pt, trim={0mm 0mm 0mm 0mm}, clip]{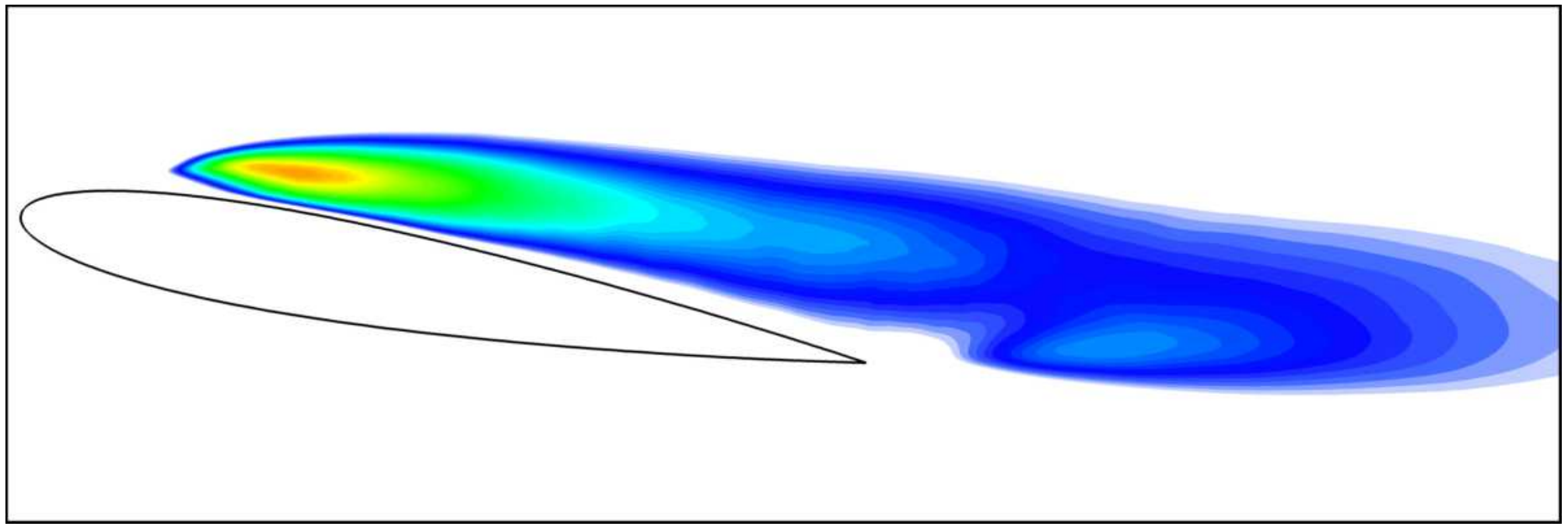}
\textit{$\alpha = 9.90^{\circ}$}
\end{minipage}
\medskip
\begin{minipage}{220pt}
\centering
\includegraphics[width=220pt, trim={0mm 0mm 0mm 0mm}, clip]{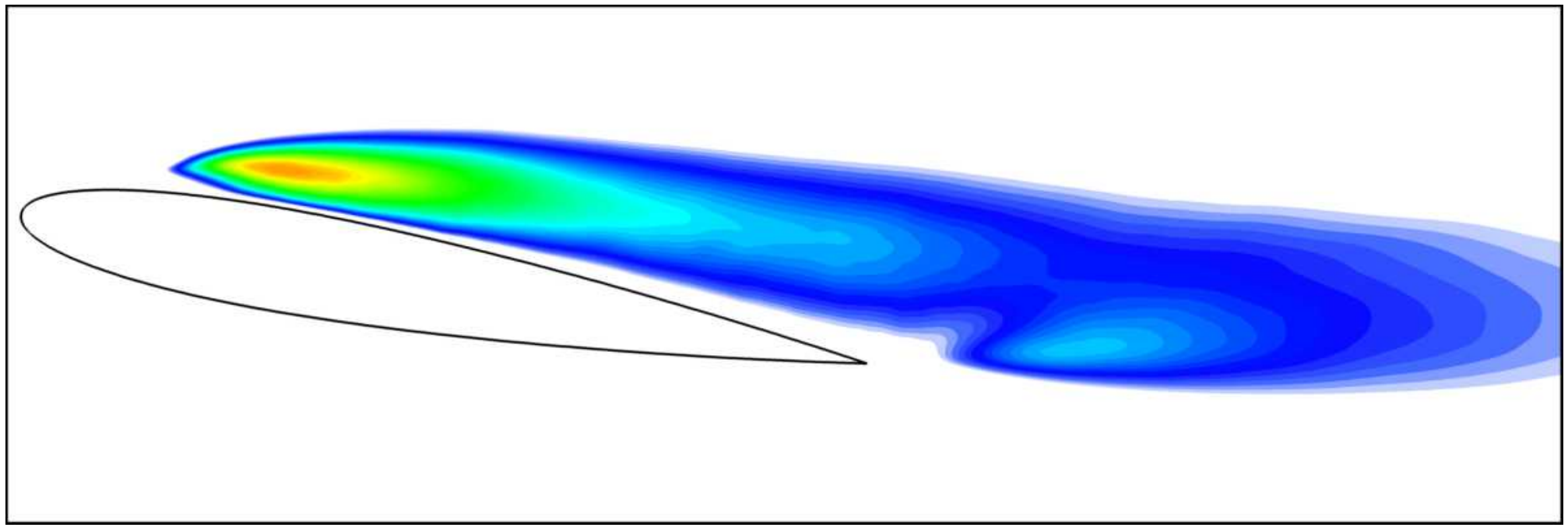}
\textit{$\alpha = 10.0^{\circ}$}
\end{minipage}
\medskip
\begin{minipage}{220pt}
\centering
\includegraphics[width=220pt, trim={0mm 0mm 0mm 0mm}, clip]{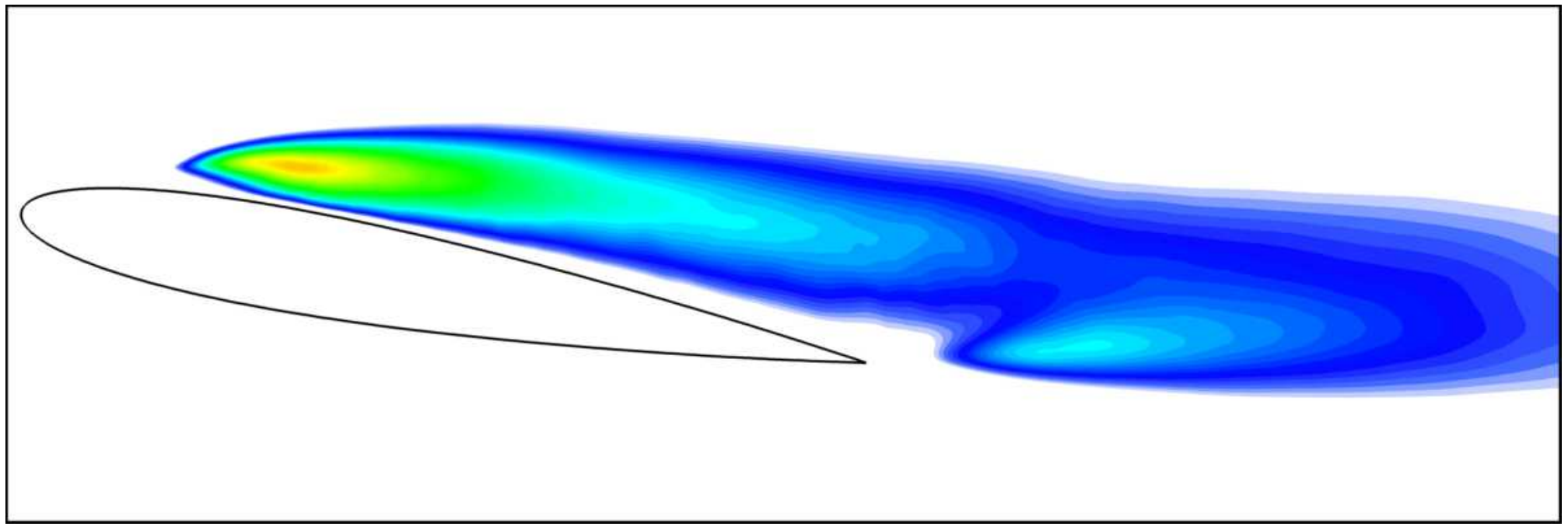}
\textit{$\alpha = 10.1^{\circ}$}
\end{minipage}
\begin{minipage}{220pt}
\centering
\includegraphics[width=220pt, trim={0mm 0mm 0mm 0mm}, clip]{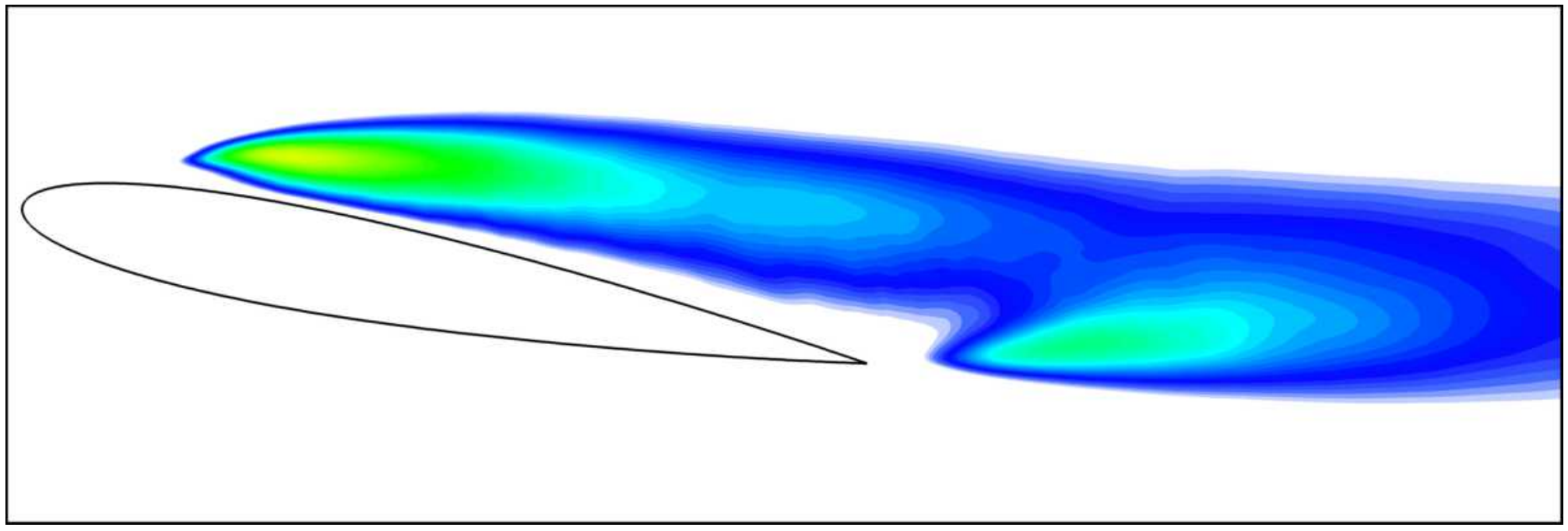}
\textit{$\alpha = 10.5^{\circ}$}
\end{minipage}
\caption{Colours map of the high-lift variance of the wall-normal velocity component, $\widehat{{v\mydprime}^2}$, for the angles of attack $\alpha = 9.25^{\circ}$--$10.5^{\circ}$.}
\label{v2_above}
\end{center}
\end{figure}

\begin{figure}
\begin{center}
\begin{minipage}{220pt}
\centering
\includegraphics[width=220pt, trim={0mm 0mm 0mm 0mm}, clip]{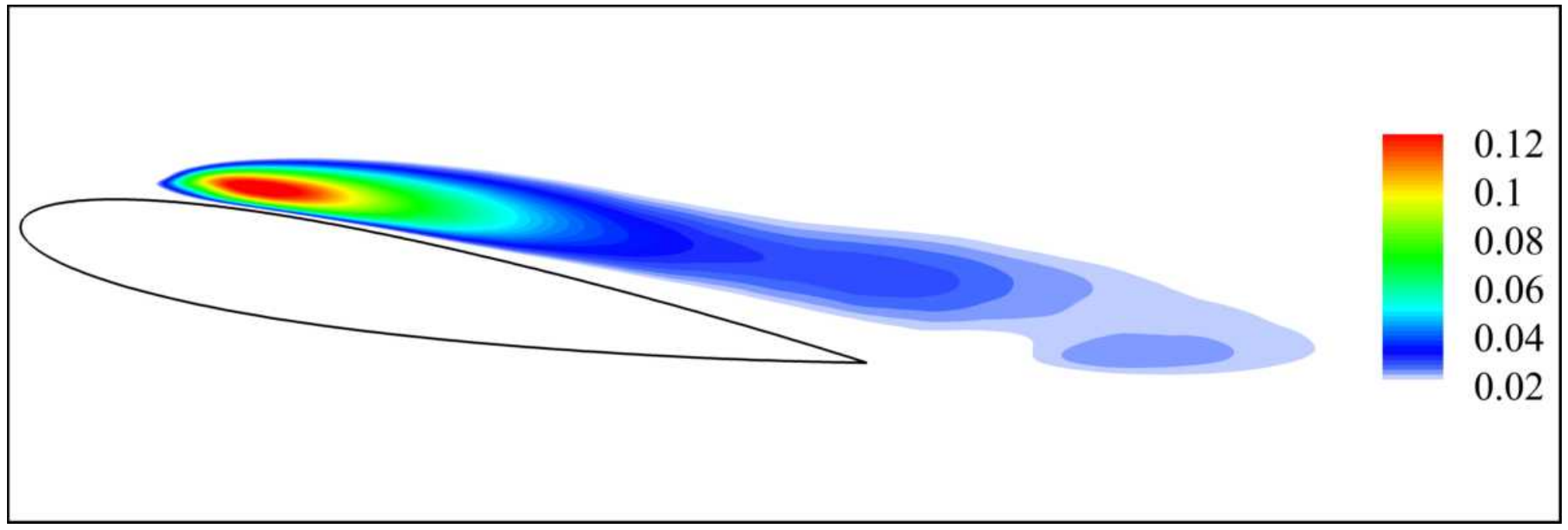}
\textit{$\alpha = 9.25^{\circ}$}
\end{minipage}
\begin{minipage}{220pt}
\centering
\includegraphics[width=220pt, trim={0mm 0mm 0mm 0mm}, clip]{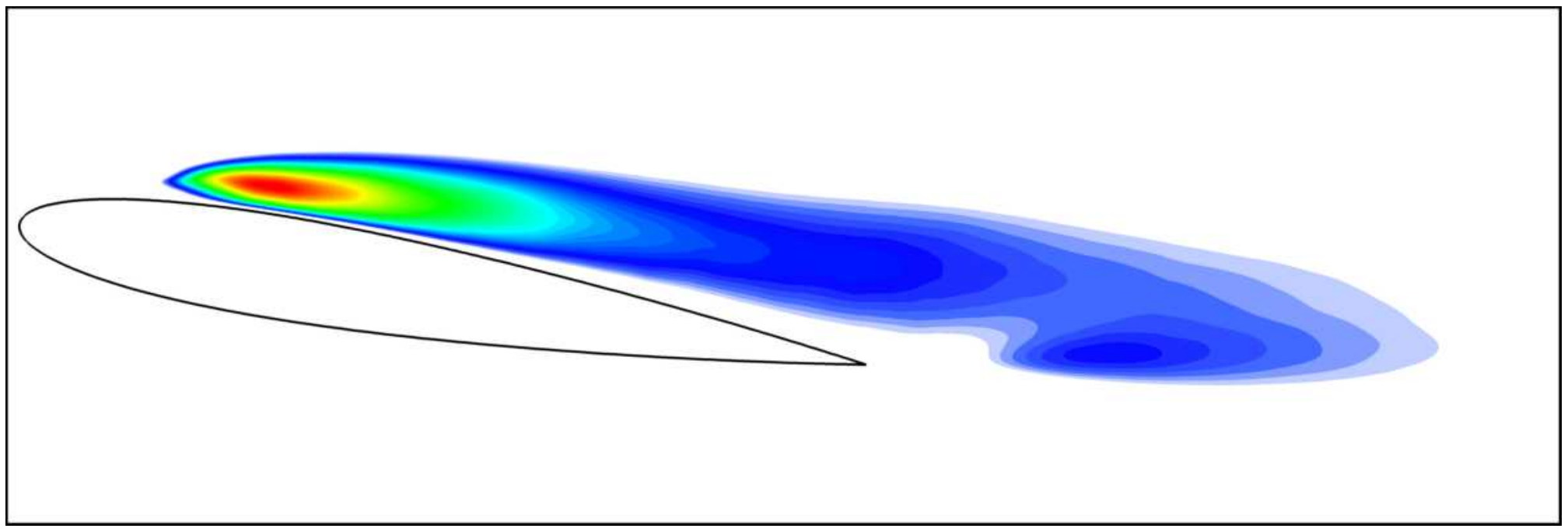}
\textit{$\alpha = 9.40^{\circ}$}
\end{minipage}
\begin{minipage}{220pt}
\centering
\includegraphics[width=220pt, trim={0mm 0mm 0mm 0mm}, clip]{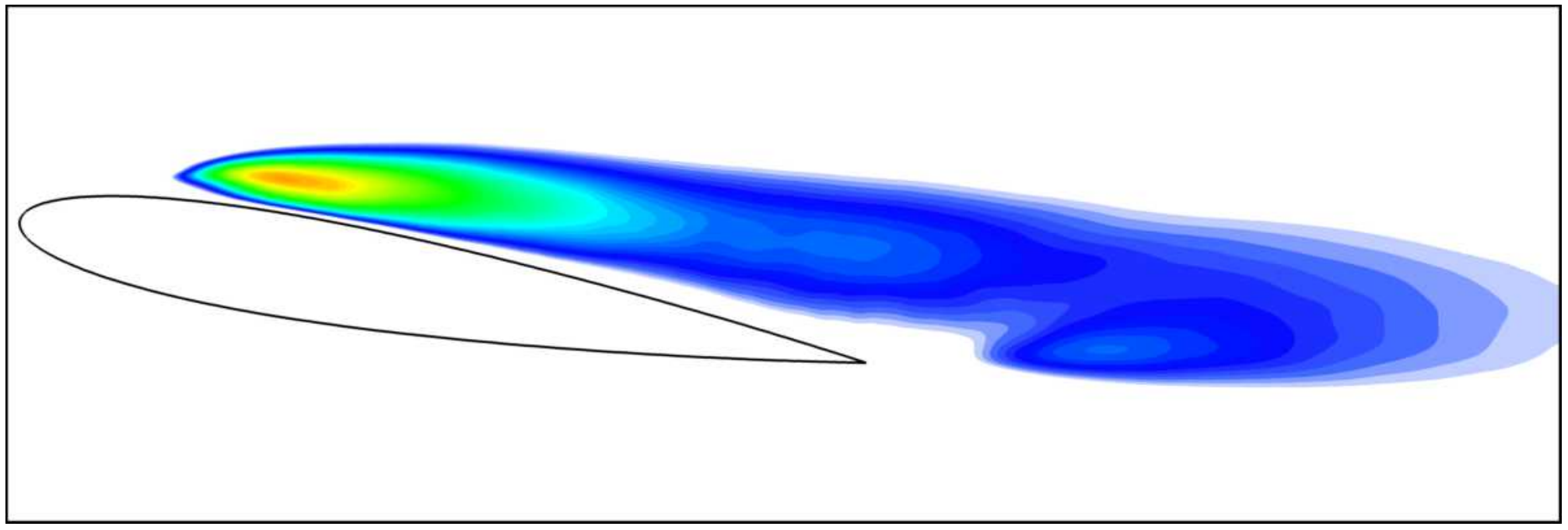}
\textit{$\alpha = 9.50^{\circ}$}
\end{minipage}
\begin{minipage}{220pt}
\centering
\includegraphics[width=220pt, trim={0mm 0mm 0mm 0mm}, clip]{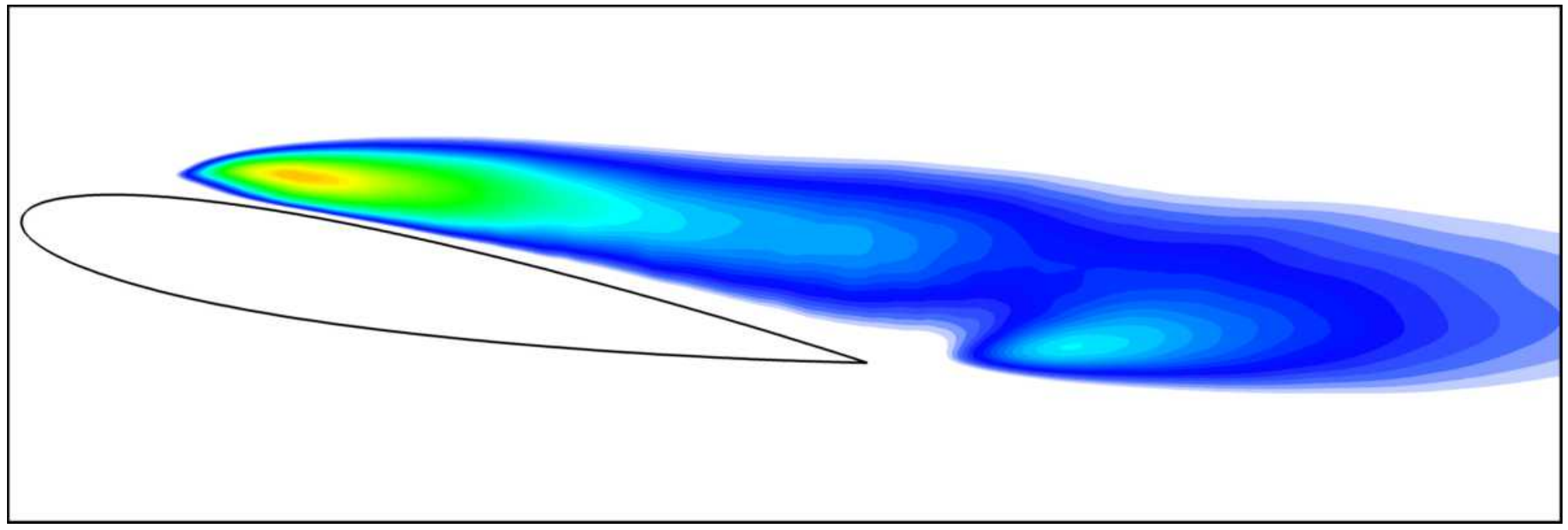}
\textit{$\alpha = 9.60^{\circ}$}
\end{minipage}
\begin{minipage}{220pt}
\centering
\includegraphics[width=220pt, trim={0mm 0mm 0mm 0mm}, clip]{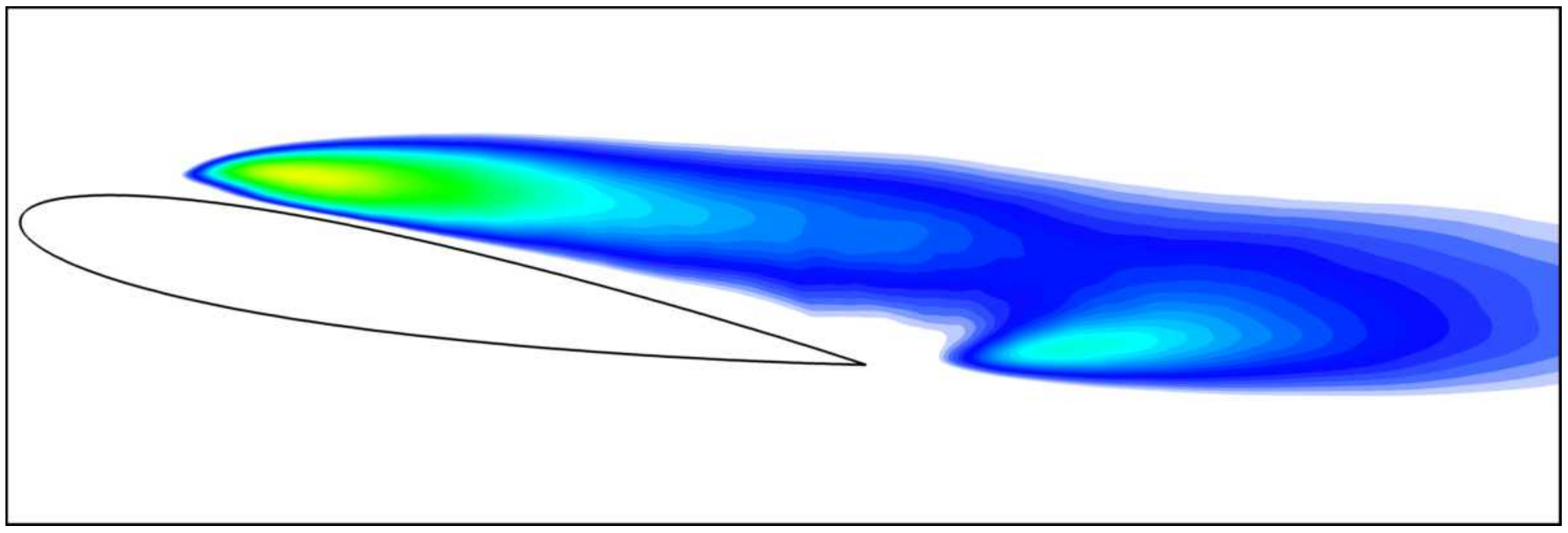}
\textit{$\alpha = 9.70^{\circ}$}
\end{minipage}
\begin{minipage}{220pt}
\centering
\includegraphics[width=220pt, trim={0mm 0mm 0mm 0mm}, clip]{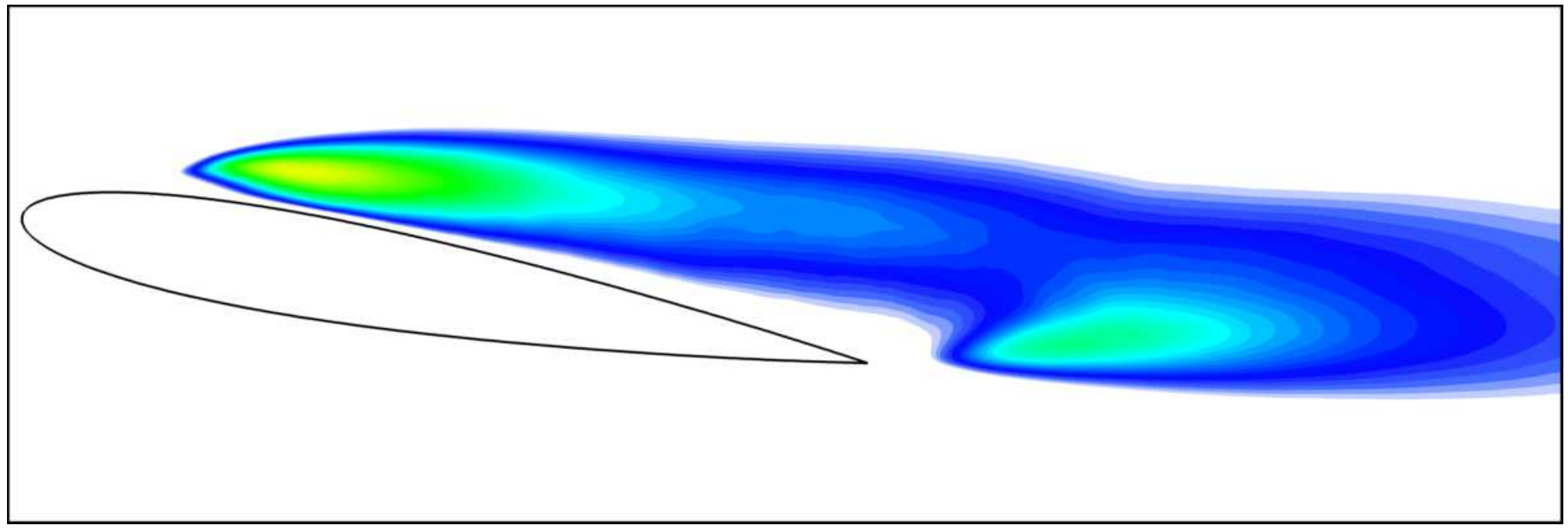}
\textit{$\alpha = 9.80^{\circ}$}
\end{minipage}
\begin{minipage}{220pt}
\centering
\includegraphics[width=220pt, trim={0mm 0mm 0mm 0mm}, clip]{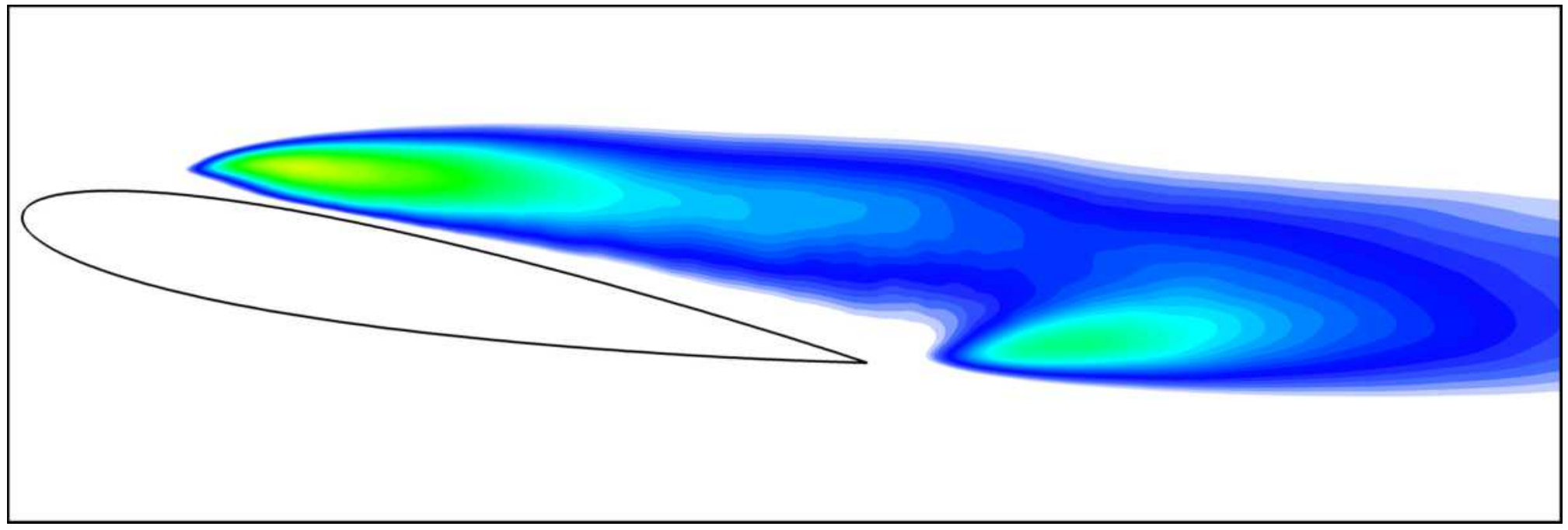}
\textit{$\alpha = 9.90^{\circ}$}
\end{minipage}
\begin{minipage}{220pt}
\centering
\includegraphics[width=220pt, trim={0mm 0mm 0mm 0mm}, clip]{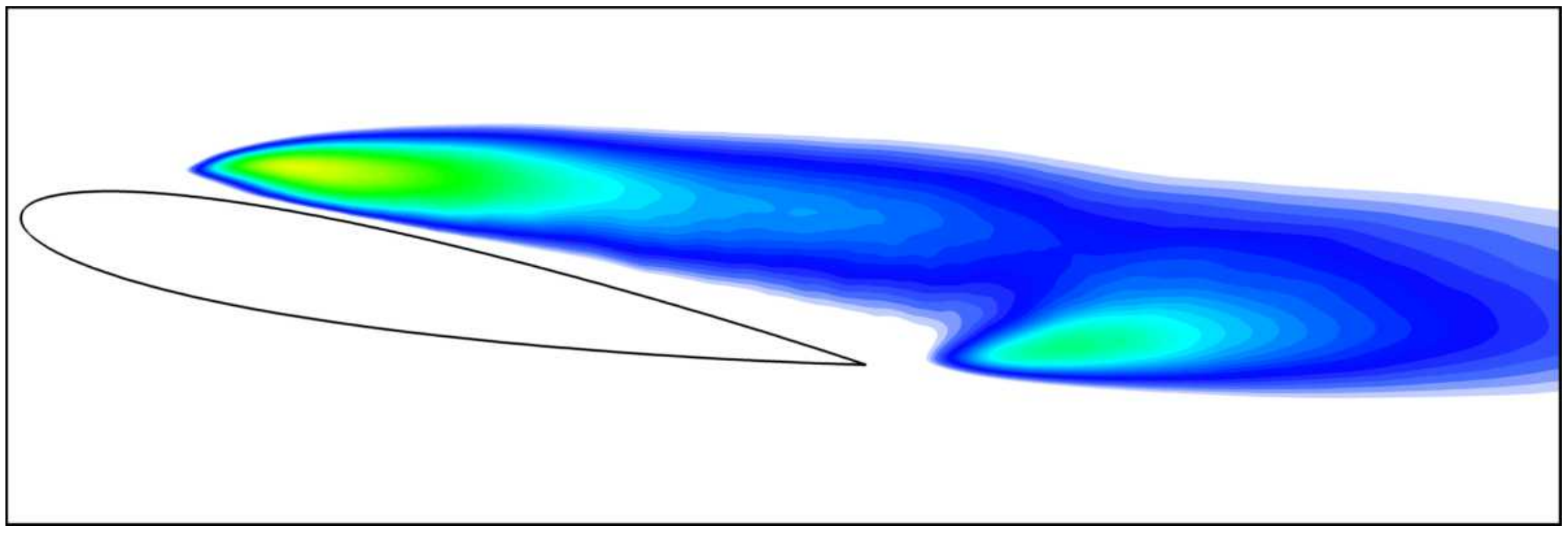}
\textit{$\alpha = 10.0^{\circ}$}
\end{minipage}
\begin{minipage}{220pt}
\centering
\includegraphics[width=220pt, trim={0mm 0mm 0mm 0mm}, clip]{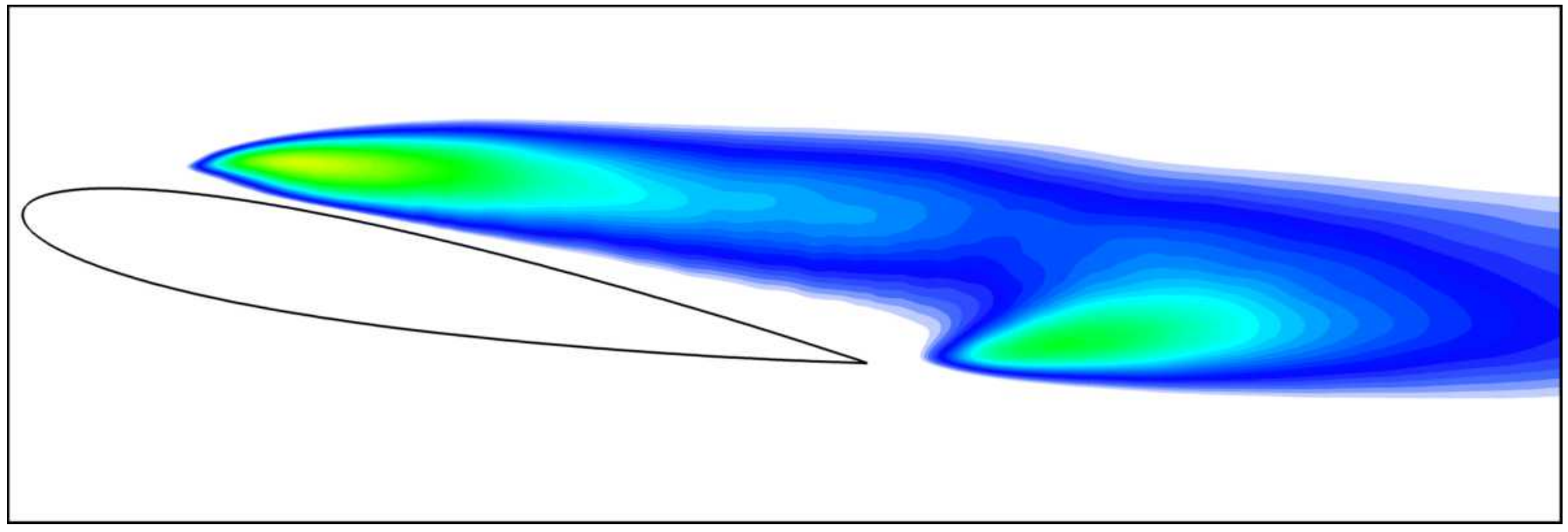}
\textit{$\alpha = 10.1^{\circ}$}
\end{minipage}
\begin{minipage}{220pt}
\centering
\includegraphics[width=220pt, trim={0mm 0mm 0mm 0mm}, clip]{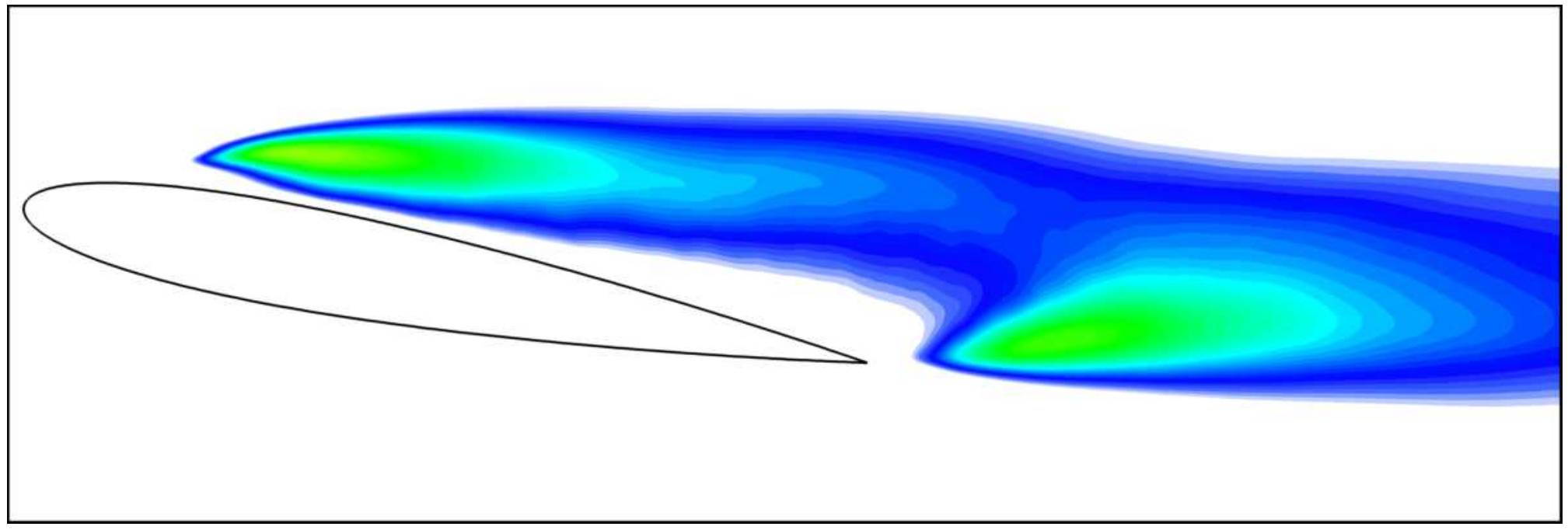}
\textit{$\alpha = 10.5^{\circ}$}
\end{minipage}
\caption{Colours map of the low-lift variance of the wall-normal velocity component, $\widecheck{{v\mydprime}^2}$, for the angles of attack $\alpha = 9.25^{\circ}$--$10.5^{\circ}$.}
\label{v2_below}
\end{center}
\end{figure}

\begin{figure}
\begin{center}
\begin{minipage}{220pt}
\includegraphics[height=125pt , trim={0mm 0mm 0mm 0mm}, clip]{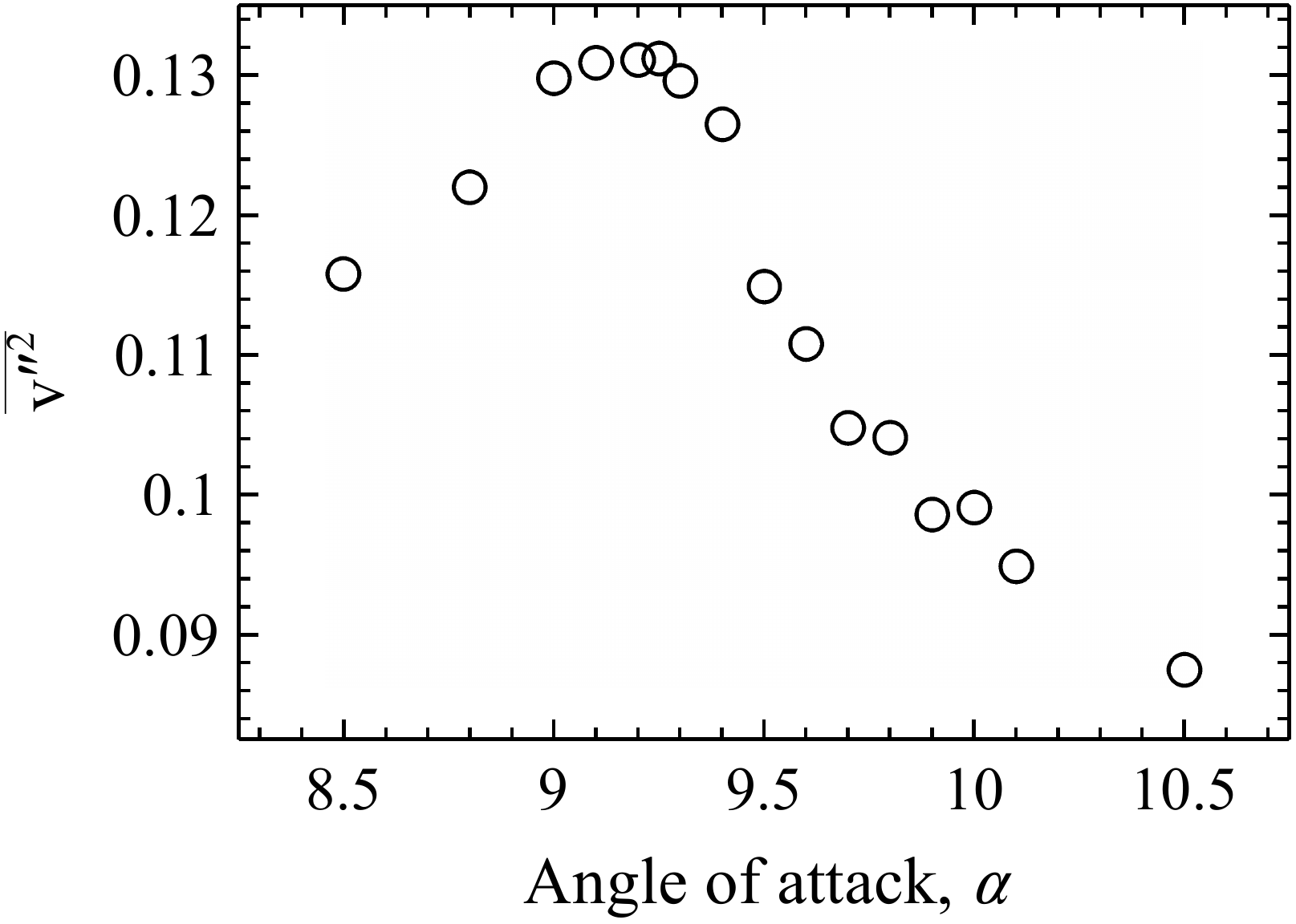}
\end{minipage}
\begin{minipage}{220pt}
\includegraphics[height=125pt , trim={0mm 0mm 0mm 0mm}, clip]{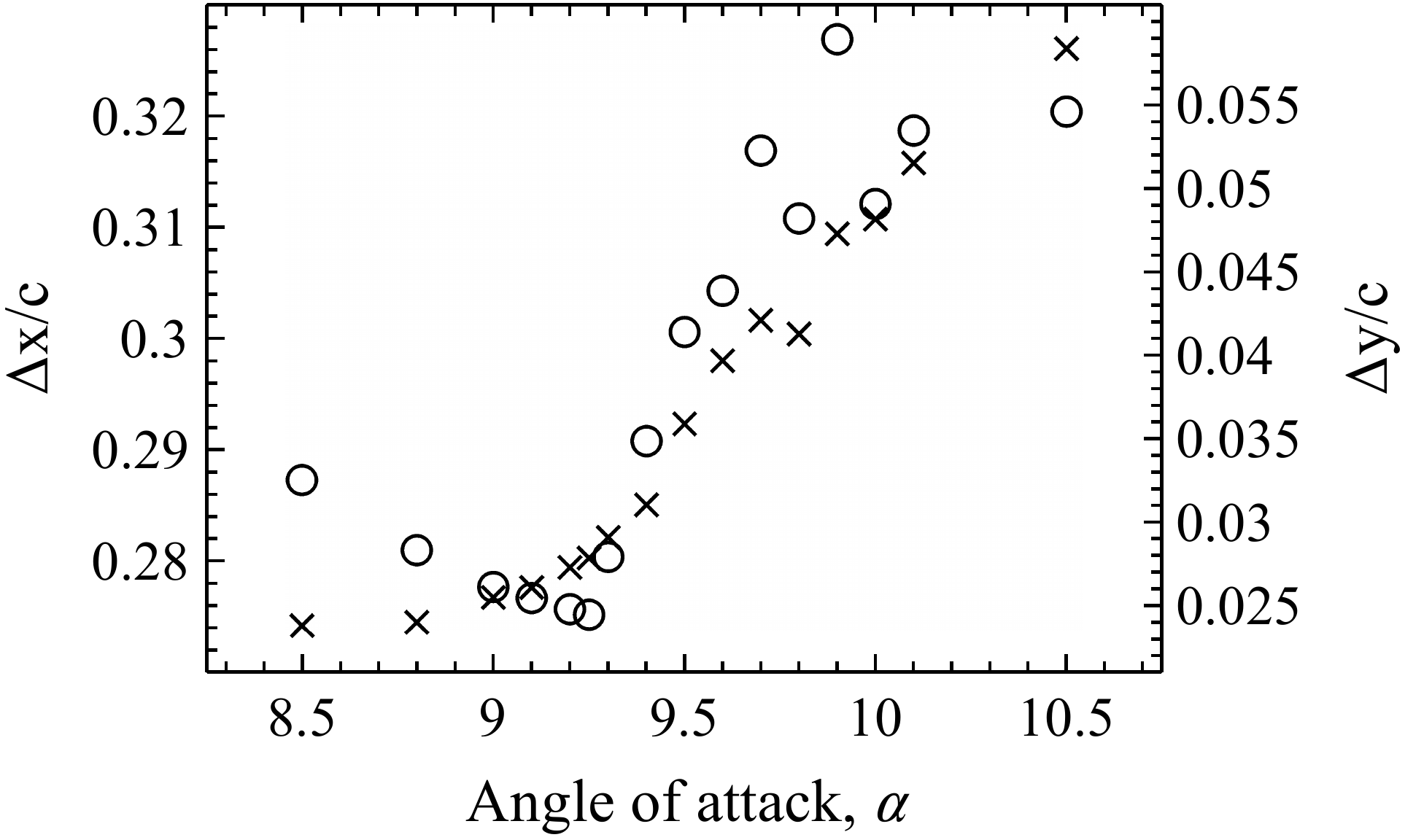}
\end{minipage}
\caption{Left: the maximum $\overline{{v\mydprime}^2}$ plotted versus the angle of attack $\alpha$. Right: the locations of the maximum $\overline{{v\mydprime}^2}$ plotted versus the angle of attack $\alpha$. Circles: $\Delta x/c$ measured from the aerofoil leading-edge, $\times$'s: $\Delta y/c$ measured from the aerofoil surface.}
\label{v2_max}
\end{center}
\end{figure}

\begin{figure}
\begin{center}
\begin{minipage}{220pt}
\centering
\includegraphics[width=220pt, trim={0mm 0mm 0mm 0mm}, clip]{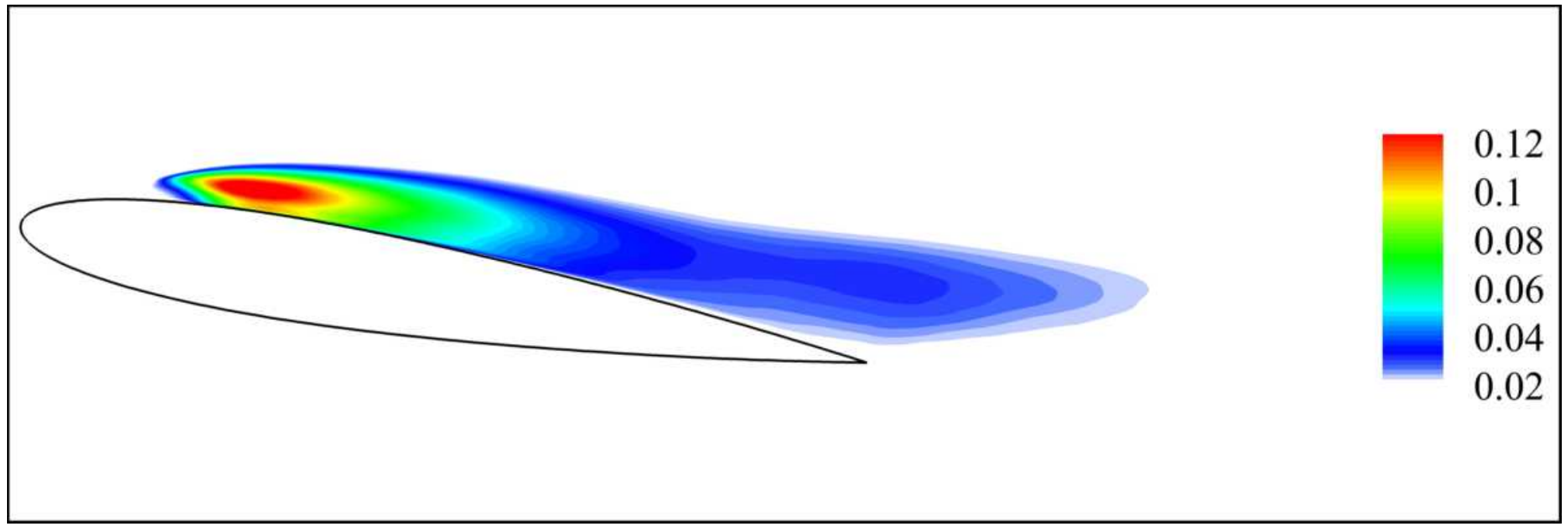}
\textit{$\alpha = 9.25^{\circ}$}
\end{minipage}
\begin{minipage}{220pt}
\centering
\includegraphics[width=220pt, trim={0mm 0mm 0mm 0mm}, clip]{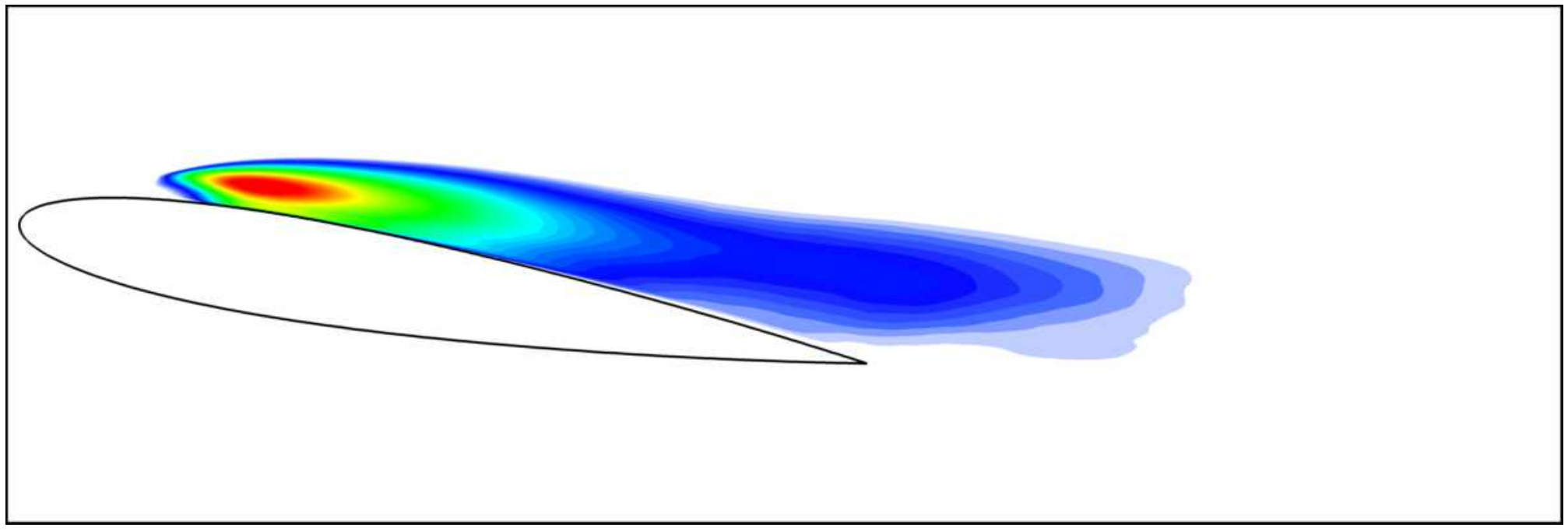}
\textit{$\alpha = 9.40^{\circ}$}
\end{minipage}
\begin{minipage}{220pt}
\centering
\includegraphics[width=220pt, trim={0mm 0mm 0mm 0mm}, clip]{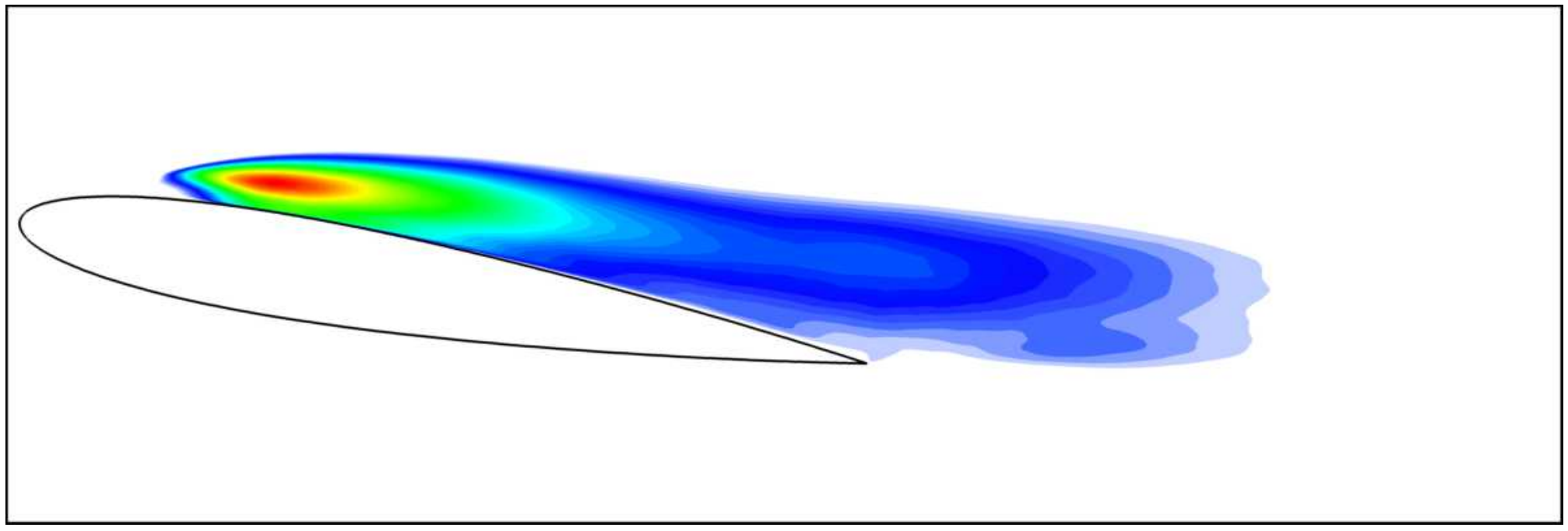}
\textit{$\alpha = 9.50^{\circ}$}
\end{minipage}
\begin{minipage}{220pt}
\centering
\includegraphics[width=220pt, trim={0mm 0mm 0mm 0mm}, clip]{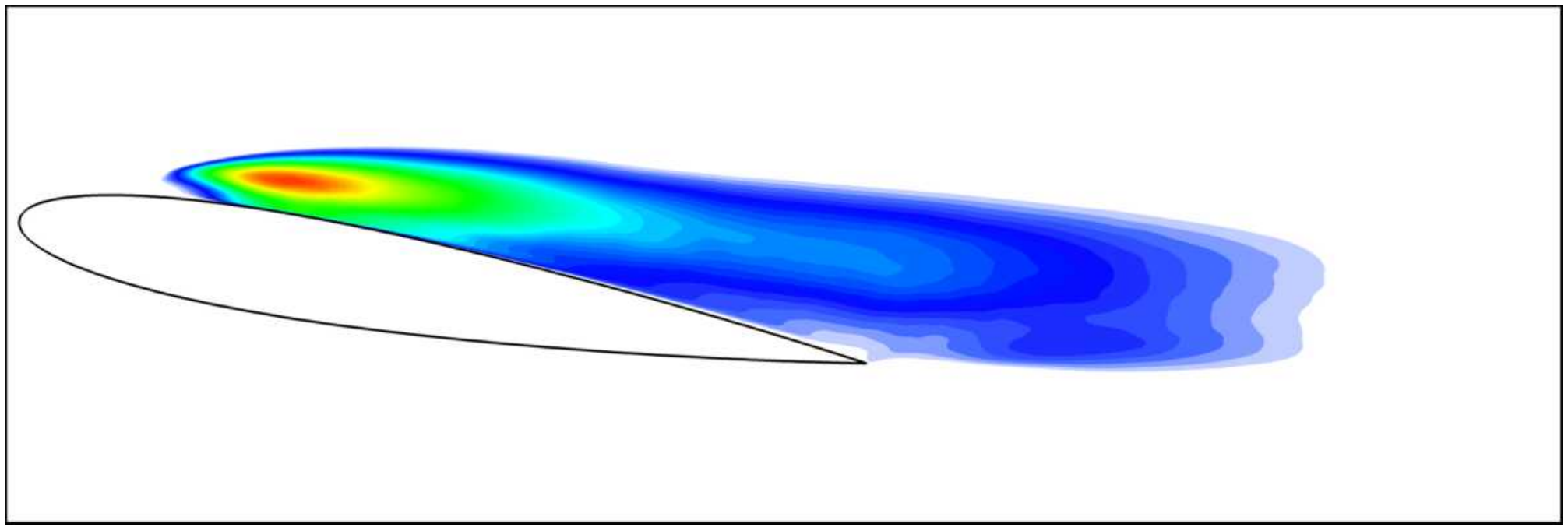}
\textit{$\alpha = 9.60^{\circ}$}
\end{minipage}
\begin{minipage}{220pt}
\centering
\includegraphics[width=220pt, trim={0mm 0mm 0mm 0mm}, clip]{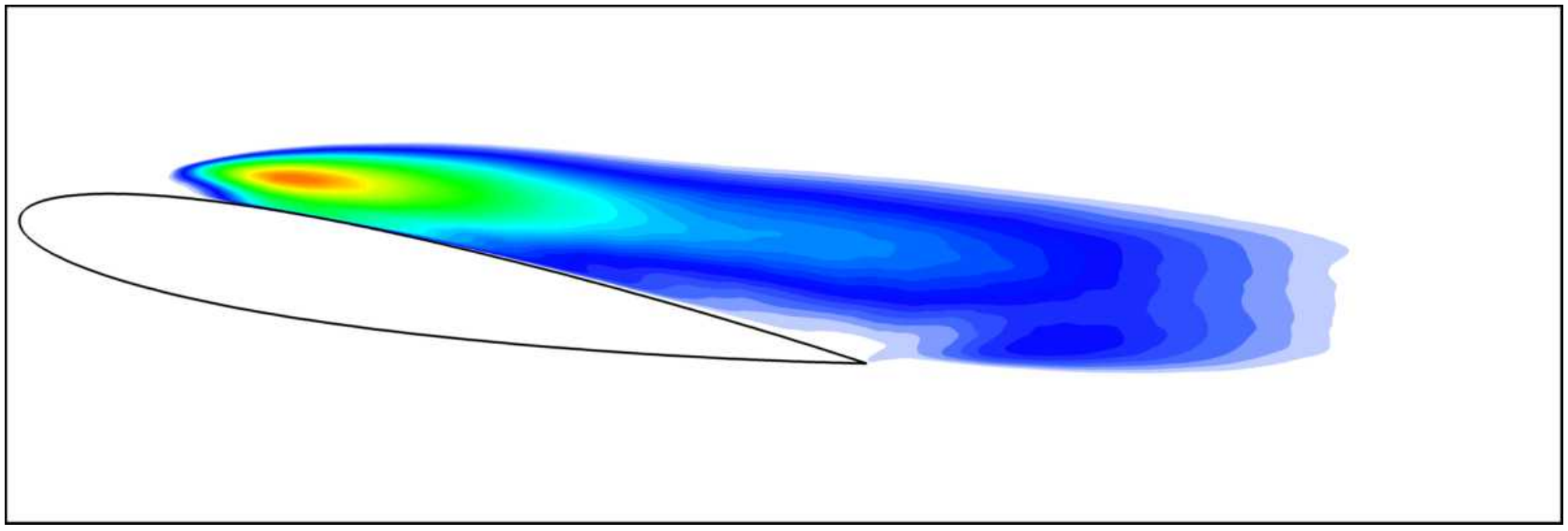}
\textit{$\alpha = 9.70^{\circ}$}
\end{minipage}
\begin{minipage}{220pt}
\centering
\includegraphics[width=220pt, trim={0mm 0mm 0mm 0mm}, clip]{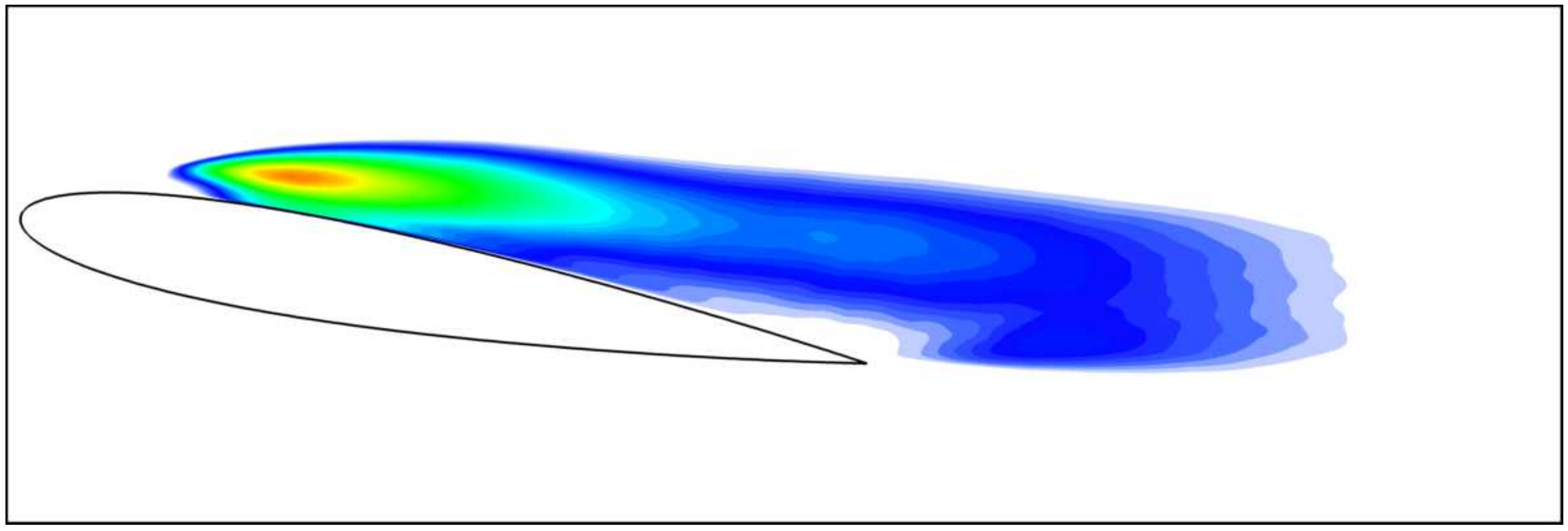}
\textit{$\alpha = 9.80^{\circ}$}
\end{minipage}
\begin{minipage}{220pt}
\centering
\includegraphics[width=220pt, trim={0mm 0mm 0mm 0mm}, clip]{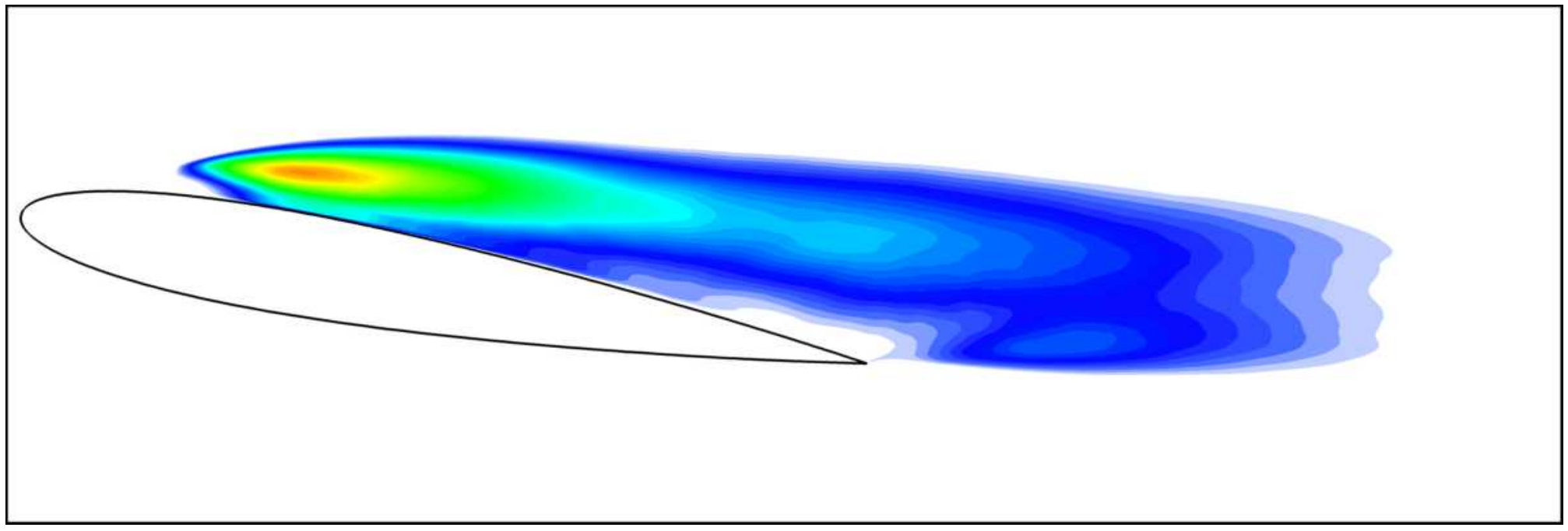}
\textit{$\alpha = 9.90^{\circ}$}
\end{minipage}
\begin{minipage}{220pt}
\centering
\includegraphics[width=220pt, trim={0mm 0mm 0mm 0mm}, clip]{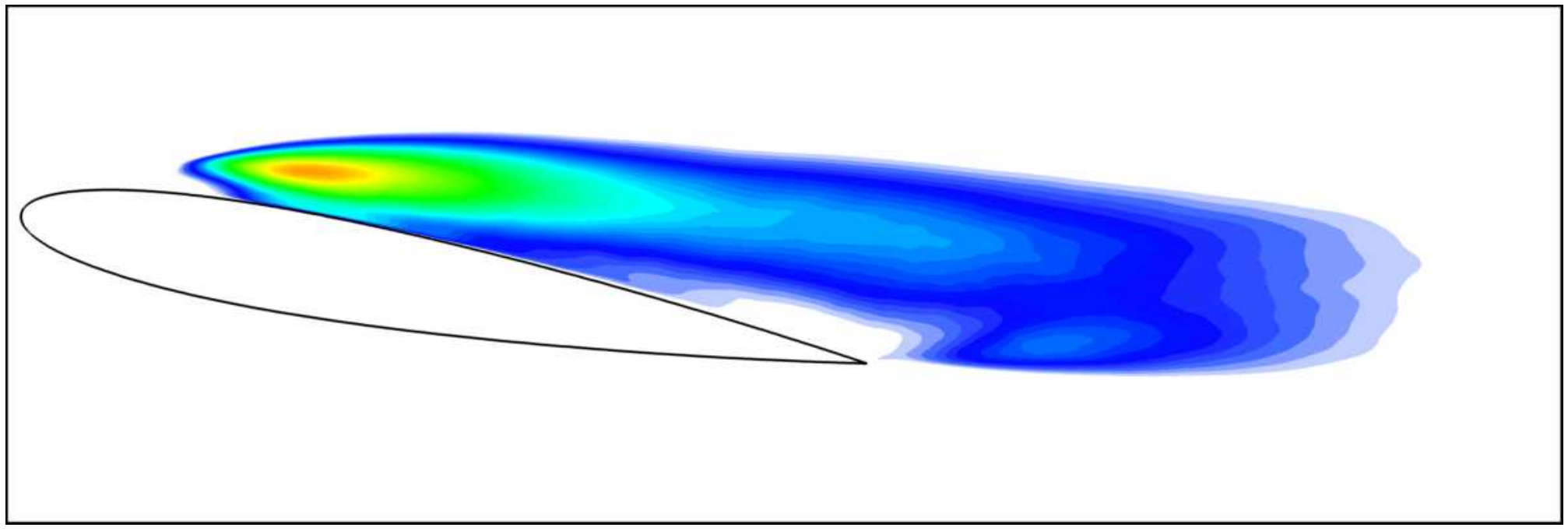}
\textit{$\alpha = 10.0^{\circ}$}
\end{minipage}
\begin{minipage}{220pt}
\centering
\includegraphics[width=220pt, trim={0mm 0mm 0mm 0mm}, clip]{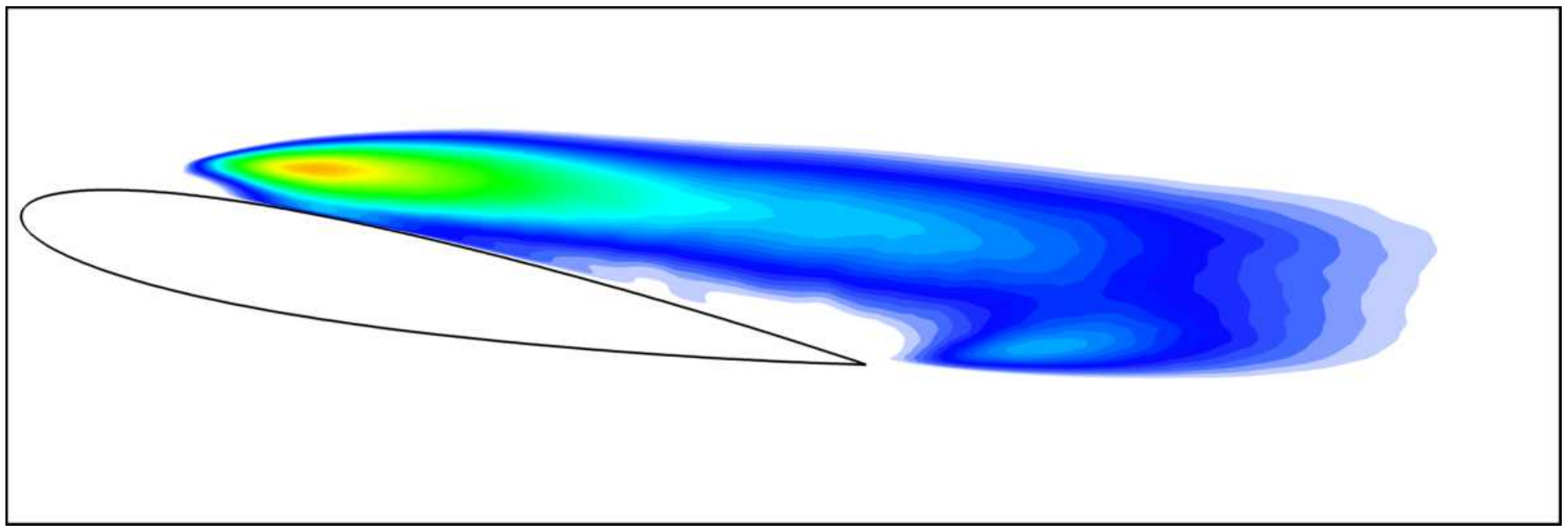}
\textit{$\alpha = 10.1^{\circ}$}
\end{minipage}
\begin{minipage}{220pt}
\centering
\includegraphics[width=220pt, trim={0mm 0mm 0mm 0mm}, clip]{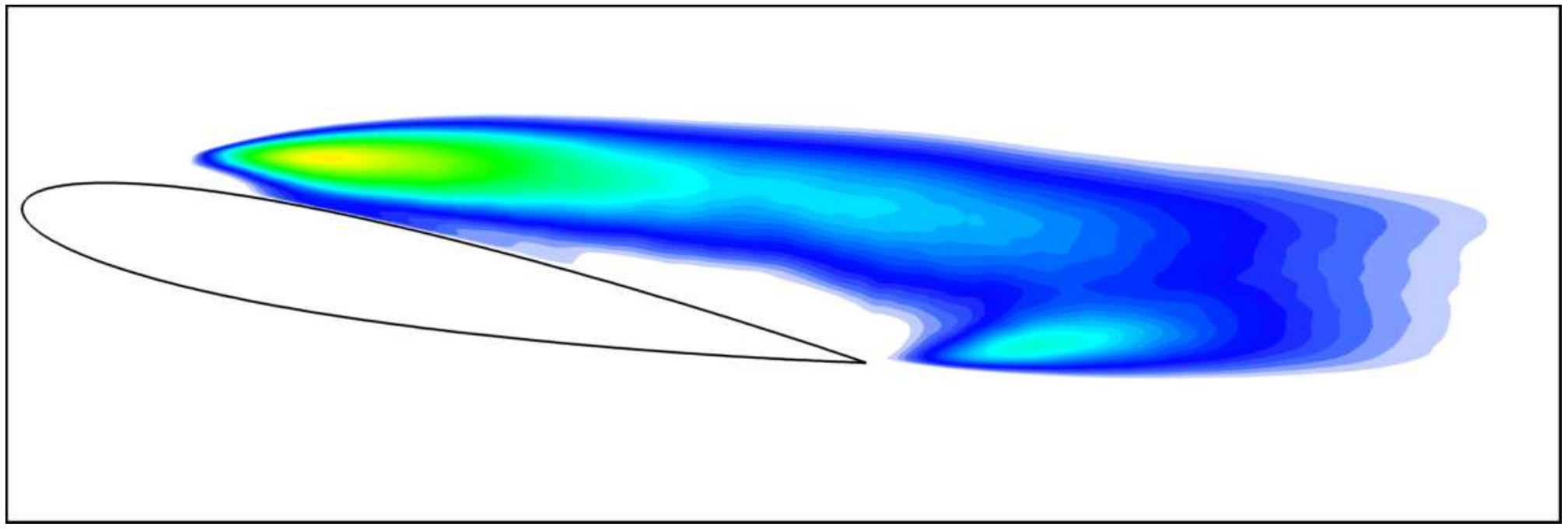}
\textit{$\alpha = 10.5^{\circ}$}
\end{minipage}
\caption{Colours map of the variance of the spanwise velocity component, $\overline{{w\mydprime}^2}$, for the angles of attack $\alpha = 9.25^{\circ}$--$10.5^{\circ}$.}
\label{w2_mean}
\end{center}
\end{figure}

\begin{figure}
\begin{center}
\begin{minipage}{220pt}
\includegraphics[height=125pt , trim={0mm 0mm 0mm 0mm}, clip]{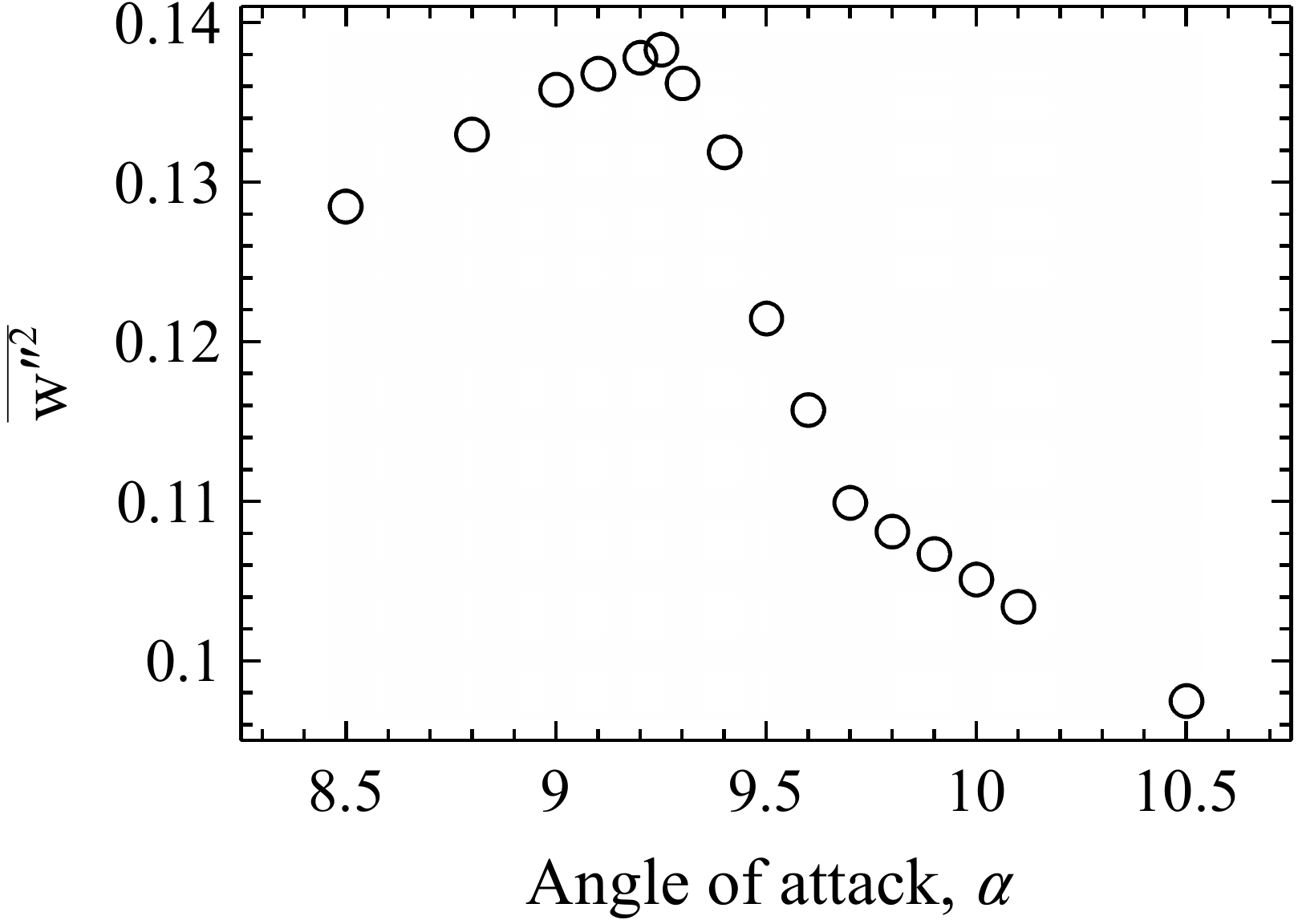}
\end{minipage}
\begin{minipage}{220pt}
\includegraphics[height=125pt , trim={0mm 0mm 0mm 0mm}, clip]{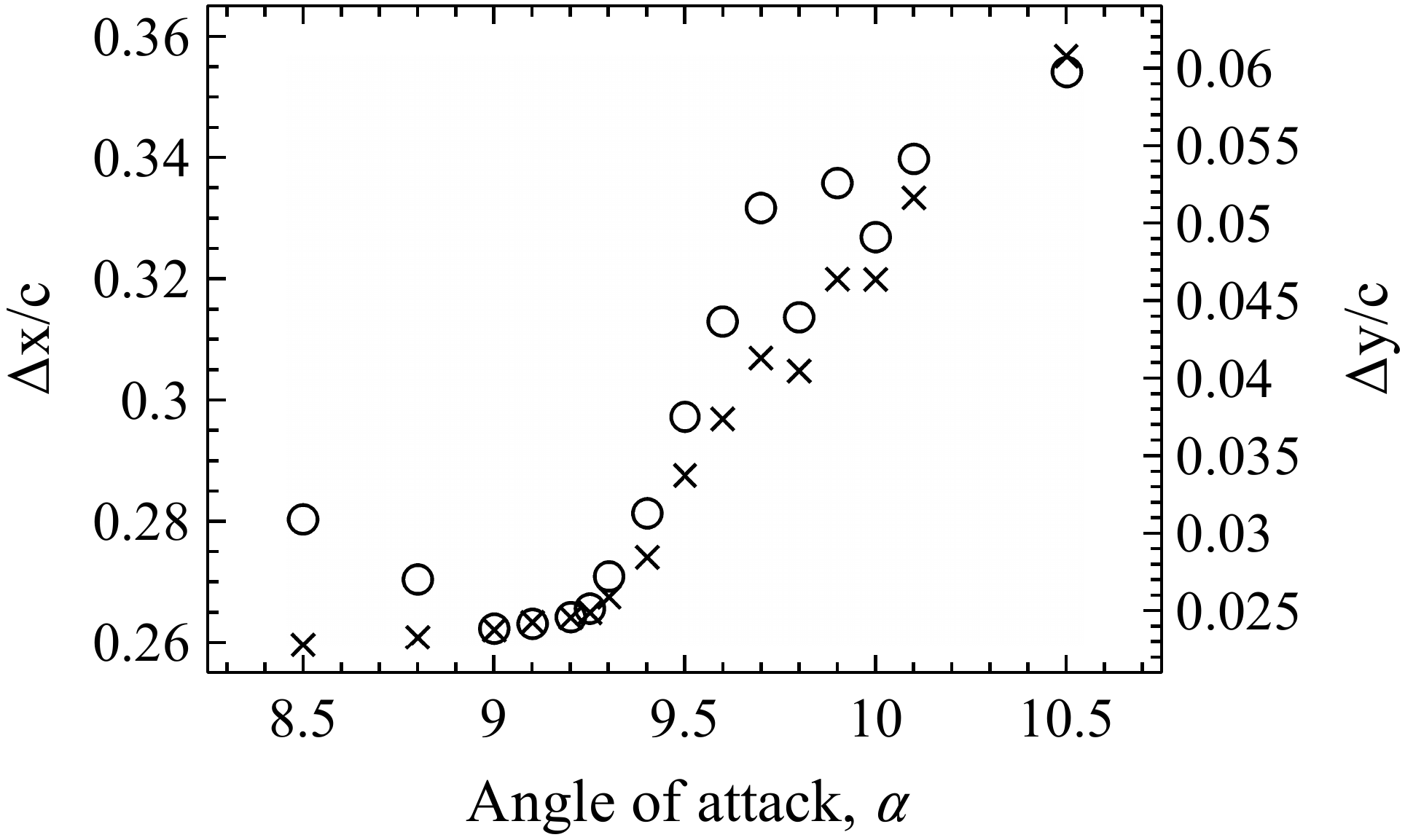}
\end{minipage}
\caption{Left: the maximum $\overline{{w\mydprime}^2}$ plotted versus the angle of attack $\alpha$. Right: the locations of the maximum $\overline{{w\mydprime}^2}$ plotted versus the angle of attack $\alpha$. Circles: $\Delta x/c$ measured from the aerofoil leading-edge, $\times$'s: $\Delta y/c$ measured from the aerofoil surface.}
\label{w2_max}
\end{center}
\end{figure}

Most of the cited studies attribute the instability of the LSB to the process of transition to turbulence along the separated shear layer. These studies start by connecting the transition process to some type of linear instability via a Tollmien-Schlichting instability in the boundary layer or a Kelvin--Helmholtz instability in the separated shear layer above the LSB. Since velocity prof\/iles in the separated region possess an inf\/lection point, amplif\/ication of the two-dimensional disturbances is often attributed to the latter type of instability. The most widely accepted theory is that during the separated phase of the LFO cycle, acoustic waves due to the intense vortex shedding at the trailing-edge propagate upstream and generate hydrodynamic disturbances, several orders of magnitude larger than the acoustic disturbances~\citet{jones2010stability}. Amplif\/ication of these disturbances within the laminar separated shear layer near the leading-edge forces earlier transition, and causing the f\/low to reattach. As the f\/low attaches, the strong vortex shedding at the trailing-edge dies down, and the acoustic feedback is cut-off. Hence, the transition is delayed, and the f\/low eventually separates. This self-sustained process repeats itself in a periodic manner.\newline
\citet{marxen2007numerical} and~\citet{marxen2008direct} studied bubble bursting and how it is related to aerofoil stall. They hypothesised that changes in the transition process and the viscous-inviscid interaction both had a major role in bubble bursting. They performed a systematic study of bubble response to incoming disturbances, and they concluded that bubble bursting occurs when incoming disturbances are switched off. They hypothesised that bursting takes place when saturated disturbances cannot drive the f\/low to reattach. Recently,~\citet{alferez2013study} studied the mechanism by which the LSB bursts and causes stall. They concluded that the bursting process is due to a change in the stability characteristics of the separated shear layer rather than a change of the instability mode from convective instability to absolute instability. Furthermore,~\citet{almutairi2013large} have shown that the location of transition moves downstream and away from the aerofoil surface during bubble bursting and upstream and close to the aerofoil surface during the reformation of the separation bubble.~\citet{almutairi2015large} and~\citet{almutairi2017dynamics} applied Dynamic Mode Decomposition (DMD) to the pressure f\/ield of the f\/low around a NACA-0012 at $\rec = 1.3\times10^5$, $\minf = 0.4$, and at the angle of attack of $11.5^{\circ}$. They concluded that the trailing-edge shedding generates acoustic waves that travel upstream and force early transition. The authors have suggested that the frequency of the trailing-edge shedding can be used to remove the LFO through periodic forcing.~\citet{eljack2017high} has shown, qualitatively, that the location of transition is connected to the trailing-edge shedding. However, in neither of these works it was established that the strong trailing-edge shedding is the necessary and suff\/icient condition for the early/late transition.\newline
Before diving into the details of the changes inside the separated shear layer and how it affects the stability of the LSB, the authors would like to clarify the nature of the kinetic energy to be discussed. Amplif\/ication of small perturbations and extraction of energy from the mean f\/low and feeding it into the f\/luctuations does not mean that the f\/low undergoes transition to turbulence. In other words, three-dimensionality and several spectral tones of the f\/luctuations do not imply that the f\/low is turbulent. Therefore, the kinetic energy will be discussed in this context unless stated otherwise. The transition location as to early and late transition will be discussed in the context of the location of the maximum turbulent kinetic energy or the location of the maximum variance of a specif\/ic f\/low variable.\newline
F\/igures~\ref{u2_mean},~\ref{u2_above}, and~\ref{u2_below} illustrate the mean, high-lift, and low-lift variance of the streamwise velocity component, $\overline{{u\mydprime}^2}$, $\widehat{{u\mydprime}^2}$, and $\widecheck{{u\mydprime}^2}$, respectively. At the angle of attack of $9.25^{\circ}$, the distribution of $\overline{{u\mydprime}^2}$ coincides with the separated shear layer and the reattachment region. The bubble is intact, and the location of the maximum $\overline{{u\mydprime}^2}$ is just downstream the maximum bubble height. As the angle of attack is increased, the maximum magnitude of $\overline{{u\mydprime}^2}$ decreases, and its location moves downstream and away from the aerofoil surface. For the case of the attached f\/low, in the mean sense, $\widehat{{u\mydprime}^2}$ have higher values compared to its corresponding $\overline{{u\mydprime}^2}$. The locations of the maximum values of $\widehat{{u\mydprime}^2}$ are closer to the aerofoil surface and are located upstream their corresponding $\overline{{u\mydprime}^2}$ locations. As for the mean separated f\/low, the magnitudes of $\widecheck{{u\mydprime}^2}$ are lower than their $\overline{{u\mydprime}^2}$ counterparts. The maximum values of $\widecheck{{u\mydprime}^2}$ has less magnitude and locates further away from the aerofoil surface and downstream their corresponding $\overline{{u\mydprime}^2}$. It is noted that as the angle of attack is increased, the f\/luctuation level of the streamwise velocity component increases signif\/icantly in the vicinity of the trailing-edge as can be seen in f\/igure~\ref{u2_below}. The oscillations are intensif\/ied during the separated phase (f\/igure~\ref{u2_below}) and die down during the attached phase (f\/igure~\ref{u2_above}). This is in total agreement with previous investigations~\citet{jones2010stability},~\citet{almutairi2013large},~\citet{almutairi2017dynamics}, and~\citet{eljack2017high}. However, does this implies that this is a suff\/icient condition for the instability of the LSB? This study found no proof that this condition being suff\/icient for the instability of the LSB. However, this is a strong indication that the instability behind the LFO is in action.\newline
F\/igure~\ref{u2_max} (left) shows the maximum values of $\overline{{u\mydprime}^2}$ plotted versus the angle of attack. As seen in the f\/igure, $\overline{{u\mydprime}^2}$ increases linearly and sharply with the angle of attack and reaches a maximum at $\alpha = 9.4^{\circ}$. After that, $\overline{{u\mydprime}^2}$ decreases linearly as the angle of attack is further increased with systematic jumps at consistent intervals. The locations of the maximum $\overline{{u\mydprime}^2}$ are plotted versus the angle of attack as illustrated in the right-hand side of f\/igure~\ref{u2_max}. $\Delta x/c$ is the distance measured from the aerofoil leading-edge in $x$ direction, and $\Delta y/c$ is the distance measured from the aerofoil surface in $y$ direction. It is noted that the location of the maximum $\overline{{u\mydprime}^2}$ moves away from the aerofoil surface as the angle of attack is increased. As for the distance in the $x$ direction, it is found that the location of the maximum $\overline{{u\mydprime}^2}$ moves slowly towards the leading-edge and becomes closest to it at $\alpha = 9.25^{\circ}$. After that, the location of the maximum $\overline{{u\mydprime}^2}$ moves downstream. Again, this is in total agreement with previous studies. However, it is discussed here as a collateral outcome of the instability rather than being the reason that triggers the instability.\newline
The distributions of the variance of the wall-normal velocity component, $\overline{{v\mydprime}^2}$, are qualitatively and quantitatively different than their corresponding streamwise component, as seen in f\/igures~\ref{v2_mean},~\ref{v2_above}, and~\ref{v2_below}. In the case of $\overline{{u\mydprime}^2}$, the variance has signif\/icant magnitude over all of the suction surface of the aerofoil including the boundary layer and the separated shear layer. At the trailing-edge, the streamwise velocity f\/luctuations have considerable amplitude. Whereas in the $\overline{{v\mydprime}^2}$ case, the f\/luctuations are signif\/icant away from the aerofoil surface and further downstream towards the trailing-edge. This is indicative that $\overline{{u\mydprime}^2}$ is the driving component in the separated f\/low in the chord-wise direction, whereas, $\overline{{v\mydprime}^2}$ act as a modulator of the reversed f\/low with no f\/lapping in the wall-normal direction. It is also noted that the level of f\/luctuations of $\overline{{v\mydprime}^2}$ at the trailing-edge is comparable to that in the vicinity of the leading-edge. The oscillating mode is mostly attributed to the f\/luctuations in the wall-normal direction.\newline
F\/igure~\ref{v2_max} (left) shows that $\overline{{v\mydprime}^2}$ reaches a maximum value at the angle of attack of $9.25^{\circ}$. This is considerably lower than the angle of maximum $\overline{{u\mydprime}^2}$. The distances of the location of the maximum $\overline{{v\mydprime}^2}$ measured in $x$ direction from the leading-edge and in $y$ direction measured from the aerofoil surface are shown in the right-hand side of f\/igure~\ref{v2_max}. These distances mimic those of $\overline{{u\mydprime}^2}$. The variance of the spanwise velocity component is similar to that of the wall-normal direction both quantitatively and qualitatively as seen in f\/igure~\ref{w2_mean}. The f\/low in this direction is periodic. Hence, the mean velocity component in this direction is zero. The only acting velocity in this direction is the spanwise perturbations. These are responsible for the break-down of the two-dimensional rolls into three-dimensional structures.\newline
The conditional time-averaging employed in the present study does not distinguish the small-scale perturbations, ``turbulence'', from oscillations induced by large-scale vortices. However, the behaviour of both $\widecheck{{v\mydprime}^2}$ and $\widecheck{{w\mydprime}^2}$ coincides with the Reynolds stress behaviour. Therefore, the f\/luctuations of both the spanwise and the wall-normal velocity components are mostly from the small scales, whereas, the f\/luctuations of the streamwise velocity component is a combination of small scales and large vortices. That is, $\widecheck{{v\mydprime}^2}$ and $\widecheck{{w\mydprime}^2}$ are extracted from the mean f\/low exclusively by the Reynolds stress. Whereas, a global f\/low oscillation extracts $\overline{{u\mydprime}^2}$ in addition to the Reynolds stress.\newline
F\/igure~\ref{w2_max} (left) illustrates the variation of the maximum $\overline{{w\mydprime}^2}$ with the angle of attack. The plot shows three distinct ranges of angles of attack. For the angles of attack $\alpha < 9.25^{\circ}$, the maximum value of $\overline{{w\mydprime}^2}$ increases linearly with the angle of attack. In this range of angles of attack, the f\/low is attached and the LSB is present and intact. The second range of angles of attack is $9.25^{\circ} \leq \alpha \leq 9.6^{\circ}$. This is the range where the LSB becomes unstable and the bursting and reformation cycle starts to develop. As seen in the f\/igure, the maximum value of $\overline{{w\mydprime}^2}$ decreases rapidly and linearly as the angle of attack is increased. In the third range, $\alpha \geq 9.7^{\circ}$, the maximum value of $\overline{{w\mydprime}^2}$ decreases slowly and linearly as the angle of attack is increased. In this range of angles of attack, the bursting and reformation cycle of the LSB becomes fully developed. The variation of the locations of the maximum $\overline{{w\mydprime}^2}$ with the angle of attack is illustrated in f\/igure~\ref{w2_max} (right). The locations are closer to the aerofoil surface compared to their $\overline{{u\mydprime}^2}$ and $\overline{{v\mydprime}^2}$ counterparts. The maximum value of $\overline{{w\mydprime}^2}$ moves upstream towards the aerofoil leading-edge until the angle of attack is $9.0^{\circ}$. After that, it moves downstream towards the trailing-edge.

\subsection{Reynolds stress}
\begin{figure}
\begin{center}
\begin{minipage}{220pt}
\centering
\includegraphics[width=220pt, trim={0mm 0mm 0mm 0mm}, clip]{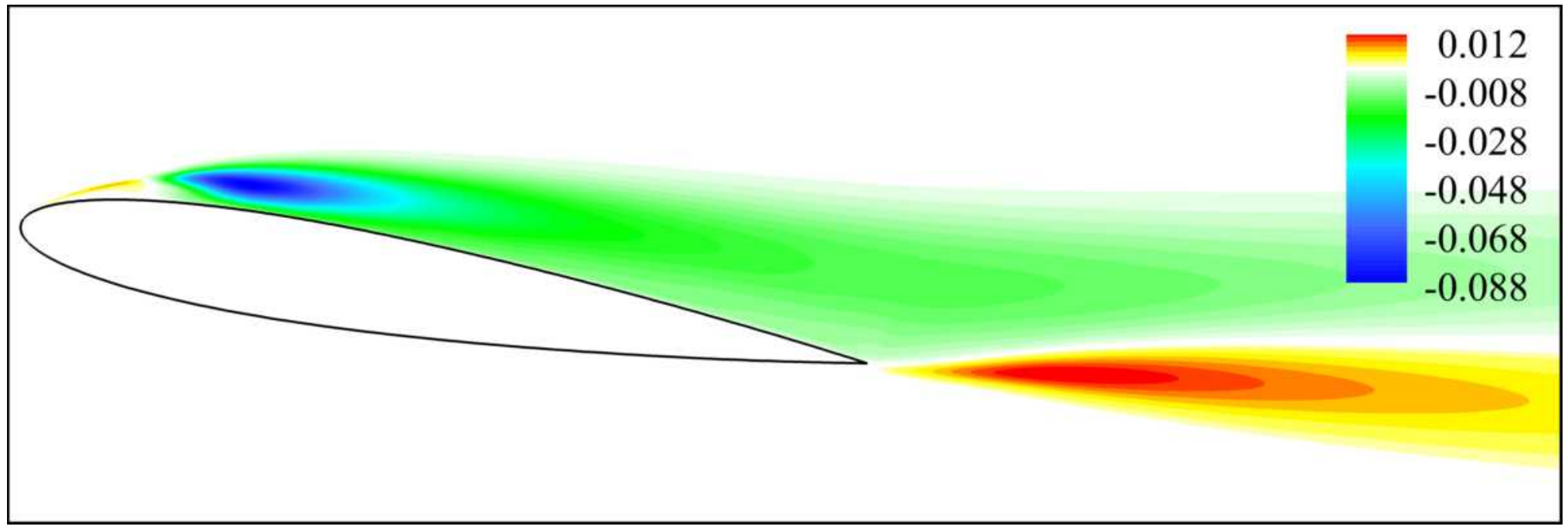}
\textit{$\alpha = 9.25^{\circ}$}
\end{minipage}
\begin{minipage}{220pt}
\centering
\includegraphics[width=220pt, trim={0mm 0mm 0mm 0mm}, clip]{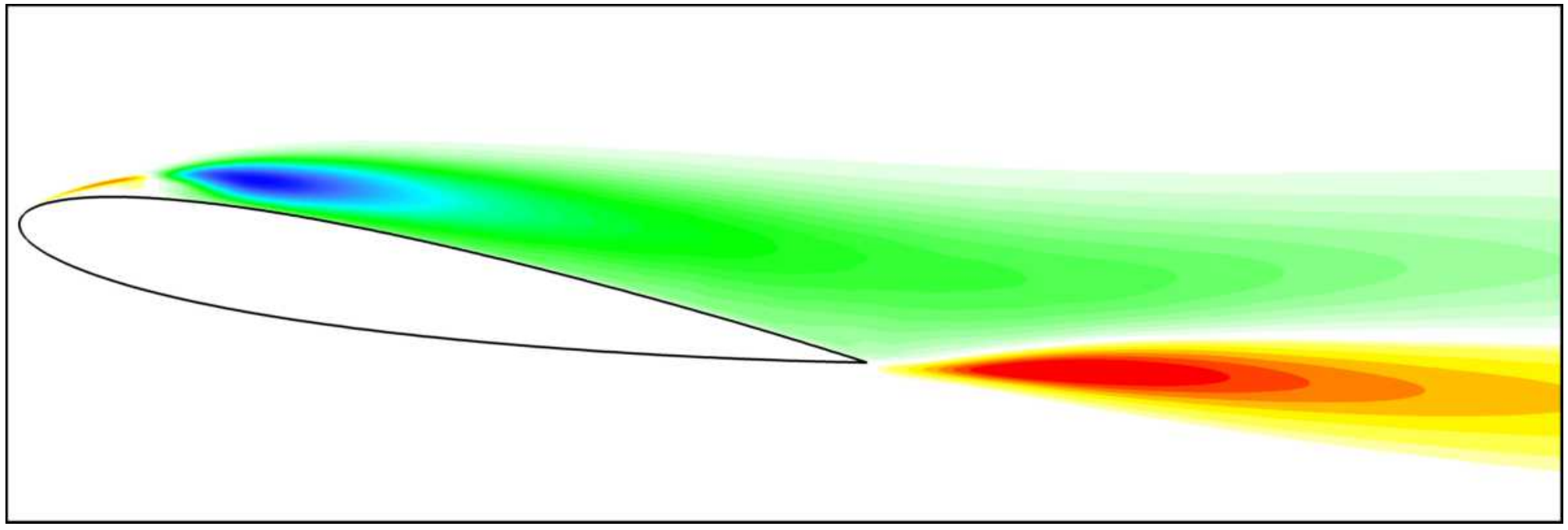}
\textit{$\alpha = 9.40^{\circ}$}
\end{minipage}
\begin{minipage}{220pt}
\centering
\includegraphics[width=220pt, trim={0mm 0mm 0mm 0mm}, clip]{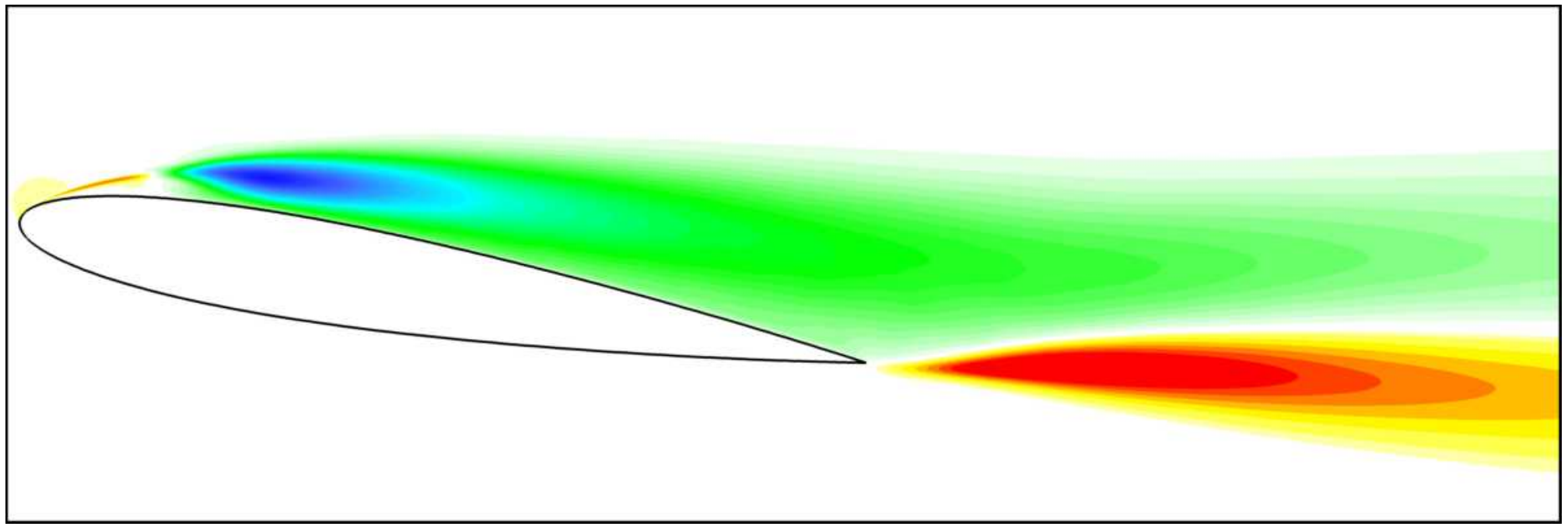}
\textit{$\alpha = 9.50^{\circ}$}
\end{minipage}
\begin{minipage}{220pt}
\centering
\includegraphics[width=220pt, trim={0mm 0mm 0mm 0mm}, clip]{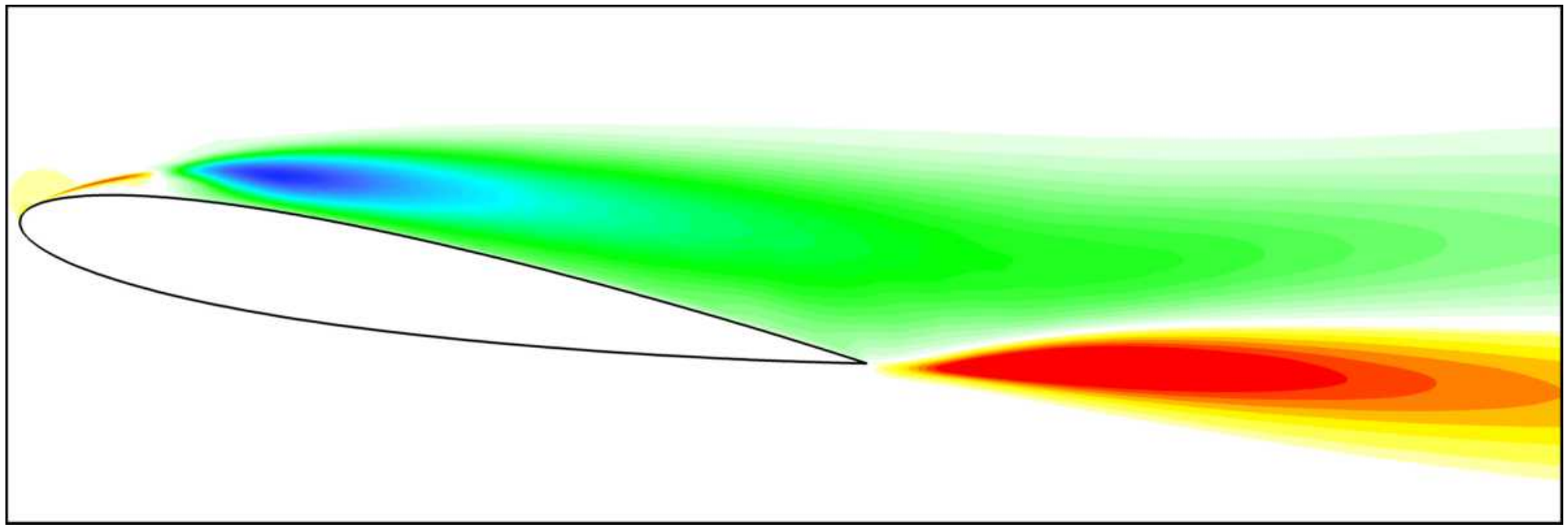}
\textit{$\alpha = 9.60^{\circ}$}
\end{minipage}
\begin{minipage}{220pt}
\centering
\includegraphics[width=220pt, trim={0mm 0mm 0mm 0mm}, clip]{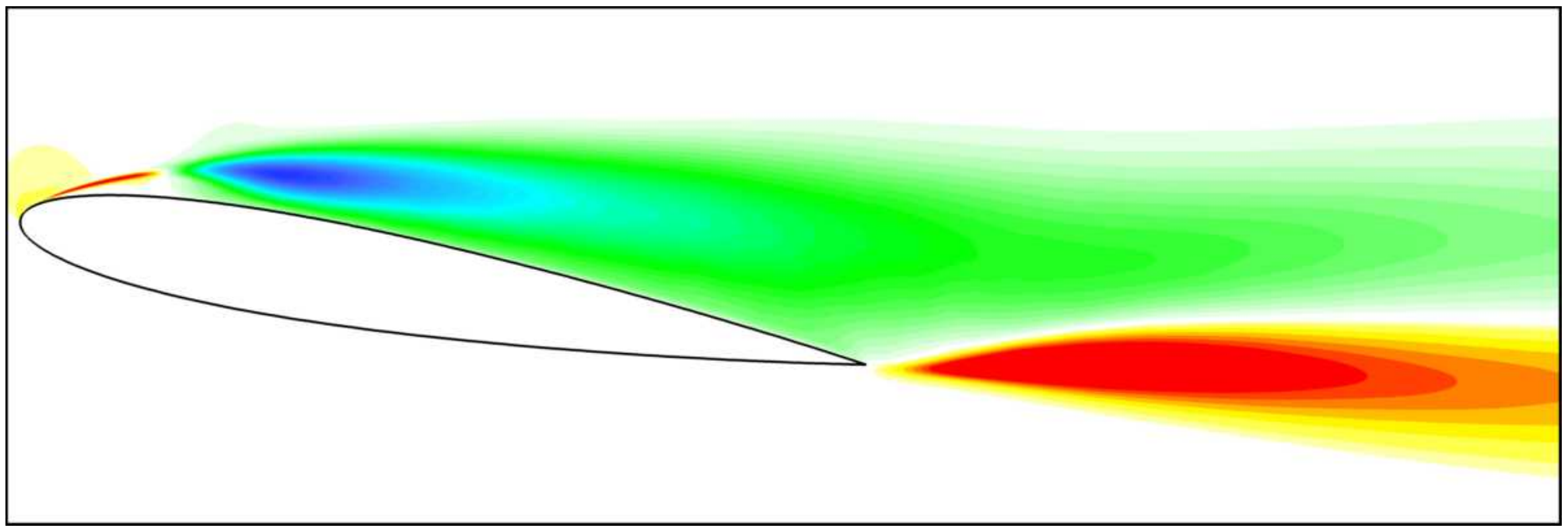}
\textit{$\alpha = 9.70^{\circ}$}
\end{minipage}
\begin{minipage}{220pt}
\centering
\includegraphics[width=220pt, trim={0mm 0mm 0mm 0mm}, clip]{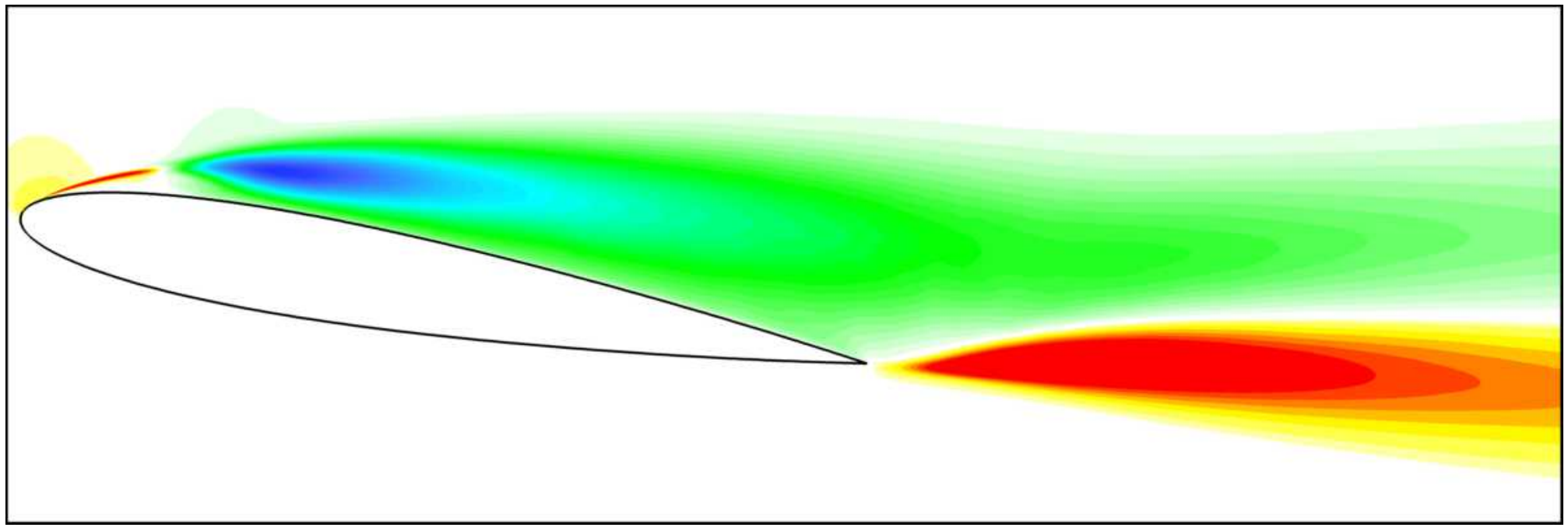}
\textit{$\alpha = 9.80^{\circ}$}
\end{minipage}
\begin{minipage}{220pt}
\centering
\includegraphics[width=220pt, trim={0mm 0mm 0mm 0mm}, clip]{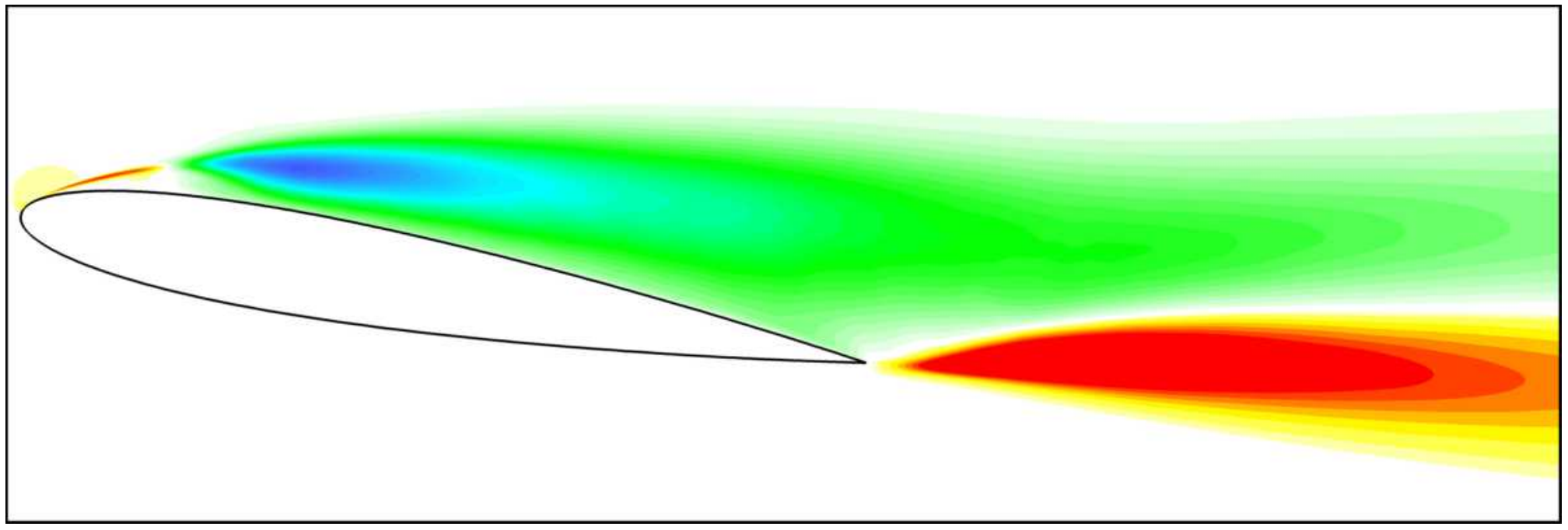}
\textit{$\alpha = 9.90^{\circ}$}
\end{minipage}
\begin{minipage}{220pt}
\centering
\includegraphics[width=220pt, trim={0mm 0mm 0mm 0mm}, clip]{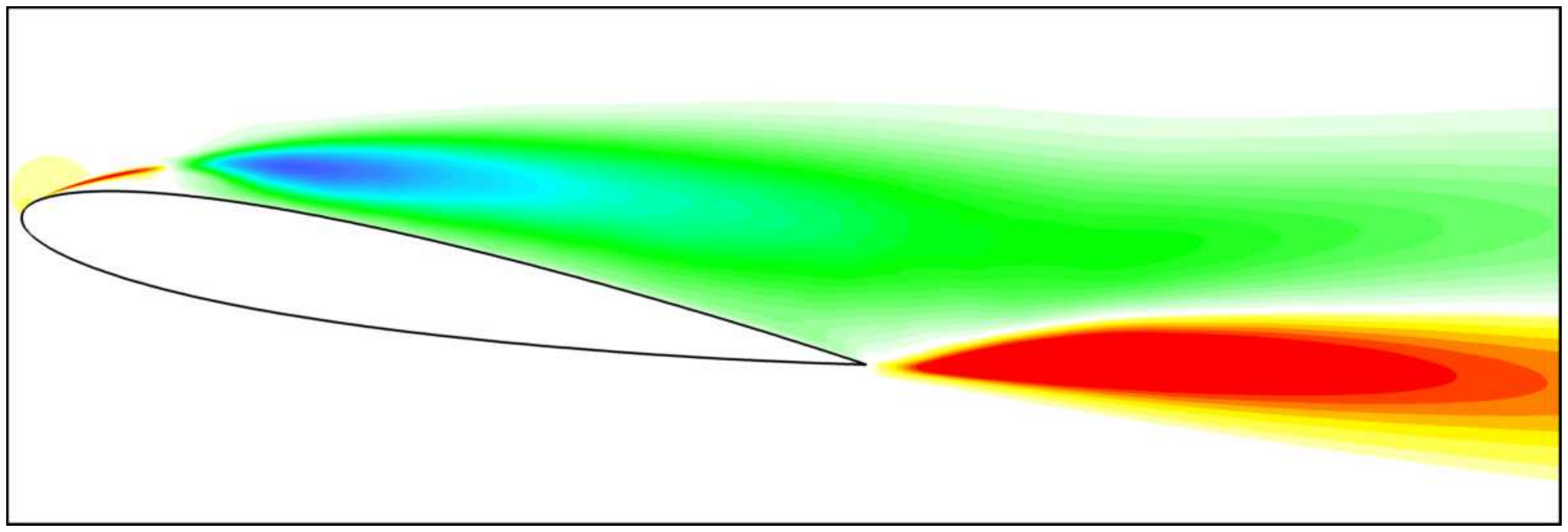}
\textit{$\alpha = 10.0^{\circ}$}
\end{minipage}
\begin{minipage}{220pt}
\centering
\includegraphics[width=220pt, trim={0mm 0mm 0mm 0mm}, clip]{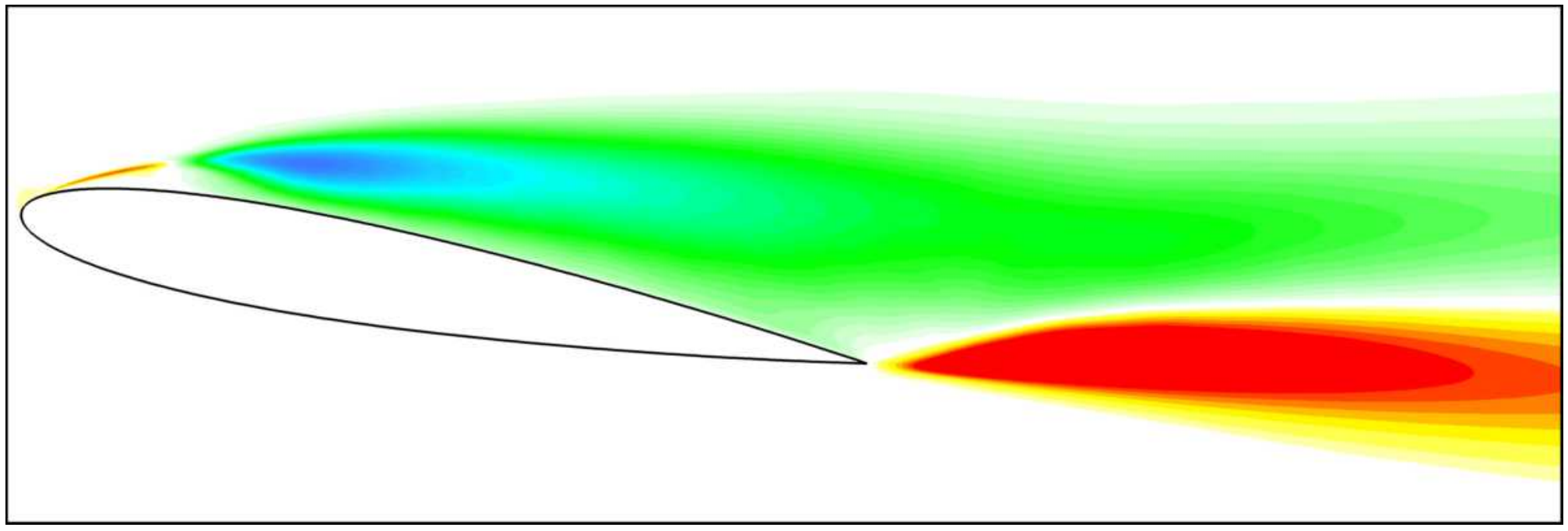}
\textit{$\alpha = 10.1^{\circ}$}
\end{minipage}
\begin{minipage}{220pt}
\centering
\includegraphics[width=220pt, trim={0mm 0mm 0mm 0mm}, clip]{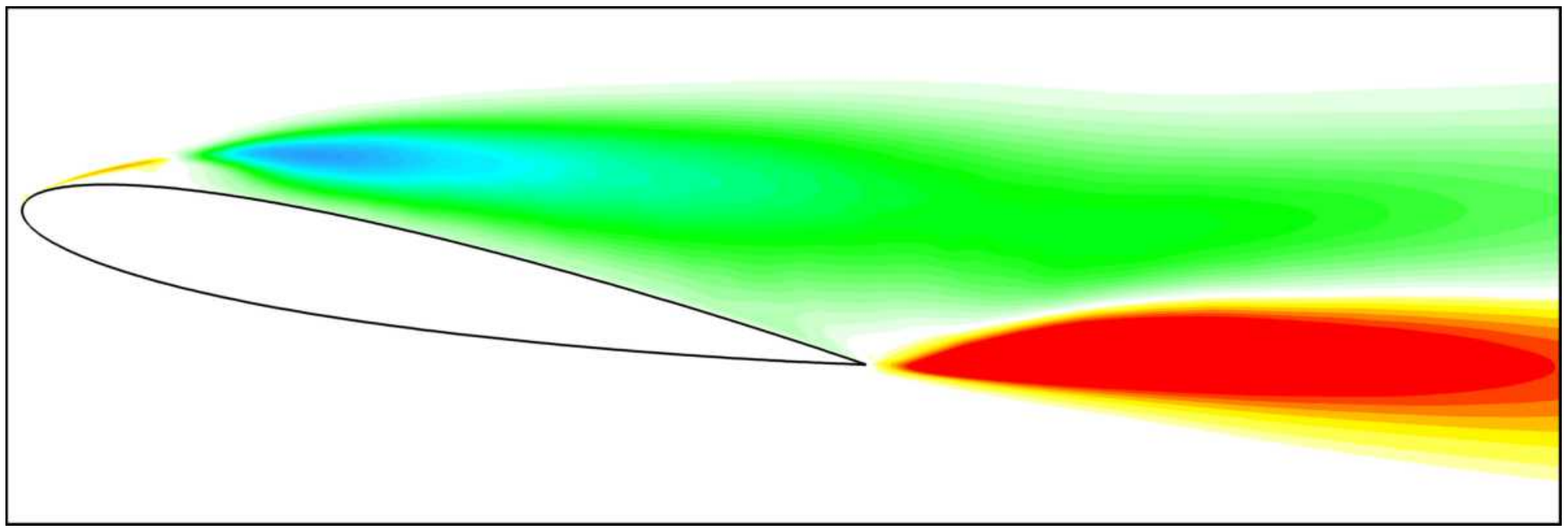}
\textit{$\alpha = 10.5^{\circ}$}
\end{minipage}
\caption{Colours map of the Reynolds stress, $\overline{{u\mydprime}v\mydprime}$, for the angles of attack $\alpha = 9.25^{\circ}$--$10.5^{\circ}$.}
\label{uv_mean}
\end{center}
\end{figure}

\begin{figure}
\begin{center}
\begin{minipage}{220pt}
\includegraphics[height=125pt , trim={0mm 0mm 0mm 0mm}, clip]{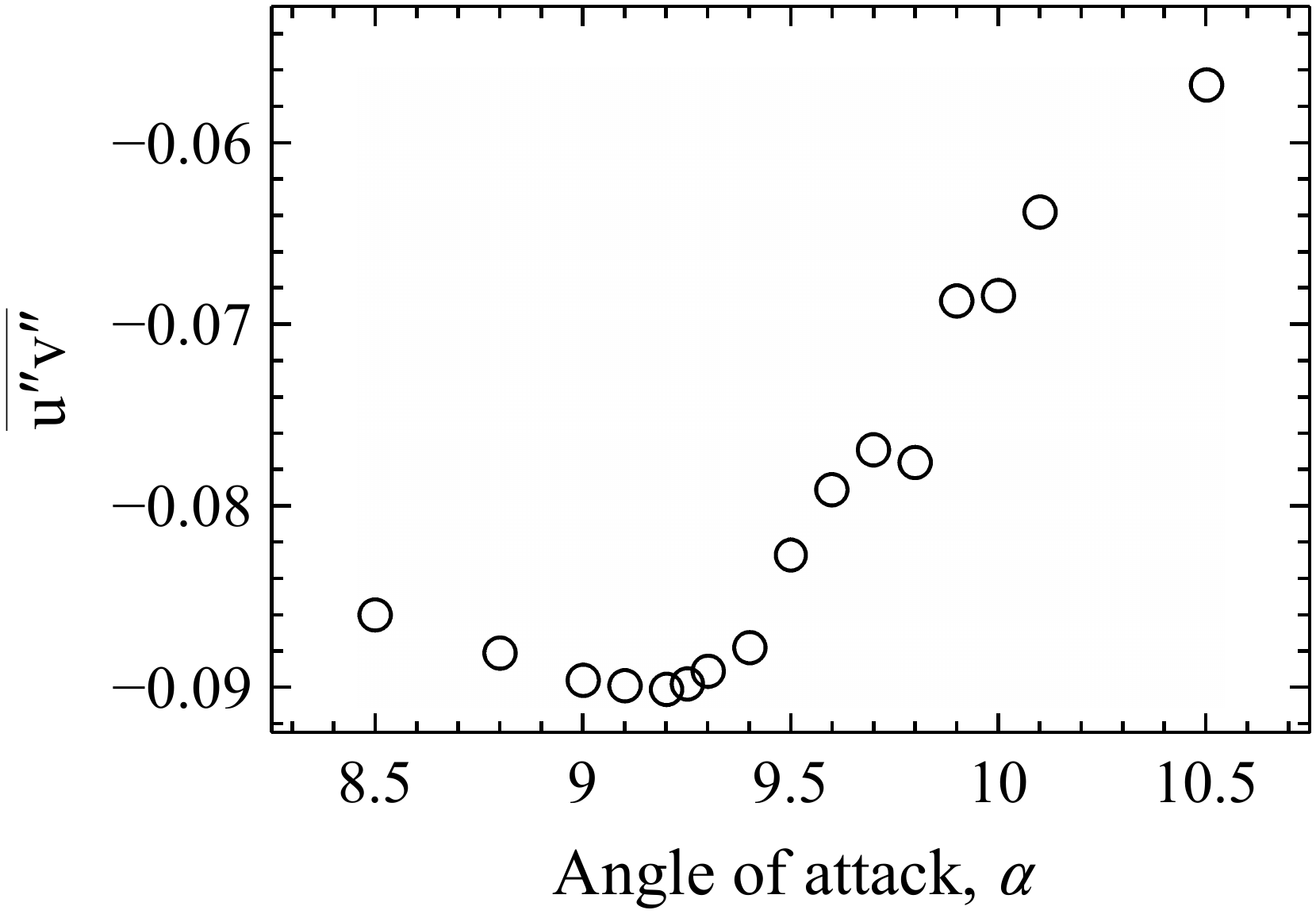}
\end{minipage}
\begin{minipage}{220pt}
\includegraphics[height=125pt , trim={0mm 0mm 0mm 0mm}, clip]{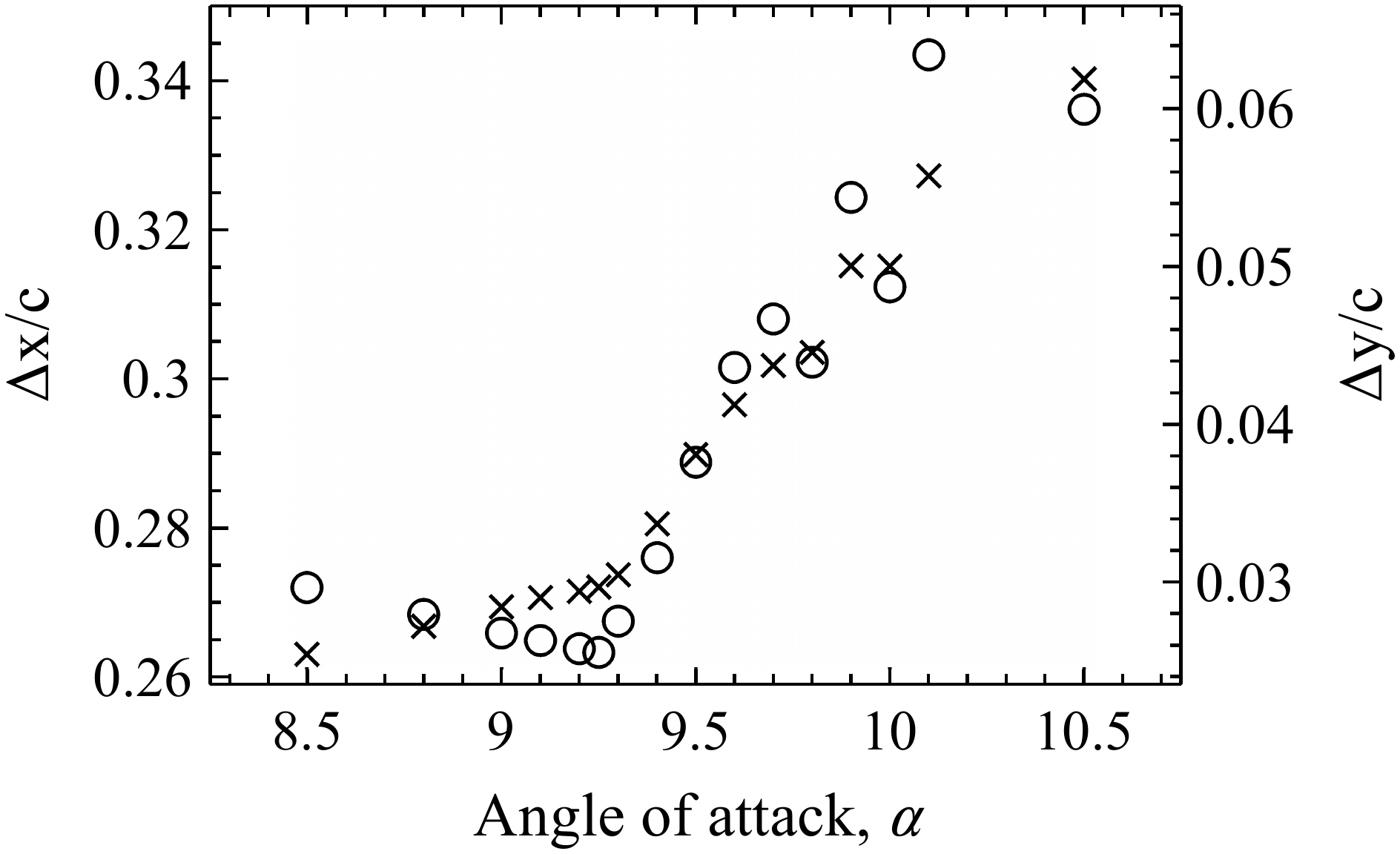}
\end{minipage}
\caption{Left: the minimum $\overline{{u\mydprime}{v\mydprime}}$ plotted versus the angle of attack $\alpha$. Right: the locations of the minimum $\overline{{u\mydprime}{v\mydprime}}$ plotted versus the angle of attack $\alpha$. Circles: $\Delta x/c$ measured from the aerofoil leading-edge, $\times$'s: $\Delta y/c$ measured from the aerofoil surface.}
\label{uv_min}
\end{center}
\end{figure}

It is well known that it is the Reynolds stress acting against the mean f\/low shear that extracts energy from the mean and feeds it into the small-scale perturbations. Also, it is well established that the Reynolds stress is proportional to the mean velocity gradient. The velocity gradient in the separated shear layer in the vicinity of the leading-edge and the shear layer at the trailing-edge are very strong. The Reynolds stress generated by the former is negative, and that produced by the latter is positive. Apparently, this is because at the trailing-edge the shear stress tends to create anticlockwise vortices, whereas, at the leading-edge, it tends to generate clockwise vortices. As the angle of attack is increased, the magnitude of the Reynolds stress in the vicinity of the leading-edge increases to a maximum value at $9.2^{\circ}$ then it drops almost linearly with the angle of attack. At the trailing-edge, the Reynolds stress increases monotonically with the angle of attack. However, the production of the turbulent kinetic energy does not only depend on the Reynolds stress, but also it depends on the mean shear. Therefore, an increase of the Reynolds stress does not guarantee an increase in the extraction of turbulent kinetic energy from the mean f\/low.\newline
F\/igure~\ref{uv_mean} shows the Reynolds stress distribution for the mean f\/low-f\/ields. The magnitude of the Reynolds stress is less than that of the high-lift and higher than that of the low-lift. This is consistent, in general, with the above discussion on turbulent kinetic energy in $\S$~\ref{sec:tke}. The left-hand side of f\/igure~\ref{uv_min} illustrates the distance of the minimum Reynolds stress in $x$ direction, from the aerofoil leading-edge, and in $y$ direction from the aerofoil surface. As seen in the f\/igure, the location of the minimum Reynolds stress moves, in $x$ direction, towards the leading-edge until it reaches the closest point to the leading-edge at the angle of attack of $9.25^{\circ}$. After that, the location of the minimum Reynolds stress moves away from the leading-edge as the angle of attack is increased. The location of minimum Reynolds stress moves away from the aerofoil surface as the angle of attack is increased.\newline
F\/igure~\ref{uv_min} (right) shows that the magnitude of the Reynolds stress reached a maximum at the angle of attack of $9.2^{\circ}$. This angle is a little bit lower than the angle at which $\overline{{v\mydprime}^2}$ and $\overline{{w\mydprime}^2}$ reached their maximum magnitude. The difference in the angle of attack of $0.05^{\circ}$ can be considered within statistical error. However, the angle of minimum Reynolds stress is much lower than that of maximum $\overline{{u\mydprime}^2}$ which is $9.4^{\circ}$. This implies that the perturbations of the wall-normal and the spanwise velocity components are extracted from the mean f\/low exclusively by the Reynolds stress. Whereas, a global f\/low oscillation generates perturbations in the streamwise velocity in addition to the Reynolds stress.

\subsection{Pressure f\/luctuations}
\begin{figure}
\begin{center}
\begin{minipage}{220pt}
\centering
\includegraphics[width=220pt, trim={0mm 0mm 0mm 0mm}, clip]{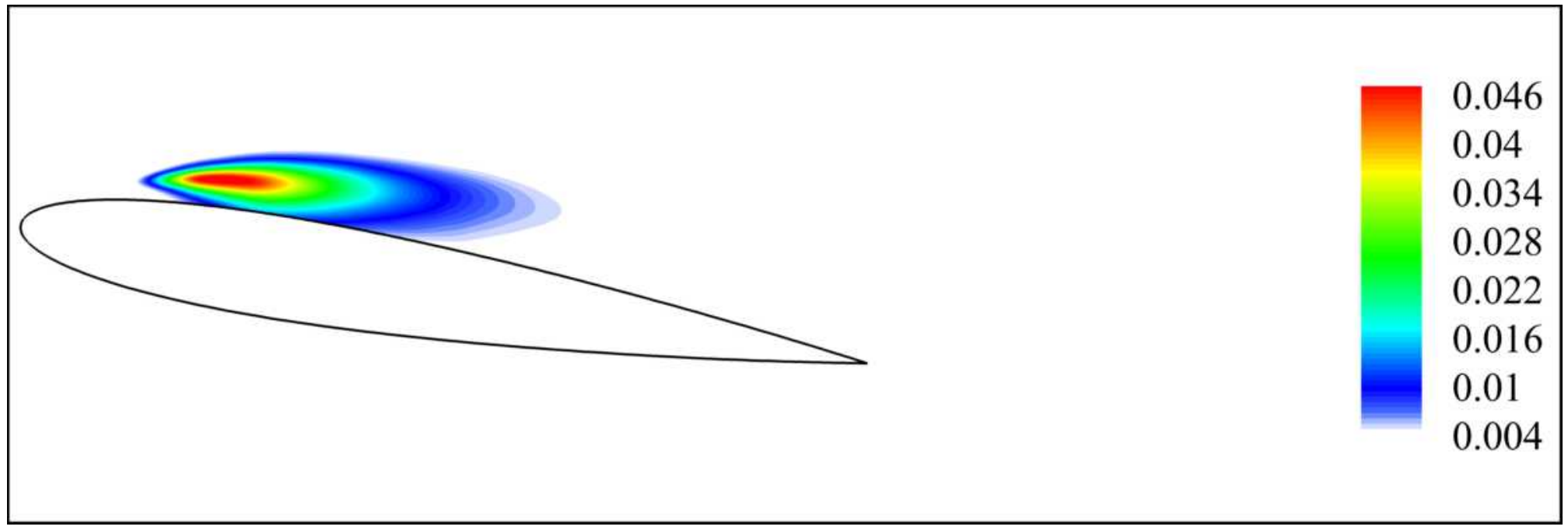}
\textit{$\alpha = 9.25^{\circ}$}
\end{minipage}
\medskip
\begin{minipage}{220pt}
\centering
\includegraphics[width=220pt, trim={0mm 0mm 0mm 0mm}, clip]{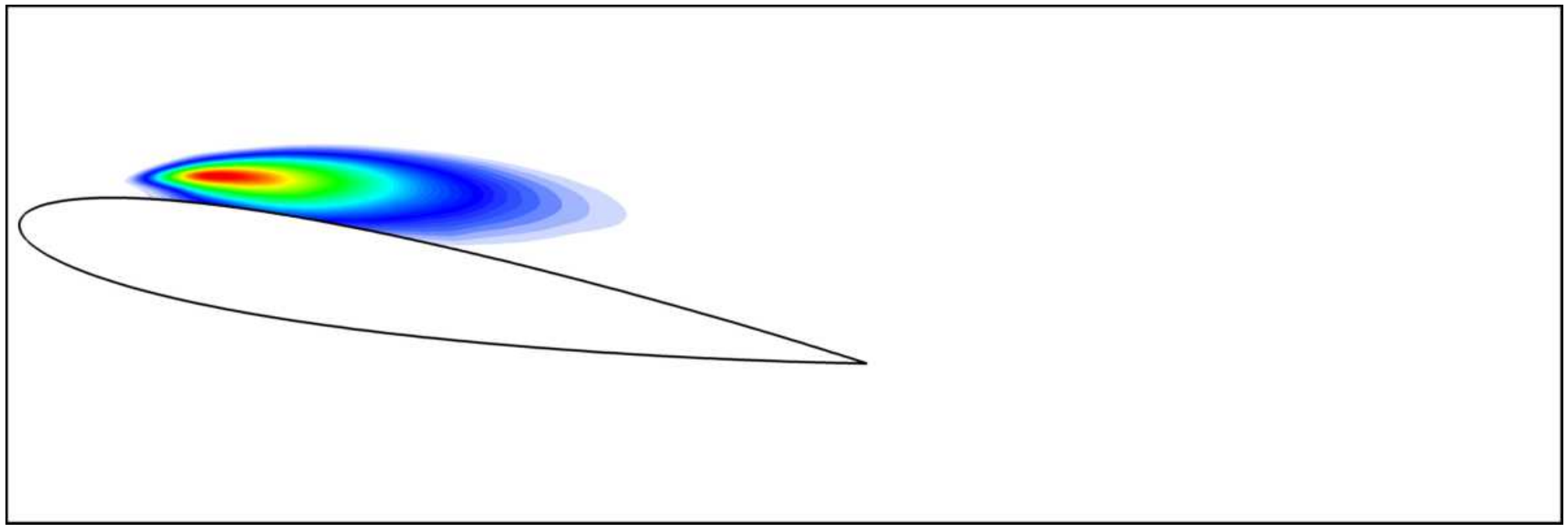}
\textit{$\alpha = 9.40^{\circ}$}
\end{minipage}
\medskip
\begin{minipage}{220pt}
\centering
\includegraphics[width=220pt, trim={0mm 0mm 0mm 0mm}, clip]{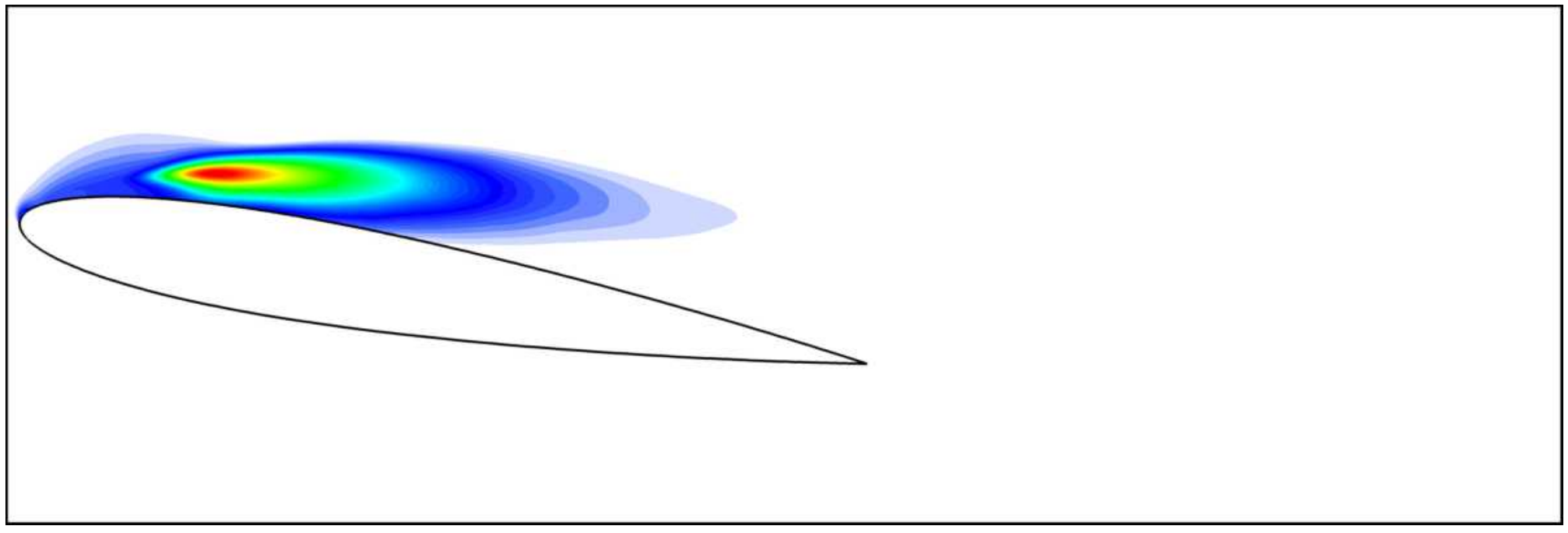}
\textit{$\alpha = 9.50^{\circ}$}
\end{minipage}
\medskip
\begin{minipage}{220pt}
\centering
\includegraphics[width=220pt, trim={0mm 0mm 0mm 0mm}, clip]{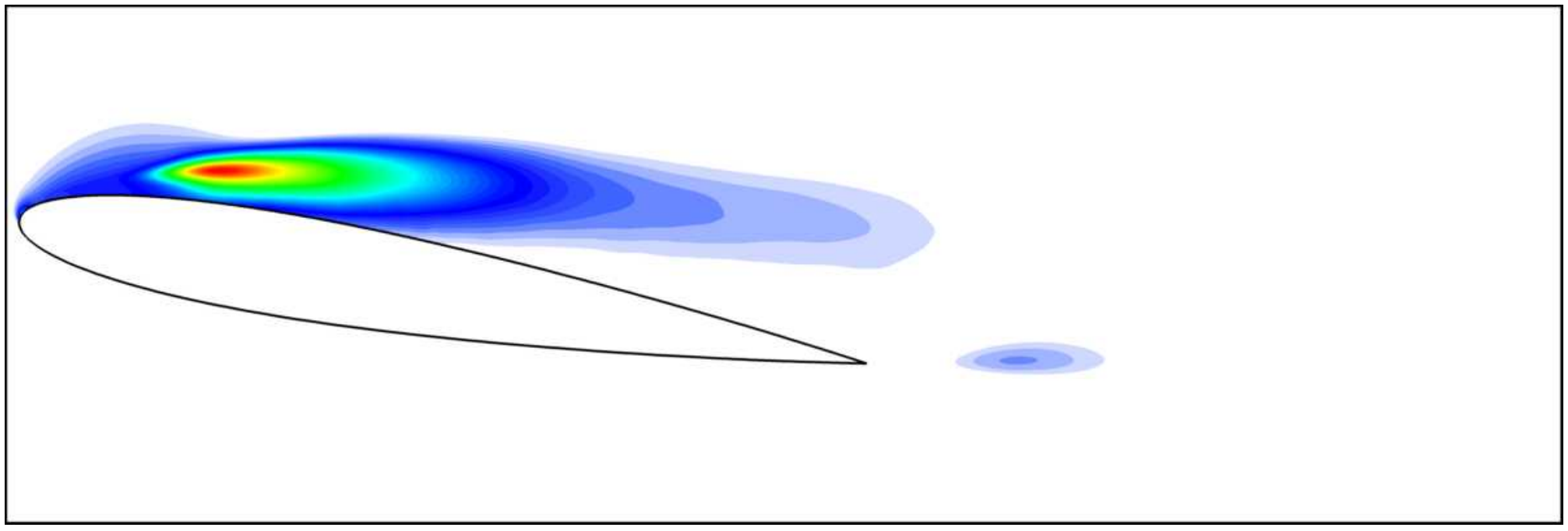}
\textit{$\alpha = 9.60^{\circ}$}
\end{minipage}
\medskip
\begin{minipage}{220pt}
\centering
\includegraphics[width=220pt, trim={0mm 0mm 0mm 0mm}, clip]{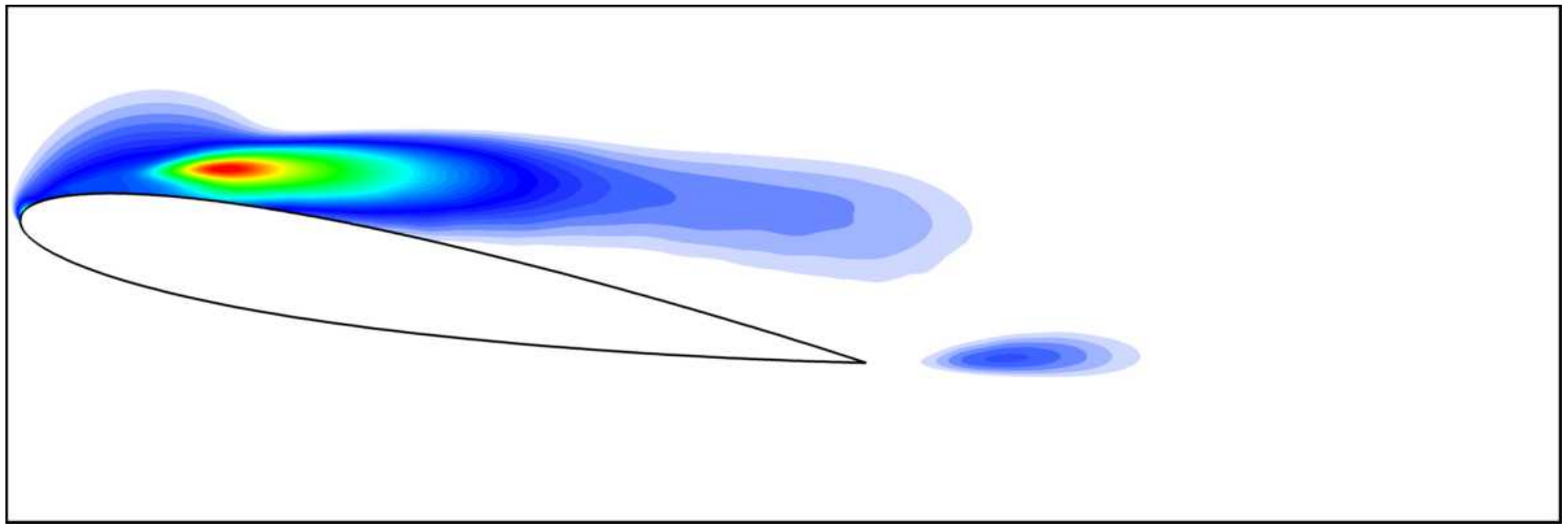}
\textit{$\alpha = 9.70^{\circ}$}
\end{minipage}
\medskip
\begin{minipage}{220pt}
\centering
\includegraphics[width=220pt, trim={0mm 0mm 0mm 0mm}, clip]{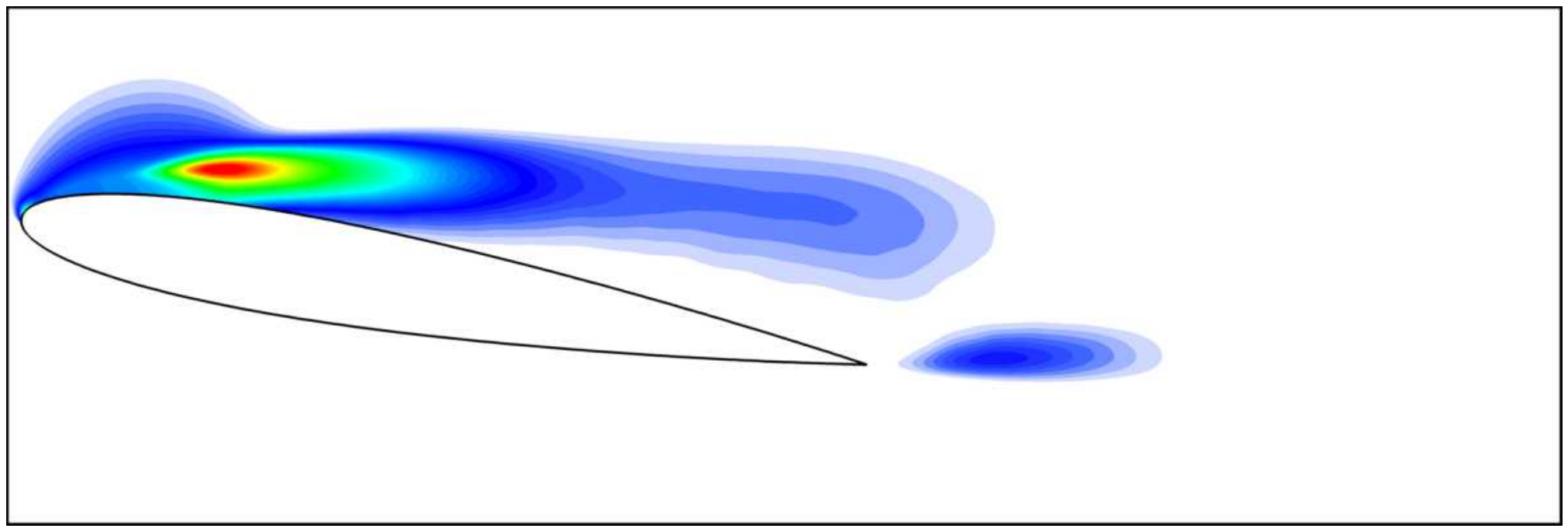}
\textit{$\alpha = 9.80^{\circ}$}
\end{minipage}
\medskip
\begin{minipage}{220pt}
\centering
\includegraphics[width=220pt, trim={0mm 0mm 0mm 0mm}, clip]{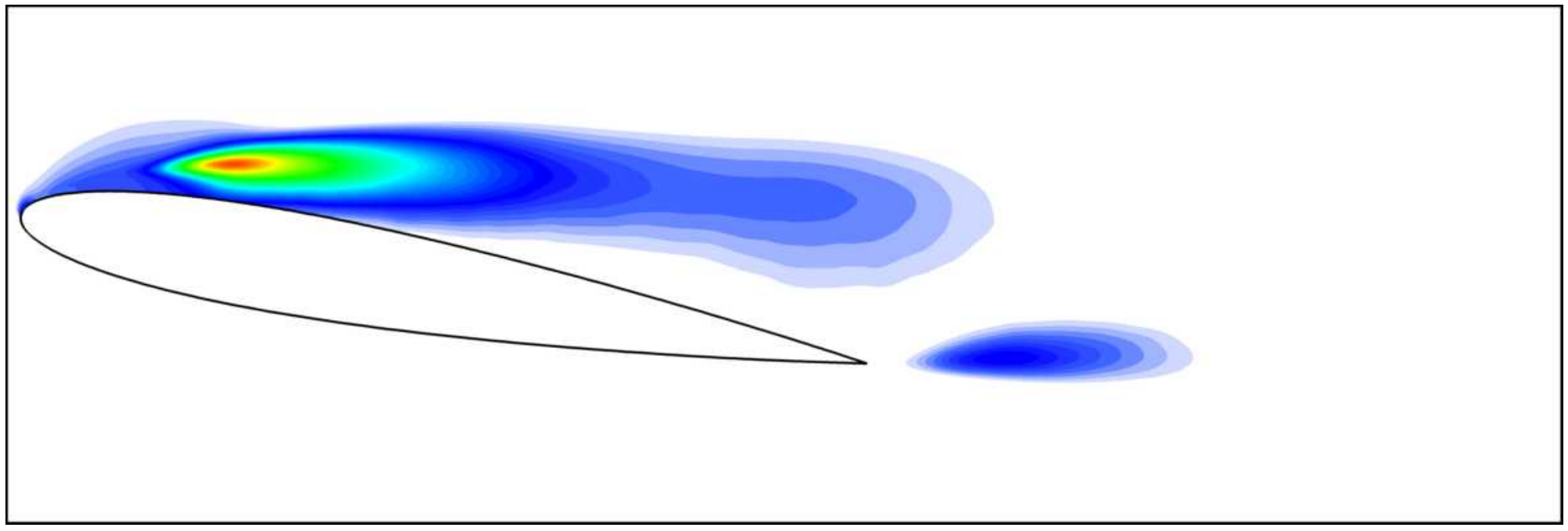}
\textit{$\alpha = 9.90^{\circ}$}
\end{minipage}
\medskip
\begin{minipage}{220pt}
\centering
\includegraphics[width=220pt, trim={0mm 0mm 0mm 0mm}, clip]{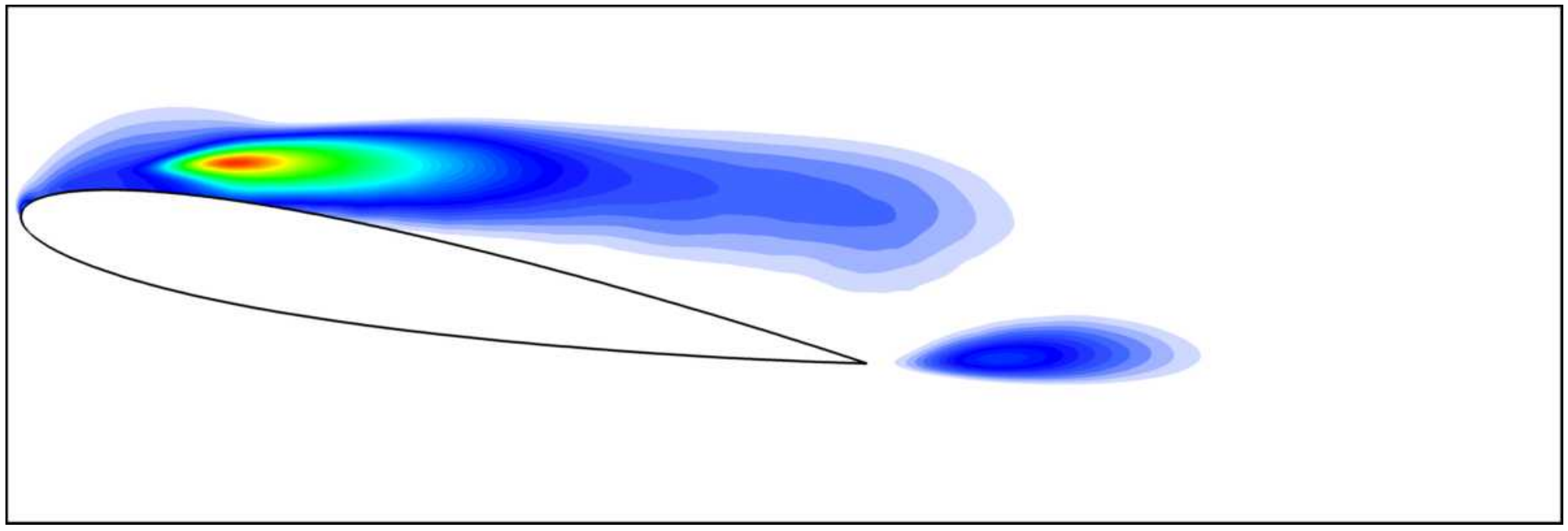}
\textit{$\alpha = 10.0^{\circ}$}
\end{minipage}
\medskip
\begin{minipage}{220pt}
\centering
\includegraphics[width=220pt, trim={0mm 0mm 0mm 0mm}, clip]{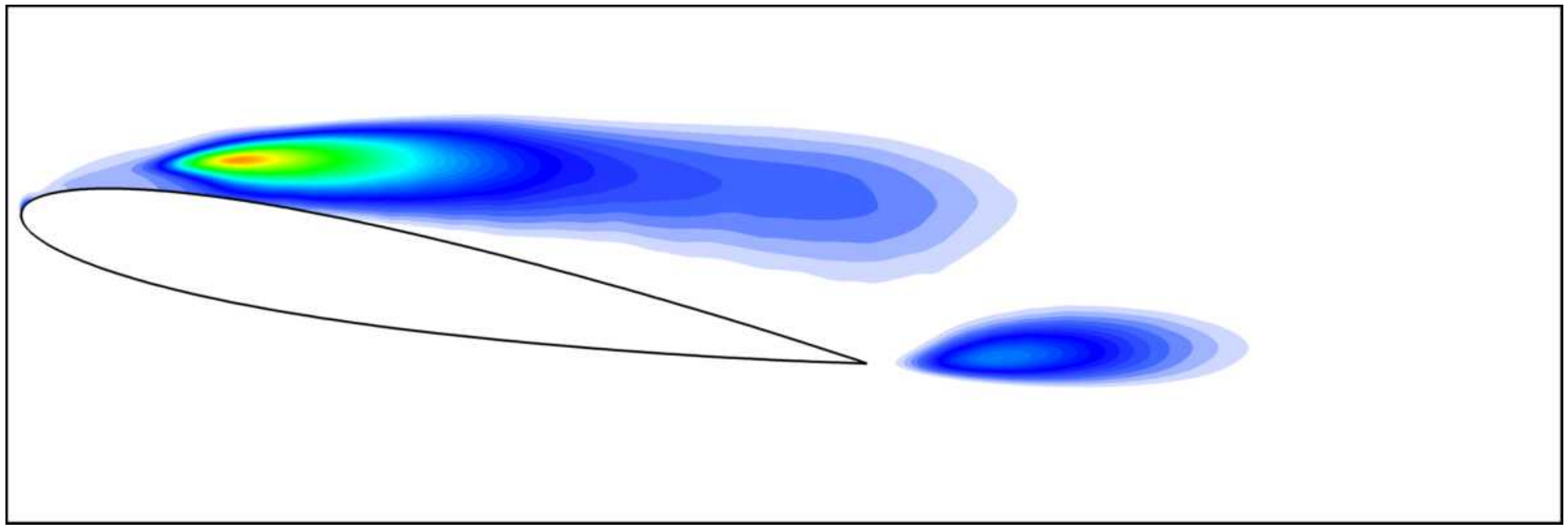}
\textit{$\alpha = 10.1^{\circ}$}
\end{minipage}
\begin{minipage}{220pt}
\centering
\includegraphics[width=220pt, trim={0mm 0mm 0mm 0mm}, clip]{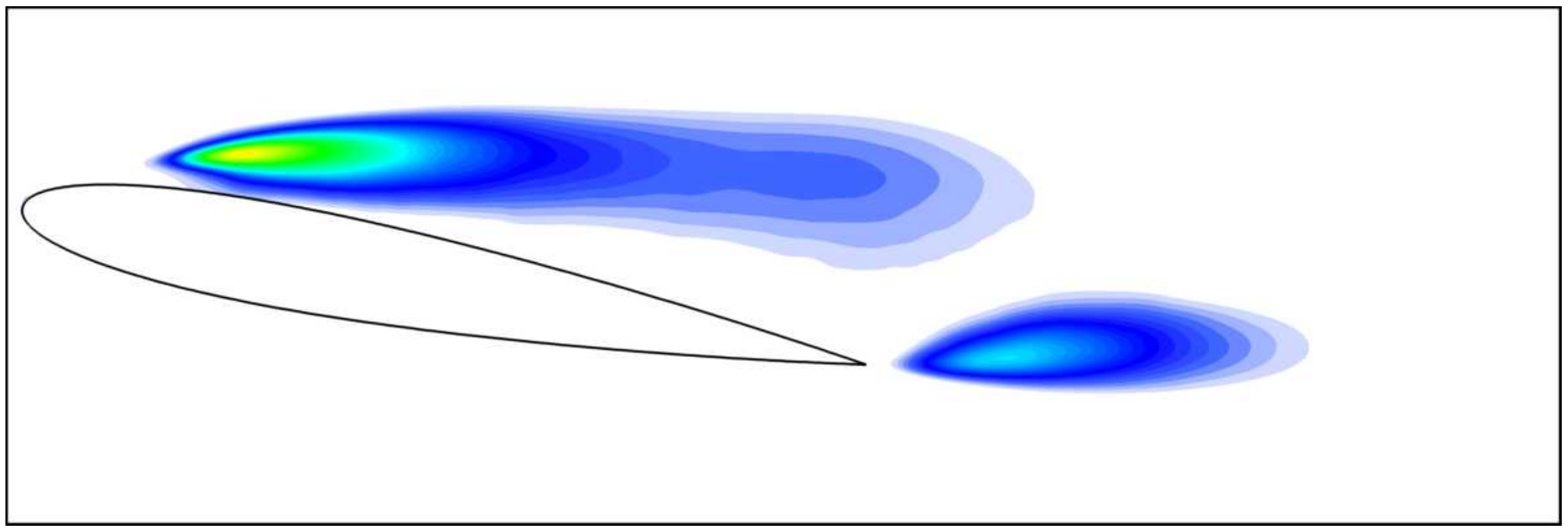}
\textit{$\alpha = 10.5^{\circ}$}
\end{minipage}
\caption{Colours map of the variance of the pressure, $\overline{{p\mydprime}^2}$, for the angles of attack $\alpha = 9.25^{\circ}$--$10.5^{\circ}$.}
\label{p2_mean}
\end{center}
\end{figure}

\begin{figure}
\begin{center}
\begin{minipage}{220pt}
\centering
\includegraphics[width=220pt, trim={0mm 0mm 0mm 0mm}, clip]{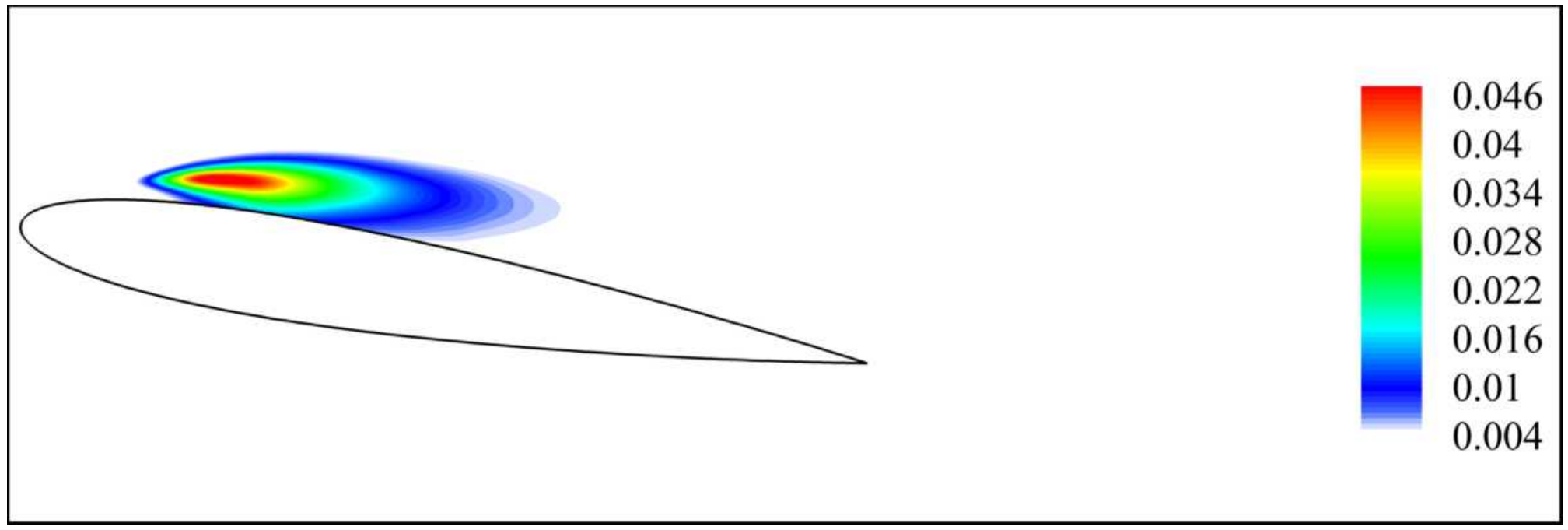}
\textit{$\alpha = 9.25^{\circ}$}
\end{minipage}
\medskip
\begin{minipage}{220pt}
\centering
\includegraphics[width=220pt, trim={0mm 0mm 0mm 0mm}, clip]{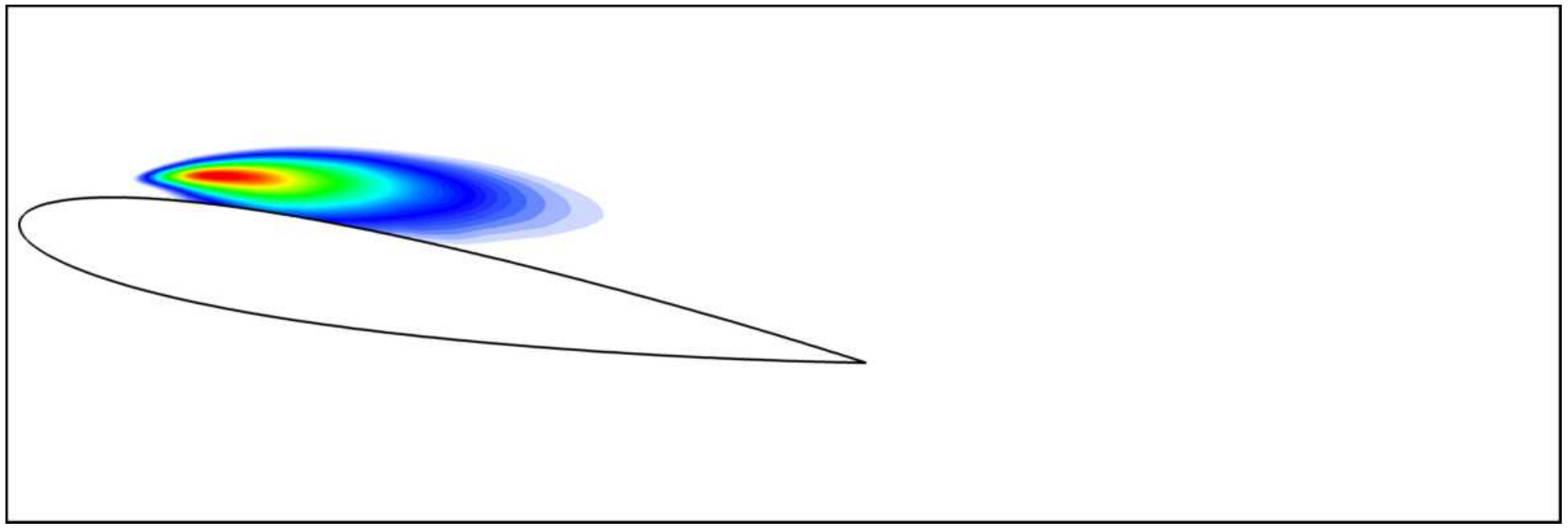}
\textit{$\alpha = 9.40^{\circ}$}
\end{minipage}
\medskip
\begin{minipage}{220pt}
\centering
\includegraphics[width=220pt, trim={0mm 0mm 0mm 0mm}, clip]{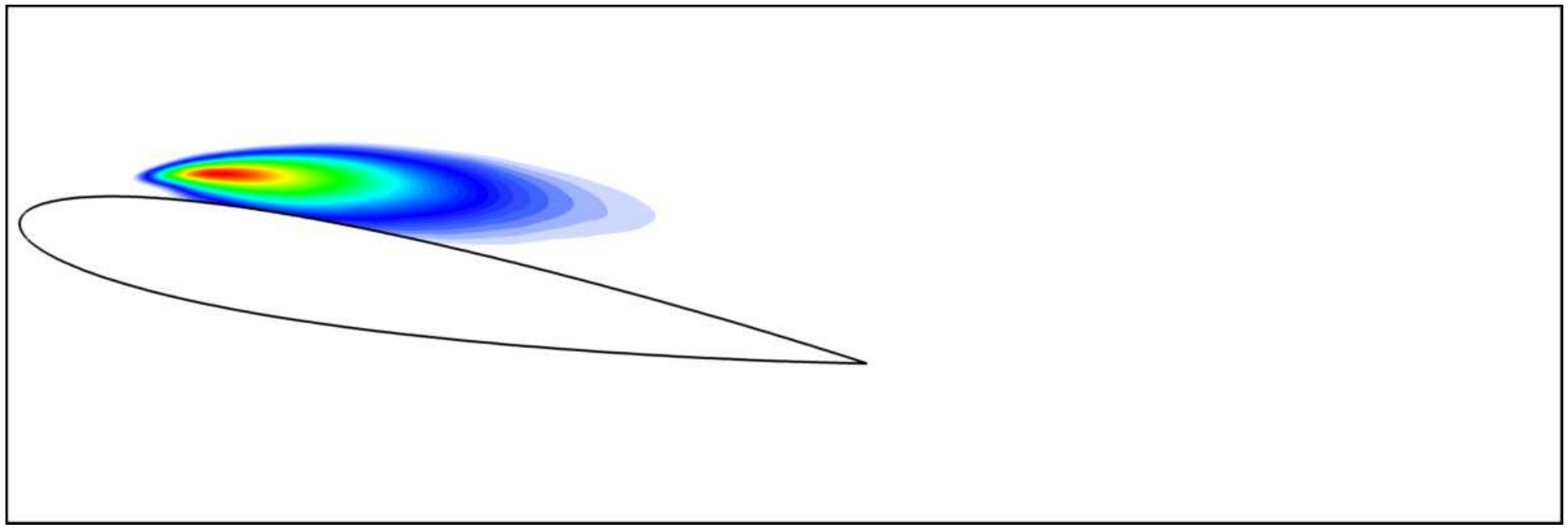}
\textit{$\alpha = 9.50^{\circ}$}
\end{minipage}
\medskip
\begin{minipage}{220pt}
\centering
\includegraphics[width=220pt, trim={0mm 0mm 0mm 0mm}, clip]{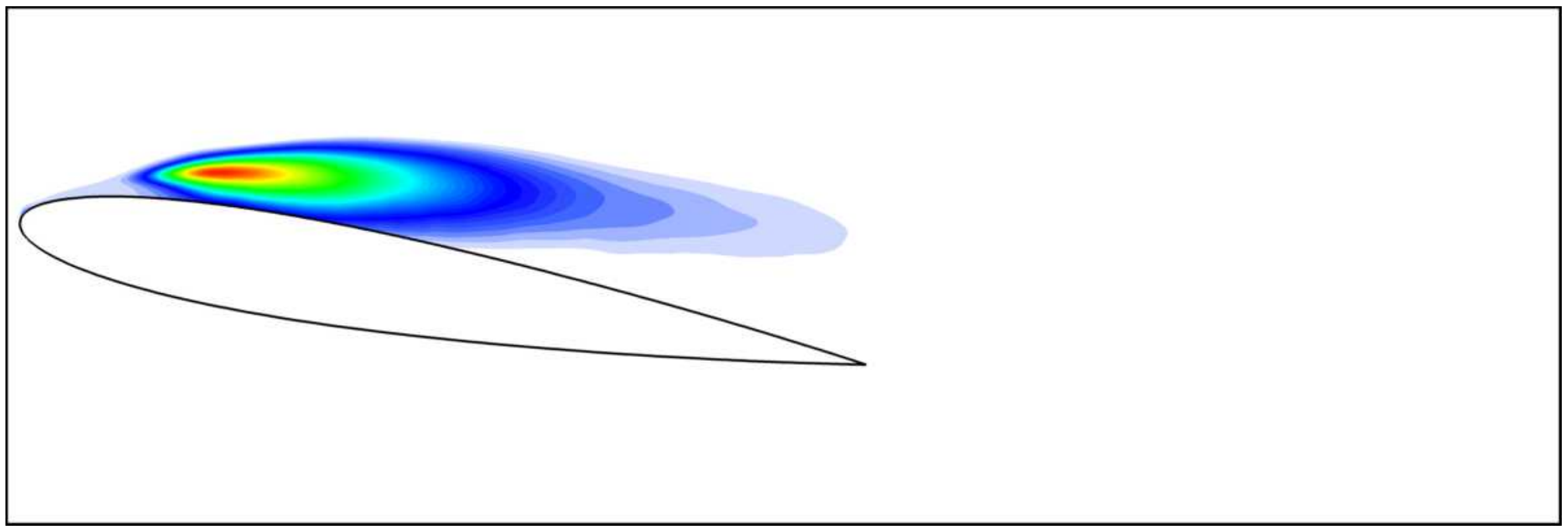}
\textit{$\alpha = 9.60^{\circ}$}
\end{minipage}
\medskip
\begin{minipage}{220pt}
\centering
\includegraphics[width=220pt, trim={0mm 0mm 0mm 0mm}, clip]{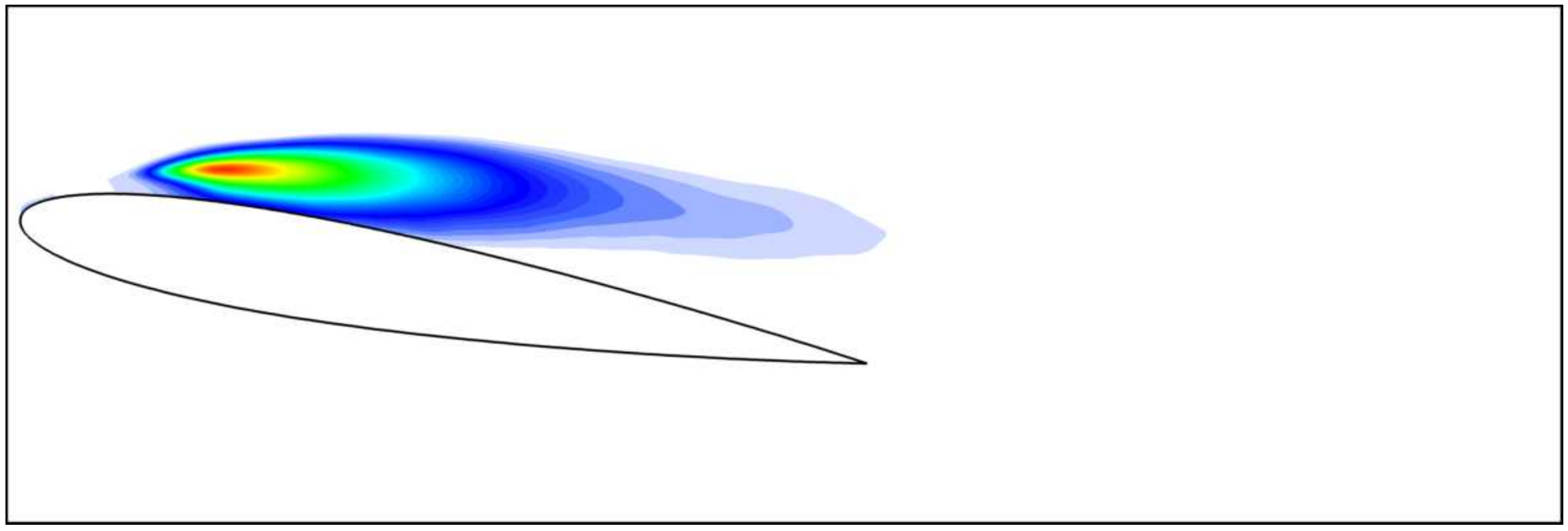}
\textit{$\alpha = 9.70^{\circ}$}
\end{minipage}
\medskip
\begin{minipage}{220pt}
\centering
\includegraphics[width=220pt, trim={0mm 0mm 0mm 0mm}, clip]{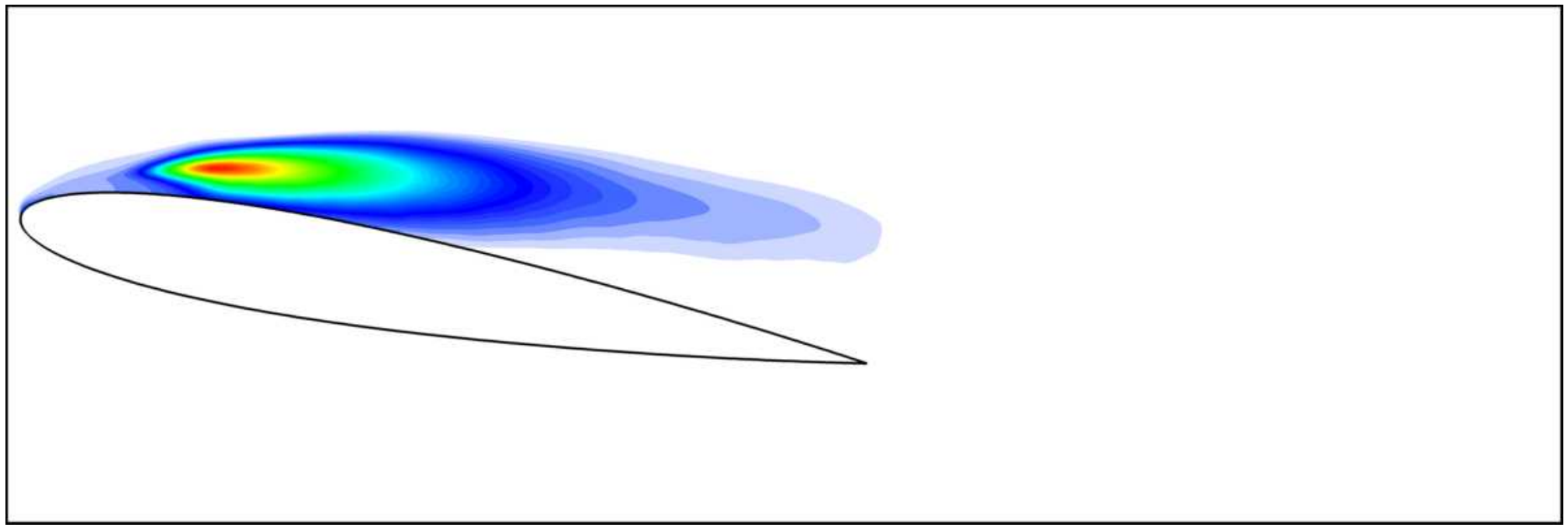}
\textit{$\alpha = 9.80^{\circ}$}
\end{minipage}
\medskip
\begin{minipage}{220pt}
\centering
\includegraphics[width=220pt, trim={0mm 0mm 0mm 0mm}, clip]{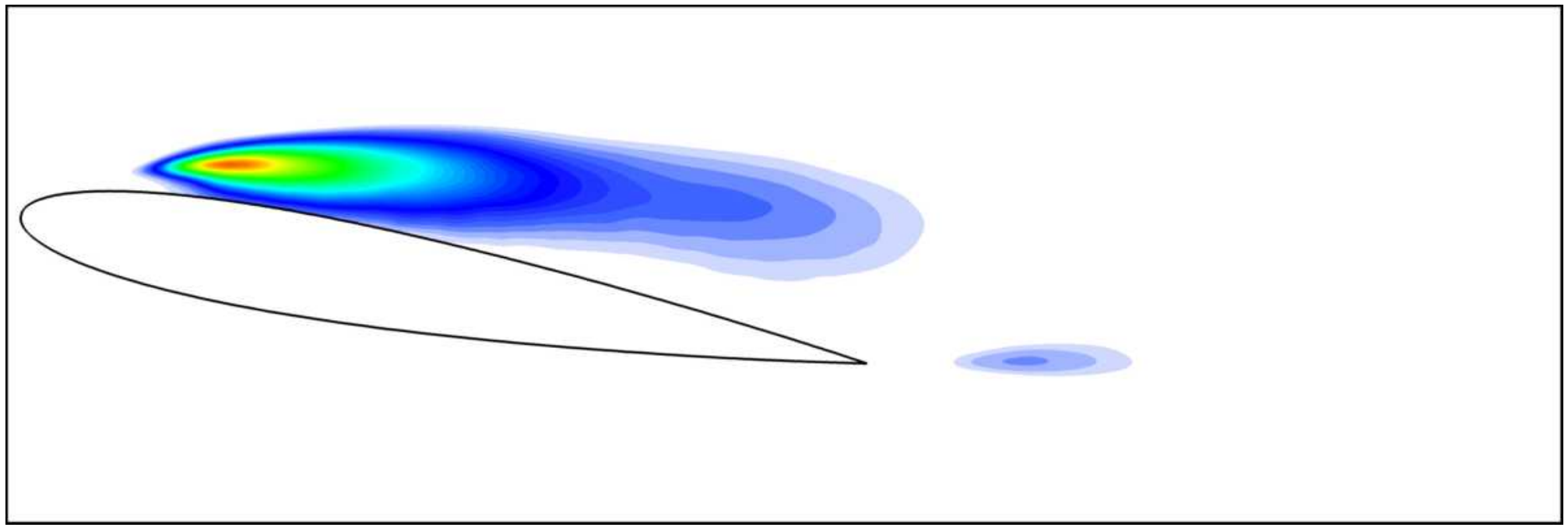}
\textit{$\alpha = 9.90^{\circ}$}
\end{minipage}
\medskip
\begin{minipage}{220pt}
\centering
\includegraphics[width=220pt, trim={0mm 0mm 0mm 0mm}, clip]{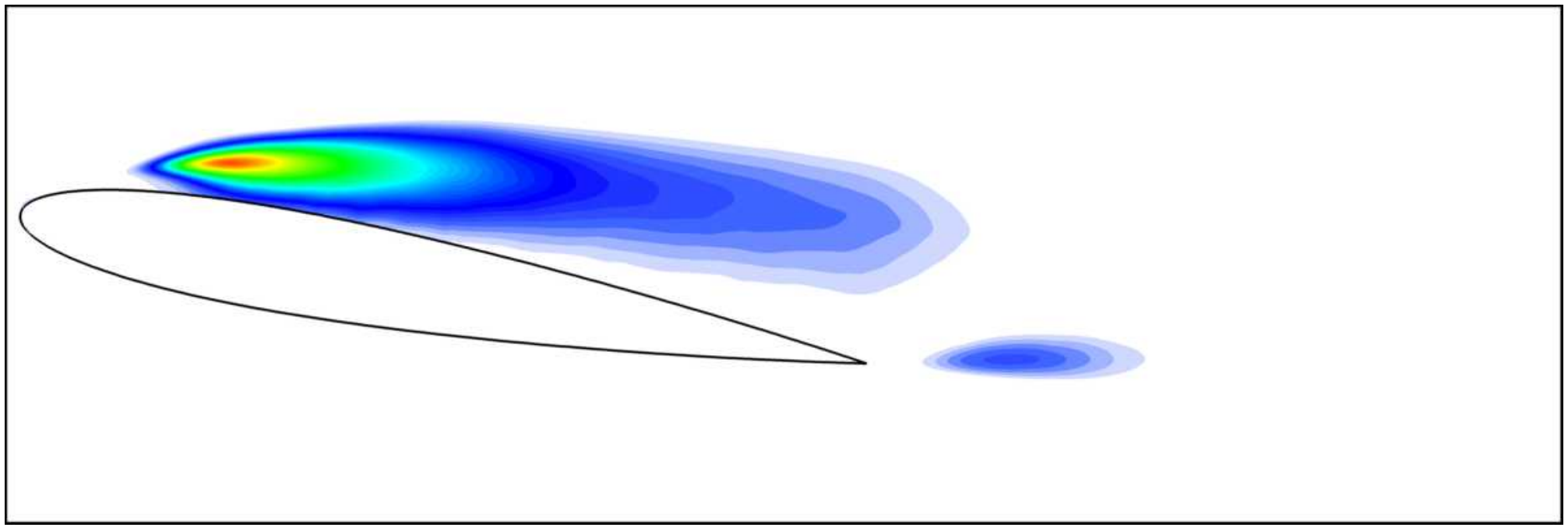}
\textit{$\alpha = 10.0^{\circ}$}
\end{minipage}
\medskip
\begin{minipage}{220pt}
\centering
\includegraphics[width=220pt, trim={0mm 0mm 0mm 0mm}, clip]{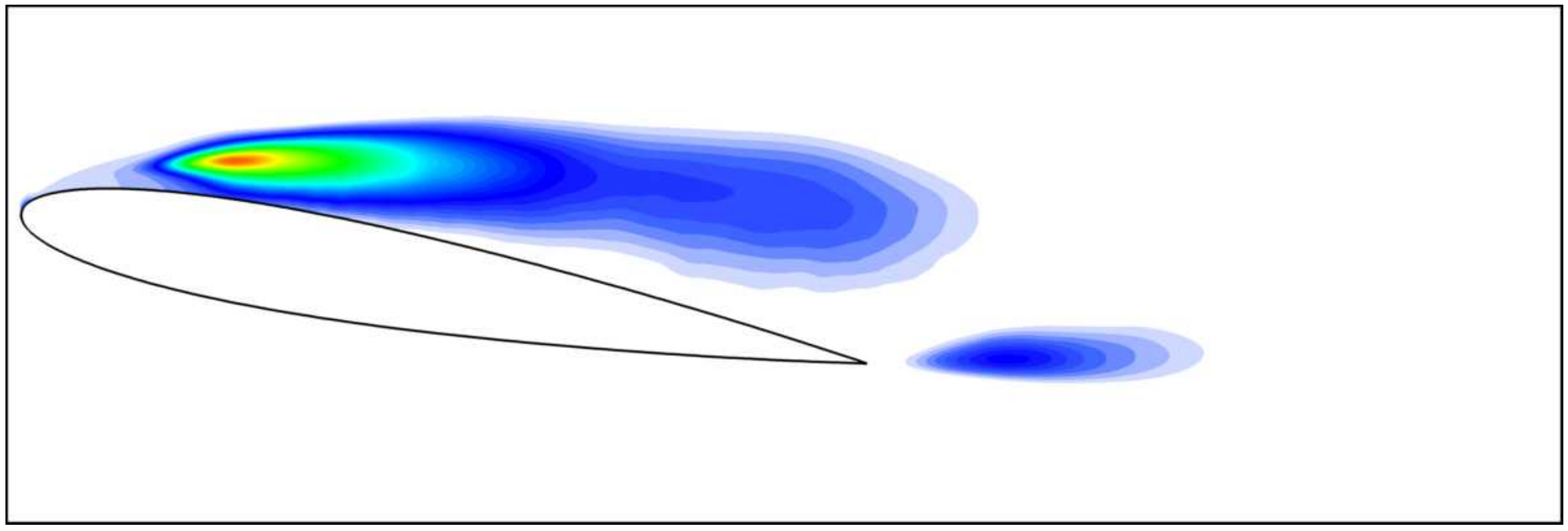}
\textit{$\alpha = 10.1^{\circ}$}
\end{minipage}
\begin{minipage}{220pt}
\centering
\includegraphics[width=220pt, trim={0mm 0mm 0mm 0mm}, clip]{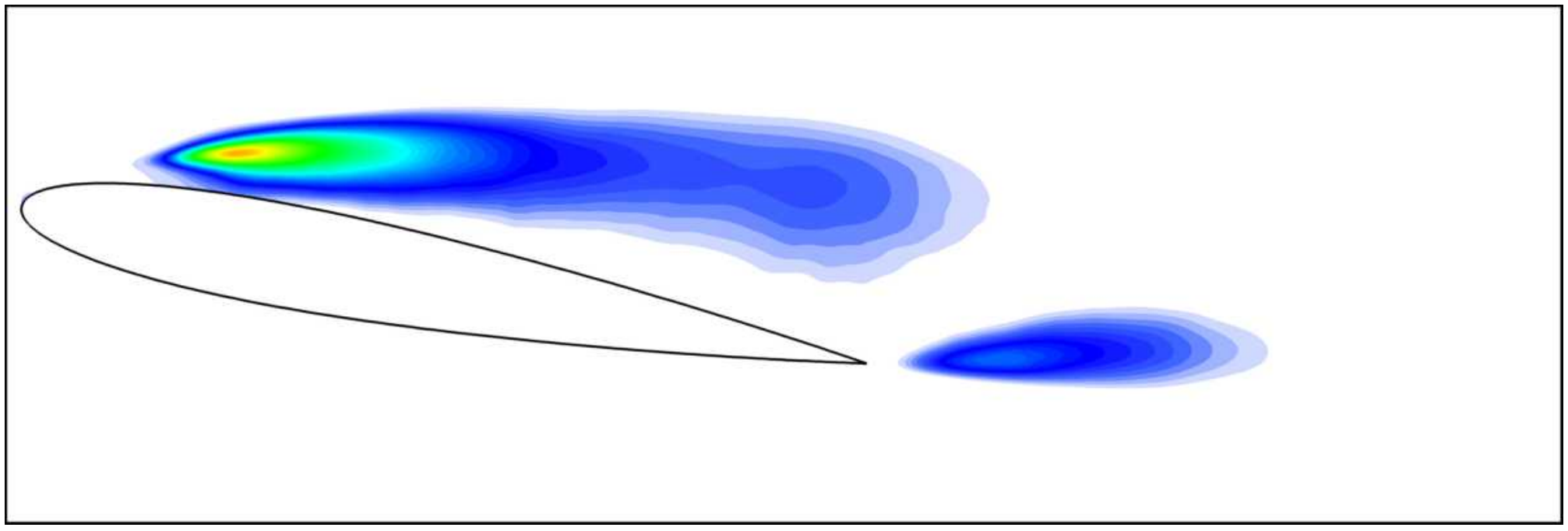}
\textit{$\alpha = 10.5^{\circ}$}
\end{minipage}
\caption{Colours map of the high-lift variance of the pressure, $\widehat{{p\mydprime}^2}$, for the angles of attack $\alpha = 9.25^{\circ}$--$10.5^{\circ}$.}
\label{p2_above}
\end{center}
\end{figure}

\begin{figure}
\begin{center}
\begin{minipage}{220pt}
\centering
\includegraphics[width=220pt, trim={0mm 0mm 0mm 0mm}, clip]{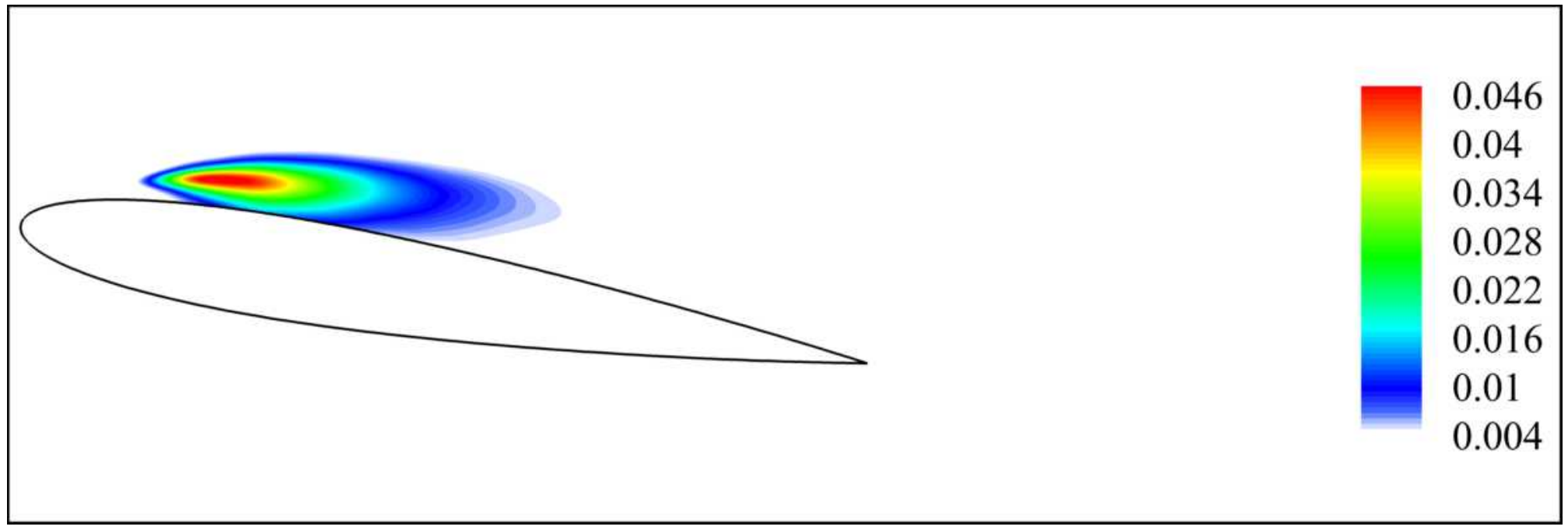}
\textit{$\alpha = 9.25^{\circ}$}
\end{minipage}
\begin{minipage}{220pt}
\centering
\includegraphics[width=220pt, trim={0mm 0mm 0mm 0mm}, clip]{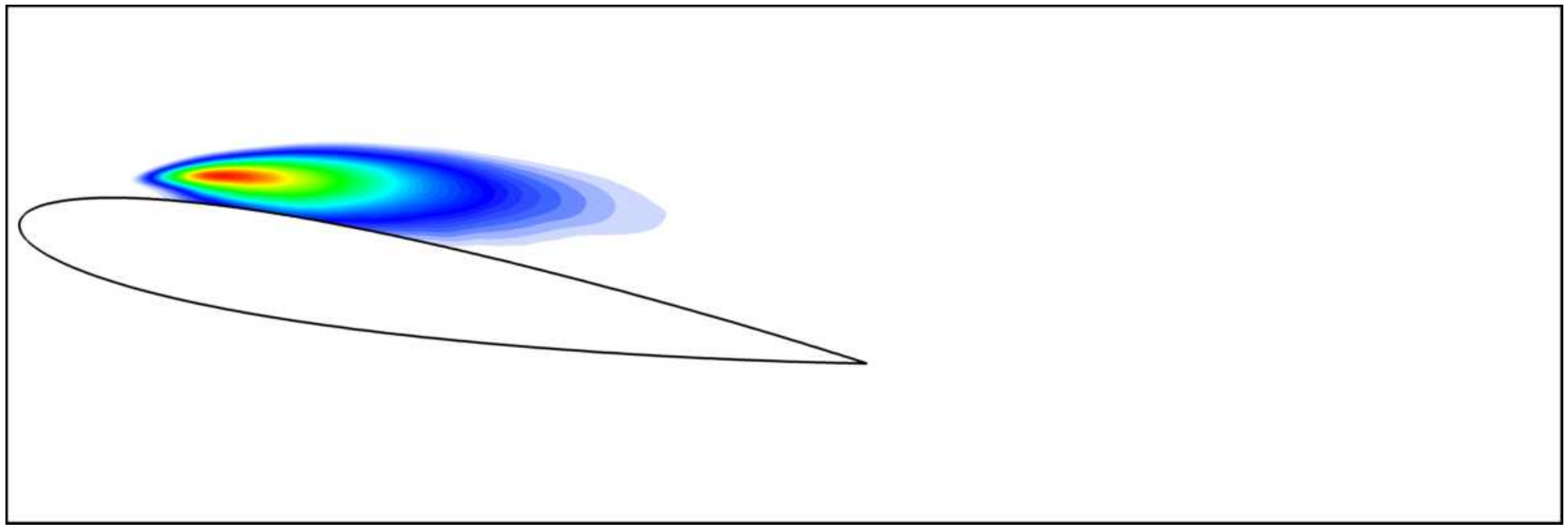}
\textit{$\alpha = 9.40^{\circ}$}
\end{minipage}
\begin{minipage}{220pt}
\centering
\includegraphics[width=220pt, trim={0mm 0mm 0mm 0mm}, clip]{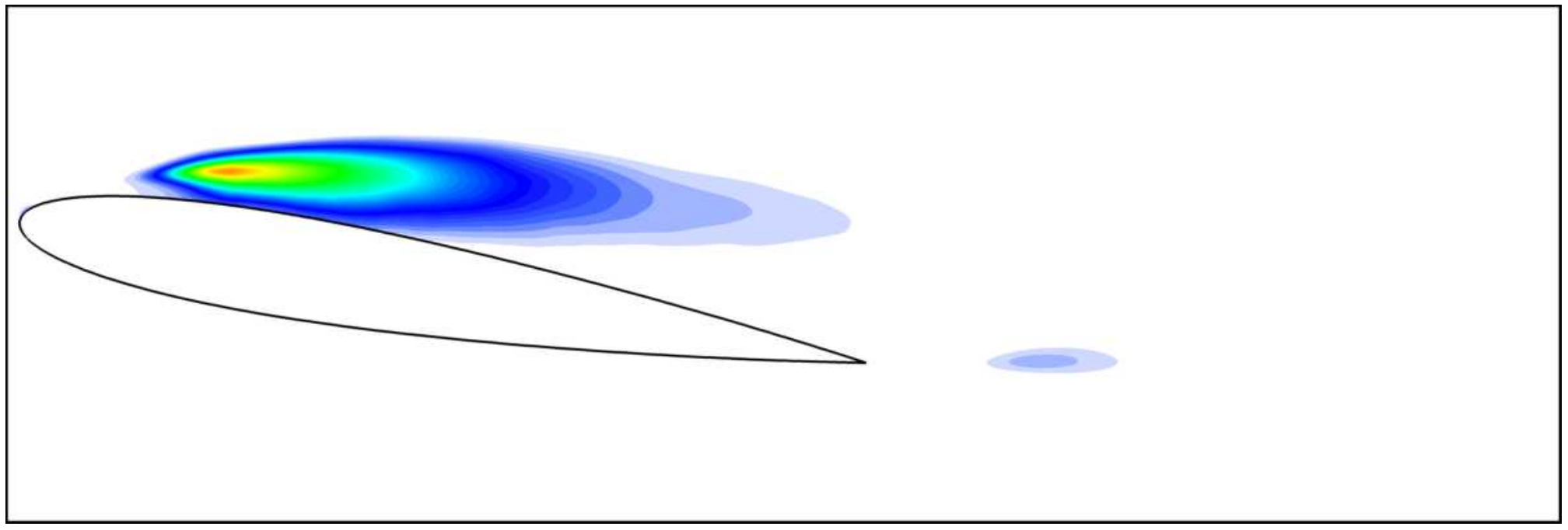}
\textit{$\alpha = 9.50^{\circ}$}
\end{minipage}
\begin{minipage}{220pt}
\centering
\includegraphics[width=220pt, trim={0mm 0mm 0mm 0mm}, clip]{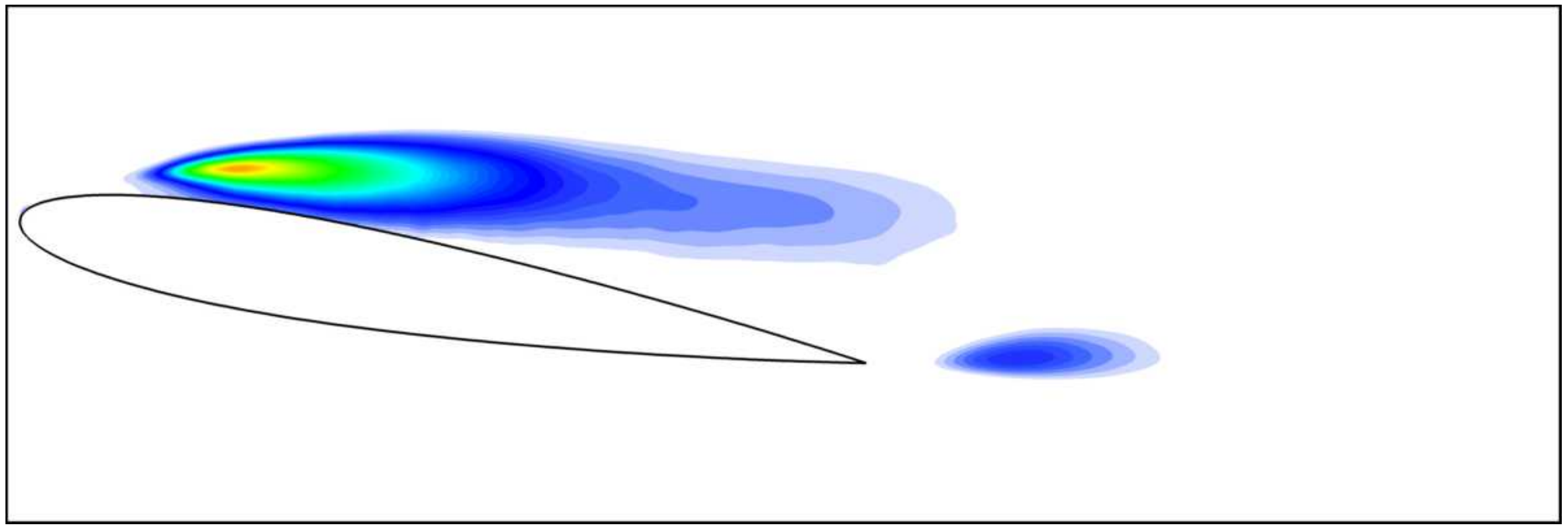}
\textit{$\alpha = 9.60^{\circ}$}
\end{minipage}
\begin{minipage}{220pt}
\centering
\includegraphics[width=220pt, trim={0mm 0mm 0mm 0mm}, clip]{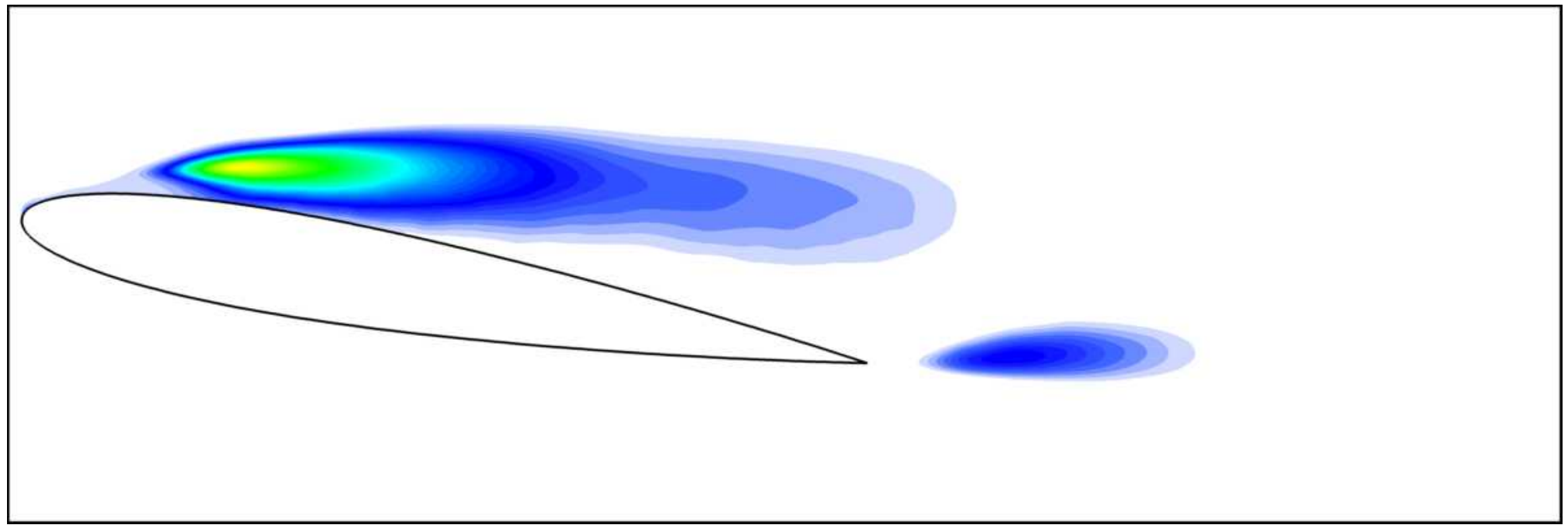}
\textit{$\alpha = 9.70^{\circ}$}
\end{minipage}
\begin{minipage}{220pt}
\centering
\includegraphics[width=220pt, trim={0mm 0mm 0mm 0mm}, clip]{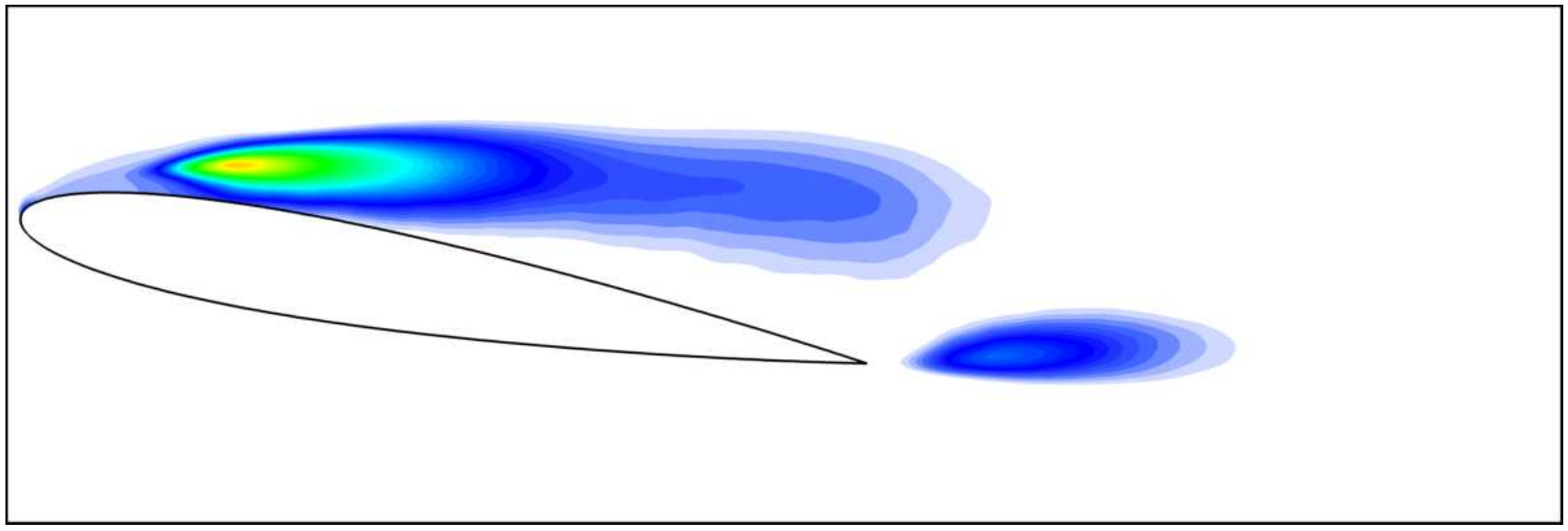}
\textit{$\alpha = 9.80^{\circ}$}
\end{minipage}
\begin{minipage}{220pt}
\centering
\includegraphics[width=220pt, trim={0mm 0mm 0mm 0mm}, clip]{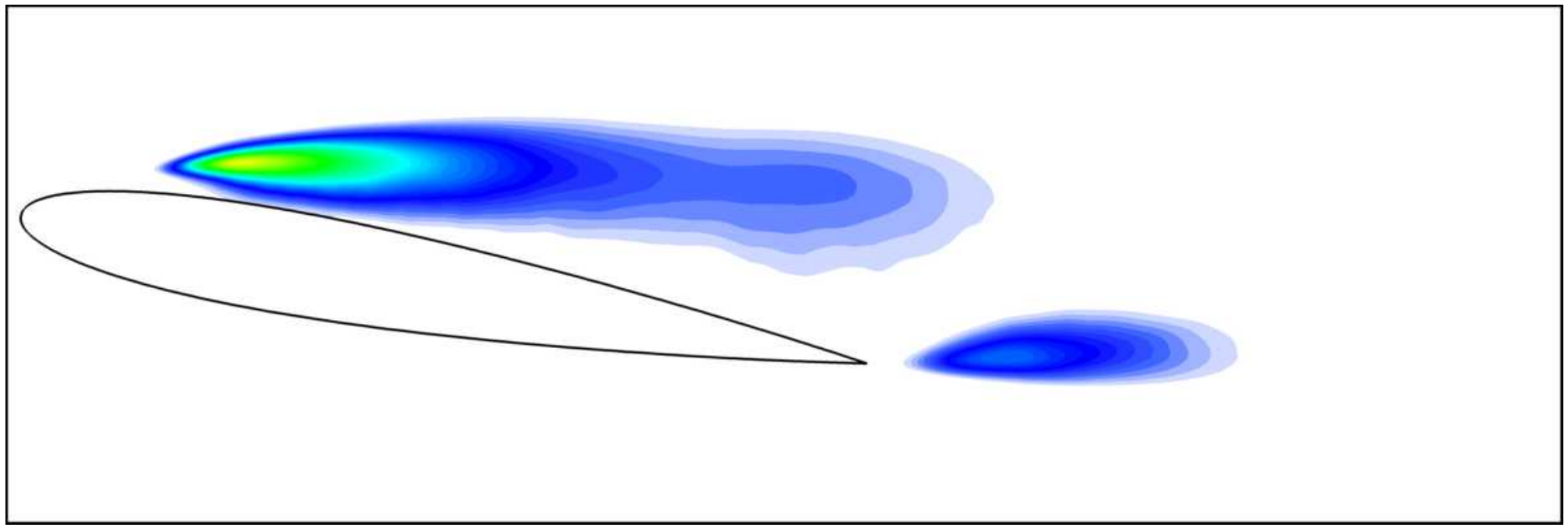}
\textit{$\alpha = 9.90^{\circ}$}
\end{minipage}
\begin{minipage}{220pt}
\centering
\includegraphics[width=220pt, trim={0mm 0mm 0mm 0mm}, clip]{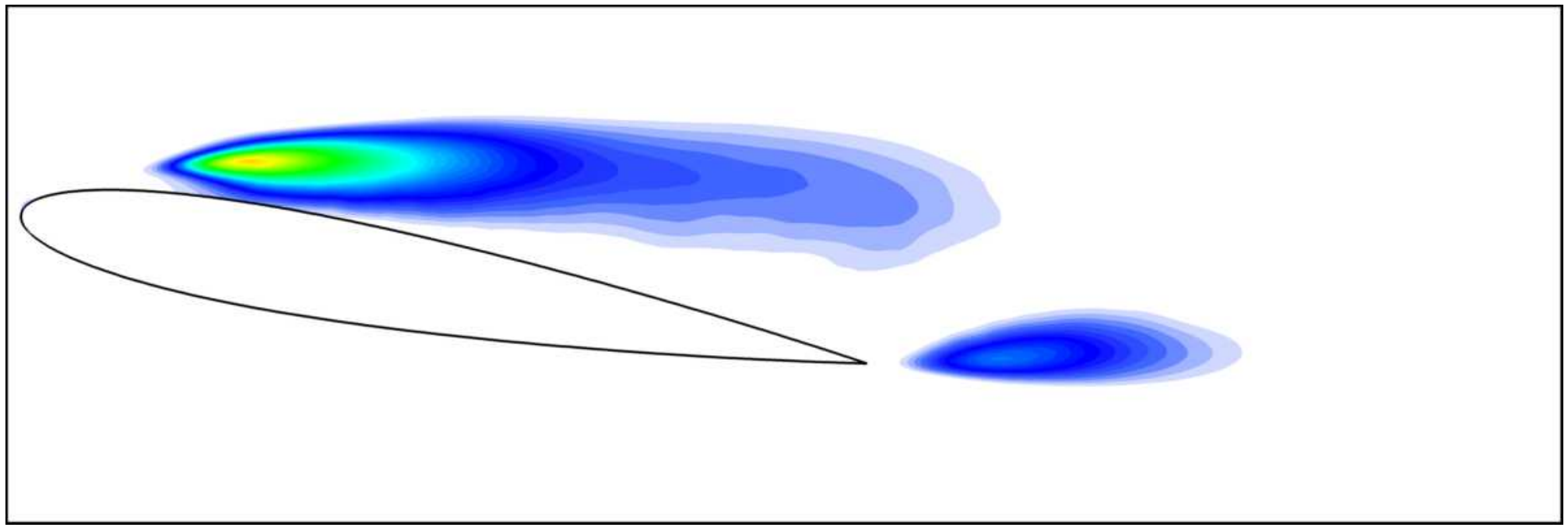}
\textit{$\alpha = 10.0^{\circ}$}
\end{minipage}
\begin{minipage}{220pt}
\centering
\includegraphics[width=220pt, trim={0mm 0mm 0mm 0mm}, clip]{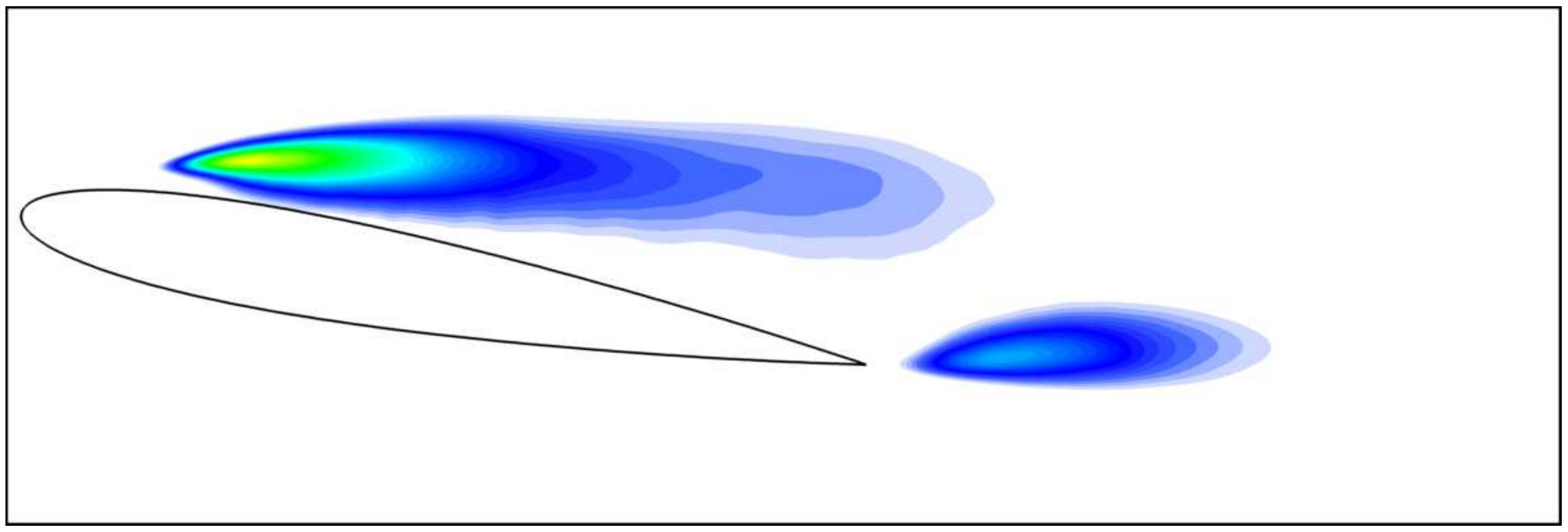}
\textit{$\alpha = 10.1^{\circ}$}
\end{minipage}
\begin{minipage}{220pt}
\centering
\includegraphics[width=220pt, trim={0mm 0mm 0mm 0mm}, clip]{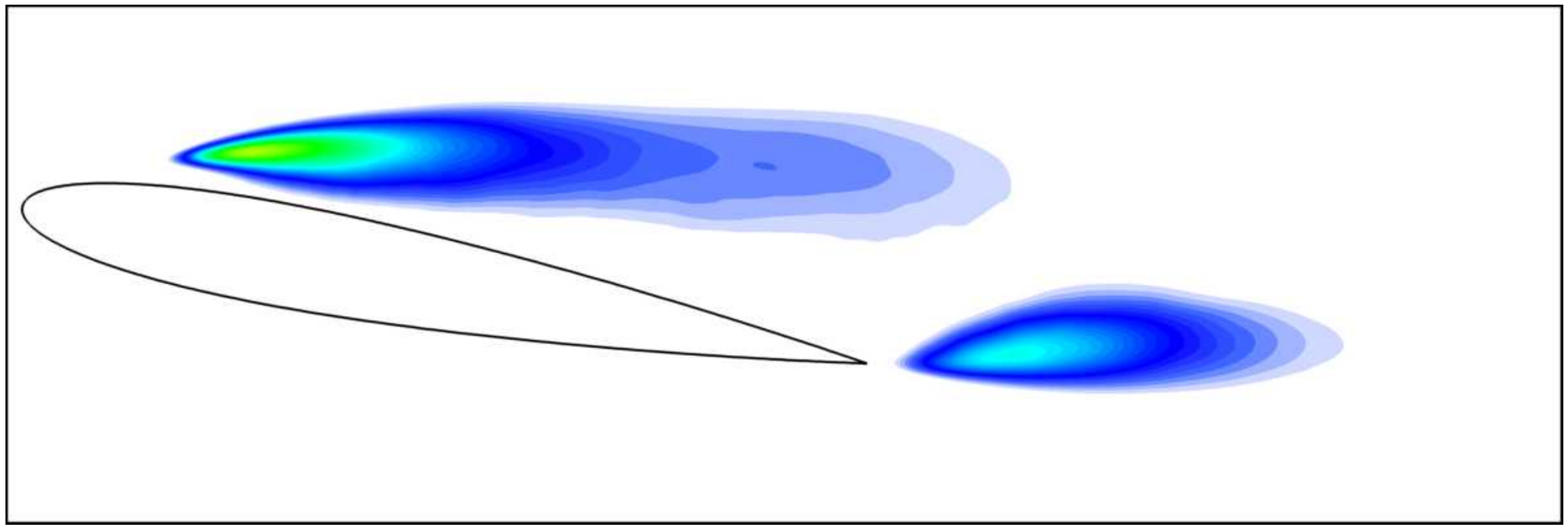}
 \textit{$\alpha = 10.5^{\circ}$}
\end{minipage}
\caption{Colours map of the low-lift variance of the pressure, $\widecheck{{p\mydprime}^2}$, for the angles of attack $\alpha = 9.25^{\circ}$--$10.5^{\circ}$.}
\label{p2_below}
\end{center}
\end{figure}

\begin{figure}
\begin{center}
\begin{minipage}{220pt}
\includegraphics[height=125pt , trim={0mm 0mm 0mm 0mm}, clip]{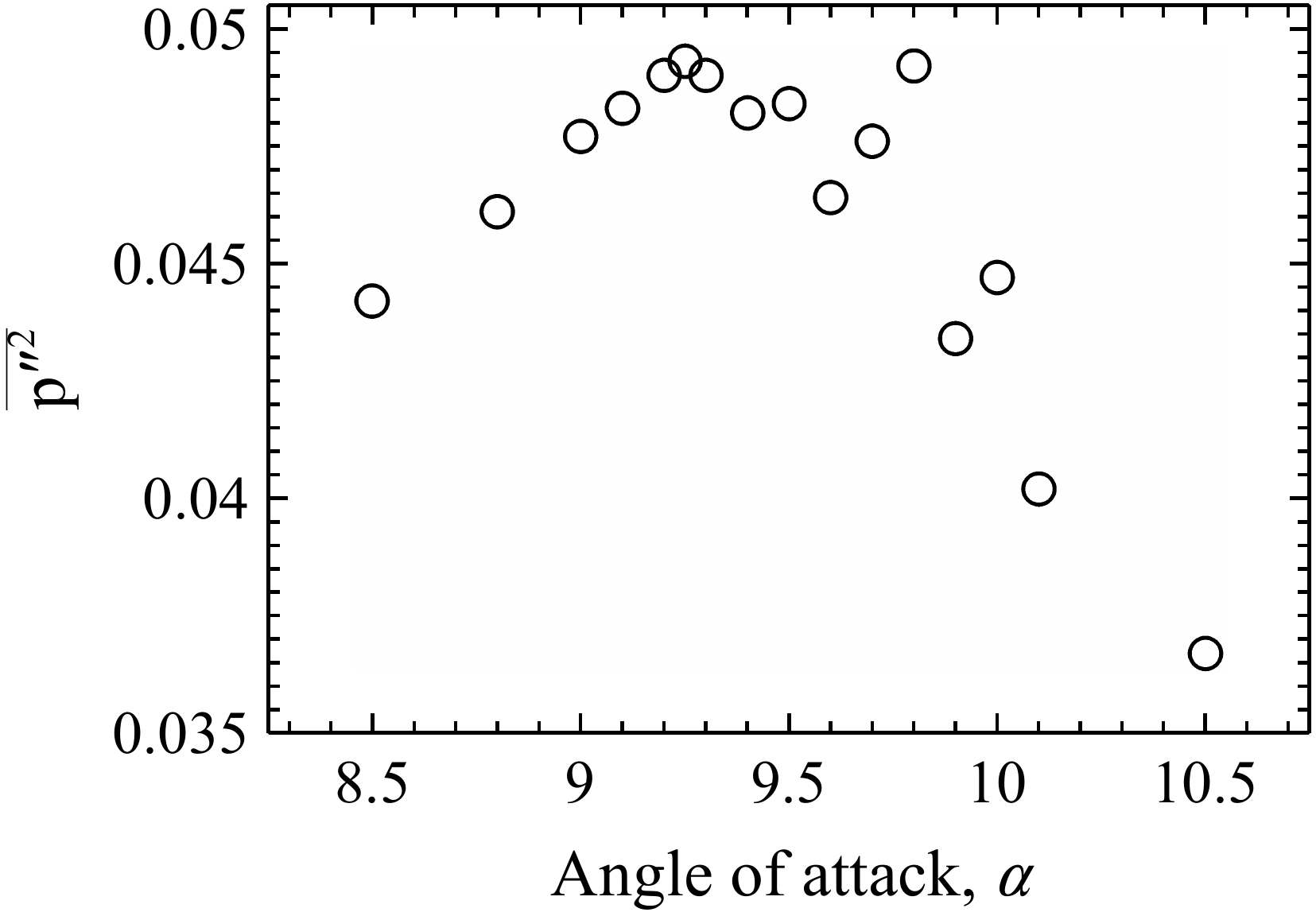}
\end{minipage}
\begin{minipage}{220pt}
\includegraphics[height=125pt , trim={0mm 0mm 0mm 0mm}, clip]{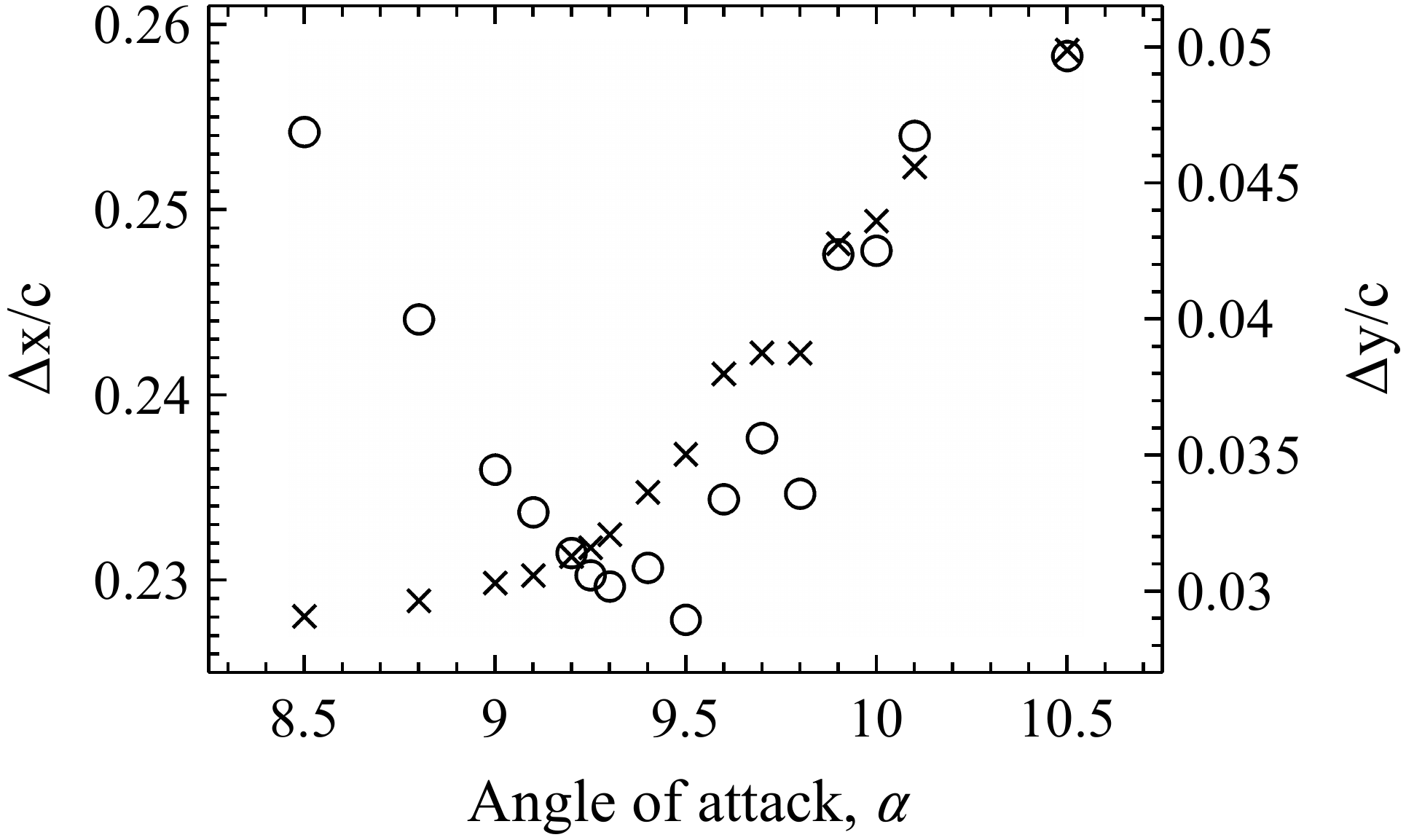}
\end{minipage}
\caption{Left: the maximum $\overline{{p\mydprime}^2}$ plotted versus the angle of attack $\alpha$. Right: the locations of the maximum $\overline{{p\mydprime}^2}$ plotted versus the angle of attack $\alpha$. Circles: $\Delta x/c$ measured from the aerofoil leading-edge, $\times$'s: $\Delta y/c$ measured from the aerofoil surface.}
\label{p2_max}
\end{center}
\end{figure}

The variance of the pressure f\/luctuations is shown in f\/igure~\ref{p2_mean}. As seen in the f\/igure, at $\alpha = 9.25^{\circ}$, $\overline{{p\mydprime}^2}$ has a signif\/icant magnitude around the location of the maximum LSB height and vanishes elsewhere. As the angle of attack is increased, $\overline{{p\mydprime}^2}$ extends downstream the bubble but away from the aerofoil surface in a fashion similar to that observed in $\overline{{v\mydprime}^2}$. However, $\overline{{p\mydprime}^2}$ has a signif\/icant magnitude upstream and in the vicinity of the separation location. In the range of angles of attack $9.4^{\circ} < \alpha < 10.5^{\circ}$ the pressure f\/luctuations intensif\/ies in the region between the leading-edge and the LSB. This is indicative that an inf\/lectional inviscid type instability is in action in this region. The variance of the pressure f\/luctuations for the mean attached f\/low and the mean separated f\/low are shown in f\/igures~\ref{p2_above} and~\ref{p2_below}, respectively. Both of $\widehat{{p\mydprime}^2}$ and $\widecheck{{p\mydprime}^2}$ are similar to their mean counterpart; however, they vanish upstream the LSB. At the trailing-edge, the pressure f\/luctuations signif\/icantly intensif\/ies in the statistically separated f\/low as seen in f\/igure~\ref{p2_below}.\newline
The prof\/ile of the maximum $\overline{{p\mydprime}^2}$ mimics that of $\overline{{u\mydprime}^2}$ but peaks at a lower angle of attack of $9.25^{\circ}$, as seen in the left side of f\/igure~\ref{p2_max}. In this regards, $\overline{{p\mydprime}^2}$ behaviour is in agreement with that of $\overline{{v\mydprime}^2}$, $\overline{{w\mydprime}^2}$, and $\overline{{u\mydprime}{v\mydprime}}$ except for the pressure f\/luctuations upstream the LSB.

\subsection{Aerodynamic coeff\/icients}
\begin{figure}
\begin{center}
\begin{minipage}{220pt   }
\centering
\includegraphics[width=220pt, trim={0mm 0mm 0mm 0mm}, clip]{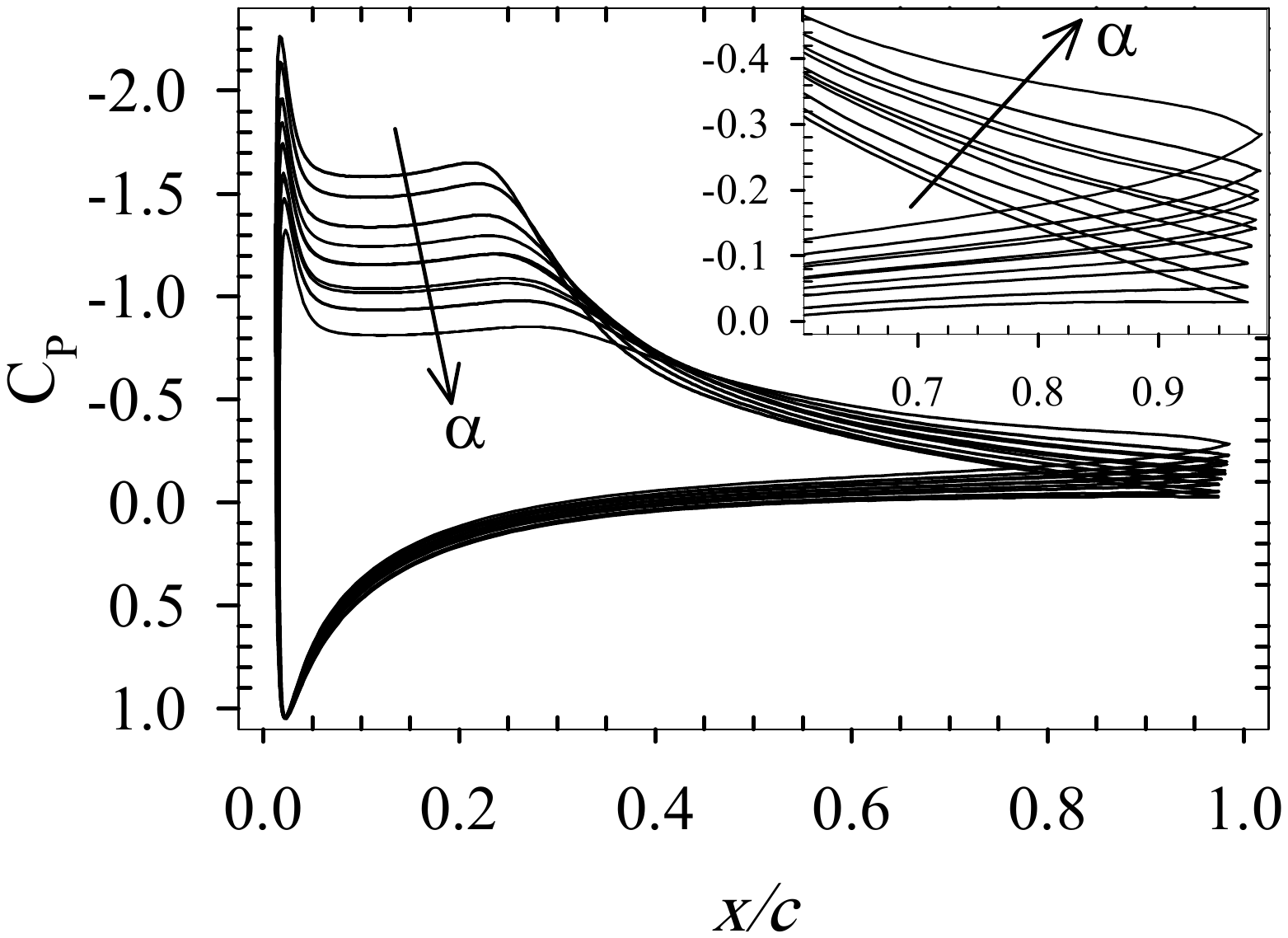}
\textit{The local pressure coefficient $\mathrm{C}_{\!_\mathrm{P}}$.}
\end{minipage}
\begin{minipage}{220pt}
\centering
\includegraphics[width=220pt, trim={0mm 0mm 0mm 0mm}, clip]{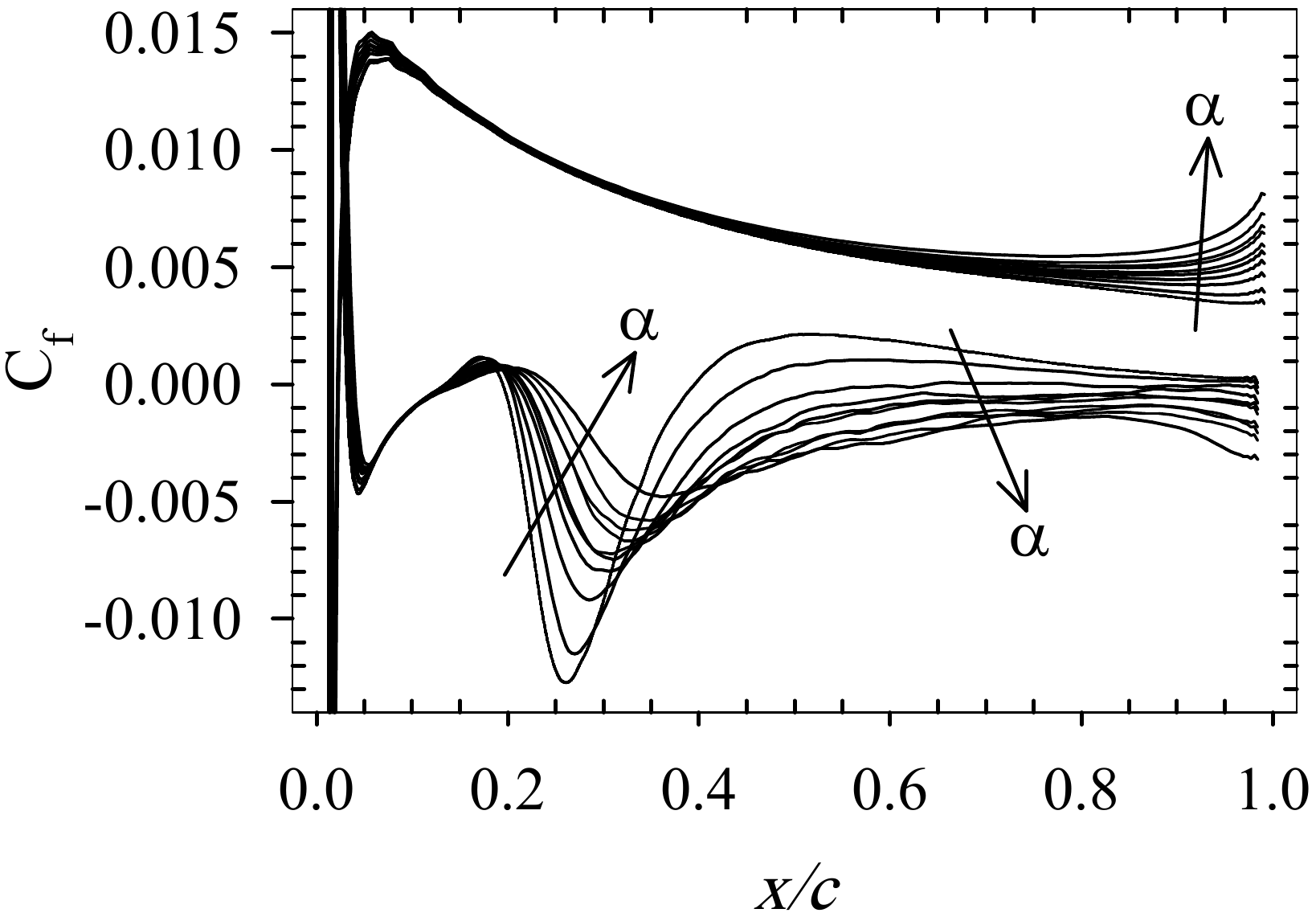}
\textit{The local skin-friction coefficient $\mathrm{C}_{\!_\mathrm{f}}$.}
\end{minipage}
\caption{Locally-time-averaged pressure and skin-friction coeff\/icients for the angles of attack $\alpha = 9.25^{\circ}$--$10.5^{\circ}$. The arrows indicate the direction in which the angle of attack, $\alpha$, increases in ascending order.}
\label{local_mean}
\end{center}
\end{figure}

\begin{figure}
\begin{center}
\begin{minipage}{220pt}
\centering
\includegraphics[width=220pt, trim={0mm 0mm 0mm 0mm}, clip]{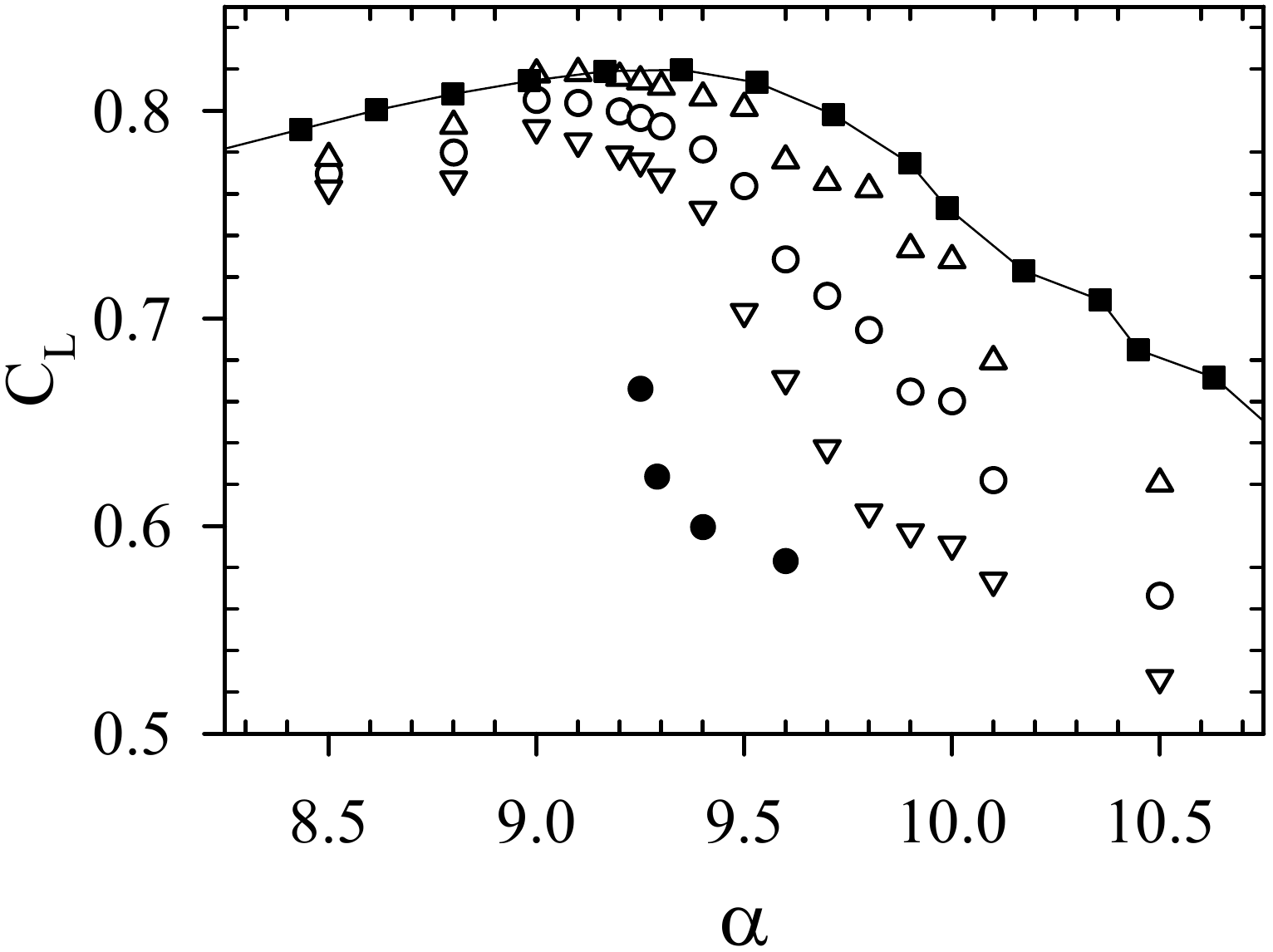}
\end{minipage}
\begin{minipage}{220pt}
\centering
\includegraphics[width=220pt, trim={0mm 0mm 0mm 0mm}, clip]{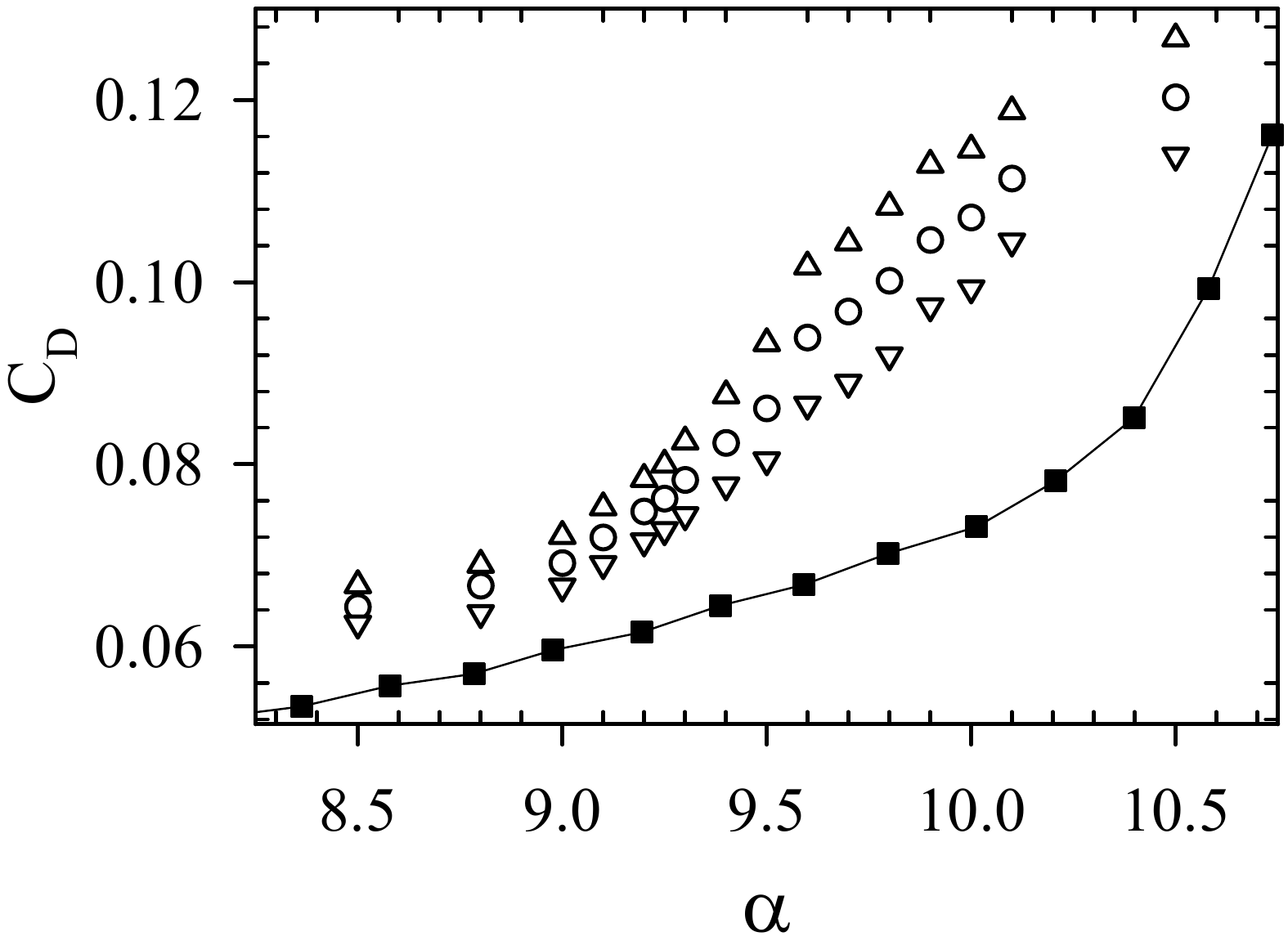}
\end{minipage}
\begin{minipage}{220pt}
\centering
\includegraphics[width=220pt, trim={0mm 0mm 0mm 0mm}, clip]{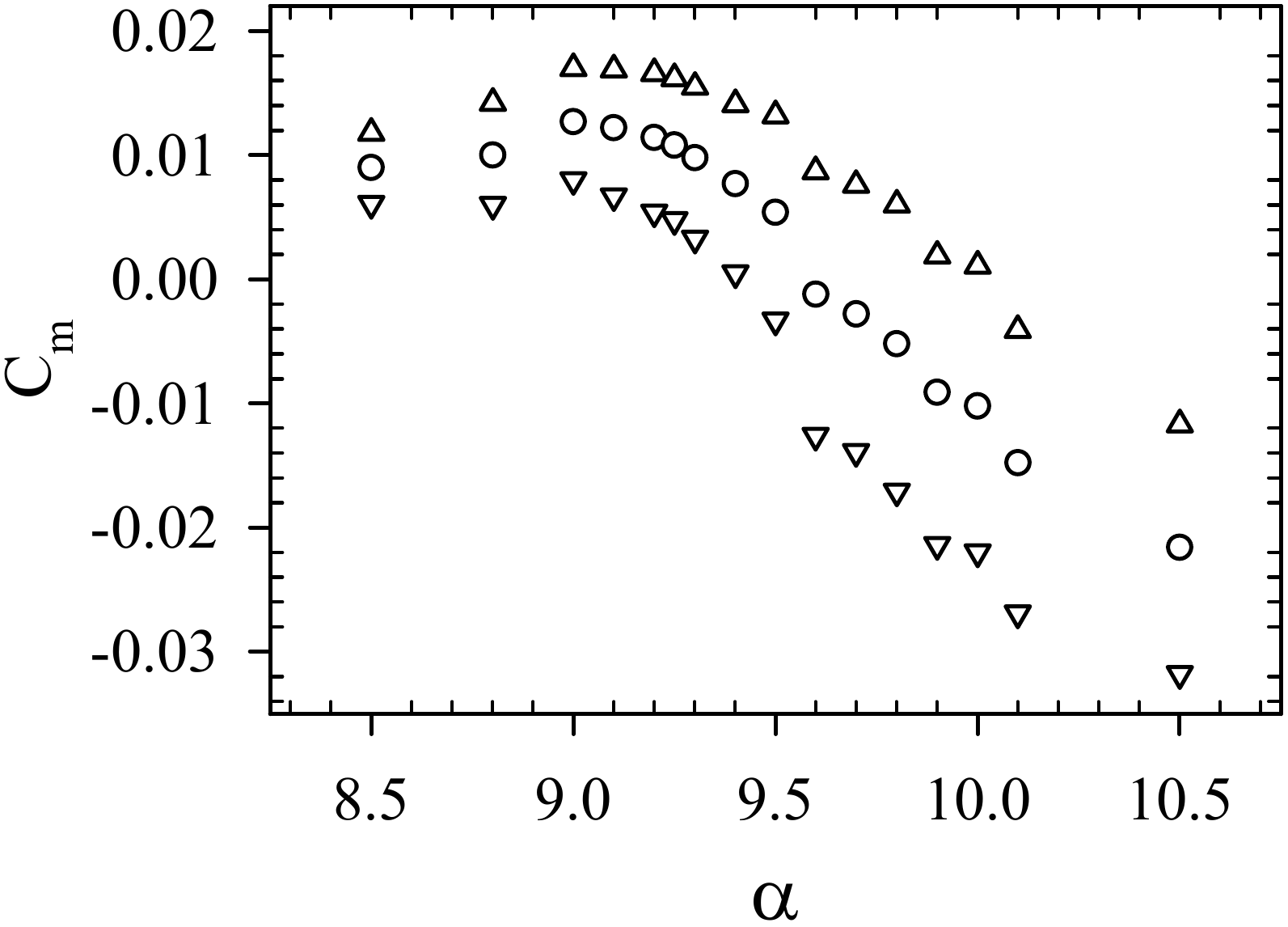}
\end{minipage}
\begin{minipage}{220pt}
\centering
\includegraphics[width=220pt, trim={0mm 0mm 0mm 0mm}, clip]{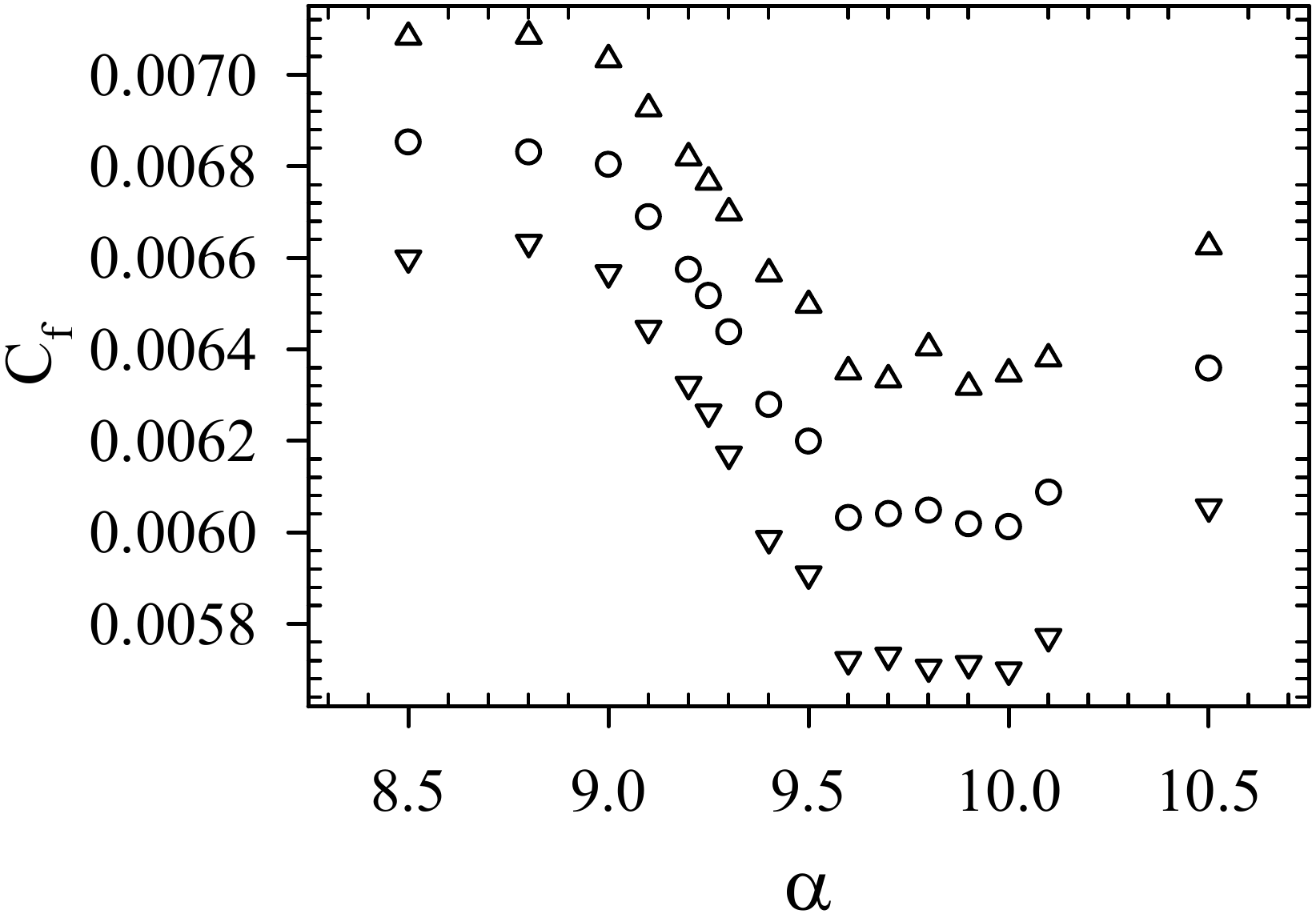}
\end{minipage}
\caption{Mean aerodynamic coeff\/icients $\mathrm{C}_{\!_\mathrm{L}}$, $\mathrm{C}_{\!_\mathrm{D}}$, $\mathrm{C}_{\!_\mathrm{m}}$, and $\mathrm{C}_{\!_\mathrm{f}}$ plotted versus the angle of attack $\alpha$. Circles: mean $\left(\overline{\cdot}\right)$, upward triangles: high-lift mean $\left(\widehat{\cdot}\right)$, downward triangles: low-lift mean $\left(\widecheck{\cdot}\right)$, solid line with filled black squares: the experimental data of~\citet{ohtake2007nonlinearity} at $\rec = 5\times10^4$ transformed to its corresponding compressible counterpart.}
\label{forces_mean}
\end{center}
\end{figure}

The locally-time-averaged pressure coeff\/icient, $\overline{\mathrm{C}_{\!_\mathrm{P}}}$, is shown in the left side of f\/igure~\ref{local_mean}. Ten prof\/iles covering the range of angles of attack $9.25^{\circ}$--$10.5^{\circ}$ are presented in the f\/igure. The zoom in window illustrates the distribution of $\overline{\mathrm{C}_{\!_\mathrm{P}}}$ in the vicinity of the trailing-edge in the range $0.6 \leq x/c \leq 1.0$. The f\/low is fully attached at $9.25^{\circ}$, in the mean sense, and fully separated at $10.5^{\circ}$ as seen in f\/igures~\ref{bubble_mean},~\ref{bubble_above}, and~\ref{bubble_below}. Near the aerofoil leading-edge, a laminar boundary layer develops, and the pressure coeff\/icient decreases drastically. The boundary layer remains laminar until the pressure gradient changes from favourable to adverse. Consequently, the laminar boundary layer detaches and travels away from the aerofoil surface to create a region of separated f\/low near the surface. The f\/low then reattaches to the aerofoil surface, and an LSB is formed. The pressure gradient seems to be favourable across the bubble in the range $0.05 < \textit{x/c} < 0.3$. Downstream the bubble the pressure gradient becomes adverse, but the f\/low has enough momentum to overcome separation. Further downstream, the pressure coeff\/icient remains almost constant. As a consequence of the bubble bursting, the pressure coeff\/icient gradually diminishes in magnitude and f\/lattens at higher angles of attack. At the vicinity of the trailing-edge the pressure gradient on the pressure surface of the aerofoil is adverse at $9.25^{\circ}$. As the angle of attack is increased, the pressure distribution becomes favourable as seen on the left-hand side of f\/igure~\ref{local_mean}. It is noted that at the trailing-edge the pressure distribution shifts upwards gradually as the angle of attack is increased. The pressure coeff\/icient is duly integrated about the aerofoil at each time-step for each angle of attack. The integrated effect of the bubble bursting on the pressure coeff\/icient indeed shows up in the aerodynamic forces, as discussed later.\newline
The locally-time-averaged skin-friction coeff\/icient $\overline{\mathrm{C}_{\!_\mathrm{f}}}$ in the same range of angles of attack is shown in the right side of f\/igure~\ref{local_mean}. In the vicinity of the leading-edge, the velocity tends to increase rapidly. Hence, a relatively high velocity gradient takes place, and the skin-friction is consequently at maximum values at all of the investigated angles of attack. This is followed by a sharp and sudden drop in the skin-friction until it crosses the $\textit{x}$--axis because of the APG and f\/low separation. The skin-friction becomes positive again around $\textit{x/c} = 0.2$; this is the region where the \textit{secondary vortex} is formed. The friction coeff\/icient then decreases again and becomes negative before it switches signs again at a location that is dependent on the angle of attack as seen in the f\/igure. In the region of negative $\overline{\mathrm{C}_{\!_\mathrm{f}}}$, the LSB is formed at each angle of attack. At higher angles of attack, the locations where $\overline{\mathrm{C}_{\!_\mathrm{f}}}$ switches signs moves further downstream as the mean bubble length increases with the angle of attack. On the suction surface of the aerofoil and in the vicinity of the trailing-edge, $\overline{\mathrm{C}_{\!_\mathrm{f}}}$ becomes negative again as a consequence of the TEB. On the pressure surface, however, $\overline{\mathrm{C}_{\!_\mathrm{f}}}$ increases in the vicinity of the trailing-edge as the angle of attack is increased.\newline
The left-hand side (top) of f\/igure~\ref{forces_mean} illustrates the averaged values of $\mathrm{C}_{\!_\mathrm{L}}$ for all of the investigated angles of attack. The circles denote the mean time-averaged lift coeff\/icient, $\overline{\mathrm{C}_{\!_\mathrm{L}}}$; the upward triangles display the high-lift time-averaged lift coeff\/icient, $\widehat{\mathrm{C}_{\!_\mathrm{L}}}$; and the downward triangles denote the low-lift time-averaged lift coeff\/icient, $\widecheck{\mathrm{C}_{\!_\mathrm{L}}}$. The values of $\mathrm{C}_{\!_\mathrm{L}}$ reaches its maximum at $\alpha=9.0^{\circ}$ and decreases with the angle of attack down to $\alpha=9.3^{\circ}$. A drastic and abrupt loss of lift then takes place in the range of angles of attack $9.3^{\circ} < \alpha < 9.8^{\circ}$. After that, the lift resumes decreasing almost linearly but faster than that observed in the range of $\alpha=9.0^{\circ}$--$9.3^{\circ}$.\newline
The solid line with filled black squares denotes the experimental data of~\citet{ohtake2007nonlinearity} at $\rec = 5\times10^4$ transformed to its corresponding compressible counterpart using the Prandtl--Glauert rule~\citet{glauert1928effect}. The filled black circles denote the LES data by~\citet{almutairi2013large}. The Prandtl--Glauert rule applies to potential flows at small angles of attack; however, it gives an estimate of how the incompressible experimental data of~\citet{ohtake2007nonlinearity} and its corresponding compressible data transformed with the Prandtl--Glauert rule compare to the compressible LES data at $\minf = 0.4$. Subsonic and incompressible potential flows have zero drag. Therefore, only $\mathrm{C}_{\!_\mathrm{L}}$ was transformed with the Prandtl--Glauert rule. The effect of compressibility can be expressed by the shift in the coeff\/icient magnitude and the angle of attack at which the coeff\/icients have maximum values. The lift coefficient compares very well to that of the LES data by~\citet{almutairi2013large}. However,~\citet{almutairi2013large} used a coarser grid (Grid--1) compared to the grid resolution used in the present study. Hence, the systematic and consistent shift in the magnitude of $\mathrm{C}_{\!_\mathrm{L}}$.\newline
The right-hand side (top) of f\/igure~\ref{forces_mean} shows the averaged values of $\mathrm{C}_{\!_\mathrm{D}}$ for all of the investigated angles of attack. The circles denote the mean drag coeff\/icient, $\overline{\mathrm{C}_{\!_\mathrm{D}}}$; the upward triangles display the high-lift drag coeff\/icient, $\widehat{\mathrm{C}_{\!_\mathrm{D}}}$; the downward triangles denote the low-lift drag coeff\/icient, $\widecheck{\mathrm{C}_{\!_\mathrm{D}}}$; and the solid line with filled black squares denotes the experimental data of~\citet{ohtake2007nonlinearity}. The drag coeff\/icient increases slowly and almost linearly in the range of $\alpha=9.0^{\circ}$--$9.3^{\circ}$. A sudden increase in the drag takes place in the range $9.3^{\circ} < \alpha < 9.7^{\circ}$. After that, it increases slowly with some degree of nonlinearity.\newline
The lift and drag coeff\/icients agree very well qualitatively and with acceptable accuracy quantitatively when compared to that of~\citet{ohtake2007nonlinearity}. The relatively small disparity between the LES data and that of the experiment is due to the fact that experimental data are sampled for relatively longer sampling time compared to that of the numerical simulation, hence, the convergence of the statistics of the experimental data to their true mean is much better than that of the LES statistics.\newline
F\/igure~\ref{forces_mean} (bottom-left) displays the averaged values of $\overline{\mathrm{C}_{\!_\mathrm{m}}}$, $\widehat{\mathrm{C}_{\!_\mathrm{m}}}$, and $\widecheck{\mathrm{C}_{\!_\mathrm{m}}}$ for all of the investigated angles of attack. As seen in the f\/igure, the moment coeff\/icient mimics the behaviour of the lift coeff\/icient. F\/igure~\ref{forces_mean} (bottom-right) shows the averaged values of $\overline{\mathrm{C}_{\!_\mathrm{f}}}$, $\widehat{\mathrm{C}_{\!_\mathrm{f}}}$, and $\widecheck{\mathrm{C}_{\!_\mathrm{f}}}$ for all angles of attack. The skin-friction remaines almost constant for angles of attack lower than $9.0^{\circ}$. After that, $\mathrm{C}_{\!_\mathrm{f}}$ drops linearly in the range $9.0^{\circ} < \alpha < 9.6^{\circ}$. F\/inally, the skin-friction f\/luctuates around a constant value for the angles of attack $\alpha \geq 9.6^{\circ}$. It is interesting to note that the difference between $\widehat{\mathrm{C}_{\!_\mathrm{f}}}$ and $\overline{\mathrm{C}_{\!_\mathrm{f}}}$ remains almost constant for all of the angles of attack. It is also noted that the difference between the mean time-averaged skin-friction coeff\/icient and the low-lift time-averaged coeff\/icient remains constant. Unlike $\mathrm{C}_{\!_\mathrm{L}}$, $\mathrm{C}_{\!_\mathrm{D}}$, and $\mathrm{C}_{\!_\mathrm{m}}$ where these differences start small at low angles of attack and tend to increase as the angle of attack is increased.

\subsection{The lift and the skin-friction spectra}
\begin{figure}
\begin{center}
\includegraphics[width=450pt, trim={0mm 0mm 0mm 0mm}, clip]{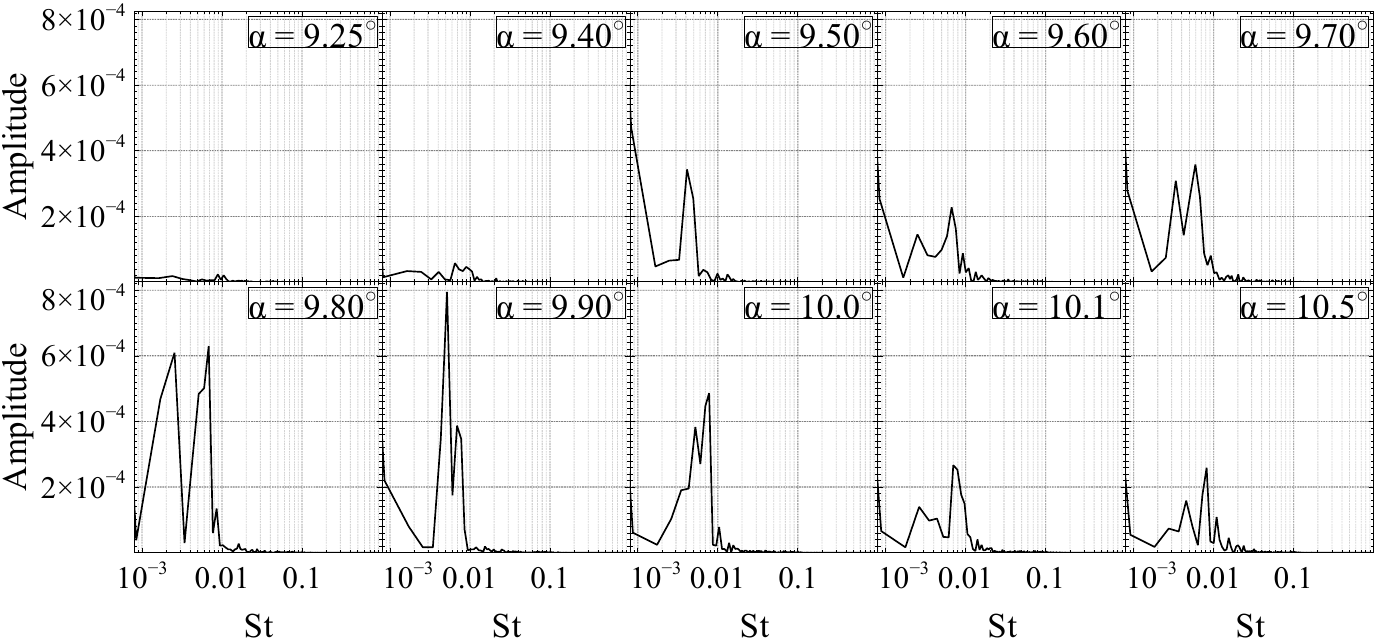}
\caption{Spectra of the lift coeff\/icient for the angles of attack $\alpha = 9.25^{\circ}$--$10.5^{\circ}$.}
\label{CL_spectra}
\end{center}
\end{figure}

\begin{figure}
\begin{center}
\includegraphics[width=450pt, trim={0mm 0mm 0mm 0mm}, clip]{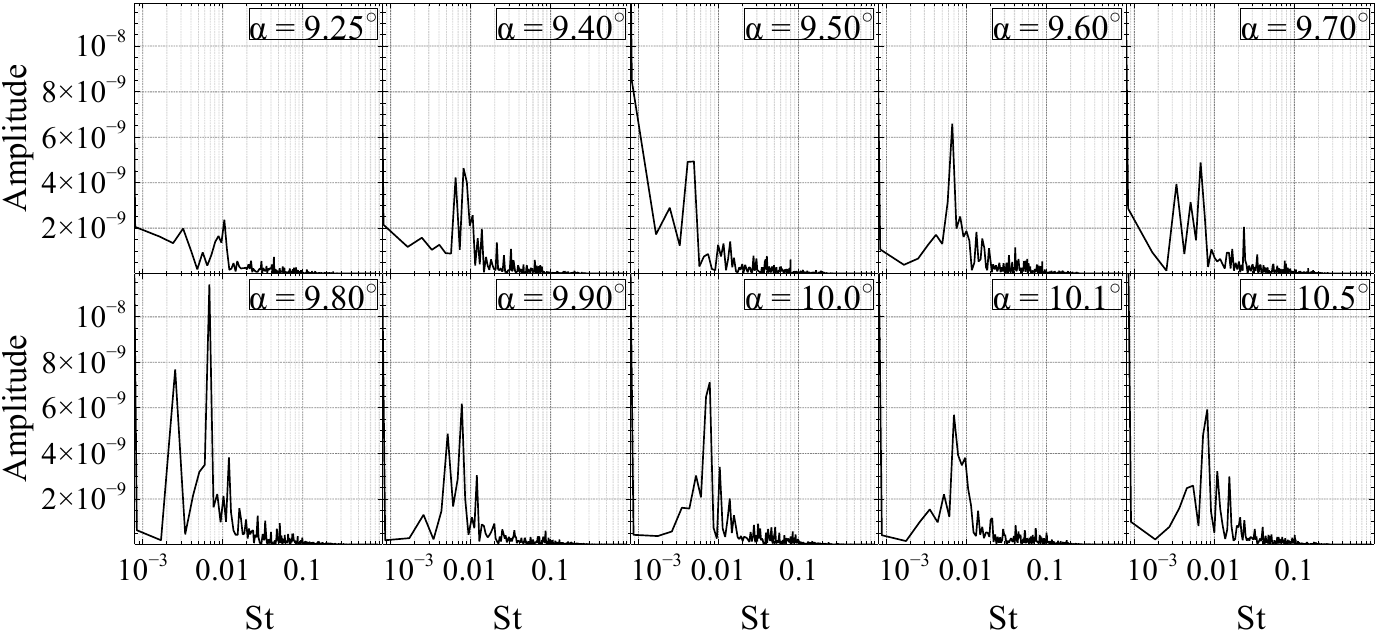}
\caption{Spectra of the skin-friction coeff\/icient for the angles of attack $\alpha = 9.25^{\circ}$--$10.5^{\circ}$.}
\label{Cf_spectra}
\end{center}
\end{figure}

\begin{figure}
\begin{center}
\begin{minipage}{220pt}
\centering
\includegraphics[width=220pt, trim={0mm 0mm 0mm 0mm}, clip]{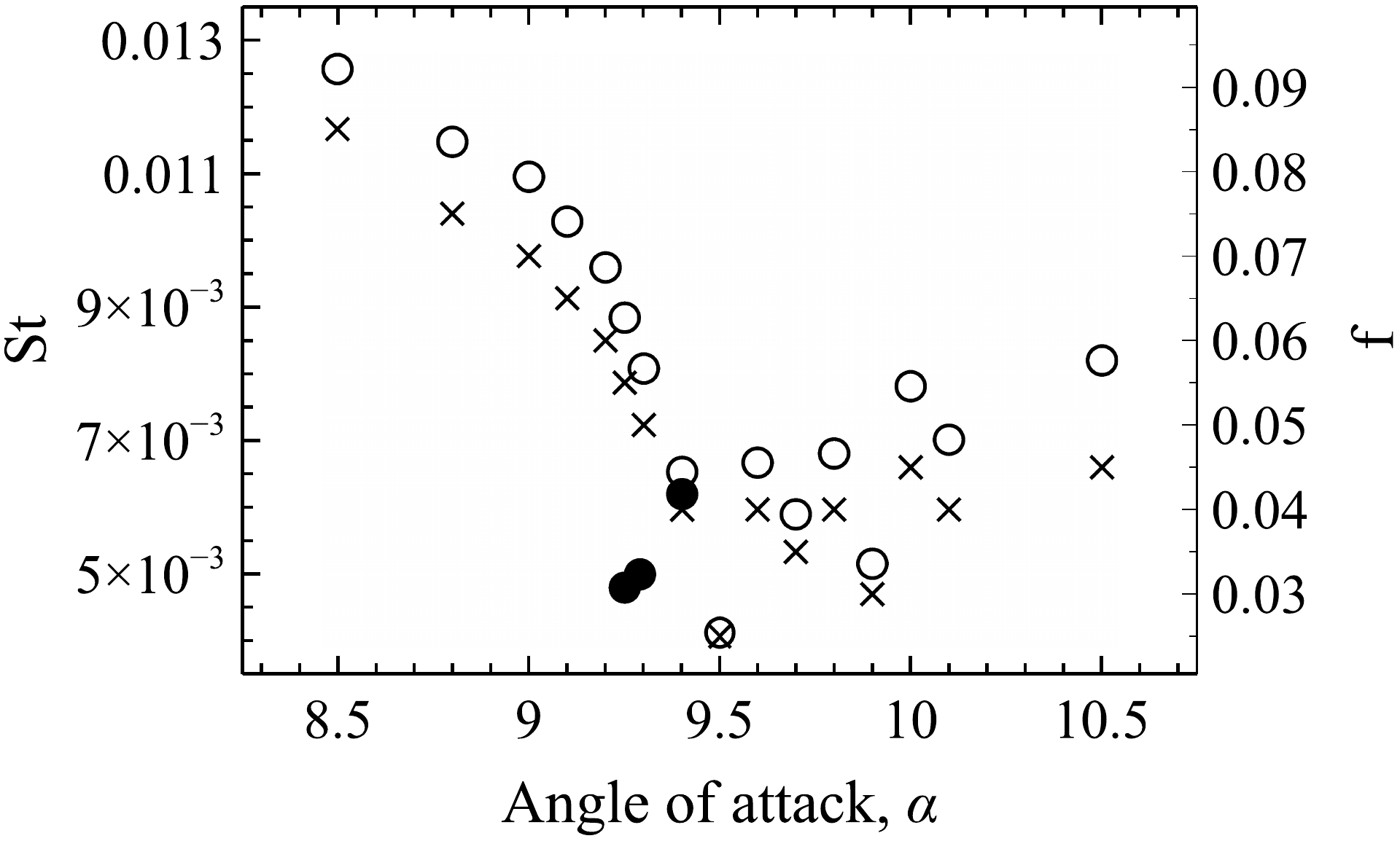}
\textit{The lift coeff\/icient.}
\end{minipage}
\begin{minipage}{220pt}
\centering
\includegraphics[width=220pt, trim={0mm 0mm 0mm 0mm}, clip]{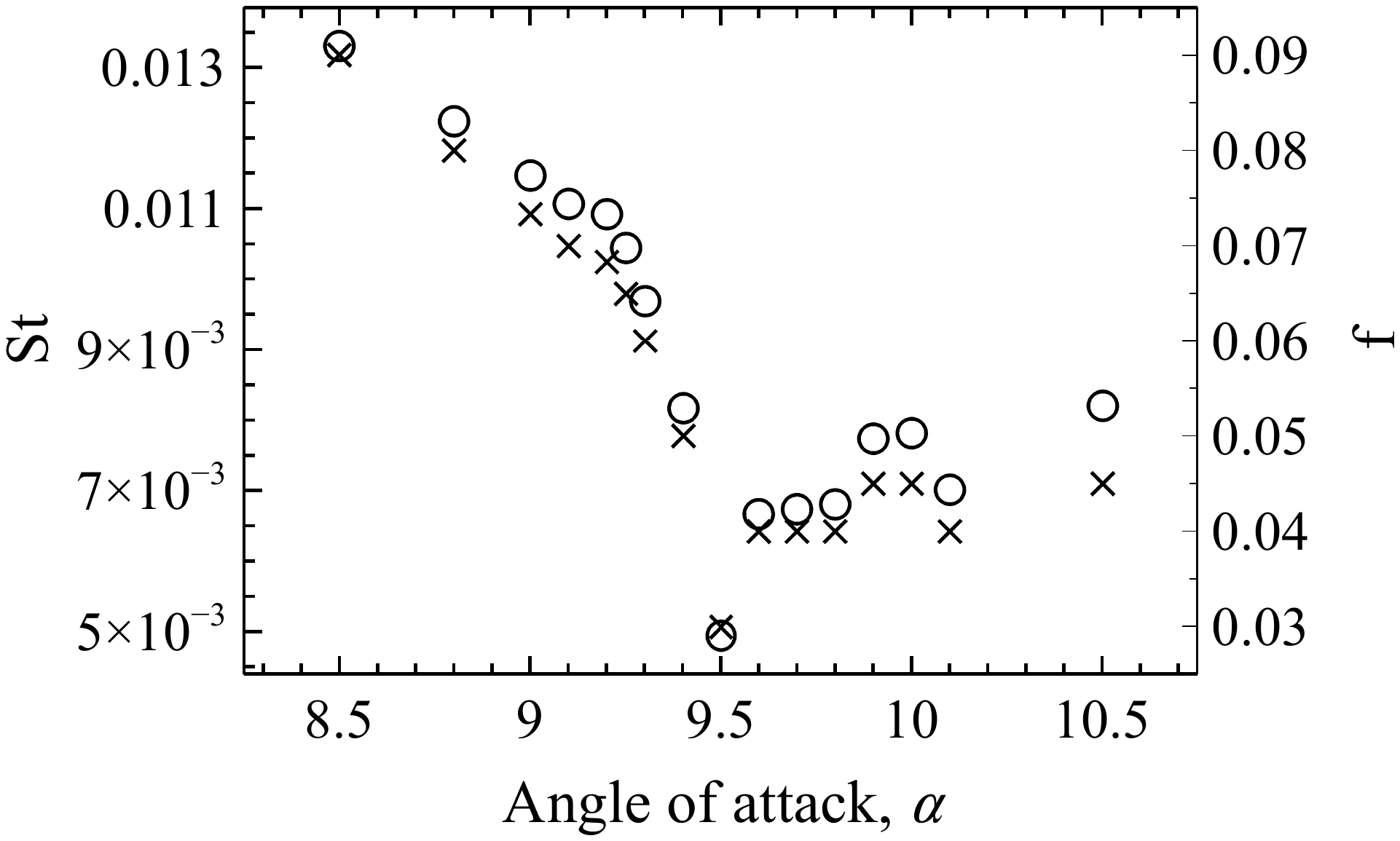}
\textit{The skin-friction coeff\/icient.}
\end{minipage}
\caption{Low-frequency Strouhal number $(St)$ and non-dimensional frequency $(f)$ plotted versus the angle of attack. Circles: Strouhal number, $\times$'s: non-dimensional frequency, filled black circles: the LES data by~\citet{almutairi2013large}.}
\label{St_cl_cf}
\end{center}
\end{figure}

Most of the variations in the pressure f\/ield are integrated into the lift coeff\/icient signal. Hence, any globally dominant f\/low feature is ref\/lected by a peak in the spectrum of $\mathrm{C}_{\!_\mathrm{L}}$. The spectra of the lift coeff\/icient were dully calculated using the fast Fourier transform (FFT) algorithm. For each of the sixteen angles of attack, the spectrum is dominated by low-frequency peak as seen in f\/igure~\ref{CL_spectra}. The $\mathrm{C}_{\!_\mathrm{L}}$ spectra peak at low-frequency with a signif\/icant amplitude at the angles of attack of $\alpha > 9.5^{\circ}$. That is, the low-frequency modes gain considerable momentum at the middle of the LFO transition process. The LFO continues to develop and gains more strength until the amplitude reaches a maximum at $\alpha = 9.9^{\circ}$. It is noted that in most of $\mathrm{C}_{\!_\mathrm{L}}$ spectrum there are two peaks at low frequencies. Each peak features a low-frequency f\/low mode. The two low-frequency modes interact with each other in a manner that relaxes the f\/low into equilibrium. The spectra of the skin-friction coeff\/icient were calculated similarly. The spectra is similar to that of the lift coeff\/icient as seen in f\/igure~\ref{Cf_spectra}.\newline
F\/igure~\ref{St_cl_cf} (left) shows the Strouhal number of the most dominant low-frequency oscillations, $St$, plotted versus the angle of attack for the lift coeff\/icient. Strouhal number decreases very fast and linearly with the angle of attack in the range of $\alpha=9.0^{\circ}$--$9.5^{\circ}$. After that, the Strouhal number increases with the angle of attack. The $\times$'s denote the non-dimensional frequency $f$. The filled black circles denote the LES data by~\citet{almutairi2013large}. However, the Strouhal number increases much faster with the angle of attack in the case of the LES data by~\citet{almutairi2013large}. This is because they used a coarser grid and consequently overestimated the size of the bubble and the rate at which the kinetic energy dissipated as discussed before. Therefore, the aerofoil in their case undergoes an early full stall at the angle of attack of $9.6^{\circ}$. The Strouhal number of the LFO for the skin-friction coeff\/icient mimics that of the lift coeff\/icient as illustrated in the right-hand side of f\/igure~\ref{St_cl_cf}.\newline
Time histories of $\mathrm{C}_{\!_\mathrm{L}}$ shows no apparent low-frequency oscillations for the angles of attack $\alpha \leq 9.25^{\circ}$. However, the spectral analysis identif\/ied a low-frequency mode at these angles. The low-frequency mode at the angles of attack $\alpha \leq 9.25^{\circ}$ may feature bubble shedding rather than bubble bursting and reformation. Whereas, the low-frequency mode at angles of attack $\alpha > 9.25^{\circ}$ features switching between fully attached and fully separated f\/lows as a consequence of bubble bursting and reformation.

\section*{Conclusions}
The effects of the angle of attack on the characteristics of the laminar separation bubble (LSB), its associated low-frequency f\/low oscillation (LFO) around a NACA-0012 aerofoil at $\rec = 5\times10^4$ and $\minf = 0.4$ was studied. In the range of the investigated angles of attack, statistics of the f\/low-f\/ield suggest the existence of three distinct angle-of-attack regimes:
\begin{enumerate}
  \item The f\/low-f\/ield and aerodynamic characteristics are not much affected by the LFO for the angles of attack $\alpha < 9.25^{\circ}$.
  \item The f\/low-f\/ield undergoes a transition regime in which the bubble bursting and reformation cycle, LFO, develops until it reaches a quasi-periodic switching between separated and attached f\/low in the range of angles of attack $9.25^{\circ} \leq \alpha \leq 9.6^{\circ}$.
  \item The f\/low-f\/ield and aerodynamic characteristics are overwhelmed by a quasi-periodic and self-sustained LFO until the aerofoil approaches the angle of a full stall for the angles of attack $\alpha > 9.6^{\circ}$.
\end{enumerate}
A trailing-edge bubble (TEB) forms at $\alpha > 9.25^{\circ}$ and grows with the angle of attack. The LSB and TEB merge and continue to deform until they form an open bubble at $\alpha = 10.5^{\circ}$. On the suction surface of the aerofoil, the pressure distribution shows that the presence of the LSB induces a gradual and continues adverse pressure gradient (APG) when the f\/low is attached. The bursting of the bubble causes a gradual and continues favourable pressure gradient (FPG) when the f\/low is separated. This is indicative that a natural forcing mechanism keeps the f\/low attached against the APG and separated despite the FPG. The length of the bubble, in the mean sense, decreases to a minimum size of $33.5\%$ of the aerofoil chord at $\alpha = 9.0^{\circ}$ then it grows in size when the angle of attack is further increased. The maximum reversed velocity (MRV) increases to a maximum of $35.6\%$ of the free-stream velocity and $22\%$ of the local free-stream velocity at $\alpha = 9.0^{\circ}$ then it decreases as the angle of attack increases. The location of the MRV is closest to the aerofoil leading-edge at the angle of attack of $9.0^{\circ}$.\newline
The variance of the pressure f\/luctuations, the wall-normal velocity f\/luctuations, and the spanwise velocity f\/luctuations have a maximum at the angle of attack of $9.25^{\circ}$. The Reynolds stress has a minimum at the angle of attack of $9.2^{\circ}$, and the variance of the streamwise velocity f\/luctuations has a maximum of $21\%$ of the free-stream velocity at the angle of attack of $9.4^{\circ}$. It is shown that perturbations of the wall-normal and the spanwise velocity components are extracted exclusively by the local velocity gradient inside the shear layer via the Kelvin--Helmholtz instability. Whereas, the f\/luctuations in the streamwise velocity component and the pressure are due to a global oscillation in the f\/low-f\/ield in addition to the velocity gradient across the shear layer. The location of maximum values of the variance of the velocity components is closest to the aerofoil leading-edge at the angle of attack of $9.25^{\circ}$ or at the inception of the LFO transition regime. The location of the minimum Reynolds stress and the maximum variance of the pressure f\/luctuations is closest to the aerofoil leading-edge at the angles of attack of $9.0^{\circ}$ and $9.5^{\circ}$, respectively. The location of the maximum reversed velocity (MRV), the locations of the maximum variance, and the location of the minimum Reynolds stress move away from the aerofoil surface as the angle of attack increases.\newline
The characterisation parameters are the bubble size; the maximum reversed f\/low (MRV); the minimum Reynolds stress; the maximum values of the variance of the velocities and the pressure; their distances from the aerofoil leading-edge and from the aerofoil surface; the aerodynamic coeff\/icients; and the Strouhal number of the LFO. In the f\/irst regime, the characterisation parameters increases as the angle of attack is increased until they approach saturation values. At the beginning of the second regime, the characterisation parameters drop sharply as the angle of attack increases. In the third regime, the characterisation parameters continue to decrease slowly with the angle of attack. The drag coeff\/icient behaviour is different; it continues to increase in the three regimes. The Strouhal number of the LFO phenomenon drops gradually in the f\/irst regime, rapidly in the second regime, and rises slowly in the third regime. However, the f\/low behaviour becomes more stochastic during the third regime, and the number of samples needed for statistical convergence might be much more than that required for the f\/irst two regimes.\newline
It is noted that the secondary bubble exists at the same location when the f\/low is attached and when it is massively separated. This is indicative that the secondary bubble plays a profound role in the underlying mechanism behind the self-sustained quasi-periodic low-frequency switching between attached and separated f\/low.\newline
The variance of pressure f\/luctuations has signif\/icant magnitude in the laminar portion of the separated shear layer. This is indicative that the instability that generates and sustains the LFO originates at this location. This challenges the traditional theory that links the instability of the LSB to the separated shear layer with its associated Kelvin--Helmholtz mechanism. The present investigation suggests that most of the observations reported in the literature about the LSB and its associated LFO are neither thresholds nor indicators for the inception of the instability, but rather are consequences of it.

\section*{Acknowledgements}
All computations were performed on \verb"Aziz Supercomputer" at King Abdulaziz university's High Performance Computing Center (\url{http://hpc.kau.edu.sa/}). The authors would like to acknowledge the computer time and technical support provided by the center.

\bibliographystyle{jfm}
\bibliography{lfo1}

\newpage
\newpage
\newpage
\newpage
\newpage
\newpage
\newpage
\newpage
\newpage
\newpage
\newpage
\newpage
\newpage
\newpage
\newpage
\newpage
\newpage

\end{document}